**Казахстанские**
**Космические**
**Исследования**

**Том3**

**Дубовиченко С.Б., Такибаев Н.Ж.**
**Чечин Л.М.**

# Ф И З И Ч Е С К И Е ПРОЦЕССЫ В ДАЛЬНЕМ И БЛИЖНЕМ КОСМОСЕ

Национальное космическое агентство
Республики Казахстан
Национальный центр космических исследований
и технологий
Астрофизический институт им. В.Г. Фесенкова

**Дубовиченко С.Б., Такибаев Н.Ж.
Чечин Л.М.**

# Ф И З И Ч Е С К И Е ПРОЦЕССЫ В ДАЛЬНЕМ И БЛИЖНЕМ КОСМОСЕ

***Космология, атмосферы звезд и планет, ядерная астрофизика***

# СОДЕРЖАНИЕ

# ПРЕДИСЛОВИЕ

Книга посвящена ряду актуальных вопросов космологии ранней Вселенной, физики атмосфер Солнца и Земли, а также ядерной астрофизике.

Она состоит из трех частей, в каждом из которых обсуждаются теоретические проблемы этих разделов физики космоса.

В первой части рассматриваются физика расширяющейся Вселенной с упором на содержание новейшей революции в космологии. Именно, сегодня надежно установлено, что во Вселенной доминирует вакуум, превосходящий по плотности барионные формы космической материи; что динамикой космического расширения управляет вакуумная антигравитация; и, наконец, что космологическое расширение ускоряется.

Открытие космического вакуума, в свою очередь, поставило ряд новых вопросов в космологии - как с расширением Вселенной меняется уравнение состояния вещества?, только ли вакуум может создавать антигравитацию?, может ли сам вакуум являться причиной фрагментации космологического субстрата?. И другие. Их обсуждение и составляет содержание первого раздела монографии.

Во второй части исследуются процессы, ведущие к электризации газовых потоков в атмосфере Солнца и генерации нейтронов в активных областях его атмосферы, вопросы электризации грозовых облаков в тропосфере Земли, формирования и развития серебристых облаков в области ее мезосферы.

Обсуждаются термодинамические и квантовомеханические механизмы накопления объемных электрических, пространственно разделенных, зарядов во встречных сталкивающихся атмосферных потоках. Особое внимание уделено обсуждениям общности и подобия ряда явлений, как на Солнце, так и на Земле с учетом различий, обусловленных масштабами процессов и влиянием внешних полей.

И, наконец, в третьей части предложены методы расчета волновых функций кластерных систем в непрерывном и дискретном спектре при заданном межкластерном взаимодействии. В ней содержится теоретическое изучение S-факторов фотоядерных реакций, которые входят в гелиевый цикл термоядерных процессов, определяющих основную долю выхода энергии звездных систем.

Первая часть написана Л.М. Чечиным, вторая - Н.Ж. Такибаевым, и третья - С.Б. Дубовиченко. Структура книги такова, что ее чтение можно начать с любого раздела. Они будут доступны в целом студентам и аспирантам физико-математических факультетов университетов, а также научным работникам, специализирующимся в области теоретической физики, физики космоса и ядерной астрофизики.

***Дубовиченко С.Б., Такибаев Н.Ж., Чечин Л.М.***

**Август 2008г.**

**Часть 1**

# КОСМОЛОГИЯ РАННЕЙ ВСЕЛЕННОЙ

***Чечин Л.М.***

***Светлой памяти моих родителей***
***- Чечина Михаила Никифоровича***
***и Чечиной Алефтины Павловны***
***- посвящаю***

## *ВВЕДЕНИЕ*

На протяжении нескольких столетий после Ньютона никто из астрономов не сомневался в том, что наша Вселенная в целом имеет статический характер. Даже А.Эйнштейн, сформулировав общую теорию относительности, сначала предложил статическую модель четырехмерной Вселенной [1].

Однако после работ А.Фридмана [2] стало ясно, что свойства Вселенной лишь в каждый заданный момент времени одинаковы во всех точках и во всех направлениях. Это свидетельствует как об отсутствии какого - либо центра мира, так и о невозможности существования во Вселенной каких - либо привилегированных направлений. Вместе с тем, модель Фридмана рассматривает давление в веществе и плотность вещества как функции времени и, тем самым, представляет собой нестационарную - расширяющуюся - модель Вселенной.

Теоретически предсказанное Фридманом расширение Вселенной было обнаружено спустя несколько лет Э. Хабблом [3]. При этом он обнаружил, что чем дальше находятся галактики, тем с большей скоростью они удаляются друг от друга. Эта закономерность математически выражается в виде $\vec{v} = H\vec{r}$, где H - постоянная величина или постоянная Хаббла.

Закон Хаббла является одним из важнейших экспериментальных доказательств о справедливости нестационарной - расширяющейся - модели Вселенной. Но если размеры Вселенной постоянно возрастают, то это означает существование момента времени, при котором вещество было сжато до плотностей, на много порядков (на восемьдесят порядков!) превышающих ядерную плотность. Это так называемое сингулярное состояние Вселенной.

Поэтому среди многих вопросов, возникших перед астрономами, был вопрос о физических характеристиках Вселенной в сингулярном состоянии. Важную роль в его исследовании сыграл Г. Гамов, выдвинувший идею «Большого взрыва» [4]. Расчеты показали, что тогда Вселенная находилась в горячем состоянии с температурой несколько миллиардов градусов и была заполнена квантами высоких энергий. Хотя позднейшие исследования существенно «подняли» температуру начального состояния Вселенной до 10(32)K, но важнейший вывод из исследований Г. Гамова остался.

Речь идет о том, что при расширении Вселенной ее температура падает и поэтому должно существовать остаточное (реликтовое) излучение, соответствующее температуре в несколько Кельвинов. Значительно позднее А. Пензиас и Р. Вильсон [5] зарегистрировали изотропное излучение, соответствующее температуре около 3K, которое стало еще одним доказательством модели горячей Вселенной.

А что представляло собой вещество во Вселенной в начальной стадии ее расширения? Наполняющая раннюю Вселенную высокотемпературная плазма состоит из электронов, протонов, некоторого количества нейтронов, фотонов и, как выяснилось в последнее время, из темного (скрытого) вещества. (Состав темного вещества и его физические свойства пока неизвестны. Поэтому исследования, проводимые в этом направлении, представляются одними из наиболее важных в современной космологии.) Согласно кинетической теории плазмы в ней необходимо возникают возмущения плотности, давления и других ее характеристик. Если возмущения вызваны гравитационной нестабильностью плазмы, то в ней начинается их рост и последующая фрагментация всего вещества во Вселенной.

Теория эволюции возмущений плотности на основе ньютоновской космологии была создана Дж. Джинсом [6] , а релятивистская теория развития возмущений в нестационарной Вселенной был предложена Е. Лифшицем [7]. Физической причиной роста возмущений плотности является гравитационное притяжение.

Действительно, если на фоне однородно распределенного вещества возникает область повышенной плотности, то она будет притягивать к себе окружающее вещество. Этот процесс будет продолжаться до тех пор, пока силы притяжения не будут уравновешены силами давления. Описанный механизм является причиной образования не только галактик и их скоплений, но и крупномасштабной структуры Вселенной.

Крупномасштабная структура Вселенной был предсказана Я. Зельдовичем (см. обзор [8]). Он обнаружил, что образующиеся объекты не всегда обладали сферической симметрией. Напротив того, они представляли собой трехмерные структуры с различными поперечными размерами, т.н. «блины» или «стенки», в которых был сконцентрирована основная доля вещества во Вселенной.

Последующие наблюдения, проведенные группой Р. Киршнера [9], подтвердили этот вывод. Поэтому третьим основным наблюдательным фактом, лежащим в основании релятивистской космологии, следует считать открытие крупномасштабной структуры Вселенной.

Четвертый основной наблюдательный тест это рас-пространенность легких химических элементов в космосе. Согласно теории горячей Вселенной, о которой уже упоминалось выше, на раннем этапе ее эволюции могли рождаться только легкие элементы - гелий, литий и их различные изотопы. При этом основной была реакция $p + n \rightarrow {}^2H + \gamma$, в которой рождался дейтерий - важнейший продукт для образования гелия и лития.

Дальнейшие теоретические исследования показали, что расчетное количество водорода (~75%) и гелия (~25%) во Вселенной достаточно хорошо согласуется с наблюдениями, что и послужило подтверждением теории первичного нуклеосинтеза.

Еще одним экспериментальным тестом релятивистской космологии является открытие анизотропии реликтового излучения. Анизотропия - это разница температуры реликтового излучения в различных направлениях на небе. Обнаружение самого реликтового излучения означало наблюдение первой мультипольной гармоники. Открытие же анизотропии реликтового излучения означало измерение ее дипольной составляющей. При этом амплитуда дипольной гармоники составила приблизительно $3(-5)$K [10]. Ее измерение позволило установить наиболее универсальную систему отсчета во Вселенной, определить величину пекулярных скоростей галактик и т.п.

Обсуждая эти эксперименты, необходимо еще раз подчеркнуть, что все они являются подтверждением нестационарной - расширяющейся - модели Фридмана. Однако здесь до сих пор не ставился вопрос о физических причинах расширения Вселенной - что именно «заставляет» ее эволюционировать, т.е. какая физическая причина приводит ее к расширению? Вместе с тем, ответ на этот вопрос существует, и он составляет содержание новейшей революции в космологии.

Согласно [11] во Вселенной доминирует вакуум, превосходящий по плотности барионные формы космической материи; динамикой космического расширения управляет вакуумная антигравитация; и, наконец, космологическое расширение ускоряется.

Открытие космического вакуума, в свою очередь, ставит ряд новых вопросов в космологии:

- как с расширением Вселенной меняется уравнение состояния вещества?
- только ли вакуум может создавать антигравитацию?
- может ли сам вакуум являться причиной фрагментации космологического субстрата?

Для более полного представления обо всем спектре важнейших проблем современной космологии можно воспользоваться литературой, указанной в ссылках [12-16].

# ***ЛИТЕРАТУРА***

# 1. ПРЕБЫВАНИЕ РАННЕЙ ВСЕЛЕННОЙ В КВИНТЭССЕНЦИАЛЬНОМ СОСТОЯНИИ

## Введение

Одной из ключевых проблем современной космологии является теоретическая интерпретация нового вида материи - квинтэссенции, ассоциированной, например, с $\Lambda$ - членом в уравнениях Эйнштейна. Ее основные наблюдаемые характеристики заключаются в том, что она изотропна и, что она не кластеризуется, т.е. остается однородной (см., например, [1,2]) на масштабах, меньших горизонта.

Важным аспектом обсуждаемой проблемы является вопрос об уравнении состояния квинтэссенции, являющейся одним из видов темной материи. Простейшее уравнение состояния выбиралось в линейном виде $p = w^2\rho$, где на показатель состояния накладывались ограничения $-1 < w^2 < -\frac{1}{3}$.

Однако наблюдательные данные о сверхновых достаточно убедительно свидетельствуют о том, что $w^2$ может быть и меньше -1 [3 - 9].

Этот факт стимулировал появление большого ряда работ, в которых для описания поля квинтэссенции стали использоваться нелинейные уравнения состояния. Так, в работе [10] оно представлено как уравнение состояния типа жидкости Тэта $p = \omega_0\rho^\gamma$, где $\omega_0 < 0$, а $\gamma > 0$; в работах [11, 12] предложено использовать уравнение состояния газа Чаплыгина

$$p = -\frac{A}{\rho^n},$$

где $A > 0$, а $n > 1$.

Для фантомных полей, также могущих быть ассоциированными с квинтэссенцией, в работе [13] (см. также [7] и цитированную в ней обширную литературу) использовалось уравнение состояния жидкости Тэта с коэффициентом $\omega_0 < -1$ и изучались его космологические следствия.

Поиск нелинейного уравнения состояния диктовался не только наблюдательными данными, но и теоретическими соображениями. Действительно, линейное уравнение состояния имеет место лишь в строго фридмановской Вселенной. Поэтому уже небольшие отклонения от нее должны приводить, например, к небаротропности давления. Тот факт, что для описания темной материи требуется нелинейное уравнение состояния, по - видимому, является и следствием того, что темная материя не кластеризуется.

## 1.1 Условие давление - доминантности в космологии

В данной работе предлагается новая теоретическая интерпретация некластеризуемости квинтэссенции. Для понимания ее сущности обсудим условие кластеризации вещества.

Кластеризация барионной материи имеет место в тех случаях, когда ее плотность подчиняется условию энерго - доминантности

$$-\rho \le p \le \rho . \qquad (1)$$

Отсюда следует, что отсутствие кластеризации в среде можно интерпретировать как невыполнение условия (1), т.е. как нарушение энерго - доминантности в космологии.

Вопрос о возможности нарушения условия энерго - доминантности в классической космологии обсуждался, напри-мер, в работе [14]. В ней было показано, что если общую теорию относительности модифицировать введением "конформной" гравитационной постоянной $G \to G \cdot \psi(\rho)$, то путем соответствующего выбора функции $\psi(\rho)$ условие энерго - доминантности в области больших плотностей может быть нарушено.

Мы, однако, не будем выходить за рамки общей теории относительности, а дадим другую трактовку некластеризуемости квинтэссенции в космологии. А именно, некластеризуемость квинтэссенции будем интерпретировать как возможность описания ее состояния посредством условия давление - доминантности, т.е. как выполнение в ней следующих неравенств

$$p < -\rho \text{ и } p > \rho . \tag{2}$$

Для нахождения уравнения состояния космологической квинтэссенции заметим, что в цитированных выше работах величина ее параметра состояния рассматривалась в весьма широких пределах: $-1.0 < w^2 < -1.3$; по другим оценкам $-1.3 < w^2 < -1.6$ и даже $-2.4 < w^2 < -1.0$ (см. работу [7]). Из приведенных оценок видно, что состояние квинтэссенции с достаточной степенью точности можно описать как малое отклонение от вакуумного состояния. Поэтому при его выводе целесообразно воспользоваться методом приближений.

Итак, запишем уравнение состояния произвольной материи в общем виде

$$p = p(\rho) \tag{3}$$

и представим ее давление и плотность энергии как

$$\begin{aligned} p &= p_0 + \delta p, \delta p > 0, \\ \rho &= \rho_0 + \delta \rho, \delta \rho > 0, \end{aligned} \tag{4}$$

где $\delta p$ и $\delta\rho$ - малые добавки к невозмущенному давлению $p_0$ и невозмущенной плотности энергии $\rho_0$, обусловленные распространением звука в среде.

В этом случае выражение (3), согласно [15], может быть разложено в ряд Тейлора

$$p = p_0 + \left(\frac{\partial p}{\partial \rho}\right)_0 \delta\rho + \frac{1}{2}\left(\frac{\partial^2 p}{\partial \rho^2}\right)_0 (\delta\rho)^2 + ... \qquad (5)$$

При дальнейшем изложении примем, что основные значения давления и плотности энергии соответствуют вакуумному состоянию

$$p_0 = -\rho_0, \qquad (6)$$

а состояние квинтэссенции будем рассматривать как сумму вакуумного состояния и малого отклонения от него. Так как

$$\left(\frac{\partial p}{\partial \rho}\right)_0 = v^2 > 0 \qquad (7)$$

есть скорость звука в среде (здесь, как обычно, используется система единиц, т.е. $\hbar = c = 1$), то для описания квинтэссенции в (5) следует ограничиться тремя основными слагаемыми. Итак, с учетом (6) давление представим в нелинейном виде

$$p = -\rho_0 + v^2\left[1 + \kappa\frac{\delta\rho}{v}\right]\delta\rho = p_0 + \delta p, \qquad (8)$$

где $\kappa = \left(\frac{\partial v}{\partial \rho}\right)_0$ показатель дисперсии среды. Если $\kappa > 0$ и $\delta\rho > 0$, то дисперсия является нормальной и выражение (6) в целом будет описывать состояние вещества, удовлетворяю-щее условию энерго - доминантности (1). Если же $\kappa < 0$ и $\delta\rho > 0$, то дисперсия будет аномальной и выражение (8), вообще говоря, может эффективно описывать уравнение состояния квинтэссенции, удовлетворяющее, по предположению, условию давление - доминантности (2).

Выбор нелинейного уравнения состояния (8), как будет видно из дальнейшего изложения, позволяет, не выходя за рамки стандартной инфляционной модели, эффективно описать динамику темной материи (квинтэссенции). Для этого рассмотрим самосогласованную задачу, которая определяет совместную эволюцию полей и Вселенной.

## 1.2 Эволюция скалярных полей во Вселенной

Запишем сразу соответствующую систему уравнений Эйнштейна и уравнений двух взаимодействующих скалярных полей

$$\left(\frac{\dot{a}}{a}\right)^2 = \frac{4\pi G}{3}\Big( \dot{\varphi}^2 + m_\varphi{}^2\varphi^2 + \dot{\psi}^2 + m_\psi{}^2\psi^2 + \frac{\lambda_\varphi}{2}\varphi^4 + \frac{\lambda_\psi}{2}\psi^4 + \nu\varphi^2\psi^2 \Big), \tag{9}$$

$$\ddot{\varphi} + 3\frac{\dot{a}}{a}\dot{\varphi} + m_\varphi{}^2\varphi + \lambda_\varphi\varphi^3 + \nu\psi^2\varphi = 0, \tag{10}$$

$$\ddot{\psi} + 3\frac{\dot{a}}{a}\dot{\psi} + m_\psi{}^2\psi + \lambda_\psi\psi^3 + \nu\varphi^2\psi = 0. \tag{11}$$

Для исследования вопроса о возможности выполнения в такой модели условия давление - доминантности положим, что массы и поля соотносятся, как $m_\varphi << m_\psi, \varphi >> \psi$, а коэффициенты самодействия удовлетворяют соотношению $\lambda_\varphi << \lambda_\psi << 1$. Поэтому период колебаний поля $\varphi$ много больше периода колебаний поля $\psi$ ($T_\varphi >> T_\psi$). Другими словами, за время изменения поля $\psi$ основное поле $\varphi$ практически не меняется и, стало быть, его можно описать условиями

$$\dot{\varphi} \approx 0, \varphi \approx \text{const}\,. \tag{12}$$

Тогда, пренебрегая самодействием полей, получаем упрощенную систему уравнений

$$\left(\frac{\dot{a}}{a}\right)^2 = \frac{4\pi G}{3}\left(m_\varphi{}^2\varphi^2 + \dot{\psi}^2 + \tilde{m}_\psi{}^2\psi^2\right), \tag{13}$$

$$\ddot{\psi} + 3\frac{\dot{a}}{a}\dot{\psi} + \tilde{m}_\psi{}^2\psi = 0\,, \tag{14}$$

которую и будем анализировать. Здесь $\tilde{m}_\psi{}^2 = m_\psi{}^2 + \nu\varphi^2$ - квадрат эффективной массы поля $\psi$, обусловленный собственной массой поля и его взаимодействием с полем $\varphi$.

Для дальнейшего исследования необходимо задать массы скалярных полей и их начальные амплитуды. При этом необходимо иметь в виду, что они не могут быть произвольными и их типичные значения обычно задаются следующим образом

$$\begin{aligned} &m_\varphi << \lambda_\varphi^{\frac{1}{2}} M_p, m_\psi << \lambda_\psi^{\frac{1}{2}} M_p, \\ &\varphi_0 \approx \lambda_\varphi^{-\frac{1}{4}} M_p, \psi_0 \approx \lambda_\psi^{-\frac{1}{4}} M_p. \end{aligned} \tag{15}$$

Имея в виду эти ограничения, при анализе системы (13) - (14) рассмотрим такой ее случай, когда

$$m_\varphi\varphi >> \tilde{m}_\psi\psi, m_\varphi\varphi >> \dot{\psi}\,. \tag{16}$$

Это может иметь место, если в рассматриваемой модели удовлетворяется условие

$$\lambda_\psi << \lambda_\varphi << 1. \qquad (17)$$

Условие (16) означает, что энергия основного поля $\varphi$ существенно больше энергии дополнительного поля $\psi$. Тогда система уравнений (13) - (14) еще больше упрощается и принимает вид

$$\begin{aligned} &\left(\frac{\dot a}{a}\right)^2 = \frac{4\pi G}{3} m_\varphi{}^2 \varphi^2, \\ &\ddot\psi + 3\frac{\dot a}{a}\dot\psi + \tilde{m}_\psi{}^2 \psi^2 = 0. \end{aligned} \qquad (18)$$

Уравнения (18), как легко видеть, сводятся к линейному дифференциальному уравнению второго порядка

$$\ddot\psi + \mathfrak{I}\dot\psi + \mathfrak{R}\psi = 0$$

с постоянными коэффициентами $\mathfrak{I} = \sqrt{12\pi G}\, m_\varphi \varphi$ и $\mathfrak{R} = \tilde{m}_\psi{}^2$. Его решение ищется в виде $\psi = \psi_0 \exp(M \cdot t)$, следствием чего выступает алгебраическое уравнение

$$M^2 + \mathfrak{I}M + \mathfrak{R} = 0$$

с корнями

$$M_{1,2} = -\frac{\mathfrak{I}}{2} \pm \sqrt{\frac{\mathfrak{I}^2}{4} - \mathfrak{R}^2}\,. \qquad (19)$$

Из (16) и (17) следует, что $m_\varphi >> \tilde{m}_\psi$. Поэтому подкоренное выражение можно разложить в ряд по малому пара-метру $\tilde{m}_\psi / m_\varphi$ и получить два решения

$$M_1 = -\frac{\tilde{m}_\psi^{\,2}}{2\sqrt{3\pi G}\, m_\varphi \varphi},$$

$$M_2 = -2\sqrt{3\pi G}\, m_\varphi \varphi\ , \tag{20}$$

второе из которых является приближенным (с нулевой точностью по $\tilde{m}_\psi / m_\varphi$ ). Поэтому искомые решения уравнения движения поля $\psi$ принимают вид

$$\psi_1 = \psi_0 \exp(M_1 \cdot t) = \psi_0 \exp\left( -\frac{\tilde{m}_\psi^{\,2} \cdot t}{2\sqrt{3\pi G}\, m_\varphi \varphi} \right), \tag{21}$$

и

$$\psi_2 = \psi_0 \exp(M_2 \cdot t) = \psi_0 \exp\left( -2\sqrt{3\pi G}\, m_\varphi \varphi \cdot t \right). \tag{22}$$

Из правой части (9) нетрудно найти добавки к плотности энергии и давлению

$$\delta\rho = \frac{1}{2}\dot{\psi}^2 + \frac{m_\psi^{\,2}}{2}\psi^2 + \frac{\nu}{2}\varphi^2\psi^2, \tag{23}$$

$$\delta p = \frac{1}{2}\dot{\psi}^2 - \frac{m_\psi^{\,2}}{2}\psi^2 - \frac{\nu}{2}\varphi^2\nu^2 . \tag{24}$$

Подставляя их в нелинейное уравнение состояния, и имея в виду введенное ограничение на характер полей (пренебрежение их самодействием), запишем основное слагаемое в выражении для дисперсии

$$\kappa \approx \kappa_0 = 2\frac{\left(\dot{\psi}^2 - \tilde{m}_\psi{}^2\psi^2\right) - v^2\left(\dot{\psi}^2 + \tilde{m}_\psi{}^2\psi^2\right)}{\left(\ \dot{\psi}^2 + \tilde{m}_\psi{}^2\psi^2\ \right)^2}, \tag{25}$$

в котором опущены слагаемые, пропорциональные коэффициенту взаимодействия.

Подставляя (21) и (22) в (25), получаем

$$\kappa_{0_{1,2}} = 2\frac{\left[\left(M_{1,2}{}^2 - \tilde{m}_\psi{}^2\right) - v^2\left(M_{1,2}{}^2 + \tilde{m}_\psi{}^2\right)\right]}{\left(M_{1,2}{}^2 + \tilde{m}_\psi{}^2\right)^2 \psi_0{}^2}\exp\left(-2M_{1,2}\cdot t\right). \tag{26}$$

Чтобы оценить знак дисперсии подставим конкретные значения (20) в (26). Для первого случая имеем

$$\kappa_{0_1} = -2\frac{\left[\left(1 - \frac{\tilde{m}_\psi{}^2}{12\pi G m_\varphi{}^2\varphi^2}\right) + v^2\left(1 + \frac{\tilde{m}_\psi{}^2}{12\pi G m_\varphi{}^2\varphi^2}\right)\right]}{m_\psi{}^2\left(1 + \frac{\tilde{m}_\psi{}^2}{12\pi G m_\varphi{}^2\varphi^2}\right)^2 \psi_0{}^2}\cdot \tag{27}$$

$$\exp\left(\frac{\tilde{m}_\psi{}^2}{\sqrt{3\pi G} m_\varphi{}^2\varphi^2}\cdot t\right)$$

Подставляя же в (26) второй корень из (20), получаем

$$\kappa_{0_2} = -2\frac{\left[\left(1 - 12\pi G\varphi^2\frac{m_\varphi{}^2}{\tilde{m}_\psi{}^2}\right) + v^2\left(1 + 12\pi G\varphi^2\frac{m_\varphi{}^2}{\tilde{m}_\psi{}^2}\right)\right]}{\tilde{m}_\psi{}^2\left(1 + 12\pi G\varphi^2\frac{m_\varphi{}^2}{\tilde{m}_\psi{}^2}\right)^2 \psi_0{}^2}\cdot \tag{28}$$

$$\exp\left(4\sqrt{3\pi G} m_\varphi \varphi\cdot t\right)$$

Помня, что в силу (16) и (17) массы полей соотносятся как $\tilde{m}_\psi << m_\varphi$, а также что $G = M_p^{-2}$, имеем следующую оценку $\dfrac{\tilde{m}_\psi^{\;2}}{G m_\varphi^{\;2} \varphi^2} \sim \dfrac{\lambda_\psi}{\sqrt{\lambda_\varphi}} << 1$. Поэтому выражение для дисперсии (27) существенно упрощается

$$\kappa_{0_1} = -2 \frac{1+v^2}{\tilde{m}_\psi^{\;2} \psi_0^{\;2}} \exp\left( \frac{\tilde{m}_\psi^{\;2}}{\sqrt{3\pi G} m_\varphi \varphi} \cdot t \right) < 0 . \tag{29}$$

Отсюда видно, что время порядка

$$T_1 \sim \frac{\sqrt{G} m_\varphi \varphi}{m_\psi^{\;2}} >> \sqrt[4]{\frac{\lambda_\varphi}{\lambda_\psi^{\;4}}} \cdot M_p^{\;-1} >> \sqrt[4]{\frac{\lambda_\varphi}{\lambda_\psi^{\;4}}} \cdot 10^{-43} c \tag{30}$$

является временным масштабом существования квинтэссенциального состояния.

Поэтому при характерных значениях постоянных самодействия полей - $\lambda_\psi \sim 10^{-14}$ и $\lambda_\varphi \sim 10^{-12}$ - показатель экспоненты на временах догорячего этапа развития Вселенной ($t \sim 10^{-37}$ с) будет порядка $10^{-5}$. Это означает, что аномальная дисперсия (27) имеет величину

$$\kappa_{0_1} \sim -2 \frac{1+v^2}{m_\psi^{\;2} \psi^2} >> -2 \frac{1+v^2}{\sqrt{\lambda_\psi}} M_p^{\;-4} \tag{31}$$

или в обычных единицах $\kappa_{0_1} >> -10^{-87}$ см$^3$/г. Полученная оценка показывает, что, несмотря на малость дисперсии (31), на начальных этапах эволюции Вселенной состояние вещества отличалось от вакуумно - подобного.

В силу тех же условий (15) и (17) имеем и такую оценку

$$G\varphi^2 \frac{m_\varphi^{\ 2}}{m_\psi^{\ 2}} \sim \frac{\sqrt{\lambda_\varphi}}{\lambda_\psi} >> 1$$

и поэтому дисперсия оказывается положительной

$$\kappa_{0_2}(t) \sim 2\frac{1 - v^2}{Gm_\varphi^{\ 2}\varphi_0^{\ 2}\psi_0^{\ 2}} \exp\left(\sqrt{12\pi G}m_\varphi\varphi \cdot t\right) \sim$$

$$\sim 2\left(1 - v^2\right)\sqrt{\frac{\lambda_\psi}{\lambda_\varphi^{\ 2}}} M_p^{\ -4} \exp\left(\sqrt{12\pi G}m_\varphi\varphi \cdot t\right) \qquad (32)$$

Состояние системы с положительной дисперсией, как это следует из (32), будет существовать на масштабе времени

$$T_2 \sim \frac{1}{\sqrt{G}m_\varphi\varphi} >> \lambda_\varphi^{\ -\frac{1}{4}} M_p^{\ -1} >> \lambda_\varphi^{\ -\frac{1}{4}} \cdot 10^{-43}\,c\,. \qquad (33)$$

Поэтому показатель экспоненты даже на начальных временах жизни Вселенной ($t > T_p$) становится очень большим ($>> 10^2$). Но несмотря на это, в силу естественного условия $v \to 1$ дисперсия (32) всегда будет стремиться к нулю, т.е. $\kappa_{0_2}(t) \to 0$.

## Заключение

Таким образом, проведенный анализ показал, что система двух гравитирующих скалярных полей, одно из которых ($\varphi$) находится в вакуумно-подобном состоянии, а второе ($\psi$) экспоненциально убывает по закону (21), может пребывать в квинтэссенциальном состоянии, описываемом уравнением (8).

# ЛИТЕРАТУРА

# 2. УРАВНЕНИЕ СОСТОЯНИЯ ВЕЩЕСТВА И ЭВОЛЮЦИЯ ВСЕЛЕННОЙ

## Введение

Для описания эволюции Вселенной, как известно, необходимо задать уравнение состояния вещества, которое предопределяет ее динамику.

Уравнение состояния вещества, как правило, задается в линейном виде [1]

$$p = w^2 \rho , \tag{1}$$

который связывает давление с плотностью среды. При этом параметр $w^2$ может принимать различные значения.

Например, если $w^2 = -1$, то такое уравнение состояния описывает вакуум; если $w^2 = 1/3$, - оно задает релятивистский газ; при $w^2 = 0$, - вещество представляет собой пылевидную материю; случай $w^2 = - 1/3$ соответствует струно - подобному состоянию вещества; при $w^2 = - 2/3$ уравнение состояния описывает доменные стенки. И наконец, если $w^2 < -1$, то оно, по предположению, описывает небарионную материю.

Численные значения этого коэффициента следуют из вида соответствующего тензора энергии импульса и условия равенства нулю его следа.

Сравнительно недавно в космологии стали использовать нелинейные уравнения состояния, которые способны более точно описать вещество во Вселенной. Так, в работе [3] оно представлено как уравнение состояния жидкости Тэта

$$p = \omega_0 \rho^{\gamma}, \tag{2}$$

где $\omega_0 < 0$, а $\gamma > 0$; в работах [4] для описания состояния $\Lambda$ - поля предложено использовать уравнение состояния газа Чаплыгина

$$p = -\frac{A}{\rho^n}. \tag{3}$$

В работе [2] для описания состояния среды во Вселенной (темной материи) предложено использовать два алгебраических параметра, связанных с масштабным фактором.

Ряд некоторых результатов в релятивистской космологии связан с использованием уравнения состояния Ван - дер - Ваальса [5]

$$p = 8\omega\frac{\rho}{3-\rho} - 3\rho^2. \tag{4}$$

Необходимость вывода общего уравнения состояния, которое приводило бы к естественной смене конкретных видов уравнения состояния во Вселенной, была подчеркнута в работе [6]. Однако, предложенное в ней идеологически корректное уравнение

$$p = \frac{4}{3}\left(1 - \frac{\rho}{\rho_{max}}\right)^{1/2}\left(1 - \frac{\rho_{min}}{\rho}\right)^{1/2}\rho \tag{5}$$

не приводит к желаемому результату.

В данной работе предлагается нестационарный тип уравнения состояния вещества, который естественным образом позволяет учесть трансформацию среды в ходе эволюции самой Вселенной

## 2.1 Вывод нестационарного уравнения состояния

Для вывода уравнения состояния вещества в очень ранней и ранней Вселенной необходимо иметь в виду, что в указанный период понятия классических пространства и времени фактически отсутствовали. Это означает, что уравнение состояния следует выводить из принципов квантовой теории. Поскольку же уравнение состояния связывает между собой давление $p$ и плотность энергии $\rho$, то для его вывода будем использовать выражения квантовой акустики, описывающие кинетику фонона. (В рамках квантовой теории поля уравнение состояния вакуума было выведено в работе [7]; см. также обзор [8].) При этом во всех получаемых результатах следует использовать условие предельного перехода к исчезающе малым значениям пространственных координат и времени, имевшее место в очень ранней Вселенной.

Пусть в первичном «сгустке» вещества, которое появилось в результате квантового туннелирования, Вселенной, возникло малое возмущение - свободный фонон. Квантовую кинетику такого фонона будем описывать уравнением Шредингера

$$H\psi = E\psi \,, \tag{6}$$

гамильтониан которого, согласно [9], имеет вид

$$H = \frac{1}{2}\int \vec{v}(r)\rho'(r)\vec{v}(r)d^3r \,. \tag{7}$$

Здесь

$$\vec{v} = \sum_k \left( \frac{\hbar\omega_k}{2\rho_0 V k^2} \right)^{1/2} \vec{k}\left[\hat{b}_k \exp(-i\omega_k t) - \hat{b}_{-k}{}^{+} \exp(i\omega_k t)\right] e^{i\vec{k}\vec{r}} \tag{8}$$

оператор скорости, а

$$\rho'(r) = \sum_k \left( \frac{\hbar \rho k^2}{2V\omega_k} \right)^{1/2} \left[ \hat{b}_k \exp(-i\omega_k t) + \hat{b}_{-k}{}^{+} \exp(i\omega_k t) \right] e^{i\vec{k}\vec{r}} \qquad (9)$$

оператор флуктуации плотности.

Кроме того, в этих выражениях $\rho_0$ - равновесная плотность среды, $V$ - объем системы (первичного «сгустка» вещества). Наконец, волновая функция свободного фонона суть

$$\psi(r,t) = \frac{1}{V^{1/2}} \sum_k \left[ \hat{b}_k \exp(-i\omega_k t) + \hat{b}_{-k}{}^{+} \exp(i\omega_k t) \right] e^{i\vec{k}\vec{r}} \ . \qquad (10)$$

В выражениях (8) - (10) $\hat{b}_{-k}{}^{+}$ и $\hat{b}_k$ представляют собой операторы рождения и уничтожения, соответственно.

Используя выражения (7) - (10), легко найти среднее значение удельной кинетической энергии фонона с учетом отмеченного выше условия предельного перехода к «нулевым» значениям пространства и времени

$$\varepsilon = \lim_{\substack{r \to 0 \\ t \to 0}} \int \overset{*}{\psi}(r,t) H \psi(r,t) d^3 r = \frac{1}{2} \overline{\rho' \cdot v^2} \ , \qquad (11)$$

где $\overline{\rho'}$ - среднее значение флуктуации плотности среды, $\overline{v}^2$ - квадрат средней скорости фонона. При этом с учетом выражения волнового вектора имеем

$$\overline{\rho'} = \rho' = \lim_{\substack{r \to 0 \\ t \to 0}} \int \overset{*}{\psi} \rho'(r) \psi d^3 r = \sum_k \left( \frac{\hbar \rho_0 k^2}{V\omega_k} \right)^{1/2} = \left( \frac{\hbar \rho_0}{V} \right)^{1/2} \sum_k \omega_k^{1/2} \ , \qquad (12)$$

а

$$\bar{v} = \lim_{\substack{r \to 0 \\ t \to 0}} \int \overset{*}{\psi} v(r) \psi d^3 r = \sum_k \left( \frac{\hbar \omega_k}{\rho_0 V} \right)^{1/2} = \left( \frac{\hbar}{\rho_0 V} \right)^{1/2} \sum_k \omega_k^{1/2} . \quad (13)$$

Опираясь на (13), можно показать, что квадрат средней скорости есть

$$\bar{v}^2 = \frac{\hbar}{\rho_0 V} \left( \sum_k \omega_k^{1/2} \right)^2 = \frac{\hbar}{\rho_0 V} \Omega_f , \quad (14)$$

где $\Omega_f$ - обобщенная частота фонона как волнового пакета.

Так как звуковая волна образуется благодаря появлению в среде, изначально описываемой равновесными давлением $p_0$ и плотностью энергии $\rho_0$, двух параметров - избыточных (недостаточных) давления $\rho'$ и плотности энергии $p'$, то полные давление и плотность энергии, как известно, представляются в виде

$$\left. \begin{aligned} p &= p_0 + p' \\ \rho &= \rho_0 + \rho' \end{aligned} \right\} . \quad (15)$$

Поэтому, считая процесс возникновения флуктуации адиабатическим, выражение (14) с учетом (15) и согласно [10] можно записать следующим образом

$$\bar{v}^2 = \frac{\hbar \Omega_f}{\rho_0 V} = \frac{p'}{\rho_0} . \quad (16)$$

Выбор положительного знака перед $p'$ объясняется тем, что в случае первичного «сгустка» вещества давление может быть направлено только вовне.

Для нахождения классического выражения избыточного давления воспользуемся выражением для избыточной плотности энергии, которое на основании (14) и (16) легко пред-

ставить как

$$\rho' = \sqrt{\rho_0 \frac{\hbar\Omega_f}{V}} = \sqrt{\rho_0 p'} . \qquad (17)$$

Но поскольку из общего уравнения состояния $p = p(\rho)$ в силу (16) следует, что

$$p' = \frac{dp}{d\rho}\rho' , \qquad (18)$$

то, объединяя (17) и (18), находим

$$\frac{p'}{\rho_0} = \left(\frac{dp}{d\rho}\right)^2 . \qquad (19)$$

Таким образом, классическую плотность кинетической энергии фонона с учетом (11) можно записать как

$$\varepsilon = \frac{1}{2}\rho' \cdot \bar{v}^2 = \frac{1}{2}\rho' \frac{p'}{\rho_0} = \frac{1}{2}\rho' \left(\frac{dp}{d\rho}\right)^2 . \qquad (20)$$

Поэтому ей соответствует плотность действия

$$s = \varepsilon\Omega_f^{-1} = \frac{1}{4\pi}\rho' \left(\frac{dp}{d\rho}\right)^2 T_f , \qquad (21)$$

где $T_f$ - период колебаний фонона.

Для дальнейшего изложения необходимо подчеркнуть, что даже в воздухе акустическое число Маха $M = \frac{v}{c_f}$, где $v$ - скорость частиц среды, а $c_f$ - скорость распространения возмущений (звука), обычно много меньше единицы. С воз-растанием же плотности среды (тем более, если речь идет о плотности первичного «сгустка» на восемьдесят порядков превышающей ядерную плотность) $M \to 0$. Это позволяет говорить о неподвижности не только среды в целом, но и о неподвижности произвольной ее области. Поэтому удельное действие для любой из областей, например, области, занимаемой фононом, будет определяться только энергией среды.

Следовательно, удельное действие фонона можно записать в классическом виде

$$s = \frac{S}{V} = \frac{Et}{V} = \rho t . \qquad (22)$$

Отсюда

$$\frac{ds}{d\rho} = t . \qquad (23)$$

Подставляя теперь (21) в (23), получаем

$$\frac{1}{2\pi}\rho' T_f {c_f}^2 \left(\frac{d^2 p}{d\rho^2}\right) = t , \qquad (24)$$

где

$${c_f}^2 = \left(\frac{dp}{d\rho}\right)$$

скорость звука.

Из (24) следует нестационарное уравнение состояния вещества в дифференциальном виде

$$\frac{d^2 p}{d\rho^2} = \frac{2\pi}{{c_f}^2 \rho'} \cdot \frac{t}{T_f} = \Gamma(t) . \qquad (25)$$

## 2.2 Виды нестационарного уравнения состояния вещества

Решение уравнения (25) имеет вид

$$p = p_0 + w_0^2 \rho + \frac{1}{2} \Gamma(t) \rho^2 \qquad (26)$$

и содержит две постоянные величины. Желая получить алгебраическое уравнение состояния в начальный момент времени, необходимо потребовать $t \to 0$. Тогда вакуумно - подобное состояние вещества возникает при $p_0 = 0$ и $w_0^2 = -1$. Таким образом, требование выполнения условия перехода к вакууму при $t \to 0$ позволяет получить следующее нестационарное уравнение состояния

$$p = -\rho + \frac{1}{2} \Gamma(t) \rho^2 . \qquad (27)$$

Начнем с несущественного переобозначения скорости звука, а именно, обозначим $c_f^2 = w^2$. (Это переобозначение, делаемое для удовлетворения выражения (1), действует только в тексте данного раздела.) Тогда в силу определения скорости звука из (26) получаем выражение

$$w^2 = w_0^2 + \frac{2\pi}{w^2 \rho'} \cdot \frac{t}{T_f} \rho , \qquad (28)$$

которое легко преобразовывается в биквадратное уравнение

$$w^4 - w_0^2 w^2 - \frac{2\pi}{\rho'} \cdot \frac{t}{T_f} \rho = 0 . \qquad (29)$$

Его решения таковы

$$w(t)^2 = -\frac{1}{2} \pm \sqrt{\frac{1}{4} + \frac{2\pi}{\rho'} \cdot \frac{t}{T_f} \rho} \,. \qquad (30)$$

Для дальнейших вычислений положим, что в (28) имеет место условие $w_0{}^2 << \frac{2\pi}{w^2\rho'} \cdot \frac{t}{T_f}\rho$. (Численные значения входящих сюда величин, могущих обеспечить выполнение этого неравенства, будут даны ниже.) Это позволяет записать приближенные значения (30) в общем виде

$$w(t)_\pm{}^2 = -\frac{1}{2} \pm \frac{1}{2}\left(1 + 4\pi\frac{t}{T_f} \cdot \frac{\rho}{\rho'}\right) \qquad (31)$$

или в виде двух решений -

$$w(t)_+{}^2 = 2\pi\frac{t}{T_f} \cdot \frac{\rho}{\rho'} \qquad (32)$$

и

$$w(t)_-{}^2 = -1 - 2\pi\frac{t}{T_f} \cdot \frac{\rho}{\rho'} \,. \qquad (33)$$

Из этих выражений видно как можно получить некоторые из приведенных выше уравнений состояний.

В самом деле, если $\frac{t}{T_f} \to 0$, то $w(t)_+{}^2 \to 0$, и из (32) в согласии с (1) следует уравнение состояния пылевидной материи. Если же при определенных значениях величин $t$, $T_f$, $\rho$ и $\rho'$ их комбинация $2\pi\frac{t}{T_f} \cdot \frac{\rho}{\rho'} \to \frac{1}{3}$, то $w(t)_+{}^2 \to \frac{1}{3}$, и из выражений (1) и (32) получается уравнение состояния реля-тивистского газа. Вероятность такого подбора четырех величин, на первый взгляд, представляется крайне малой. Однако если отношения $\frac{t}{T_f}$ и $\frac{\rho}{\rho'}$ приближенно будут находиться в обратной зависимости, то вероятность требуемого значения становится вполне определенной. Так, если считать $\frac{t}{T_f} \propto 10^{-5} \div 10^{-6}$ (отношение времени доминирования релятивистского вещества ко времени существования Вселенной), то, при величине отношения $\frac{\rho}{\rho'}$ такого же порядка (см. ниже), действительно можно получить требуемую величину $w(t)_+{}^2$.

Более того, если $\frac{t}{T_f} \to 0$, то уже из (33) и (1) следует уравнение состояния вакуума. И наконец, при $\frac{t}{T_f} \neq 0$ из (33) следует, что $w(t)_-{}^2 < -1$. Уравнение же состояния с таким значением параметра описывает, согласно (1), состояние темной энергии (см. например работу [11]).

Анализируя эти результаты, видно, что результатом эволюции Вселенной является естественная смена уравнений состояния среды. Действительно, в выражения (32) - (33) входит частота $\Omega_f$, являющаяся обобщенной частотой фонона. Но поскольку фонон представляет собой волновой пакет, то ему соответствует набор частот и, следовательно, набор периодов колебаний. Поэтому чтобы обеспечить основную последовательность смены уравнений состояния среды (вакуум $\to$ релятивистский газ $\to$ пыль) в выражениях (32) - (33) под $T_f$ следует понимать разные периоды колебаний мод фонона.

Приведем некоторые численные оценки, могущие более точно охарактеризовать обсуждаемый физический процесс.

Начальную плотность среды, очевидно, следует считать не больше планковской, т.е. $\rho \leq 10^{94}\ \text{г}/\text{см}^3$. Что касается возмущающей плотности $\rho'$, то ее величину можно выбрать в достаточно широких пределах. В качестве примера ограничимся следующим соображением.

Известно, что при планковских плотностях квантово - гравитационные эффекты становятся настолько большими, что квантовые флуктуации метрики начинают превосходить значения самой метрики и описание Вселенной в терминах классического пространства - времени становится невозможным. С другой стороны, квантовые поправки к уравнениям Эйнштейна становятся заметными при температуре $T \geq 10^{17}\,\text{Гэв}$, которой соответствует плотность среды $\rho' \propto 10^{90}\ \text{г}/\text{см}^3$ [1]. Эту плотность уже можно рассматривать в качестве добавочной плотности к $\rho_0$ или плотности фонона. Отсюда следует упоминавшаяся выше оценка - $\frac{\rho}{\rho'} \propto 10^{-4} \div 10^{-5}$. В принципе плотность возмущения может быть еще меньше, и даже стать бесконечно малой величиной. С физической точки зрения это означает, что сколь угодно малое случайное возмущение в начальном «сгустке» среды неизбежно приведет к раздуванию Вселенной. Поэтому состояние «сгустка» среды является неустойчивым.

Итак, для планковского объема $V \propto 10^{-99}\,\text{см}^3$ плотность действия $s \propto 10^{72}\ \text{эрг}\cdot\text{сек}/\text{см}^3$. Но так как при $c_a \propto c = 3\cdot 10^{10}\ \text{см}/\text{сек}$ плотность энергии фонона $\varepsilon \propto 10^{111}\ \text{эрг}/\text{см}^3$, то соответствующая ему частота $\omega \propto \frac{\varepsilon}{s} \propto 10^{39}\,\text{сек}^{-1}$. Поэтому период колебаний фонона как квазичастицы составляет величину $T_f \propto 10^{-39}\,\text{сек}$, которая близка ко времени начала разогрева Вселенной $T_h \propto 10^{-37}$ сек. Отсюда следует, что в начальный момент времени $t \propto t_P \propto 10^{-43}$ сек отношение $\frac{t_P}{T_f} \propto 10^{-4}$ и, следовательно, с указанной степенью точности правую часть в (25) можно считать стремящейся к нулю. Поэтому с точностью не меньшей чем $10^{-4}$ состояние Вселенной в начальный момент времени будет описываться вакуумным состоянием.

## Заключение

Таким образом, для описания эволюции Вселенной впервые предложено нелинейное нестационарное уравнение состояния вещества. Оно позволило естественным образом описать трансформацию вещества в процессе эволюции Вселенной, предсказывая основную последовательность смены его уравнений состояния - вакуум $\rightarrow$ релятивистский газ $\rightarrow$ пыль.

При этом предсказан будущий момент времени в эволюции Вселенной, при котором плотность энергии будет равняться нулю, так что состояние среды будет удовлетворять условию давление - доминантности.

Показано, что взаимодействующие скалярные поля допускают нестационарное уравнение состояния, но с определенным значением константы взаимодействия. Кроме того, такие поля приводят к весьма широкому спектру возмущений плотности, которые достаточны для образования всей наблюдаемой структуры Вселенной.

Анализ самосогласованной задачи о влиянии скалярного поля на параметры Вселенной и обратно, показал, что нестационарное уравнение состояния фактически выполняется на протяжении достаточно длительного времени эволюции Вселенной.

# ЛИТЕРАТУРА

# 3. РАЗВИТИЕ ВОЗМУЩЕНИЙ ВО ВСЕЛЕННОЙ, ОПИСЫВАЕМОЙ НЕСТАЦИОНАРНЫМ УРАВНЕНИЕМ СОСТОЯНИЯ

## Введение

Вопросам эволюции флуктуаций барионной материи посвящена значительная литература (см., например, [1]), но физическая причина их роста до конца не исследована. Дело заключается в том, что традиционный подход к этой проблеме основан на изучении гравитационной неустойчивости барионного вещества во Вселенной.

Однако наблюдательные данные последнего десятилетия убедительно доказали существенное количественное превосходство во Вселенной небарионной субстанции над барионной материей [2].

Поэтому возникает естественный вопрос о том, может ли сама небарионная субстанция (например, темная энергия) явиться причиной образования космических структур во Вселенной? Различные аспекты этой проблемы были рассмотрены в работах [3, 4]. Среди них отметим анализ антигравитационной - в частности, вакуумной - неустойчивости барионного космологического субстрата. При этом в статьях [5] было показано, что вакуум сам может порождать объекты типа карликовых галактик.

Целью работы является продолжение исследований по развитию возмущений барионной материи, обусловленных небарионной субстанцией. Конкретно - здесь речь идет об анализе роста возмущений барионной материи в ходе эволюции Вселенной с нестационарным уравнением состояния.

Нестационарное уравнение состояния небарионной субстанции, по - видимому, впервые было введено в [6]. Его физический смысл заключается в том, что свойства небарионной материи также должны меняться вместе с эволюцией Вселенной. При этом состояние небарионной материи, видимо, уже становится отличным от чисто вакуумного, а на самых ранних этапах развития Вселенной допускает приближение к фантомной темной энергии [7].

Влияние нестационарного уравнения состояния на эволюцию структур во Вселенной было рассмотрено, например, в работе [8]. В ней для шести типов нестационарных уравнений состояния численным методом проанализирован рост ее возмущений на фоне темной материи, описываемой относительной плотностью $\Omega_D = 0.3$. Общий результат заключается в том, что хотя возмущения плотности барионной материи растут, но они растут медленнее, чем масштабный фактор. Поэтому отношение $\frac{\delta(t)}{a(t)} < 1$ (в единицах $г/см^4$).

В отличие от цитированной работы [8], здесь используется лишь один вид нестационарного уравнения состояния. Но при этом применяется общий вид уравнения, описывающего эволюцию возмущений только в случае барионной материи (Присутствием темной материи мы пренебрегаем). Кроме того, мы учитываем факт, что в процессе эволюции Вселенной меняется уравнение состояния не только небарионной, но и барионной материи.

И, наконец, здесь дано аналитическое решение поставленной задачи, что позволяет явным образом представить временную эволюцию масштабного фактора Вселенной, а также оценить изменение плотности возмущений барионного вещества в зависимости от соотношения ее кинетической и потенциальной энергий для получения выводов космогонического характера для различных эпох Вселенной.

## 3.1 Эволюция масштабного фактора

Уравнения Эйнштейна, описывающие эволюцию масштабного фактора, имеют вид [1]

$$\ddot{a} = -\frac{4\pi}{3}G(\rho_{nb} + 3p_{nb})a, \qquad (1)$$

$$H^2 + \frac{k}{a^2} = \left(\frac{\dot{a}}{a}\right)^2 + \frac{k}{a^2} = \frac{8\pi}{3}G\rho_{nb}. \qquad (2)$$

Из уравнений (1) и (2) следует закон сохранения энергии, который записывается

$$\dot{\rho}_{nb}a^3 + 3(\rho_{nb} + p_{nb})a^2\dot{a} = 0. \qquad (3)$$

В этой системе $\rho_{nb}$ - плотность небарионной материи, $p_{nb}$ - ее давление; $G$ - гравитационная постоянная. Здесь же $H = \frac{\dot{a}}{a}$ - постоянная Хаббла, которая, как отмечалось выше, зависит от времени. Кроме того, в (2) k - кривизна пространства, которая равна 1 для закрытой, 0 для плоской и - 1 для открытой моделей Вселенной, соответственно.

Для определения того, как Вселенная эволюционирует во времени, необходимо задать уравнение состояния небарионного вещества, которое связывает между собой его плотность энергии и давление. Для адиабатических процессов оно задается в виде $p_{nb} = \omega \cdot \rho_{nb}$, где $\omega$ - параметр состояния (постоянный в модели Фридмана). Для известных видов небарионной материи, например квинтэссенции, вакуума, фантомной энергии, он принимает значения $-1\langle\omega\langle 1/3$, $-1$, $\omega\langle -1$, соответственно [2].

Согласно постановке задачи мы используем параметризацию Шевалье - Поларски - Линдера [6]

$$\omega(a) = \omega_0 + \omega_1 (1 - \frac{a}{a_0}) \frac{a}{a_0}, \qquad (4)$$

которая описывает широкий класс нестационарных уравнений состояния. Здесь $\omega_0 = -1.3$, а $\omega_1$ равняется $4$ , либо $-2$; $a_0$ - фиксированный масштабный фактор, $a$ - его текущее значение.

Решая уравнение (3) с (4), получаем

$$\rho_{nb} = \rho_0 x^{-3\kappa} \exp[\frac{3}{2}\omega_1 (x-1)^2], \qquad (5)$$

где $x = \dfrac{a}{a_0}$, а $\kappa = 1 + \omega_0$ - постоянная величина. Подставляя необходимые параметры в (5), получим неоднородное дифференциальное уравнение второго порядка следующего вида

$$\frac{d^2 x}{dt^2} = Cx^2 \exp[\frac{3}{2}\omega_1 (x-1)^2] \cdot (-3 + 3\omega_1 x - 3\omega_1 x^2) , \qquad (6)$$

с постоянной $C = -\dfrac{4\pi}{3} G\tilde{\rho}_0$.

Для решения приведенного уравнения будем рассматривать случай, когда $\dfrac{a}{a_0} \langle\langle 1$. Такое условие, при соответствующем выборе фиксированного масштабного фактора $a_0$, выполняется как в очень ранней Вселенной, так и на более поздних этапах ее эволюции. Тогда (6) упрощается и записывается следующим образом

$$\frac{d^2 x}{dt^2} = -3Cx^2 \exp[\frac{3}{2}\omega_1 - 3\omega_1 x]. \qquad (7)$$

Дифференциальные уравнения приведенного типа решаются заменой - $\frac{dx}{dt} = p$, $\frac{d^2x}{dt^2} = p\frac{dp}{dx}$. В результате получаем уравнение с разделенными переменными

$$pdp = Fx^2 \exp[-3\omega_1 x]dx\,, \qquad (8)$$

в которое введена новая постоянная величина

$$F = -3C\exp(3/2\omega_1)\,.$$

Вычисляя квадратуры и возвращаясь к первоначальным переменным, находим решение

$$t = \frac{1}{\sqrt{2F}}\frac{2}{3\omega_1}\exp(3/2\omega_1 x)\,. \qquad (9)$$

Обратив (9), имеем

$$x = \frac{2}{3\omega_1}\ln\chi t\,, \qquad (10)$$

где

$$\chi = \frac{3\omega_1\sqrt{F}}{\sqrt{2}}\,.$$

Дифференцируя (10) и пользуясь определением постоянной Хаббла, легко находим ее выражение

$$H = \frac{\dot{x}}{x} = \frac{1}{t\ln\chi t}\,. \qquad (11)$$

Используя явный вид переменной $x$, получаем следую-щую зависимость масштабного фактора от времени

$$a = \frac{2a_0}{3\omega_1} \ln \chi t \,. \tag{12}$$

## 3.2 Рост плотности возмущений во Вселенной

Запишем общее нерелятивистское уравнение, описывающее эволюцию возмущений плотности барионной материи [1, 4], -

$$\ddot{\delta} + 2H\dot{\delta} + (v_s^2 k^2 - 4\pi G \rho_b)\delta = 0 \,, \tag{13}$$

где $v_s$ - скорость звука в барионной материи, $\vec{k}$ - волновой вектор, а $\rho_b$ - плотность барионной материи. Для его дальнейшего анализа сделаем два замечания.

В выражении (13) в скобках присутствуют два слагаемых, первое из которых описывает внутреннюю энергию барионной материи, а второе - ее внешнюю (гравитационную) энергию. При этом соотношение между этими видами энергий в ходе эволюции Вселенной меняется.

Далее, в процессе эволюции Вселенной, строго говоря, меняется не только уравнение состояния небарионной материи, но и барионной материи. Поэтому выражение для плотности барионной материи также не является постоянным, а зависит от времени.

В очень ранней Вселенной вещество имеет релятивистский характер, и поэтому $v_s \sim 1$; в более поздние эпохи вещество становится нерелятивистским и $v_s \to 0$. Кроме того, учтем, что согласно выражению (12) и условию $a_0 = t_0$, волновой вектор, уменьшается как

$$k^2 \propto a^{-2} = \frac{9\omega_1^{\;2}}{4t_0^{\;2}\ln^2\chi t}.$$

Закон сохранения (3) с учетом наличия не только небарионной, но и барионной материи, обобщается очевидным образом -

$$(\dot{\rho}_{nb} + \dot{\rho}_b)a = -3[(\rho_{nb} + \rho_b) + (p_{nb} + p_b)]\dot{a}\,. \tag{14}$$

Но так как барионное вещество и небарионная субстанция не взаимодействуют между собой, то входящие сюда переменные величины являются независимыми. Поэтому для барионной материи получаем эволюционное уравнение

$$\dot{\rho}_b = -K \cdot H \cdot \rho_b\,, \tag{15}$$

где $K$ - коэффициент, зависящий от ее уравнения состояния (для релятивистского газа $K = 4$, для пыли $K = 3$). Его решение, полученное с учетом (11), имеет вид $\rho_b = \tilde{\rho}_0 \ln^{-K}\chi t$, где $\tilde{\rho}_0 = \mathrm{const}$.

Подставляя эти величины в (20), получаем дифференциальное уравнение второго порядка с переменными коэффициентами

$$\ddot{\delta} + P(t)\dot{\delta} + Q(t)\delta = 0\,, \tag{16}$$

которые равны

$$P(t) = \frac{2}{t\ln\chi t}, \quad Q(t) = \frac{9\omega_1^2}{4t_0^2\ln^2\chi t} - \frac{4\pi G\tilde{\rho}_0}{\ln^K\chi t}\,,$$

соответственно.

Для решения этого уравнения введем новую функцию z, связанную с $\delta$ соотношением $\delta = u(t)z$. Согласно стандарт-ной методике [8], продифференцируем указанное соотношение два раза и полученные значения подставим в уравнение (22). В итоге получаем выражение

$$u\ddot{z} + (2\dot{u} + P(t)u)\dot{z} + (\ddot{u} + P(t)\dot{u} + Q(t)u)z = 0 . \qquad (17)$$

Приравняем к нулю коэффициент при первой производной $\dot{z}$

$$2\dot{u} + P(t)u = 0 . \qquad (18)$$

Отсюда легко найти значения u и ее производных. Подставляя их в (17) и производя необходимые преобразования, получаем следующее уравнение

$$\ddot{z} + I(t)z = 0 , \qquad (19)$$

где $I(t) = -\frac{1}{4}P^2(t) - \frac{1}{2}\dot{P}(t) + Q(t)$. Подставляя в него значения P(t), Q(t) получаем

$$\ddot{z} + \frac{1}{t^2}\left(\frac{1}{\ln\chi t} + t^2\left[\frac{9\omega_1^{\,2}}{4t_0^{\,2}\ln^2\chi t} - \frac{4\pi G\rho_0}{\ln^K\chi t}\right]\right)z = 0 . \qquad (20)$$

Обозначив выражение в круглых скобках как $\Omega(t)$, получаем нелинейное уравнение Эйлера [8]

$$\ddot{z} + \frac{\Omega(t)}{t^2}z = 0 . \qquad (21)$$

Исследуем эволюцию барионной материи на временном интервале $t_1 = 10^{-36}\,с < t < t_2 = 10^{-6}\,с$. Правомерность такой постановки обусловлена тем, что одним из наиболее актуальных вопросов современной космологии, согласно [9], являет-ся изучение процесса формирования возмущений барионного вещества в самых ранних эпохах Вселенной. Поэтому рассмотрим поведение функции $\Omega(t)$, характерное в эпоху очень ранней Вселенной, положим здесь $K = 4$ и применим теорему сравнения для нахождения максимально допустимого значения z.

Выбор нижней границы обусловлен тем, что с этого момента времени $t_1 \sim 10^{-36}$c наступает стадия рождения барионного вещества. Верхняя граница $t_2 = t_0 \sim 10^{-6}$c задана из соображений выполнения принятого выше условия $\frac{a}{a_0} = \frac{t}{t_0} \langle\langle 1$ с учетом численного значения входящей в $\Omega(t)$ постоянной $\chi$ ($\chi \sim 10^{41}c^{-1}$ для $\omega_1 = 4$), а также на основании оценок завершения адронной эры [1].

Пусть имеет место условие

$$\frac{9\omega_1^{\ 2}}{4t_0^{\ 2}\ln^2\chi t} >> \frac{4\pi G\rho_0}{\ln^4\chi t},$$

означающее преобладание кинетической энергии барионного вещества над потенциальной энергией. Тогда, пренебрегая в круглых скобках уравнения (20) третьим слагаемым, находим, что

$$\Omega(t) = \frac{1}{\ln\chi t} + \frac{9\omega_1^{\ 2}t^2}{4t_0^{\ 2}\ln^2\chi t}. \tag{22}$$

Анализ этой функции показывает, что на выбранном временном промежутке она лишь монотонно убывает. Поэтому ее максимальное значение имеет место в нижней границе интервала. Численные оценки показывают, что $\Omega(t_1) \approx 10^{-1} \div 10^{-2} = \text{const}_1$.

В соответствии с теоремой сравнения и общей теорией дифференциальных уравнений [8], решение уравнения Эйлера с постоянной функцией $\Omega(t_n)$ (если она меньше $\left|\frac{1}{2}\right|$ для заданного значения времени $t_n$) можно записать в следующем приближенном виде $z \sim C_1 t + C_2$, где $C_1$ и $C_2$ - новые постоянные величины. Так что возмущающая функция эволюционирует не быстрее чем

$$\delta(t) = \frac{u_0}{\ln \chi t}\left(C_1 t + C_2\right). \tag{23}$$

Пусть удовлетворяется обратное условие

$$\frac{9\omega_1^{\,2}}{4t_0^{\,2}\ln^2 \chi t} << \frac{4\pi G\tilde{\rho}_0}{\ln^4 \chi t},$$

которое имеет место в случае преобладания в барионном веществе потенциальной энергии над кинетической. Тогда

$$\Omega(t) = \frac{1}{\ln \chi t} - \frac{4\pi G\tilde{\rho}_0}{\ln^4 \chi t}t^2. \tag{24}$$

Анализ этой функции показывает, что на выбранном временном промежутке она также монотонно убывает, как и в первом случае. Поэтому ее максимальное значение тоже будет в нижней границе интервала. Вычисления показывают, что для начальной плотности $\tilde{\rho}_0 \sim 10^{73}\,г/см^3$ (что примерно соответствует моменту времени $t \sim 10^{-36}\,c$), второе слагаемое намного меньше первого и поэтому функция $\Omega(t_1) \approx 10^{-1} \div 10^{-2} = const_2$.

Следовательно, как и в предыдущем случае, возмущение плотности барионного вещества изменяется во времени не быстрее чем

$$\delta(t) = \frac{u_0}{\ln \chi t}\left(C_1 t + C_2\right).$$

Наконец рассмотрим случай, когда имеет место соотношение

$$\frac{9\omega_1^{\,2}}{4t_0^{\,2}\ln^2 \chi t} \approx \frac{4\pi G\rho_0}{\ln^4 \chi t}.$$

С физической точки зрения оно описывает устойчивое состояние возмущения в барионном веществе. Тогда уравнение (13) примет вид

$$\ddot{\delta} + 2H\dot{\delta} = \ddot{\delta} + \frac{2}{t\ln \chi t}\dot{\delta} = 0. \qquad (25)$$

Его приближенное решение также можно представить в виде

$$\delta(t) \sim \frac{t}{\ln \chi t}.$$

Далее, требование о том, чтобы волновой вектор был пропорционален масштабному фактору, т.е. $k \propto a$, дает возможность оценить рост возмущений на масштабах горизонта. Но для того, чтобы сделать заключение о роли первичных возмущений в барионной материи для формирования объектов галактического типа, необходимо использовать условие их устойчивости.

Другими словами, в уравнении (20) следует положить

$$v_s^2 k^2 - 4\pi G\rho_b = 0. \qquad (26)$$

Отсюда с учетом решения уравнения (21), находим длину волны возмущений

$$\lambda = \frac{2\pi}{k} = \lambda_J \ln^2 \chi t, \qquad (27)$$

где $\lambda_J$ - длина волны Джинса. Поэтому масса возмущения, оцененная стандартным образом, имеет вид

$$M(t) = \frac{4}{3}\pi\lambda(t)^3\delta(t) \approx \frac{4}{3}\pi\lambda_J{}^3\delta_0 \ln^7 \chi t. \qquad (28)$$

Следовательно, в начальный момент времени ($t_1 \approx 10^{-36}$ с) затравочная масса будет иметь величину $M \approx 10^7 M_J$, что на семь порядков больше массы Джинса.

Рассмотрим теперь поведение функции $\Omega(t)$ на временном интервале $t_1 = 10^2\,c < t < t_2 = 10^{18}\,c$ (эпоха нейтральной материи) и также применим теорему сравнения для нахождения максимально допустимого значения z.

Сначала найдем $\chi$ для данного интервала времени. Для этого подставим в соотношение (12) $t = t_2 = 10^{18}\,c$, и из требования $a = a_0$ получаем величину искомого параметра - $\chi \approx 10^{-15}\,c^{-1}$.

В эпоху нейтральной материи барионное вещество во Вселенной можно описывать в виде пылевидной материи. Поэтому здесь $K = 3$. В уравнении (20), следовательно,

$$\Omega(t) = \frac{1}{\ln\chi t} - t^2\left[\frac{9\omega_1{}^2}{4t_0{}^2\chi t} - \frac{4\pi G\rho_0}{\ln^3 \chi t}\right].$$

Как и для предыдущего временного интервала, рассмотрим три варианта соотношения между потенциальной и кинетической энергиями.

Пусть

$$\frac{9\omega_1^2}{4t_0^2 \ln^2 \chi t} >> \frac{4\pi G\rho_0}{\ln^3 \chi t},$$

тогда

$$\Omega(t) = \frac{1}{\ln \chi t} + \frac{9\omega_1^2 t^2}{4t_0^2 \ln^2 \chi t}.$$

Анализ поведения этой функции на временном интервале $t_1 = 10^2\,c < t < t_2 = 10^{18}\,c$ показывает, что она на нем только убывает. Поэтому ее максимальное значение находится в верхней границе выбранного интервала; при этом $\Omega(t) \approx -3.0 \cdot 10^{-2}$.

Следовательно,

$$\delta(t) = \frac{u_0}{\ln \chi t}(C_1 t + C_2).$$

Если

$$\frac{9\omega_1^2}{4t_0^2 \ln^2 \chi t} << \frac{4\pi G\rho_0}{\ln^3 \chi t},$$

то

$$\Omega(t) = \frac{1}{\ln \chi t} - \frac{4\pi G\rho_0}{\ln^3 \chi t} t^2.$$

Аналогично предыдущему случаю она тоже убывает на интервале $t_1 = 10^2\,c < t < t_2 = 10^{18}\,c$; ее максимальное значение при величине $\rho_0 \sim 10^4\,г/см^3$ оценивается следующим образом - $\Omega(t_1 = 10^2) \approx -3.0 \cdot 10^{-2}$. Поэтому искомое решение совпадает с решением (25).

В случае

$$\frac{9\omega_1^2}{4t_0^2\ln^2\chi t}\approx\frac{4\pi G\rho_0}{\ln^3\chi t}$$

решение также совпадает с решением (23), поскольку, как показывает анализ, основной вклад в величину $\Omega(t)$ дает первое слагаемое.

## Заключение

Проведенное исследование показало, что во всех рассмотренных эпохах эволюции Вселенной рост возмущений плотности барионной материи, обусловленный выбранным видом нестационарного уравнения состояния небарионной материи (4) и соответствующим ему релятивистским режимом расширения (11), одинаков. При этом плотность изменяется по закону

$$\delta(t)=\frac{u_0}{\ln\chi t}\left(C_1t+C_2\right),$$

означающем менее интенсивный ее рост в сравнении со степенным ( $\delta(t)\sim t$ ) темпом роста возмущений барионной материи в рамках обычной фридмановской космологии со стационарным уравнением состояния небарионной материи [1]. Поэтому, как и в работе [7], выполняется соотношение $\frac{\delta(t)}{a(t)}<1$.

Но затравочная масса (28) оценивается существенно большей величиной, а именно величиной порядка $10^7 M_J$. Поэтому ее дальнейшая эволюция, обусловленная уже гравитационным притяжением, будет вести к более эффективному образованию объектов галактического типа.

## ЛИТЕРАТУРА

# 4. АНТИГРАВИТАЦИОННАЯ НЕУСТОЙЧИВОСТЬ КОСМИЧЕСКОГО СУБСТРАТА В НЬЮТОНОВСКОЙ КОСМОЛОГИИ

## Введение

Новейшие достижения современной космологии - открытие вакуума [1], темной материи и темной энергии [2], ископаемых и невидимых галактик [3] - существенно меняют саму постановку задачи об образовании наблюдаемых структур Вселенной. Если традиционным подходом в проблеме происхождения, например, галактик является исследование гравитационной неустойчивости барионного вещества, то факт существенного преобладания во Вселенной небарионного («темного») вещества (95%) над барионной материей (5%) приводит к необходимости исследования антигравитационной (например, вакуумной) неустойчивости космической среды.

Более конкретно это означает - не может ли сам вакуум выступать в качестве того фактора, который приводит к формированию крупномасштабных структур во Вселенной? Однако до сих пор вопрос о роли вакуума в формировании наблюдаемых крупномасштабных структур сводился, например, к исследованию эволюции флуктуаций скалярного поля, основное состояние которого обеспечивало вакуум [4]; к изучению остановки роста возмущений в стандартной космологической модели [5] под воздействием вакуума и т.д.

В отличие от таких работ, в статье рассматривается вакуум как основная причина развития возмущений в космологическом субстрате. Для определенности постановки задачи мы пренебрегаем гравитационным самодействием среды, а рассматриваем лишь влияние вакуума на нее ее динамику. Причем исследование антигравитационной неустойчивости космической среды проведем здесь в рамках ньютоновской космологии.

## 4.1 Уравнения гидродинамики на фоне вакуума

Запишем уравнения гидродинамики барионной материи (космической среды) на фоне вакуума в ньютоновом приближении. Для этого учтем, что на произвольно выделенный в среде элементарный объем со стороны вакуума действует только дополнительная сила $F_V$. Итак, согласно [6] имеем

$$\frac{\partial \rho_m}{\partial t} + \mathrm{div}(\rho_m \vec{v}) = 0, \qquad (1)$$

$$\frac{\partial \vec{v}}{\partial t} + (\vec{v}\,\mathrm{grad})\vec{v} + \frac{1}{\rho_m}(\mathrm{grad}P_m + F_V) = 0. \qquad (2)$$

Что касается $P_m$, то это обычное давление, оказываемое самой барионной материей на выделенный объем среды. Соотношение между этими видами давлений может быть различным; оно определяется характерными масштабами распределения барионной материи. В работе [7], например, показано, что на расстояниях порядка одного мегапарсека доминирует барионная материя, так что $P_m > P_v$; на расстояниях от одного до десяти мегапарсек они приблизительно равны друг другу; на расстояниях же более десяти мегапарсек преобладает вакуум, т.е. $P_v > P_m$. В дальнейшем нас будет интересовать случай, когда вакуум является основным возмущающим фактором структурной эволюции Вселенной

Для исследования уравнений (1) - (2) заметим, что сам вакуум описывается уравнением состояния $p_v + \rho_v = 0$. Поэтому он создает антигравитацию, а его эффективная грави-тирующая энергия $\rho_G$ является отрицательной [7], т.е.

$$\rho_G = \rho_v + 3p_v = -2\rho_v . \tag{3}$$

Выделим теперь в пространственно бесконечной среде (барионной материи), двигающейся на фоне вакуума, шаровой слой из двух концентрических сфер радиусами $r_1$ и $r_0$. Тогда на любую частицу среды внутри слоя, расположенную на расстоянии $r$ $(r_0 > r > r_1)$ от его центра, будет действовать гравитационная сила

$$F_v = -\frac{4\pi G\rho_G}{r^2}\int_{r_1}^{r}\xi^2 d\xi = \frac{8\pi G\rho_v}{r^2}\int_{r_1}^{r}\xi^2 d\xi . \tag{4}$$

Она легко вычисляется из классической теории притяжения с учетом эффективной энергии вакуума (3). Полагая $r_1 = 0$ (т.е. переходя к шару радиуса $r$) и производя элементарное интегрирование в (4), получаем выражение гравитационной силы вакуума

$$F_v = -\frac{8\pi G\rho_v}{3} r .$$

Подставляя ее в (1) - (2), находим уравнения гидродинамики барионной материи на фоне вакуума

$$\frac{\partial\rho_m}{\partial t} + \mathrm{div}(\rho_m\vec{v}) = 0, \tag{5}$$

$$\frac{\partial\vec{v}}{\partial t} + (\vec{v}\,\mathrm{grad})\vec{v} + \frac{1}{\rho_m}\mathrm{grad}P_m - \frac{8\pi G\rho_v}{3\rho_m}\vec{r} = 0. \tag{6}$$

Решение уравнений (5) - (6) будем искать методом теории возмущений, считая, как и в теории Джинса, невозмущенным такое состояние барионной материи, когда она опи-сывается условиями $\rho_{m_0} = \text{const}$, $P_{m_0} = \text{const}$ и $\vec{v}_0 = 0$. Возмущенное же решение удобно представить в виде плоских волн, наложенных на основное решение.

Поэтому запишем его следующим образом

$$\rho_m(\vec{r}, t) = \rho_{m_0} + \delta\rho_m = \rho_{m_0}\left[1 + \delta(t)\cdot\sin(\vec{k}\vec{r})\right], \tag{7}$$

$$\vec{v}(\vec{r}, t) = 0 + \vec{w}(t)\cdot\cos(\vec{k}\vec{r})\,. \tag{8}$$

Что касается давления барионной материи, то в силу общего вида ее уравнения состояния $P_m = P_m(\rho_m)$, оно также может быть разложено в ряд

$$P_m = P_{m_0} + \left(\frac{\partial P_m}{\partial \rho_m}\right)_0 \delta\rho_m + \cdots = P_{m_0} + b^2\delta\rho_m + \cdots, \tag{9}$$

где $b$ - скорость звука в барионной материи.

Наконец, в виду постановки задачи примем еще следующее ограничение

$$\frac{\rho_v}{\rho_m} << 1,$$

а также положим, что все добавки имеют одинаковый порядок, т.е.

$$\delta(t) \sim \frac{w(t)}{v(r,t)} \sim \frac{\rho_v}{\rho_m}.$$

При всех этих условиях получаем такую систему уравнений для нахождения добавок первого порядка к невозмущенному состоянию барионной материи

$$\left.\begin{aligned} &\frac{d\delta(t)}{dt} - \vec{k}\vec{w}(t) = 0 \\ &\cos(\vec{k}\vec{r})\frac{d\vec{w}(t)}{dt} + \vec{k}b^2\cos(\vec{k}\vec{r})\delta(t) - \frac{8\pi G\rho_v}{3\rho_{m_0}}\vec{r} = 0 \end{aligned}\right\}. \qquad (10)$$

Система уравнений (10) эквивалентна линейному неоднородному дифференциальному уравнению

$$\frac{d^2\delta(t)}{dt^2} + A^2\delta(t) = \Phi, \qquad (11)$$

в котором коэффициенты равны следующим выражениям

$$A^2 = k^2b^2,$$

$$\Phi = \frac{8\pi G\rho_v}{3\rho_{m_0}} \cdot \frac{(\vec{k}\vec{r})}{\cos(\vec{k}\vec{r})}. \qquad (12)$$

## 4.2 Формирование первичных возмущений барионного вещества

Для интегрирования этого уравнения необходимо учесть расширение барионного вещества, порождаемое антигравитационным фоном вакуума. Другими словами, необходимо еще в явном виде знать зависимость $\vec{r} = \vec{r}(t)$. Это позволит задать правую часть уравнения (11) в виде конкретной функции от времени $t$.

Например, для произвольной космологической модели легко видеть, что, расстояние между любой парой точек барионной материи, в соответствии с законом расширения Хаббла $\dot{\vec{r}} = H\vec{r}$, эволюционирует следующим образом

$$\vec{r}(t) = \vec{r}_0 \exp Ht, \qquad (13)$$

где $\vec{r}_0$ - заданный начальный масштаб распределения барионной материи. Поэтому коэффициенты (12) принимают вид

$$A = kb,$$

$$\Phi = \Phi(t) = \frac{8\pi G\rho_v}{3\rho_{m_0}} \cdot \frac{(\vec{k}\vec{r}_0)\exp Ht}{\cos(\vec{k}\vec{r}_0 \cdot \exp Ht)}. \tag{14}$$

Физически наиболее интересным является непериодическое решение уравнения (11), поскольку именно оно дает возможность оценить эволюцию возмущений плотности барионной материи за космологически значимое время.

Общий вид этого решения записывается следующим образом

$$\delta(t) = \frac{8\pi G\rho_v}{3\rho_{m_0}} \cdot \frac{(\vec{k}\vec{r}_0)}{kb} \int_0^t \frac{\exp H\tau}{\cos(\vec{k}\vec{r}_0 \cdot \exp H\tau)} \sin kb(t-\tau)d\tau, \tag{15}$$

но его точное интегрирование представляется весьма проблематичным. Поэтому заметим, что, так как при любом значении времени $\cos(\vec{k}\vec{r}_0 \cdot \exp H\tau) \le 1$, то возмущение плотности барионной материи будет не меньше величины

$$\delta_{min}(t) = \frac{8\pi G\rho_v}{3\rho_{m_0}} \cdot \int_0^t \exp H\tau \cdot \sin kb(t-\tau)d\tau. \tag{16}$$

Интегрирование (16) дает

$$\delta_{min}(t) = \frac{8\pi G\rho_v}{3\rho_{m_0}} \cdot \frac{\exp Ht}{H^2 + k^2b^2} \cdot \left(kb - H\sin kbt - kb\cos kbt\right). \tag{17}$$

Следовательно, после каждого полного периода колебаний T возмущение плотности барионной материи увеличивается в

$$\delta_{min}(T) = \frac{8\pi G\rho_v}{3\rho_m} \cdot kb \cdot \frac{\exp HT}{H^2 + k^2b^2} \qquad (18)$$

раз. Что касается возмущения скорости барионного вещества, то за тот же период колебаний она возрастает в

$$w(T) = \frac{8\pi G\rho_v}{3\rho_m} \cdot Hb \cdot \frac{\exp HT}{H^2 + k^2b^2}$$

число раз. Численные оценки этих значений таковы. Представляет также интерес исследование уравнения (11) в конкретно заданной космологической модели. Рассматривая в рамках ньютоновской космологии эволюцию однородной и изотропной космологической модели в случае $\rho_m < \rho_c$, где $\rho_c$ - критическая плотность Вселенной, можно показать [6], что искомая зависимость имеет линейный вид

$$r(t) = r_0\sqrt{\frac{8\pi G}{3}(\rho - \rho_c)} \cdot t \qquad (19)$$

Здесь $r_0$ - упоминавшееся выше заданное значение размеров шарового распределения барионной материи. Вводя выражение (19) в значения коэффициентов (12), получаем для рассматриваемой космологической модели

$$A^2 = k^2b^2, \qquad (20)$$

$$\Phi(t) = \left(\frac{8\pi G}{3}\right)^{3/2} \frac{\vec{k}\vec{r}_0}{\cos\left(\vec{k}\vec{r}_0\sqrt{\frac{8\pi G}{3}}(\rho_c - \rho_m)\cdot t\right)} \left[\frac{\rho_v{}^2}{\rho_m}\left(\frac{\rho_c}{\rho_m} - 1\right)\right]^{1/2} \cdot t$$

Решение уравнения (11) с учетом только непериодических слагаемых имеет вид

$$\delta(t) = \left(\frac{8\pi G}{3}\right)^{3/2} \cdot \left[\frac{\rho_v^{\,2}}{\rho_m}\left(\frac{\rho_c}{\rho_m} - 1\right)\right]^{1/2} \cdot$$
$$\frac{\vec{k}\vec{r}_0}{kb} \int_0^t \frac{\sin kb(t-\tau)}{\cos\left(\vec{k}\vec{r}_0\sqrt{\frac{8\pi G}{3}(\rho_c - \rho_m)} \cdot t\right)} \cdot dt \qquad (21)$$

Вычисление этого интеграла приводит к следующему результату

$$\delta(t) = \left(\frac{8\pi G}{3}\right)^{3/2} \cdot \left[\frac{\rho_v^{\,2}}{\rho_m}\left(\frac{\rho_c}{\rho_m} - 1\right)\right]^{1/2} \cdot \qquad (22)$$
$$\frac{\vec{k}\vec{r}_0}{kb^2} \cdot \sin kbt \cdot (-1)^{kb/2\aleph} \cdot \ln \tan\left(\frac{\pi}{4} + \frac{t}{2}\right)$$

где

$$\aleph = \vec{k}\vec{r}\sqrt{\frac{8\pi G}{3}(\rho_c - \rho_m)} \; .$$

## Заключение

Из (22) видно, что возмущение плотности барионного вещества, вообще говоря, неограниченно возрастает со временем. Поэтому антигравитационная неустойчивость барионной космической материи приводит к эффекту нарастания амплитуды ее (космической материи) колебаний и, в конечном итоге, к процессу формирования галактик и их скоплений.

# ЛИТЕРАТУРА

# 5. ДИНАМИКА СТОХАСТИЧЕСКИХ КОСМИЧЕСКИХ СТРУН В РАННЕЙ ВСЕЛЕННОЙ

## Введение

Проблема рождения галактик связана с проблемой возникновения начальных неоднородностей в космологическом субстрате. Обычно для решения этого вопроса привлекается концепция космических струн, которые как считается появились на самых ранних этапах рождения Вселенной. В силу их огромной линейной плотности масс $(10^{22}\text{г}/\text{см})$ они действительно могли притягивать к себе окружающее вещество и являться зародышами галактик. Но для этого космические струны должны обладать регулярными свойствами.

В соответствии с моделью Смита - Виленкина [1] космические струны приобретают регулярный характер на временах $t_1 \approx 100\text{сек}$ и, следовательно, с этой эпохи начинается формирование устойчивых неоднородностей в среде. Цель настоящей работы заключается в том, чтобы показать, что устойчивые неоднородности в среде возникают раньше - на временах $t_2 \approx 30\text{сек}$, при которых космические струны обладают как стохастическим так и регулярными свойствами.

## 5.1 Тензор энергии - импульса нитевидной материи со стохастическими возмущениями

Рассмотрим нитевидную материю, состоящую из бесконечного числа невзаимодействующих нитей. При этом каждая нить при своем движении заметает гиперповерхность, которую можно параметризовать двумя переменными: вре-

мени - подобной $\tau$ и пространственно - подобной $\rho$.

Далее будем рассматривать случай, когда лишь параметр $\rho$ подвержен стохастическим возмущениям. Это означает, что от параметра $\rho$ мы должны перейти к параметру $\tilde{\rho}$ такому, что

$$\tilde{\rho} = \rho + \rho'(x^0) = \rho\left(1 + \rho'(x^0)\big/\rho\right) = \rho\left(1 + z(x^0)\right) \quad (1)$$

где $z(x^0)$ - безразмерная стохастическая функция времени. В соответствии с (1), следовательно, получаем с линейной точностью модифицированный пространственно - временной вектор

$$\tilde{l}^{\alpha} = dx^{\alpha}\big/d\tilde{\rho} \approx l^{\alpha}\left(1 - z(x^0)\right), \quad (2)$$

где $z(x^0) << 1$, и обычный времени - подобный вектор

$$u^{\alpha} = dx^{\alpha}\big/d\tau. \quad (3)$$

Для построения динамики нитевидной материи будем исходить из принципа действия

$$S = \int (\sqrt{-g\mu})dV_4, \quad (4)$$

где $g$ - детерминант метрического тензора, а $\mu$ линейная плотность масс. Вычисляя вариацию действия, получаем

$$\delta S = -\int \mu\left(u^{\alpha}u_{\beta} - \tilde{l}^{\alpha}\tilde{l}_{\beta}\right)\nabla_{\alpha}\delta x^{\beta}dV_4 = 0. \quad (5)$$

Отсюда легко получить закон сохранения

$$\nabla_\chi \tilde{T}^{\alpha\beta} = 0, \tag{6}$$

где

$$\tilde{T}^{\alpha\beta} = \mu(u^\alpha u^\beta - \tilde{l}^\alpha \tilde{l}^\beta) \approx \mu(u^\alpha u^\beta - l^\alpha l^\beta) - 2\mu l^\alpha l^\beta z(x^0) \tag{7}$$

тензор энергии - импульс нитевидной материи со стохастическими добавками.

## 5.2 Гравитационное поле космической струны, подверженной стохастическим возмущениям

Теперь из выражения (7) нетрудно найти тензор энергии - импульса уединенной космической струны. Для этого необходимо умножить (7) на $\delta$ - функцию Дирака. Итак, имеем

$$\tilde{T}^{\alpha\beta} = \tilde{T}^{\alpha\beta}\delta_3(\vec{x} - \vec{x}') \tag{8}$$

Линеаризированные уравнения Эйнштейна

$$\partial_4 \tilde{h}_{\alpha\beta} = -16\pi(\tilde{T}^{\alpha\beta} - {}^1\!/_2\, \tilde{g}^{\alpha\beta}\tilde{T}) \tag{9}$$

имеют, следовательно, такие стохастические решения

$(\tilde{h}_{\alpha\beta} = h_{\alpha\beta} + \frown{h}_{\alpha\beta})$,

где в покомпонентной записи

$$\frown{h}_{00} = -4\gamma\mu \int \frac{z(x^0)}{|\vec{x} - \vec{x}'|}\delta_3(\vec{x} - \vec{x}')dV = 8\gamma\mu \cdot z(x^0) \cdot \ln\frac{r}{r_0} \quad , \tag{10'}$$

$$\widehat{h}_{ab} = -8\gamma\mu \int \frac{z(x^0)}{|\vec{x}-\vec{x}'|}\delta_3(\vec{x}-\vec{x}')dV = -16\gamma\mu \cdot z(x^0)\cdot \ln\frac{r}{r_0} \quad , \qquad (10'')$$

$$\widehat{h}_{33} = -12\gamma\mu \int \frac{z(x^0)}{|\vec{x}-\vec{x}'|}\delta_3(\vec{x}-\vec{x}')dV = -24\gamma\mu \cdot z(x^0)\cdot \ln\frac{r}{r_0} \qquad (10''')$$

$(a, b = 1,2)$ .

Поэтому полный интервал гравитационного поля, порожденного космической струной со стохастически возмущенной линейной плотностью масс имеет вид

$$ds^2 = \left(1 + 8\gamma\mu \cdot z(x^0)\cdot \ln\frac{r}{r_0}\right)(dx^0)^2 +$$

$$\left(1 - 8\gamma\mu \cdot z(x^0)\cdot \ln\frac{r}{r_0} + 16\gamma\mu \cdot z(x^0)\cdot \ln\frac{r}{r_0}\right)\left((dx^1)^2 + (dx^2)^2\right) -$$

$$\left(1 + 24\gamma\mu \cdot z(x^0)\cdot \ln\frac{r}{r_0}\right)(dx^3)^2 . \qquad (11)$$

## 5.3 Уравнения движения пробной нити в гравитационном поле космической струны, подверженной стохастическим возмущениям

Выпишем уравнения движения космической струны в произвольном гравитационном поле

$$\frac{D^2x^\alpha}{d\tau^2} - \frac{D^2x^\alpha}{d\rho^2} = 0 \qquad (12)$$

и для простоты положим, что

$$x^0 = x^0(\tau) \quad , \tag{13 a}$$

$$x^k = x^k(\tau, \rho) \quad . \tag{13 b}$$

Решение уравнений (12) будем искать методом пост - ньютоновского приближения путем разложения в ряд по степеням $\frac{v}{c}$, где $v$ - скорость движения струны как целого, $c$ - скорость света. В соответствии с этим методом векторы $u^\alpha$ и $l^\alpha$ могут быть разложены следующим образом

$$u^0 = 1 + \underset{2}{u^0} + \underset{4}{u^0} + \cdots , \tag{14 a}$$

$$u^k = \underset{1}{u^k} + \underset{3}{u^k} + \cdots \tag{14 b}$$

и

$$l^0 = \underset{1}{l^0} + \underset{3}{l^0} + \cdots , \tag{15 a}$$

$$l^k = \underset{0}{l^k} + \underset{2}{l^k} + \cdots . \tag{15 b}$$

Следовательно, в наинизшем - нулевом – приближении для нулевой компоненты из (12) находим уравнение

$$\frac{d^2 x^0}{d\tau^2} = 0 \tag{16}$$

с простейшим решением

$$x^0 = \tau \ . \tag{17}$$

Для пространственных компонент уравнение (12) имеет второй порядок малости, так что с учетом (17) имеем

$$\frac{d^2x^k}{dx^{0^2}}+4\frac{\gamma\mu}{r}\eta^k-4\frac{\gamma\mu}{r}\cdot z(x^o)\cdot\eta^k=0\,. \qquad (18)$$

Ниже будет дано возможное космологическое приложение этого уравнения движения. Поэтому ограничимся простейшим типом движения нити - круговым движением. В этом случае $r=R_0=\mathrm{const}$ и уравнение (18) распадется на два уравнения гармонических колебаний, подверженных стохастическим возмущениям

$$\frac{d^2x^k}{dt^2}+4\frac{\gamma m}{R_0^{\ 2}}x^k=4\frac{\gamma m}{R_0^{\ 2}}\cdot z(t)\cdot x^k \qquad (19)$$

где $k=1{,}2$.

## 5.4 Решение усредненных стохастических уравнений движения

Введя два обозначения

$$\omega^2=4\frac{\gamma m}{R_0^{\ 2}},\ f^k(x)=3\omega^2x^k\ , \qquad (20)$$

можно записать уравнение (19) в стандартном математическом виде

$$\frac{d^2x^k}{dt^2}+\omega^2x^k=3f^k(x)\cdot z(t)\quad . \qquad (21)$$

В силу своей линейности это уравнение распадается на пару одинаковых по виду стохастических уравнений

$$\frac{d^2x}{dt^2}+\omega^2x=3f(x)\cdot z(t), \qquad (22)$$

описывающих одномерные колебания частицы. В дальнейшем предположим, что стохастическая сила описывает процесс с независимыми приращениями всех аргументов. Они называются «белым шумом». Тогда уравнение (21) может быть переписано как

$$\frac{d^2x}{dt^2}+\omega^2x=f(x)\frac{d\zeta(t)}{dt} \quad , \qquad (23)$$

где $\zeta(t)-$ «белый шум».

Решение этого уравнения эквивалентно решению стохастических дифференциальных уравнений Ито

$$dx(t)=y(t)dt, \qquad (24')$$

$$dy(t)=\omega^2dt+f(x)d\zeta \quad . \qquad (24'')$$

Это решение представляет собой марковский процесс $\{x(t),y(t)\}$ в фазовом пространстве динамической системы. Плотность распределения этого процесса $W(x_0,y_0,t_0;x,y,t)$ удовлетворяет уравнению Колмогорова - Фоккера - Планка

$$\frac{\partial W}{\partial t}+\frac{\partial(yW)}{\partial x}-\frac{\partial(\omega^2xW)}{\partial y}=\frac{1}{2}\frac{\partial^2(f^2(x)W)}{\partial y^2} \qquad (25)$$

с соответствующим ее начальным значением

$$W(x_0,y_0,t_0;x,y,t_0)=\delta(x-x_0)\delta(y-y_0)$$

и условием нормировки

$$\int_{-\infty}^{\infty}\int_{-\infty}^{\infty} W(x_0, y_0, t_0; x, y, t)dxdy = 1 \quad . \tag{26}$$

Переходя к новым переменным

$$x = A\cos\varphi, \quad y = -\omega A\sin\varphi, \quad \varphi = \omega t + \theta$$

и используя принцип усреднения, уравнение Колмогорова - Фоккера - Планка принимает вид

$$\frac{\partial W_0}{\partial t} + \frac{\partial}{\partial A}\left(A(A)W_0\right) + \frac{\partial}{\partial \theta}\left(B(A)W_0\right) = \\ \frac{1}{2}\left\{\frac{\partial^2}{\partial A^2}\left(\Sigma(A)W_0\right) + 2\frac{\partial^2}{\partial A\partial\theta}\left(H(A)W_0\right) + \frac{\partial^2}{\partial\theta^2}\left(E(A)W_0\right)\right\}, \tag{27}$$

где $A, B, \Sigma, H, E$ - полностью вычисляемые коэффициенты (диффузии, переноса и смешанный).

При анализе колебательной системы со стохастическими возмущениями важную роль играет стационарная плотность распределения. Поэтому в дальнейшем мы будем рассматривать именно такую плотность. С учетом этого допущения и после вычисления необходимых величин, получаем окончательную форму уравнения Колмогорова - Фоккера - Планка

$$\frac{\partial}{\partial A}\left(\overline{A}(A)W_0\right) = \frac{1}{2}\left\{\frac{\partial^2}{\partial A^2}\left(\overline{\Sigma}(A)W_0\right) + \frac{\partial^2}{\partial\theta^2}\left(\overline{E}(A)W_0\right)\right\} \quad . \tag{28}$$

Из (28) следует, что усредненную стационарную плотность распределения можно представить в виде суммы, члены которой зависят от $A$ и $\theta$ по отдельности. Итак, мы ищем решение (28) в виде

$$W_0 = W'(A_0, A) + W''(\theta_0, \theta) \quad . \tag{29}$$

Такое представление позволяет написать два выражения

$$W'(A_0, A) = \sigma' \cdot (A - A_0) \quad , \qquad (30')$$

$$W''(\theta, \theta_0) = \sigma'' \cdot (\theta - \theta_0) \quad , \qquad (30'')$$

в которых постоянные величины $\sigma'$ и $\sigma''$ могут быть определены из условий нормировки.

Итак, в качестве результата мы получили, что фазовые плотности распределений $W'$ и $W''$ линейно зависят от амплитуды $A$ и фазы $\theta$.

## 5.5 Приближенное решение стохастического уравнения движения

Введя обозначение $\nu^2 = 4\frac{\gamma}{R_0^{\;2}}$, можно записать уравнение (22) в следующей форме

$$\frac{d^2x}{dt^2} + \omega^2 x = 3\nu^2 \cdot m(t) \cdot x \, , \qquad (31)$$

где $m(t) = m \cdot z(t)$ стохастическая функция времени. Примем, что $m(t)$ может быть представлена набором периодических стохастических функций времени

$$m(t) = \sum_n ({}^A m_n \cos\Omega_n t + {}^B m_n \sin\Omega_n t) \quad . \qquad (32)$$

Более того, положим, что ${}^A m_n$ и ${}^B m_n$ являются стохастическими функциями с одинаковыми дисперсией и нормальным гауссовым распределением

$$f\left({}^{\Xi}m_n\right)=\frac{1}{\sqrt{2\pi}\sigma_m}\exp\left\{-\frac{({}^{\Xi}m_n-{}^{\Xi}m_0)^2}{2\sigma_m{}^2}\right\}, \qquad (33)$$

где $\Xi = A, B.$

Согласно принятому предположению (2), будем искать решение (31) - (32) в виде $x = x_0 + x'$, где $x_0$ - невозмущенная координата, а $x'$ - малая добавка к ней $(x' << x_0)$.

Поэтому из (31) получаем частное невозмущенное решение

$$x_0(t) = R_0 \cos\omega t \quad . \qquad (34)$$

Подставляя (34) в (31), и рассматривая случай резонанса $(\omega = \Omega)$ и одномодового приближения $(\omega = \Omega)$, получаем уравнение для определения возмущения

$$\frac{d^2x'}{dt^2} + \omega^2 x' = \frac{1}{2}\nu^2 m R_0 [1 + 2\cos\omega t] \quad . \qquad (35)$$

Его физически интересное решение таково

$$\overset{*}{x}'(t) = \frac{1}{2}\nu^2 m \frac{R_0}{\omega^2}\left(1 - \frac{1}{3}\cos 2\omega t\right) \quad . \qquad (36)$$

Используя (36) можно вычислить среднюю величину добавки

$$< \overset{*}{x}'(m) >= \int_{-\infty}^{\infty} \overset{*}{x}'(m) f(m) dm = \frac{\nu^2 R_0}{2\omega^2} m_0 (1 - \frac{1}{3}\cos 2\omega t) \qquad (37)$$

и среднее по периоду T ее стохастическое возмущение

$$\overline{<\overset{*}{x}'(m)>} = \frac{\nu^2 R_0}{2\omega^2 T} m_0 \int_0^T (1 - \frac{1}{3}\cos 2\omega t)dt = \frac{2}{3}\frac{m_0}{m} R_0 = \delta R_m . \quad (38)$$

## 5.6 Возмущения плотности субстрата, порожденные хаотическим движением космической нити

Типичное расстояние между космическими струнами в рамках модели Смита - Виленкина оценивается как $\xi \propto k^{-1/2} t$, где $k$ коэффициент порядка 20. Пусть $R_0 \propto \xi$, тогда можно ввести величину $\eta_m = \delta R_m / R_0$, которая описывает меру хаотизации системы космических струн, обусловленную стохастичностью распределения «амплитуды» масс.

Ранее было подчеркнуто, что на временных масштабах $t_1 \approx 30$сек космические струны имели явный броуновский характер, а на временах $t_2 \approx 100$сек - они почти выпрямились. Поэтому мерой хаотизации в модели Смита - Виленкина является величина $\eta = \xi_1 / \xi_2 = t_1 / t_2 \propto 0.3$.

Пусть $\eta_m \le \eta$. Это означает, что первоначально безразмерная космическая нить в силу стохастических возмущений приобретает эффективный поперечный размер

$$\delta R \cong \eta R_0 . \quad (39)$$

## Заключение

Итак, благодаря эффективному поперечному размеру нить может породить во внешнем субстрате возмущения с длиной волны $(\lambda_{pert} \propto \delta R)$. Легко посчитать, что $\delta R \propto 10^{12}\,\text{см}$. Сопоставляя эту длину волны с диной волны Джинса ($\lambda_J \propto 10^{11}\,\text{см}$), имеем $\lambda_{pert}/\lambda_J \propto 10$. Это означает, что хаотические флуктуации космической нити могут порождать во внешнем субстрате длинные волны возмущения. А они, в свою очередь, будут приводить субстрат в неустойчивое состояние, т.е. расщеплять его на отдельные фрагменты, из которых позднее будут формироваться галактики [2].

## ЛИТЕРАТУРА

# 6. ГРАВИТАЦИОННОЕ ВЗАИМОДЕЙСТВИЕ ДВУХ ОСЦИЛЛИРУЮЩИХ КОСМИЧЕСКИХ СТРУН

## Введение

После пионерской статьи Виленкина [1], в которой была выведена метрика массивной прямолинейной космической струны, появилось много работ, обобщающих этот результат. Так, в [2,3] предложена метрика массивной вращающейся космической струны; в [4,5] - невращющейся струны конечных размеров; в [6] - вращюющейся струны конечных размеров.

Кроме того, были выведены метрики прямолинейной [7, 8] и круговой заряженной струны; сверхмассивной [10] и сверхпроводящей [11] космических струн; космической струны с кинками [12], полых космических струн [13] и т.д.

Вместе с тем, при выводе этих метрик не принималось во внимание одна из важнейших особенностей поведения космических струн. Речь идет о том, что эволюция космической струны как протяженного объекта должна сопровождаться ее колебаниями. Особенно если речь идет о динамике космической струны с массами на концах [14]. Поэтому на ранних стадиях эволюции Вселенной реальная космическая струна это осциллирующая струна.

Осцилляции могут быть причиной множественных процессов перекоммутации, пересечений и разрывов струн. Следовательно, осцилляции должны приводить к появлению соответствующих добавок в компонентах метрического тензора, которые должны содержать периодические слагаемые.

Частный случай такого типа метрик был выведен Вачаспати [15,16]. Она описывает пространственно -

временной интервал вблизи космической струны, вдоль которой со скоростью света распространяется слабая волна возмущения - космическая струна с бегущей вдоль нее волной. Позднее простой физический процесс - отклонение лучей света - в такой метрике был рассмотрен в работе [17].

Основная цель настоящего раздела состоит в выводе метрики слабо осциллирующей, но не излучающей космической струны другого типа - метрики космической струны с наложенными на нее стоячими колебаниями. Кроме того, в ней кратко обсуждаются некоторые физические процессы вблизи такой струны.

Среди них динамика открытой пробной космической нити, совершающей в таком поле вынужденные колебания. Природа таких колебаний существенно отличается от ранее изучавшихся механизмов. В самом деле, колебания космических струн рассматривались в контекстах динамического трения при их движении в среде [18], их пересечений [19, 20], коллапса [21] и других причин.

Наиболее детальное изучение колебаний космических струн было сделано при анализе их гравитационного излучения [22 - 24]. Поэтому решения уравнений движения космических струн выбирались в виде бегущих волн различной структуры [25, 26]. Это позволило найти периодические вклады в компоненты метрического тензора и изучить спектральное распределение гравитационной энергии.

В настоящей главе колебания космической нити возникают в результате воздействия внешней периодической силы, что ведет не к затуханию колебаний, а наоборот - к их усилению. Соответственно, и гравитационное излучение космической нити будет линейно возрастать со временем.

## 6.1 Пространство - время вблизи космической струны, осциллирующей в виде стоячих волн

Зададим тензор энергии - импульса нитевидной материи [27]

$$T^{\alpha\beta} = \sum_{a} \overset{a}{\mu}\left( \overset{a}{u}^{\alpha}\, \overset{a}{u}^{\beta} - \overset{a}{l}^{\alpha}\, \overset{a}{l}^{\beta} \right)\delta_3\left( \vec{x} - \vec{\overset{a}{\xi}} \right) \qquad (1)$$

и уравнение движения a - й нити

$$\frac{D\overset{\alpha}{u}^{\alpha}}{d\overset{\alpha}{\tau}} - \frac{D\overset{\alpha}{l}^{\alpha}}{d\overset{\alpha}{\rho}} = 0, \qquad (2)$$

которое следует из (1) в силу закона сохранения энергии. В уравнении (2)

$$\overset{a}{u}^{\alpha} = \frac{d\overset{a}{x}^{\alpha}}{d\overset{a}{\tau}}$$

и

$$\overset{\alpha}{l}^{\alpha} = \frac{d\overset{a}{x}^{\alpha}}{d\overset{a}{\rho}}$$

времени - подобный и пространственно - подобный векторы, характеризующие движение и ориентацию космической струны как целой.

Кроме того, наложим обычное условие ортонормировки на эти векторы

$$\left(\overset{\alpha}{u}^{\alpha} \pm \overset{\alpha}{l}^{\alpha}\right)^2 = 0\,.$$

Для дальнейшего исследования заметим, что в [1] все вычисления проводились в рамках линеризированной теории гравитации Эйнштейна, т.е. в предположении, что $g_{\mu\nu} = \delta_{\mu\nu} + h_{\mu\nu}$, где $h_{\mu\nu}$ малые добвки к псевдоэвклидовому фону. Оставаясь в рамках такого приближения, урвнения общей теории относительности можно записать в виде

$$h_{\mu\nu} = -16\pi\gamma\left(T_{\mu\nu} - \frac{1}{2}\delta_{\mu\nu}T\right). \tag{3}$$

Подставляя сюда тензор энергии - импульса (2) получаем в компонентах (для N струн, неподвижных как целое)

$$h_{00} = 0, \tag{4}$$

$$h_{kl} = 16\pi\gamma\sum_a \overset{a}{\mu}\left(\delta_{kl} + \overset{a}{l}_k \overset{a}{l}_l\right)\delta_3\left(\vec{x} - \overset{a}{\vec{\xi}}\right). \tag{5}$$

Из (4) имеем частное решение $h_{00} = 0$. Что касается уравнения (5), то его решение имеет форму запаздывающего потенциала

$$h_{kl} = -4\gamma\sum_a \int \overset{a}{\mu}\frac{\delta_{kl} + \overset{a}{l'}_k \overset{a}{l'}_l}{\left|\vec{x} - \vec{x}'\right|}\delta_3\left(\vec{x} - \overset{a}{\vec{\xi}}\right)dV', \tag{6}$$

где $\overset{a}{l'}_k = \overset{a}{l'}_k\left(\vec{x}', x'^0\right)$.

В соответствии с постановкой задачи будем искать метрический тензор вблизи уединенной космической струны. Разлагая выражение (6) в ряд по степеням $\lambda = \left|\vec{x} - \vec{x}'\right| / x^0$ и сохраняя лишь главное слагаемое, для случая $a = 1$ имеем

$$h_{kl} = -4\gamma \int \mu \frac{\delta_{kl} + l'_k l'_l}{|\vec{x} - \vec{x}'|} \delta_3 (\vec{x} - \vec{x}') dV', \tag{7}$$

где теперь $l'_k = l'_k (\vec{x}', x^0)$. Для нахождения явного вида (7), вычислим вектор $l_k$.

В нашей предыдущей статье [28] было показно, что

$$l^k = l_0^k + \sum_{n=1}^{\infty} \frac{\pi n}{L} \left( A_n^k \cos \frac{\pi n}{L} \tau + B_n^k \sin \frac{\pi n}{L} \tau \right) \cos \frac{\pi n}{L} \rho, \tag{8}$$

где $A_n^k$ и $B_n^k$ набор амплитуд, L дина струны. Поэтому потенциал (7) может быть разложен на две части $\left(h_{kl} = {}^{(1)}h_{kl} + {}^{(2)}h_{kl}\right)$, которые имеют вид

$$^{(1)}h_{kl} = -4\gamma \int \mu \frac{\delta_{kl} + l_0^k l_0^l}{|\vec{x} - \vec{x}'|} \delta_3 (\vec{x} - \vec{x}') dV' \tag{9}$$

и

$$\begin{aligned} ^{(2)}h_{kl} = -4\gamma\delta_{kl} \int \mu \sum_{n=1}^{\infty} \frac{\pi^2 n^2}{L^2} \Bigg( A_n^m \cos \frac{\pi n}{L} \tau + \\ + B_n^m \sin \frac{\pi n}{L} \tau \Bigg)^2 \frac{\cos^2 \frac{\pi\ n}{L} \rho'}{|\vec{x} - \vec{x}'|} \delta_3 (\vec{x} - \vec{x}') dV' \end{aligned}, \tag{10}$$

соответственно.

Для вычисления этих интегралов положим, что струна расположена вдоль оси z, а возмущающие волны остаются в плоскости {y, z}. Если $\mu = \mu\delta(z - z')$ линейная плотность массы струны, то из (9) получаем стандартное выражение

$$^{(1)}h_{kl} = -8\ln\frac{\left(x^2 + y^2\right)^{1/2}}{r_0}\delta_{kl}, \tag{11}$$

где $r_0$ - постоянная величина, имеющая смысл размера поперечного сечения струны.

Что касается выражения (10), то в силу (8), условия ρ=z вдоль космической струны и выбора плоскости колебаний, оно может быть представлено как

$$^{(2)}h_{22} = -8\gamma\tilde{M}\ln\frac{y}{r_0}, \tag{12}$$

где зависящая от времени линейная плотность масс такова

$$\tilde{M} = \mu\sum_{n=1}^{\infty}\frac{\pi^2 n^2}{L^2}\left(\overset{*}{A}_n\cos\frac{\pi n}{L}wx^0 + \overset{*}{B}_n\sin\frac{\pi n}{L}wx^0\right)^2 \cdot$$
$$\cos^2\frac{\pi n}{L}z'\delta(z - z'). \tag{13}$$

(Для простоты записи мы здесь ввели обозначения $\overset{*}{A}_n = A_n^2$ и $\overset{*}{B}_n = B_n^2$).

Следовательно, пространственно - временной интервал космической струны с наложенным на него семейством стоячих волн, есть

$$dS^2 = dx^{02} - \left[1 + 8\gamma\tilde{\mu}\ln\frac{\left(x^2 + y^2\right)^{1/2}}{r_0}\right]dx^2 -$$
$$-\left[1 + 8\gamma\tilde{\mu}\ln\frac{\left(x^2 + y^2\right)^{1/2}}{r_0} + 8\gamma\tilde{M}\ln\frac{y}{r_0}\right]dy^2 - dz^2. \tag{14}$$

В силу того, что полученная выше метрика существенно анизотропна, то заранее ясно, что и все физические процессы в ней также будут анизотропными. Например, угол отклонения луча света вдоль осей x и y будет различным.

Периодический характер этой метрики будет существенно влиять и на колебательные процессы, происходящие вблизи осциллирующей космической струны. Действительно, пробная нить в пространстве - времени (14) будет совершать вынужденные колебания. Найдем уравнение этих колебаний.

## 6.2 Открытая пробная нить в гравитационном поле массивной осциллирующей струны

Запишем уравнение движения пробной нити в заданном пространстве - времени в компонентах

$$\left.\begin{array}{l} \dfrac{d^2x^0}{d\tau^2}-\dfrac{d^2x^0}{d\rho^2}+\Gamma^0_{00}\left(u^0u^0-l^0l^0\right)+ \\ \quad +2\Gamma^0_{0m}\left(u^0u^m-l^0l^m\right)+ \\ \quad \Gamma^0_{mn}\left(u^mu^n-l^ml^n\right)=0 \end{array}\right\}, \qquad (15)$$

$$\left.\begin{array}{l} \dfrac{d^2x^k}{d\tau^2}-\dfrac{d^2x^k}{d\rho^2}+\Gamma^k_{00}\left(u^0u^0-l^0l^0\right)+ \\ \quad +2\Gamma^k_{0m}\left(u^0u^m-l^0l^m\right)+ \\ \quad \Gamma^k_{mn}\left(u^mu^n-l^ml^n\right)=0 \end{array}\right\}. \qquad (16)$$

Решение уравнений (16) будем искать в виде

$$x^k = x^k_0 + \xi^k, \qquad (17)$$

где $\xi^k$ - малые добавки к основному невозмущенному смещению $x^k_0$. Подставляя (17) в (16), получаем однородное ги-

перболическое уравнение

$$\frac{d^2x_0^k}{d\tau^2} - \frac{d^2x_0^k}{d\rho^2} = 0, \tag{18}$$

которое описывает свободные колебания нити в виде стоячих волн.

Для вывода явного вида возмущенного уравнения движения необходимо подставить в (15) и (16) соответствующие символы Римана - Кристоффеля, а также векторы $u^m$ и $1^m$. В дальнейшем будем интересоваться в метрическом только теми слагаемыми, которые имеют периодический характер, т.е. слагаемыми в $g_{22}$ - компоненте. В соответствии со сказанным, требуемые символы Римана - Кристоффеля таковы

$$\begin{gathered}\Gamma_{22}^0 = \Gamma_{02}^2 = -2\gamma\mu\sum_{m=1}^{\infty}\frac{\pi^3 m^3}{\Lambda^3}\left[\left(A_m^2 - B_m^2\right)\sin\frac{\pi m}{\Lambda}\tau - \right. \\ \left. 2A_m B_m \cos 2\frac{\pi m}{\Lambda}\tau\right]\cos^2\frac{\pi m}{\Lambda}z' \cdot \ln\frac{y}{r_0},\end{gathered} \tag{19}$$

$$\Gamma_{22}^2 = 2\gamma\mu\sum_{m=1}^{\infty}\frac{\pi^2 m^2}{\Lambda^2}\left[\left(A_m \cos\frac{\pi m}{\Lambda}\tau + B_m \sin\frac{\pi m}{\Lambda}\tau\right)^2 \cos^2\frac{\pi m}{\Lambda}z'\right]\frac{1}{y}. \tag{20}$$

В силу линейности уравнений Эйнштейна (3) и решения уравнений движения струны, можно записать векторы $u^\mu$ и $1^\mu$ в виде их главных слагаемых

$$\left.\begin{aligned} u^0 &= \frac{1}{\omega}, \\ u^2 &= v_0^2 - \sum_{n=1}^{\infty}\frac{\pi n}{L}\left(a_n \sin\frac{\pi n}{L}\tau - b_n \cos\frac{\pi n}{L}\tau\right)\sin\frac{\pi n}{L}\rho; \end{aligned}\right\} \tag{21}$$

$$\left.\begin{aligned} &1^0 = 0, \\ &1^2 = \sum_{n=1}^{\infty} \frac{\pi n}{L}\left( a_n \cos\frac{\pi n}{L}\tau + b_n \sin\frac{\pi n}{L}\tau \right)\cos\frac{\pi n}{L}\rho. \end{aligned}\right\} \qquad (22)$$

Полагая для простоты $n = 1$ и обозначая $\xi^2 = y, p = z$, уравнение (15) с учетом (19), (21) и (22) принимает вид

$$\frac{d^2x^0}{d\tau^2} = 2\gamma\mu\sum_{m=1}^{\infty}\frac{\pi^3 m^3}{\Lambda^3 L}\left[\begin{aligned}&\left(A^2{}_m - B^2{}_m\right)\sin 2\frac{\pi m}{L}\tau - \\ &2A_m B_m \cos 2\frac{\pi m}{L}\tau\end{aligned}\right]\cdot$$

$$\cos^2\frac{\pi m}{L}z\cdot\left[\begin{aligned}&2v_0\left( a\sin\frac{\pi}{L}\tau - b\cos\frac{\pi}{L}\tau\right)\sin\frac{\pi}{L}z \\ &+\frac{1}{L}\left( a\cos\frac{\pi}{L}\tau + b\sin\frac{\pi}{L}\tau\right)^2\cdot\cos^2\frac{\pi}{L}\tau\end{aligned}\right]\cdot \qquad (23)$$

$$\left[\ln\frac{a}{r_0} + \ln\cos\frac{\pi}{L}\tau + \ln\sin\frac{\pi}{L}\tau\right]$$

Совершенно аналогично уравнение (16) может быть записано как

$$\frac{d^2y}{d\tau^2} - \frac{d^2y}{dz^2} = 4\gamma\mu\sum_{m=1}^{\infty}\frac{\pi^4 m^3}{\Lambda^3 L}\left[\begin{aligned}&\left(A_m^2 - B_m^2\right)\sin 2\frac{\pi m}{L}\tau - 2A_m B_m\cdot \\ &\cos 2\frac{\pi m}{L}\tau\end{aligned}\right]$$

$$\cos^2\frac{\pi m}{\Lambda}z\left[\ln\frac{a}{r_0} + \ln\cos\frac{\pi}{L}\tau + \ln\sin\frac{\pi}{L}z\right]\frac{1}{w}\left( a\sin\frac{\pi}{L}\tau - b\cos\frac{\pi}{L}\tau\right)\cdot$$

$$\sin\frac{\pi}{L}z + 2\gamma\mu\sum_{m=1}^{\infty}\frac{\pi^4 m^3}{\Lambda^2 L}\left( A_m\cos\frac{\pi m}{\Lambda}\tau + B_m\sin\frac{\pi m}{\Lambda}\tau\right)^2\cos^2\frac{\pi m}{\Lambda}z\cdot$$

$$\left[2v_0\left(a\sin\frac{\pi}{L}\tau-b\cos\frac{\pi}{L}\tau\right)\sin\frac{\pi}{L}z-\frac{\pi}{L}\left(a\sin\frac{\pi}{L}\tau-b\cos\frac{\pi}{L}\tau\right)^2\cdot\right.$$
$$\sin^2\frac{\pi}{L}\tau+$$
$$\left.+\frac{\pi}{L}\left(a\cos\frac{\pi}{L}\tau+b\sin\frac{\pi}{L}\tau\right)^2\cos^2\frac{\pi}{L}z\right]\frac{1}{\left(a\cos\frac{\pi}{L}\tau+b\sin\frac{\pi}{L}\tau\right)\sin\frac{\pi}{L}z}. \quad (24)$$

Здесь необходимо подчеркнуть, что в (23) и (24) опущены члены, которые не описывают взаимного влияния космической нити и космической струны. Кроме того, во избежание недоразумений y - я компонента скорости здесь обозначена как $v_0$. Таким образом, уравнения (23) - (24) описывают колебания пробной нити, которая движется вдоль оси $y$ к массивной космической струне со скоростью $v_0$.При этом космическая струн и пробная нить ориентированы как целое вдоль оси *z*.

Теперь оценим порядки величин, входящих в правые части (23) и (24). Из (24) видно, что одна часть слагаемых имеет порядок $\frac{A^2}{\Lambda^2}$, тогда как другая $\frac{A^2}{\Lambda^2}\frac{a}{L}$; из (25) - одна часть слагаемых имеет не только порядок $\frac{A^2}{\Lambda^2}\frac{a}{L}$, но и порядок $\frac{A^2a^2}{\Lambda^2L^2}$. Пренебрегая слагаемыми порядка $\frac{A^2}{\Lambda^2}\frac{a}{L}$ и выше, легко видеть, что уравнение (23) сводится к простейшему виду

$$\frac{d^2x^0}{d\tau^2}=0, \quad (25)$$

которое имеет известное решение

$$x^0 = \frac{\tau}{w} \ . \qquad (26)$$

Что касается уравнения (24), то, полагя m=1 и B=b=0, учитывая (26) и в силу сделанных выше упрощений, получаем

$$\frac{d^2 y}{dx^{0^2}} - w^2 \frac{d^2 y}{dx^2} = F\left(x^0; z\right) \qquad (27)$$

с возмущающей силой

$$F\left(x^0; z\right) = 4\gamma\mu \cdot v_0 \frac{\pi}{\Lambda^2 L} A^2 \cos^2 \frac{\pi w}{\Lambda} x^0 \cdot$$

$$\cos^2 \frac{\pi}{\Lambda} z \cdot \sin \frac{\pi}{L} z \frac{\sin \frac{\pi w}{L} x^0}{\cos \frac{\pi w}{L} x^0} \ . \qquad (28)$$

В дальнейшем будем рассматривать случай, когда размеры космической струны и космической нити удовлетворяют соотношению $\Lambda = 2L$.

В результате решение урвнений движения космической струны может быть записно в виде суммы двух слагаемых

$$y\left(x^0, z\right) = y_1\left(x^0, z\right) + y_2\left(x^0, z\right), \qquad (29)$$

которые равны

$$y_1\left(x^0, z\right) = \frac{1}{2} \gamma\mu \cdot v_0 \frac{\pi w^2 A^2}{L^3} \cdot I\left(x^0\right) \cdot \sin \frac{\pi}{L} z, \qquad (30)$$

$$y_2\left(x^0,z\right)=\frac{1}{4}\gamma\mu\cdot v_0\frac{\pi w^2A^2}{L^3}\cdot I\left(x^0\right)\cdot\sin 2\frac{\pi}{L}. \qquad (31)$$

При этом

$$I\left(x^0\right)=\int_0^{x^0}\cos^2\frac{\pi\omega}{2L}\chi\cdot\frac{\sin\frac{\pi\omega}{L}\chi}{\cos\frac{\pi\omega}{L}\chi}\sin\frac{\pi w}{2l}\left(x^0-\chi\right)d\chi. \qquad (32)$$

Вычисляя этот интеграл, нетрудно убедиться в том, что он представляет собой сумму двух слагемых - периодического и непериодического. В дальнейшем мы сохраним лишь непериодическое слагаемое. Поэтому

$$\left.\begin{aligned}y_1\left(x^0,z\right)&=\frac{1}{8}\gamma\mu\cdot v_0\frac{wA^2}{L^2}\cdot x^0\cdot\cos\frac{\pi w}{L}x^0\cdot\sin\frac{\pi}{L}z,\\ y_2\left(x^0,z\right)&=\frac{1}{16}\gamma\mu\cdot v_0\frac{wA^2}{L^2}\cdot x^0\cdot\cos\frac{\pi w}{L}x^0\cdot\sin 2\frac{\pi}{L}z.\end{aligned}\right\}. \qquad (33)$$

Это решение описывает периодические колебания с линейно по времени зависящей амплитудой. Следовательно, космическая нить, движущаяся в поле массивной космической струны будет испытывать вынужденные колебания с постоянно растущей амплитудой.

## 6.3 Гравитационное излучение от сильно осциллирующей космической нити

Оценим мощность гравитационного излучения от сильно колеблющейся космической нити, находящейся во внешнем поле массивной осциллирующей космической струны.

Известно, что полная мощность гравитационного излучения описывается формулой

$$-\frac{d\varepsilon}{dx^0} = 4\pi d\overline{\mathfrak{I}}\,. \tag{34}$$

Здесь $d\overline{\mathfrak{I}}$ - усредненная по всем направлениям интенсивность излучения в телесный угол $dO$

$$d\overline{\mathfrak{I}} = |S|d\sigma = |S|R_0^2 dO, \tag{35}$$

где $|S|$ - модуль вектор Пойнтинга. Пространственные компоненты потока энергии выражаются через компоненты псевдотензора энергии - импульса следующим образом

$$S^k = -it^{0k}. \tag{36}$$

Имея в виду эти замечания, можно вычислить излучение гравитционной энергии сильно осциллирующей космической нити. Обознчя мплитуды колебаний как

$$\left.\begin{aligned} \tilde{A}_1(x^0) &= \frac{1}{8}\gamma\mu \cdot v_0 \cdot \frac{wA^2}{L^2} \cdot x^0 \\ \tilde{A}_2(x^0) &= \frac{1}{16}\gamma\mu \cdot v_0 \cdot \frac{wA^2}{L^2} \cdot x^0 \end{aligned}\right\}, \tag{37}$$

Можно переписать возмущающие слагаемые координат колеблющейся нити в виде

$$\left.\begin{gathered} \xi^0 = 0, \\ y = \xi^2 = \tilde{A}_1(x^0)\cos\frac{\pi w}{L}x^0 \cdot \sin\frac{\pi}{L}z \\ + \tilde{A}_2(x^0)\cos\frac{\pi w}{L}x^0 \cdot \sin 2\frac{\pi}{L}z \\ \xi^1 = \xi^3 = 0 \end{gathered}\right\}. \tag{38}$$

Теперь легко найти соответствующие волновые добавки к векторам $u^{\mu}$ и $l^{\mu}$. Но в соответствии с общим выражением пространственных компонент запаздывающих потенциалов, необходимо знать добавки только к вектору $l^{k}$. Обозначая их $\ell^{k}$, из (5) находим соответственно

$$\left.\begin{aligned} &\ell^{0}=0, \\ &\ell^{2}=\left(\tilde{A}_{1}\left(x^{0}\right)\frac{\pi}{L}\cos\frac{\pi}{L}z+2\tilde{A}_{2}\left(x^{0}\right)\frac{\pi}{L}\cos 2\frac{\pi}{L}z\right)\cdot \\ &\cos\frac{\pi\omega}{L}x^{0} \\ &\ell^{1}=\ell^{3}=0. \end{aligned}\right\}. \tag{39}$$

Отсюда видно, что компоненты гравитационного потенциала зависят не только от $z$, но и от времени $x^{0}$. Запишем их в форме

$$\delta h_{22}=-4\gamma\overset{*}{\mu}\frac{\pi^{2}}{L^{2}}\int\left[\tilde{A}_{1}\left(x^{0\prime}\right)\cos\frac{\pi}{L}z'+2\tilde{A}_{2}\left(x^{0\prime}\right)\cos 2\frac{\pi}{L}z'\right]^{2}\cdot \frac{\cos^{2}\frac{\pi w}{L}x^{0\prime}}{\left|x-x'\right|}\delta_{3}\left(\vec{x}-\vec{x}'\right)dV', \tag{40}$$

где $\overset{*}{\mu}$ - линейная плотность масс пробной космической нити. Обозначая

$$\overset{*}{\rho}\left(x^{0\prime},z'\right)=\overset{*}{\mu}\frac{\pi^{2}}{L^{2}}\left(\tilde{A}_{1}\left(x^{0\prime}\right)\cos\frac{\pi}{L}z'+2\tilde{A}_{2}\left(x^{0\prime}\right)\cos 2\frac{\pi}{L}z'\right)^{2}\cdot \cos^{2}\frac{\pi w}{L}x^{0\prime}, \tag{41}$$

найденный потенциал сводится к стандарной форме

$$\delta h_{22} = -4\gamma \int \frac{\overset{*}{\rho}\left(x^{0'}, z'\right)}{\left|\vec{x} - \vec{x}'\right|} \delta_3\left(\vec{x} - \vec{x}'\right) dV' \cdot \tag{42}$$

Разлагая этот потенциал в ряд по параметру $\lambda$, мы определим слагаемое, описывающее гравитационное поле в волновой зоне $\left(R_0 \rangle\rangle L\right)$

$$\delta\tilde{h}_{22} = -8\gamma \tilde{D}_{ij}\left(x^0; z\right) \frac{\partial^2}{\partial x_i \partial x_j} \ln \frac{y}{\Delta} = \\ - 8\gamma\Delta \ln \frac{y}{\Delta} \cdot \frac{d^2 \tilde{D}_{ij}}{dx^{0^2}} \cdot n_i n_j, \tag{43}$$

где

$$\tilde{D}_{ij}\left(x^0, z\right) = \frac{1}{\Delta} \int \overset{*}{\rho}\left(x^0, z\right)\left(x^i x^j - \frac{1}{3}\delta_{ij} r^2\right) dV, \tag{44}$$

а $\Delta$ - число, характеризующее поперечные размеры нити. Подставляя сюда (41), получаем выражение

$$\tilde{D}_{ij}\left(x^0, z\right) = \overset{*}{\mu} \frac{\pi^2}{\Delta L^2}\left(\tilde{A}_1\left(x^0\right)\cos\frac{\pi}{L} z + 2\tilde{A}_2\left(x^0\right)\cos 2\frac{\pi}{L} z\right)^2 \cdot \\ \cos^2 \frac{\pi w}{L} x^0 \cdot D_{ij} = \frac{1}{\Delta} \overset{*}{D}_{ij}, \tag{45}$$

где

$$D_{ij} = \int \overset{*}{\mu}\left(x_i x_j - \frac{1}{3}\delta_{ij} r\right)^2 dV - \tag{46}$$

квдрупольный момент.

Для вычисления излучения гравитционной энергии будем использовть псевдотезор Ландау - Лифшица. Тогда, согласно [29], вдоль оси $x$

$$t^{01} = \frac{1}{64\pi\gamma}\left(\frac{d\delta\tilde{h}_{22}}{dx^0}\right)^2 . \qquad (47)$$

Подставляя сюда потенциал (43), найдем, что для произвольной точки $R_0$ в волновой зоне

$$t^{01} = \frac{\gamma}{\pi}\Delta^2 \cdot \frac{1}{\Delta^2}\ln^2\frac{R_0}{\Delta}\cdot\frac{d^3\tilde{D}_{ij}}{dx^{0^3}}\cdot\frac{d^3\tilde{D}_{kl}}{dx^{0^3}}\cdot n_i n_j n_k n_l . \qquad (48)$$

Поэтому интенсивность излучения гравитационной энергии в угол $dO$ равна

$$d\Im = \frac{\gamma}{\pi}\Delta^2 \cdot \frac{R_0^2}{\Delta^2}\cdot\ln^2\frac{R_0}{\Delta}\cdot\frac{d^3\tilde{D}_{ij}}{dx^{0^3}}\cdot\frac{d^3\tilde{D}_{kl}}{dx^{0^3}}\cdot n_i n_j n_k n_l \cdot dO. \qquad (49)$$

Для последующих преобрзовний зметим, что соглсно [30]

$$\frac{R_0}{\Delta}\ln\frac{R_0}{\Delta} = \sum_{k-1}^{\infty}\left(1-\frac{\Delta}{R_0}\right)^k \cdot\sum_{m=1}^{k}\frac{1}{m}. \qquad (50)$$

Но в силу того, что $\frac{\Delta}{R_0}\langle\langle 1$, с точностью до главного члена, получем

$$-\frac{dE}{dx^0} = \frac{4\gamma}{45}\cdot\frac{d^3\overset{*}{D}_{ij}}{dx^{0^3}}\cdot\frac{d^3\overset{*}{D}_{ij}}{dx^{0^3}}. \qquad (51)$$

И наконец, последне выржение может быть преобрзовно в более привычную форму

$$-\frac{dE}{dx^0} = G\gamma \overset{*}{\mu}{}^2 \quad , \qquad (52)$$

где $G$ - полностью вычисляемая величина.

## Заключение

Таким образом, найденное выражение численным коэффициентом отличается как от аналогичных формул, описывающих гравитационное излучение небесных тел [29], так и излучение массивных космических струн [22, 23]. Но по порядку оно меньше, чем эти виды гравитационного излучения в силу того, что принято условие $\overset{*}{\mu} < \mu$. Поэтому оно играет роль добавочного слагаемого в полной энергии гравитационного излучения от двух космических струн. Но эти добавки в силу (61) линейно растут во времени. Поэтому гравитационное излучение взаимодействующих космических струн может быть более интенсивным, чем это было вычислено ранее [22 - 24].

## *ЛИТЕРАТУРА*

# 7. ВЫТЯГИВАНИЕ КОСМИЧЕСКОЙ СТРУНЫ В ПРИСУТСТВИИ ВАКУУМНОЙ ДОМЕННОЙ СТЕНКИ

## Введение

Калибровочные теории со спонтанно нарушенной симметрией предсказывают существование топологически устойчивых структур трех типов - монополи, струны и доменные стенки, имеющие пространственную размерность ноль, один и два, соответственно. Из них наиболее динамично эволюционируют космические струны, которые первоначально выглядели как броуновские траектории и лишь на более поздних этапах эволюции под действием натяжения выпрямлялись и начинали двигаться как целое со скоростью, близкой к скорости света. Это приводит к их пересечениям, перехлестам, к образованию замкнутых петель, к интенсивному гравитационному излучению и другим динамическим процессам [1 - 6].

Из этого сценария видно, что важнейшую роль в эволюции космических струн играет их выпрямление, которое, подчеркнем еще раз, происходит за счет увеличения натяжения в результате расширения Вселенной.

Между тем, в реальной Вселенной взаимодействие происходит не только между струной и Метагалактикой, между самими космическими струнами [7], но и между другими топологическими дефектами вакуума - монополями [8], стенками [9] и т.д. В дальнейшем будем интересоваться лишь взаимодействием между струнами и доменными стенками, на которое было обращено внимание в [10]. Исследование их взаимодействия, как оказывается, важно не только в общетеоретическом плане, но и в космологическом аспекте. Дело в том, что это гравитационное взаимодействие, как оказывается, носит характер отталкивания и поэтому должно приводить к вытягиванию космических струн.

## 7.1 Колебания космической струны в гравитационном поле доменной стенки

Итак, рассмотрим движение космической струны в гравитационном поле массивной доменной стенки. Полагая, что она расположена в плоскости $\{y, x\}$, соответствующий четырехмерный интервал запишем в конформно - плоском виде [3]

$$ds^2 = \left(1 - 4\pi \frac{\gamma\sigma}{c^2} x^1\right)\left(dx^{0^2} - dx^{1^2} - dx^{2^2} - dx^{3^2}\right), \qquad (1)$$

где $\sigma$ - поверхностная плотность доменной стенки.

Запишем теперь общерелятивистское уравнение движения космической струны

$$\frac{du^\mu}{d\tau} - \frac{dl^\mu}{d\rho} + \Gamma^\mu_{\alpha\beta}\left(u^\alpha u^\beta - l^\alpha l^\beta\right) = 0, \qquad (2)$$

в котором использованы стандартные обозначения для времени - подобного

$$u^\mu = \frac{dx^\mu}{d\tau}$$

и пространственно - подобного

$$l^\mu = \frac{dx^\mu}{d\rho}$$

векторов, описывающих динамику струны.

Кроме того, $\tau$ и $\rho$ - величины, задающие параметризацию гиперповерхности, которую при своем движении заметает струна.

Переходя к координатным выражениям, исходное уравнение движения с учетом типа внешнего гравитационного поля перепишем в виде следующей системы

$$\frac{du^0}{d\tau} - \frac{dl^0}{d\rho} = 2\Gamma^0_{01}\left(l^0 l^1 - u^0 u^1\right) \quad , \tag{3}$$

$$\frac{du^1}{d\tau} - \frac{dl^1}{d\rho} = \Gamma^1_{00}\left(l^0 l^0 - u^0 u^0\right) + \tag{4}$$

$$+ \Gamma^1_{11}\left(l^1 l^1 - u^1 u^1\right) + \Gamma^1_{22}\left(l^2 l^2 - u^2 u^2\right) + \Gamma^1_{33}\left(l^3 l^3 - u^3 u^3\right) \quad ,$$

$$\frac{du^2}{d\tau} - \frac{dl^2}{d\rho} = 2\Gamma^2_{12}\left(l^1 l^2 - u^1 u^2\right) \quad , \tag{5}$$

$$\frac{du^3}{d\tau} - \frac{dl^3}{d\rho} = 2\Gamma^3_{13}\left(l^1 l^3 - u^1 u^3\right) \quad . \tag{6}$$

Для последующего анализа уравнений (3) - (6) примем, что струна расположена вдоль оси $x^1$, а расчеты будем проводить методом последовательных приближений, считая, что в основном приближении струна совершает свободные колебания. В дальнейшем все величины, относящиеся к свободному колебанию, будем снабжать индексом «(0)».

Потребуем, чтобы

$$x^0 = x^0(\tau).$$

Тогда в силу определения $l^\alpha$ имеем $l^0_{(0)}$. Уравнение (3),

следовательно, примет вид

$$\frac{du^0}{d\tau} - \frac{dl^0}{d\rho} = 0. \qquad (7)$$

Его решение $x^0 = \tau$, выбранное ради удобства, представляет собой известную временную параметризацию в теории струны [11].

Проанализируем теперь уравнения (3) - (6), имея в виду, что нижеследующие условия задают параметры движения струны в основном приближении

$$\left.\begin{array}{l} u^0_{(0)} = u^0_{(0)}\left(1,0,0,u^3_{(0)}\right) \\ l^0_{(0)} = l^0_{(0)}\left(0,1,0,l^3_{(0)}\right) \end{array}\right\}. \qquad (8)$$

Итак, в силу (8) уравнение (5) принимает вид однородного гиперболического уравнения

$$\frac{du^2}{d\tau} - \frac{dl^2}{d\rho} = 0. \qquad (9)$$

Его решение естественно выбрать в виде $x^2 = 0$. Далее, с учетом ограничений (8) уравнение (4) суть

$$\frac{du^1}{d\tau} - \frac{dl^1}{d\rho} = -\Gamma^1_{00} + \Gamma^1_{11} + \Gamma^1_{33}\left({l^3_{(0)}}^2 - {u^3_{(0)}}^2\right). \qquad (10)$$

И, наконец, рассмотрим уравнение (6). Учитывая параметры движения струны в основном приближении (8), получаем

$$\frac{du^3}{d\tau} - \frac{dl^3}{d\rho} = 2\Gamma^3_{13} l^1_{(0)} l^3_{(0)}. \qquad (11)$$

Наша дальнейшая задача - решение уравнений (10) и (11). Перепишем их следующим образом, произведя для простоты замену $x^1 \to x, x^3 \to z$, -

$$\frac{d^2x}{dx^{0\,2}} = -\Gamma^1_{00} + \Gamma^1_{11} + \Gamma^1_{33}\left(l^{3\,\,2}_{(0)} - u^{3\,\,2}_{(0)}\right), \tag{12}$$

$$\frac{d^2z}{dx^{0\,2}} - \frac{d^2z}{dx^2} = 2\Gamma^3_{13} l^3_{(0)}. \tag{13}$$

Рассмотрим сначала уравнение (13). Учитывая приближенный характер искомого решения, положим в нем

$$z = Z_{(0)} + \varsigma,$$

где основное смещение $Z_{(0)}$ описывается решением однородного уравнения, а добавка $\varsigma$ - решением неоднородного уравнения.

Таким образом, уравнение для $Z_{(0)}$ таково

$$\frac{d^2Z_{(0)}}{dx^{0\,2}} - \frac{d^2Z_{(0)}}{dx^2} = 0. \tag{14}$$

Его решение в одномодовом приближении представляет собой стоячую волну

$$Z_{(0)}\left(x, x^0\right) = \left(A\cos\frac{\pi}{L}x^0 + B\sin\frac{\pi}{L}x^0\right)\cdot\sin\frac{\pi}{L}x, \tag{15}$$

где $L$ - длина струны.

Отсюда легко найти первый неизвестный параметр, описывающий положение струны

$$l^3_{(0)} = \frac{dZ_{(0)}}{dx} = \frac{\pi}{L}\left(A\cos\frac{\pi}{L}x^0 + B\sin\frac{\pi}{L}x^0\right)\cdot\cos\frac{\pi}{L}x. \qquad (16)$$

Поскольку в нашей метрике $\Gamma^3_{13} = -2\frac{\pi\gamma\sigma}{c^2}$, то уравнение для поправки $\varsigma$ принимает вид

$$\frac{d^2\varsigma}{dx^{0\,2}} - \frac{d^2\varsigma}{dx^2} = -4\frac{\pi\gamma\sigma}{c^2}\cdot\frac{\pi}{L}\left(A\cos\frac{\pi}{L}x^0 + B\sin\frac{\pi}{L}x^0\right)\cos\frac{\pi}{L}x. \qquad (17)$$

Для его решения рассмотрим уравнение

$$\frac{d^2\varsigma}{dx^{0\,2}} - \frac{d^2\varsigma}{dx^2} = f\left(x^0, x\right) = \Phi\left(x^0\right)\cdot\cos\frac{\pi}{L}x, \qquad (18)$$

в котором

$$\Phi\left(x^0\right) = -4\frac{\pi^2\gamma\sigma}{c^2 L}\left(A\cos\frac{\pi}{L}x^0 + B\sin\frac{\pi}{L}x^0\right). \qquad (19)$$

Для однозначного решения (18) введем начальные условия

$$\left.\begin{aligned} &\varsigma(0, x) = \phi\cdot\cos\frac{\pi}{L}x, \\ &(\varsigma)'_{x^0}(0, x) = \psi\cos\frac{\pi}{L}x \end{aligned}\right\}. \qquad (20)$$

и будем искать его в виде гармонического колебания

$$\varsigma\left(x^0, x\right) = \varsigma\left(x^0\right)\cdot\cos\frac{\pi}{L}x \qquad (21)$$

с зависящей от времени амплитудой. В соответствии с [12] функцию $\varsigma\left(x^0\right)$ представим в виде

$$\varsigma\left(x^0\right)=\varsigma_1\left(x^0\right)+\varsigma_2\left(x^0\right), \tag{22}$$

где $\varsigma_1\left(x^0\right)$ есть общее решение линейного однородного уравнения, а $\varsigma_2\left(x^0\right)$ - частное решение неоднородного уравнения. В дальнейшем будем интересоваться только решением $\varsigma_2\left(x^0\right)$, которое имеет вид

$$\begin{gathered}\varsigma_2\left(x^0\right)=\frac{L}{\pi}\int_0^{x^0}\cos\frac{\pi}{L}\left(x^0-\theta\right)\Phi(\theta)d\theta= \\ -4\frac{\pi\gamma\sigma}{c^2}\int_0^{x^0}\cos\frac{\pi}{L}\left(x^0-\theta\right)\cdot\begin{pmatrix}A\cos\frac{\pi}{L}\theta+ \\ B\sin\frac{\pi}{L}\theta\end{pmatrix}d\theta.\end{gathered} \tag{23}$$

Вычисляя необходимые интегралы и оставляя лишь непериодические выражения, получаем следующее решение для $\varsigma_2\left(x^0\right)=\varsigma\left(x^0\right)$

$$\varsigma\left(x^0\right)=-A\left(x^0\right)\cdot\cos\frac{\pi}{L}x^0-B\left(x^0\right)\cdot\sin\frac{\pi}{L}x^0, \tag{24}$$

где

$$A=A\frac{2\pi\gamma\sigma}{c^2}\cdot x^0, B=B\frac{2\pi\gamma\sigma}{c^2}x^0. \tag{25}$$

Что касается полного решения $\varsigma\left(x^0,x\right)$, то оно таково

$$\varsigma\left(x^0,\rho\right)=-\left(A\left(x^0\right)\cdot\cos\frac{\pi}{L}x^0+B\left(x^0\right)\cdot\sin\frac{\pi}{L}x^0\right)\cos\frac{\pi}{L}x. \qquad (26)$$

Переходим теперь к решению уравнения (10). Поскольку входящие в него символы Римана - Кристоффеля приближенно равны

$$\Gamma^1_{00}=-\Gamma^1_{11}=-\Gamma^1_{33}=-2\frac{\pi\gamma\sigma}{c^2},$$

то оно принимает вид

$$\frac{d^2x}{dx^{0\,2}}=2\frac{\pi\gamma\sigma}{c^2}\left(2+l^{3\;2}_{(0)}-u^{3\;2}_{(0)}\right). \qquad (27)$$

Для нахождения решения этого уравнения необходимо вычислить второй неизвестный параметр, определяющий движение струны. Из (15) находим

$$u^3_{(0)}=\frac{dz_{(0)}}{dx^0}=-\frac{\pi}{L}\left(A\sin\frac{\pi}{L}x^0-B\cos\frac{\pi}{L}x^0\right)\cdot\sin\frac{\pi}{L}x. \qquad (28)$$

Подставляя (28) в (27) и учитывая ранее найденное выражение (16), получаем

$$\frac{d^2x}{dx^{0\,2}}=2\frac{\pi^3\gamma\sigma}{c^2}\left[\begin{array}{l}2+\frac{1}{L^2}\left(A\cos\frac{\pi}{L}x^0+B\sin\frac{\pi}{L}x^0\right)^2\cdot\\ \cos^2\frac{\pi}{L}x-\end{array}\right. \qquad (29)$$

$$\left.-\frac{1}{L^2}\left(A\sin\frac{\pi}{L}x^0-B\cos\frac{\pi}{L}x^0\right)^2\right]\cdot\sin^2\frac{\pi}{L}x.$$

Решение этого уравнения также ищем методом последо-вательных приближений, полагая $x = X_{(0)} + \xi$, где $X_{(0)}$ - решение уравнения движения в отсутствии доменной стенки. Оно имеет вид

$$\frac{d^2 X_{(0)}}{dx^{0\,2}} = 0.$$

Поскольку в отсутствии стенки движения струны по оси x нет, то следует положить $X_{(0)} = 0$.

Уравнение для добавки $\xi$, следовательно, принимает вид

$$\frac{d^2\xi}{dx^{0\,2}} = 2\frac{\pi^3\gamma\sigma}{c^2}\left[2 + \frac{1}{L^2}\left(A\cos\frac{\pi}{L}x^0 + B\sin\frac{\pi}{L}x^0\right)^2\right]. \qquad (30)$$

Его решение, как и прежде, представляется в виде суммы периодического и непериодического слагаемых. Ограничиваясь только непериодическим членом, в результате простого интегрирования получаем искомое смещение

$$\xi = \left[4 + \frac{1}{2L^2}\left(A^2 + B^2\right)\right]\frac{\pi^3\gamma\sigma}{c^2}\cdot x^{0\,2}, \qquad (31)$$

которое квадратично зависит от времени.

Проанализируем полученные результаты. Из формулы (26) следует, что амплитуда колебаний струны вдоль оси $z$ со временем уменьшается, поскольку $\varsigma\langle 0$. С другой стороны, из формулы (31) видно, что $\xi\rangle 0$ и, следовательно, длина волны вдоль оси со временем увеличивается.

Отсюда можно сделать следующий вывод - влияние доменной стенки на колебания космической струны состоит в ее вытягивании, приводящем к распрямлению струны. Несмотря на то, что полученные решения являются приближен-ными и имеют смысл на ограниченных временных интервалах, не вызывает сомнения факт распрямления космической струны в гравитационном поле доменной стенки. Этот вывод представляет собой новый физический механизм распрямления космических струн, полученный в добавление к хорошо известным результатам [13].

## 7.2 Энергия натяжения космической струны в гравитационном поле вакуумной доменной стенки

Рассчитаем теперь энергию, переданную струне доменной стенкой, обратив особое внимание на ее зависимость от времени.

Действие для релятивистской струны, как известно, имеет вид [11, 13]

$$S = -\mu \int_{\tau_1}^{\tau_2} L d\tau, \tag{32}$$

в котором функция Лагранжа описывается выражением

$$L = \int_{\rho_1(\tau)}^{\rho_2(\tau)} \sqrt{-g} \sqrt{\left(u^{\alpha} l_{\alpha}\right)^2 - \left(u^{\alpha}\right)^2 \left(l^{\beta}\right)^2} d\rho. \tag{33}$$

Имея в виду, что длина струны равна L, и учитывая в (33) лишь добавки первого порядка малости, в выбранной калибровке получаем

$$L = L_0 + \delta\tilde{L}, \tag{34}$$

где

$$L_0 = \int_0^L \sqrt{-g}\sqrt{\left(u^{\alpha}_{(0)} l_{\alpha(0)}\right)^2 - \left(u^{\alpha}_{(0)}\right)^2 \left(l^{\beta}_{(0)}\right)^2}\, d\rho, \tag{35}$$

а

$$\delta\tilde{L} = \int_0^L \sqrt{-g}\Big\{\Big[u^{\alpha}_{(0)} l_{\alpha(0)}\left(u^{\beta}_{(0)}\lambda_{\beta(0)} + l^{\beta}_{(0)} v_{\beta}\right)\left(u^{\alpha}_{(0)}\right)^2 l^{\beta}_{(0)}\lambda_{\beta} - \left(l^{\alpha}_{(0)}\right)^2 u^{\beta}_{(0)} v_{\beta}\Big]\cdot$$

$$\sqrt{\left(u^{\alpha}_{(0)} l_{\alpha(0)}\right)^2 - \left(u^{\alpha}_{(0)}\right)^2 \left(l^{\beta}_{(0)}\right)^2}^{\,1/2}\Big\} d\rho. \tag{36}$$

В приведенных выражениях использованы следующие обозначения

$$u^{\alpha} = u^{\alpha}_{(0)} + v^{\alpha}, \tag{37}$$

где $v^{\alpha}$ и $\lambda^{\alpha}$ - малые поправки к векторам, описывающим скорость и ориентацию струны в пространстве, соответственно.

Для рассматриваемой нами динамической модели с учетом параметров движения (8) и принятой калибровки имеем в компонентах

$$\begin{aligned} u^{\alpha} &= u^{\alpha}\left(1, v^1, 0, u^3_{(0)} + v^3\right) \\ l^{\alpha} &= l^{\alpha}\left(0, 1, 0, l^3_{(0)} + \lambda^3\right) \end{aligned}. \tag{38}$$

Опираясь на выражения (25), (26), (30) и учитывая, как всегда, лишь непериодические слагаемые, находим

$$v^1 = \frac{d\xi}{dx^0} = \frac{\pi^3\gamma\sigma}{L^2 c^2}\left(A^2 + B^2\right)\cdot x^0, \tag{39}$$

$$v^3 = \frac{d\varsigma}{dx^0} = 2\frac{\pi^2\gamma\sigma}{Lc^2}\cos\frac{\pi}{L}x\cdot x^0\begin{pmatrix} A\sin\frac{\pi}{L}x^0 - \\ B\cos\frac{\pi}{L}x^0 \end{pmatrix} \qquad (40)$$

и

$$\lambda^1 = \frac{d\xi}{dx} = 0, \qquad (41)$$

$$\lambda^3 = \frac{d\varsigma}{dx} = 2\frac{\pi^2\gamma\sigma}{Lc^2}\sin\frac{\pi}{L}x\cdot x^0\left(A\cos\frac{\pi}{L}x^0 + B\sin\frac{\pi}{L}x^0\right). \qquad (42)$$

Что касается $u^3_{(0)}$ и $l^3_{(0)}$, то они, напомним, описываются формулами (28) и (16), соответственно.

В силу громоздкости получаемых выражений будем рассматривать случай малых колебаний, т.е. будем считать $\frac{A}{L} \langle\langle 1$ и $\frac{B}{L} \langle\langle 1$, и удерживать только члены, пропорциональные первым степеням этих выражений.

Это ограничение приводит к следующим значениям

$$u^\alpha_{(0)} l_{\alpha(0)}\left(u^\beta_{(0)}\lambda_\beta + l^\beta_{(0)}v_\beta\right) - \left(u^\alpha_{(0)}\right)^2 l^\beta_{(0)}\lambda_\beta - \left(l^\alpha_{(0)}\right)^2 u^\beta_{(0)}v_\beta \approx \qquad (43)$$

$$\approx -\frac{\pi^3\gamma\sigma}{L^2c^2}\sin 2\frac{\pi}{L}x\cdot\begin{bmatrix} \left(A^2 - B^2\right)\cos 2\frac{\pi}{L}x^0 + \\ 2AB\sin 2\frac{\pi}{L}x^0 \end{bmatrix},$$

а

$$\sqrt{\left(u^{\alpha}_{(0)} l_{\alpha(0)}\right)^2 - \left(u^{\alpha}_{(0)}\right)^2 \left(l^{\beta}_{(0)}\right)^2} \approx$$

$$\approx 1 + \frac{\pi^2}{2L^2}\left(A\cos\frac{\pi}{L}x^0 + B\sin\frac{\pi}{L}x^0\right)^2 \cos^2\frac{\pi}{L}x. \qquad (44)$$

Поскольку для метрики (1)

$$\sqrt{-g} \approx 1 - 8\frac{\pi\gamma\sigma}{L^2c^2}x,$$

то, подставляя это выражение вместе с (44) в (35), получаем

$$L_0 = L + \delta L_1, \qquad (45)$$

где

$$\delta L_1 = -8\frac{\pi\gamma\sigma}{c^2}\left[1 + \frac{\pi^2}{L^2}\left(A\cos\frac{\pi}{L}x^0 + B\sin\frac{\pi}{L}x^0\right)^2\right] \cdot \int\limits_0^L x \cdot \cos^2\frac{\pi}{L}x dx. \qquad (46)$$

Что касается добавки $\delta\tilde{L} = \delta L_2$, то она, в силу (36), (43) и (44), имеет вид

$$\delta L_2 = -4\frac{\pi^2\gamma\sigma}{Lc^2} \cdot x^0 \left[\left(A^2 - B^2\right)\cos 2\frac{\pi}{L}x^0 + 2AB\sin 2\frac{\pi}{L}x^0\right] \cdot \int\limits_0^{L/4} \sin 2\frac{\pi}{L} x d\left(\frac{\pi}{L}x\right). \qquad (47)$$

Из двух полученных к лагранжиану добавок наибольший интерес представляет слагаемое (47), поскольку оно не явля-ется периодической функцией времени. Вычисляя интеграл в (47), находим окончательно

$$\delta L_2 = 2\frac{\pi^2\gamma\sigma}{Lc^2}\cdot x^0\begin{bmatrix}\left(A^2+B^2\right)\cos 2\frac{\pi}{L}x^0 + \\ 2AB\sin 2\frac{\pi}{L}x^0\end{bmatrix}. \qquad (48)$$

Соответствующая же добавка к энергии равна добавке к функции Гамильтона, которая, в свою очередь, равна поправке к лагранжиану с противоположным знаком, т.е.

$$\delta\varepsilon = \delta H = -\delta L_2 = 2\frac{\pi^2\gamma\sigma}{Lc^2}\cdot x^0\begin{bmatrix}\left(B^2-A^2\right)\cos 2\frac{\pi}{L}x^0 - \\ 2AB\sin 2\frac{\pi}{L}x^0\end{bmatrix}. \qquad (49)$$

## Заключение

Полученный результат дает ответ на вопрос о причинах вытягивания космической струны. Согласно (49) присутствие доменной стенки приводит к возрастанию по квазипериодическому закону ее энергии натяжения. Другими словами, доменная стенка накачивает струну дополнительной энергией, которая и увеличивает ее натяжение [12].

## *ЛИТЕРАТУРА*

# *8. НЕЛИНЕЙНЫЕ УРАВНЕНИЯ ДВИЖЕНИЯ РЕЛЯТИВИСТСКИХ СТРУН*

## Введение

Космические струны представляют собой одномерные конфигурации скалярного поля, возникшие в результате его фазовых переходов на ранних этапах эволюции Вселенной. Космические струны играют важную роль в формировании наблюдаемой части Вселенной (большой Метагалактики), являясь затравочными объектами при зарождении галактик.

Эволюция космической струны в пространстве - времени $V_4$, в свою очередь, определяется гиперповерхностью $x^{\alpha} = x^{\alpha}(\tau, \rho)$, на которой введены времени - подобный параметр $\tau$ и пространственно - подобный параметр $\rho$. Действие релятивистской струны имеет вид

$$S = -\mu \int_{\tau_1}^{\tau_2} L \cdot d\tau , \qquad (1)$$

где лагранжиан представляет собой интеграл вдоль длины струны

$$L = \int \sqrt{(u^{\alpha} \ell^{\beta})^2 - (u^{\alpha})^2 (\ell^{\beta})^2}\, d\rho . \qquad (2)$$

Здесь векторы $u^{\alpha} = \dfrac{dx^{\alpha}}{d\tau} = \dot{x}^{\alpha}$ и $\ell^{\alpha} = \dfrac{dx^{\alpha}}{d\rho} = x'^{\alpha}$ описывают скорость и ориентацию струны, соответственно.

Для нахождения уравнений движения введем, сле-дуя [2], простейшую неинвариантную параметризацию, в наших обозначениях имеющую вид

$$\tau = x^0, \rho = z. \qquad (3)$$

Это означает, что струна расположена вдоль оси $z$, ее колебания совершаются в плоскости $\{x, z\}$. Тогда уравнение движения струны в пространстве - времени $E_4$ имеет вид [2 ]

$$\frac{\partial}{\partial x^0}\left[\dot{x}(1-\dot{x}^2+x'^2)^{\frac{-1}{2}}\right]-\frac{\partial}{\partial z}\left[x'(1+x'^2-\dot{x}^2)^{\frac{-1}{2}}\right]=0. \qquad (4)$$

Полагая колебания струны малыми, так что выполняется условие $x'^2 << 1$и $\dot{x}^2 << 1$, из (4) получаем линейное уравнение движения

$$\ddot{x} - x'' = 0. \qquad (5)$$

Между тем, такое требование выполняется не всегда и поэтому космическая струна, вообще говоря, должна описываться нелинейными уравнениями. Последнее утверждение, впрочем, непосредственно следует и из (4).

## 8.1 Нелинейные уравнения движения релятивистских струн

Желая учесть нелинейные члены, будем считать $\left(x'^2\right)^2 << 1$, $\left(\dot{x}^2\right)^2 << 1$ и, следовательно, оставлять члены, кубические по различным производным $x$. В результате из (4) получаем следующее нелинейное уравнение движения

$$\ddot{x} = x''(1 - x'^2 + \dot{x}^2) + 2\dot{x}x'\dot{x}' \; . \qquad (6)$$

В довершение к сделанным предположениям будем иметь в виду, что в пост - ньютоновском приближении $\dot{x}^2 << x'^2$. Поэтому из (6) получаем искомое нелинейное уравнение движения космической струны

$$\ddot{x} - x''(1 - x'^2) = 0. \qquad (7)$$

С другой стороны, в работе [3] предложено следующее нелинейное уравнение движения космической струны в плоском пространстве - времени

$$\frac{d^2x^\alpha}{d\tau^2} - \frac{d^2x^\alpha}{d\rho^2}\left(1 - \frac{\varepsilon^2}{2}\frac{dx^\beta}{d\rho}\frac{dx_\beta}{d\rho}\right) = 0\,, \qquad (8)$$

где $\varepsilon^2$ - малый параметр. Оно получено благодаря изначальному учету в тензоре энергии - импульса струны нелинейных членов по пространственно - подобному вектору $1^\alpha$.

Выбирая упоминавшуюся параметризацию (3) и считая, что колебания струны, по - прежнему, происходят в плоскости $\{x, z\}$, из (8) получаем

$$\ddot{x} - x''[1 + \frac{\varepsilon^2}{2}(1 + x'^2)] = 0. \qquad (9)$$

Заметим, что если $\varepsilon \to 0$, то из уравнения (9) следует стандартное линейное уравнение (5).

Сопоставляя (7) и (9) видно, что они близки по своей структуре. Возникает, следовательно, естественный вопрос - возможна ли репараметризация уравнения (9), которая приводила бы его к виду (7).

Анализируя (7) и (9), видно, что искомая параметризация может касаться только пространственной переменной $z$. Итак, проведем замену

$$z \to z_0 = z + \zeta(z)\,, \tag{10}$$

в которой малую добавку $\zeta(z)$ будем считать линейной функцией $z$, и примем, что отношение $\frac{\zeta(z)}{z} \propto \varepsilon^2$. Тогда имеем

$$x' = \frac{dx}{dz} = \frac{dx}{dz_0}\left(1 + \zeta'_z\right) = \underbrace{x'}_{0}(1 + \zeta'_z)\,,$$

$$x'' = \frac{d^2x}{dz^2} = \frac{d^2x}{{dz_0}^2}\left(1 + 2\zeta'_z\right) = \underbrace{x''}_{0}(1 + 2\zeta'_z)\,,$$

$$x'^2 = \left(\frac{dx}{dz}\right)^2 = \left(\frac{dx}{dz_0}\right)^2\left(1 + 2\zeta'_z\right) = \underbrace{x'^2}_{0}(1 + 2\zeta'_z)\,,$$

где значок «$\underbrace{x'}_{0}$» символизирует дифференцирование по переменной $z_0$. Подставляя эти выражения в (9), с требуемой точностью получаем уравнение

$$\ddot{x} - \underbrace{x''}_{0}\left(1 + 2\zeta'_z + \frac{\varepsilon^2}{2} + \frac{\varepsilon^2}{2}\underbrace{x'^2}_{0}\right) = 0 \quad . \tag{11}$$

Сравнивая теперь (11) с (7), нетрудно видеть, что для их совпадения необходимо соблюдение условий

$$1 + 2\zeta'_z + \frac{\varepsilon^2}{2} = 1\,, \frac{\varepsilon^2}{2} = -1 \quad . \tag{12}$$

Отсюда легко найти искомую функцию $\zeta(z)$, а именно $\zeta(z) = \frac{1}{2} z$. Таким образом, совершая репараметризацию

$$x^0 \to x^0, \; z \to z_0 = \frac{3}{2} z, \tag{13}$$

уравнение (9) приближенно может быть приведено к виду (7).

## 8.2 Эволюция нелинейной космической струны

Для дальнейшего анализа нелинейных уравнений движения найдем решение уравнения (7). Согласно [4] его точным решением является линейная функция

$$x(x^0; z) = A \cdot x^0 + B \cdot z + C \tag{14}$$

с постоянными величинами $A$, $B$ и $C$.

Решение (14) описывает поступательное движение струны как целой. Поэтому оно не детерминировано нелинейным характером уравнения движения, поскольку, как нетрудно убедиться, решение (14) удовлетворяет и линейному уравнению движения (5). Таким образом, решение (14) не учитывает нелинейного характера уравнения движения (7).

Для того, чтобы учесть такую нелинейность, рассмотрим приближенное периодическое (колебательное) решение этого уравнения. Итак, положим

$$x = X + \xi, \tag{15}$$

где $\xi$ - малая добавка порядка $\left(\frac{A}{L}\right)^3 \approx \left(\frac{B}{L}\right)^3$. Выбранный порядок малости диктуется естественным требованием $\frac{A}{L} < 1$ и $\frac{B}{L} < 1$, означающим малость амплитуд колебаний $A$ и $B$ в сравнении с длиной волны $L$.

Подставляя (15) в (7), получаем уравнение свободных колебаний

$$\ddot{x} - x'' = 0 \qquad (16)$$

с решением в одномодовом приближении

$$x = \left( A \cdot \cos\frac{\pi}{L} x^0 + B \cdot \sin\frac{\pi}{L} x^0 \right) \sin\frac{\pi}{L} z . \qquad (17)$$

Что касается уравнения для возмущенной добавки, то оно имеет вид неоднородного волнового уравнения

$$\ddot{\xi} - \xi'' = f(x^0; z) = X'' \cdot X'^2 , \qquad (18)$$

где с учетом (17) возмущающая сила

$$f(x^0; z) = -\left(\frac{\pi}{L}\right)^4 \left( A \cdot \cos\frac{\pi}{L} x^0 + B \cdot \sin\frac{\pi}{L} x^0 \right)^3 \cdot \cos^2\frac{\pi}{L} z \cdot \sin\frac{\pi}{L} z \qquad (19)$$

Для однозначного решения уравнения (18) с правой частью (19) введем также общие начальные условия

$$\xi(0, z) = \varphi(z) , \xi'(0, z) = \psi(z) \qquad (20)$$

и однородное граничное условие

$$\xi(x^0,0) = \xi(x^0,L) = 0. \qquad (21)$$

Решение уравнения (18), как известно, представляет собой сумму двух слагаемых, одно из которых описывает свободные колебания, а другое - вынужденные колебания. Интересуясь лишь вынужденными колебаниями, перепишем (19) в виде $f(x^0,z) = f_1(x^0,z) + f_2(x^0,z)$, где

$$f_1(x^0,z) = -\frac{1}{4}\left(\frac{\pi}{L}\right)^4 \left(A \cdot \cos\frac{\pi}{L}x^0 + B \cdot \sin\frac{\pi}{L}x^0\right)^3 \cdot \sin\frac{\pi}{L}z = \Phi_1(x^0) \cdot \sin\frac{\pi}{L}z , \qquad (22)$$

а

$$f_2(x^0,z) = -\frac{1}{4}\left(\frac{\pi}{L}\right)^4 \left(A \cdot \cos\frac{\pi}{L}x^0 + B \cdot \sin\frac{\pi}{L}x^0\right)^3 \cdot \sin 3\frac{\pi}{L}z = \Phi_2(x^0) \cdot \sin 3\frac{\pi}{L}z . \qquad (23)$$

Соответственно этому представлению решение запишем в виде

$$\xi(x^0,z) = \xi_1(x^0) \cdot \sin\frac{\pi}{L}z + \xi_2(x^0) \cdot \sin 3\frac{\pi}{L}z, \qquad (24)$$

где

$$\xi_1(x^0) = \frac{L}{\pi}\int_0^{x^0} \sin\frac{\pi}{L}(x^0 - \tau) \cdot \Phi_1(\tau)d\tau, \qquad (25)$$

а

$$\xi_2(x^0)=\frac{L}{\pi}\int_0^{x^0}\sin 3\frac{\pi}{L}(x^0-\tau)\int\cdot\Phi_2(\tau)d\tau. \qquad (26)$$

Вычисляя необходимые интегралы и оставляя лишь те из них, которые имеют непериодический характер, получаем

$$\xi_1(x^0)=\frac{3}{32}\left(\frac{\pi}{L}\right)^3\left(A^2+B^2\right)\cdot x^0\cdot\left(A\cdot\cos\frac{\pi}{L}x^0-B\cdot\sin\frac{\pi}{L}x^0\right)=$$

$$=\left(A_1(x^0)\cdot\cos\frac{\pi}{L}x^0-B_1(x^0)\cdot\sin\frac{\pi}{L}x^0\right) \qquad (27)$$

и

$$\xi_2(x^0)=-\frac{3}{288}\left(\frac{\pi}{L}\right)^3(A^2+B^2)\cdot x^0\cdot\left(\begin{array}{l}A\cdot\cos 3\frac{\pi}{L}x^0-\\ B\cdot\sin 3\frac{\pi}{L}x^0\end{array}\right)=$$

$$=\left(A_2(x^0)\cdot\cos 3\frac{\pi}{L}x^0-B_2(x^0)\cdot\sin 3\frac{\pi}{L}x^0\right) \quad . \qquad (28)$$

Выражения (27) и (28) представляют собой квазипериодические колебания, поскольку все амплитуды линейно растут со временем. Последнее обстоятельство означает, что нелинейная струна обладает свойством самодействия уже в плоском пространстве - времени. Что касается взаимодействия линейных космических струн, то он, как известно [5], возможен только во внешних гравитационных полях.

## Заключение

Проведенный анализ, таким образом, показывает, что для описания динамики нелинейных космических струн можно воспользоваться достаточно общим уравнением движения (7). Нелинейный характер движения струны приводит к эффекту самодействия, следствием которого является монотонное возрастание амплитуды ее колебаний.

## *ЛИТЕРАТУРА*

**Часть 2**

# СОЛНЕЧНЫЕ И ЗЕМНЫЕ АТМОСФЕРНЫЕ ЯВЛЕНИЯ

***Такибаев Н.Ж.***

***Посвящаю моей маме -***
***Г.А. Тлеубергеновой, декану физфака***
***КазПИ им. Абая в 1968-1970 г.г.***

Во второй части дан краткий обзор особенностей атмосферных процессов на Солнце и Земле. Исследуются процессы, ведущие к электризации встречных газовых потоков в хромосфере Солнца и грозовых облаков в тропосфере Земли, формированию и развитию серебристых облаков в области ее мезосферы.

Рассмотрение ведется на основе метода ионизационного равновесия - теории Саха', примененного не только к равновесным (или квазиравновесным) процессам ионизации газов на Солнце, но и к процессам нуклеации и гидратации паров воды в атмосфере Земли. Это дает возможность связать макроскопические процессы и явления с характеристиками квантовых состояний микрочастиц (атомов, молекул и ионов) и их взаимодействий.

Обсуждаются вопросы генерации нейтронов в атмосферных ядерных процессах на Солнце. Предложен новый механизм генерации атмосферных нейтронов - процесс захвата электрона протоном в $H_2^+$-системе, вызванный потоком жесткого γ-облучения.

Особое внимание уделяется общности и подобию целого ряда электрических явлений на Солнце и на Земле, с учетом различий, связанных с масштабами процессов и влиянием внешних полей.

# ВВЕДЕНИЕ

Научные исследования атмосферных явлений приобретают все большую актуальность в последние годы. Это касается как земной атмосферы, так и атмосфер Солнца, звезд и крупных планет (см., например, [1-7]).

Интерес к атмосферным явлениям на нашей планете - планете Земля, вполне очевиден. Атмосфера - это среда, которая обеспечивает и сохраняет условия нашей жизни. Она является динамичной и изменчивой - ее климатические условия зависят от географического положения региона и периодически меняются в течение суток и месяцев. Известны и более длительные периоды колебаний атмосферных условий, например, годы засух, годы высокого и низкого уровня рек и озер, десятилетия наступления пустынь и т.д. Напомним о ледниковых периодах в истории Земли.

Климатические проблемы сейчас, в век глобальных информационных систем, приобретают общемировой характер. Широко обсуждаются планетарные угрозы такие, как глобальное потепление, опустынивание земель, рост кризисных ситуаций в большинстве регионов (наводнений, ураганов, снежных заносов, лесных пожаров и т.п.).

Все большей проблемой становятся вредные промышленные выбросы в атмосферу. Обсуждаются вопросы воздействия человека на природу и атмосферу. В отмеченном плане исследования атмосферных явлений приобретают первоочередное значение.

Наша Земля, как известно, находится под «властью» Солнца, которое дает нам свет, тепло и жизнь. Периоды активности Солнца во многом определяют земные климатические изменения.

Солнечно-земные связи - одно из самостоятельных научных направлений в физике космоса. Они изучаются как наземными аппаратными средствами, так и приборами, установленными на высотных шарах-зондах, спутниках и космических станциях. Исследуются процессы, происходящие на Солнце и его атмосфере, солнечное излучение и солнечные выбросы материи, их воздействие на земную атмосферу и состояние ближнего к нам космоса [7-12].

Накоплена обширная база данных, развиты теории, описывающие Солнце как динамическую систему и объясняющие многие явления, происходящие на Солнце и его атмосфере. Выдвигаются различные гипотезы, касающиеся еще не понятых событий и явлений. Они исследуются учеными и, возможно, в ближайшее время некоторые из них будут разгаданы.

В данной работе дается краткий обзор, и исследуются некоторые вопросы атмосферных явлений на Солнце и Земле [8-12]. Они касаются процессов образования заряженных газовых потоков и объемных газоплазменных формирований на Солнце, мезосферных серебристых облаков и грозовых тропосферных облаков в атмосфере Земли.

Предлагается механизм генерации нейтронов в активных областях атмосферы Солнца. В его основе лежат два последовательных процесса: образование в солнечных «молниях» ионной молекулы водорода с наведенным дипольным моментом - $H_2^+(d)$ и захват электрона одним из протонов в этой трех частичной системе при его облучении потоком жесткого γ-облучения, возникающего от «убегающих электронов».

Особое внимание уделяется общности и подобию целого ряда электроразрядных явлений на Солнце и на Земле, с учетом различий, связанных с масштабами процессов, влиянием внешних полей и т.п.

Солнце и ее планеты имеют тесную взаимосвязь, как в плане истории образования и эволюции во времени, так и продолжающегося взаимного влияния. Особенно сильным оказывается воздействие Солнца на ближайшие к ней планеты. Земля среди этих планет занимает особое место. Поверхность Земли защищена атмосферой и магнитными полями, что в сочетании с ее температурным режимом создает оптимальные условия для появления жизни.

Атмосфера Земли будет рассматриваться нами не только защитный слой, но и как уникальный физический прибор, или как природная исследовательская лаборатория, в которой можно наблюдать и изучать многие загадочные атмосферные явления.

Например, наблюдения и научный анализ процессов формирования на Земле грозовых облаков, молний, спрайтов, и т.п. можно сопоставить с подобными явлениями на Солнце. Понимание этих процессов, позволит создать основы научного прогноза, как солнечных атмосферных явлений, так и погодных изменений на Земле.

В свою очередь, рассматривая Солнце и ее атмосферу как более масштабную исследовательскую лабораторию, можно продвинуться в понимании физики атмосферных процессов и на других звездах.

Отметим, что анализ энергетических процессов, в нашем случае, электроразрядных процессов в атмосферах Солнца и Земли, является не только очень сложной, но и очень интересной и важной самостоятельной научной задачей.

# 1. ФИЗИЧЕСКИЕ ХАРАКТЕРИСТИКИ АТМОСФЕР СОЛНЦА И ЗЕМЛИ

В этой главе мы дадим краткий обзор основных характеристик атмосфер Солнца и Земли. Отметим существенные их отличия и ряд общих свойств

Сначала коснемся структуры атмосфер этих двух близких и для нас самых важных космических тел. Затем опишем процессы, происходящие в их атмосферах, солнечно-земных связи и космические связи, в частности, воздействия космических лучей, приходящих на Землю от Солнца и из глубин дальнего космоса.

## 1.1 Солнце и Земля - общие характеристики

Хотя характеристики Солнца, Земли и других планет можно найти в любом справочнике по астрономии или физике (см., например, [7-16]), мы для цельности нашего изложения приведем некоторые общие данные, чтобы иметь возможность их сравнить между собой и сопоставить с масштабом процессов, в них протекающих.

**Солнце** - это типичная звезда нашей Галактики, относящаяся к карликам спектрального класса G2. Оно представляет собой вращающийся плазменный шар с экваториальным радиусом $\approx 6{,}96 \cdot 10^5$ км. и массой $\approx 1{,}99 \cdot 10^{33}$ г. Из анализа спектра солнечного излучения следует, что солнечное вещество имеет следующий состав - на 1000 атомов вещества приходится: ≈ 900 атомов водорода, ≈ 99 атомов гелия и только 1 атом одного из всех других элементов.

Вращение Солнца имеет дифференциальный характер: экваториальная зона вращается быстрее, чем высокоширотные области. Средний период на экваторе $\sim 25{,}38$ суток, энергия вращения $\sim 2{,}4 \cdot 10^{42}$ эрг (или $\sim 1{,}5 \cdot 10^{48}$ МэВ). Эф-фективная температура поверхности Солнца равна 5780 °К.

Солнце - центральное тело Солнечной системы, является главным поставщиком энергии в этой системе. Оно же является основным источником газовой составляющей вещества в межпланетной среде.

**Земля** имеет массу $\approx 5{,}98 \cdot 10^{27}$ г. , т.е. в треть миллиона раз меньше солнечной, радиус почти в 109 раз меньше солнечного, но плотность вещества почти в 4 раза большую, чем средняя солнечная. Расстояние от Солнца $\sim 149{,}6$ млн. км. Температура поверхности Земли достигает $\sim 58$ °C в пустынях Африки и до $-90$ °C на Антарктиде.

Существование жизни на поверхности Земли стало возможным благодаря ее основным физическим характеристикам: массе, гелиоцентрическому расстоянию и быстрому вращению вокруг своей оси.

К важным характеристикам Земли, как планеты, следует отнести ее химический состав, наличие атмосферы, твердой коры и твердой поверхности, поверхностных водных пространств и пресной воды во всех трех состояниях: газообразном, жидком и твердом. Земля имеет магнитное поле и радиационные пояса, простирающиеся далеко за пределы земной атмосферы. Эти исключительные особенности определили единственно возможный путь эволюции живого и неживого вещества Земли в Солнечной системе. У других планет физические условия существенно отличаются от земных.

**Структура Солнца** (Рис.1.1). Внутренние слои Солнца представляют собой плотно сжатую горячую плазму - ядро. Под действием гравитации Солнце, как и любая звезда, стремится сжаться. Этому сжатию противодействует перепад давления, возникающий из-за высокой температуры и плотности внутренних слоев Солнца ($\approx$ 160 г/см$^{-3}$). Высокая температура в центральных областях Солнца (~ 15 млн. К$^{o}$) может длительно поддерживаться только ядерными реакциями синтеза. Эти реакции и являются основным источником энергии Солнца. В результате этих реакций 4 ядра водорода (протоны) сливаются в ядро гелия, излучаются нейтрино и выделяется огромная энергия $\approx$ 26 МэВ.

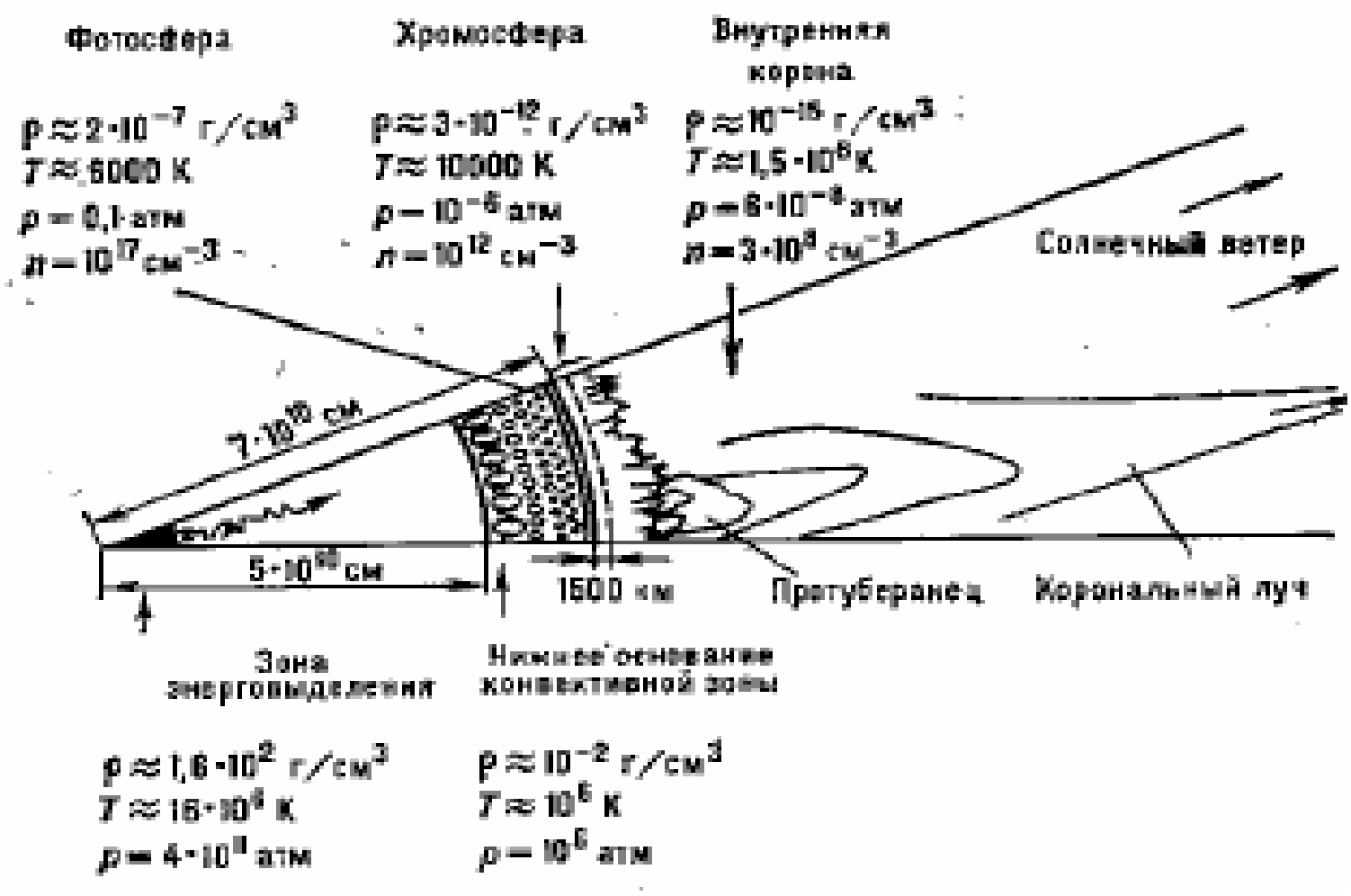

Рис. 1.1 Схема строения Солнца и его атмосферы. Здесь: r - плотность, T - температура, p - давление, n - число частиц в 1 см$^3$ [16].

В соответствии с планковским спектром излучения следует, что при температурах, характерных для центральной области Солнца, максимум энергии излучения будет приходиться на рентгеновский диапазон.

Из центра Солнца такое электромагнитное излучение, многократного поглощаясь и переизлучаясь на ионах и электронах в плазме с высокой степенью ионизации, пройдет долгий путь и достигнет поверхности за время ~ 1 млн. лет. При этом спектр излучения существенно измениться - его максимум сдвинется вниз, достигнув тепловой энергии среды. Далее в 200 раз больший путь от поверхности Солнца до Земли свет пройдет всего за время ~ 8 минут.

В отличие от фотонов, солнечные нейтрино, возникающие в ядерных реакциях в центре Солнца, достигают Земли практически мгновенно, поскольку они почти не взаимодействуют с солнечной средой. Регистрация солнечных нейтрино дает информацию о текущем состоянии ядерных реакций внутри Солнца.

Несовпадение результатов теоретических предсказаний с данными крупных и длительных экспериментов, касающихся интенсивности солнечных нейтрино, вызывает много вопросов. Теоретические результаты по разным оценкам превышают данные анализа экспериментов на величину от 15% до 30 %. Имеется ряд интересных гипотез о составе, структуре и температурных характеристиках внутренних слоев Солнца, объясняющих такие противоречия.

Основная гипотеза состоит в том, что имеются перемешивания (или осцилляции во времени) трех сортов нейтрино между собой. Т.е. компоненты (или сорта) нейтрино переходят друг в друга, и длина волны таких осцилляций близка по порядку величины к расстоянию между Солнцем и Землей. Таким образом, мы на Земле фиксируем только часть электронных нейтрино, образовавшихся на Солнце, а остальные «тяжелые» сорта нейтрино становятся для нас как бы невидимыми. Напомним, что нейтрино различаются на электронные, мюонные и тауонные.

Есть и другие гипотезы. Одна из них касается масштабных во времени конвекции горячих и «холодных» внутренних слоев вещества Солнца. По этой гипотезе температура внутри Солнца сейчас стала ниже, чем была миллионы лет назад, т.е. спектры излучения фотонов и нейтрино относятся к разным периодам жизни Солнца.

Другие гипотезы отмечают важность резонансных мод реакций синтеза, зависимость которых от температуры является более чувствительной вблизи энергий, отвечающих подпороговым резонансным реакциям.

Есть даже предположения, что солнечное ядро является железистым, а не гелиевым. Т.е. история солнечной системы, ее звезды и планет, шло по единому сценарию, и когда случился взрыв первичной протозвезды, то произошел раскол ее большого железистого ядра. Вокруг центральной части этого железистого ядра образовалось Солнце, и ядерные реакции продолжились. Разбросанные вокруг осколки этого ядра образовали планеты, в которых ядерные процессы замедлились, и температуры существенно снизились.

Выше ядра Солнца расположена его зона излучения. Здесь продукты ядерных реакций, прежде всего высокоэнер-гичные фотоны, сталкиваясь с ионами и свободными электронами, перерассеиваются и порождают вторичное излучение.

Над зоной излучения находится конвективная зона (~ 200 тыс. км.), в которой вещество, нагреваясь, поднимается и отдает тепло поверхностным слоям, затем опускается вниз и процесс повторяется. Скорости конвективных движений в глубине зоны малы (~ 1 м/с), но они достигают в тонком верхнем слое зоны больших скоростей (~ 2 км/с).

Над конвективной зоной Солнца располагается его атмосфера.

**Структура Земли.** Как ни покажется странным, но глубинные слои Земли изучены еще очень слабо [16-18]. Процессы, происходящие в центре Земли, нам практически не известны.

Согласно сейсмическим данным, недра Земли разделяются на три основных области: ядро, мантию и кору. Ядро Земли состоит из жидкого внешнего ядра (2885- 4980 км), переходной зоны (4980-5120 км) и твердого внутреннего ядра (5120-6371 км).

В интервале глубин 35-2885 км расположена силикатная оболочка, или мантия Земли. Сейсмическая граница на глубине 2775 км между мантией и ядром была открыта в 1914 Б. Гутенбергом (B. Gutenberg). Эта граница - наиболее резкая граница раздела в недрах Земли. Она сильно отражает и преломляет сейсмические волны. Непропускание ядром Земли поперечных волн означает, что модуль сдвига ядра равен нулю, т. е. ядро Земли - жидкое [17-20].

Кора отделена от мантии резкой сейсмической границей, на которой скорости сейсмических продольных волн и плотность скачкообразно возрастают. Эту границу называют границей Мохоровичича (граница Мохо).

Основные типы земной коры - океанический (толщина с учётом слоя воды ~10 км) и материковый (толщина в горных районах до нескольких десятков км); в зонах перехода от материка к океану кора имеет промежуточный тип. Эффективная толщина коры принимается равной 35 км.

Современные модели Земли выделяют литосферу -

наружную зону, включающую в себя кору и верхнюю зону мантии приблизительно до глубины 70 км. Литосфера расколота примерно на 10 больших плит, на их границах расположены основные очаги землетрясений.

Под жёсткой литосферой расположен слой повышенной текучести - астеносфера. Из-за ее малой вязкости литосферные плиты плавают в "астеносферном океане", находясь в изостатическом равновесии. Теплота внутри планеты сохранилась частично со времен ее формирования и дополнилась при последующей гравитационной дифференциации вещества на силикатную мантию и железное ядро. В состав Земли, кроме Fe (34,6%), O (29,5%), Si (15,2%), Mg (12,7%), входят в меньшем количестве многие другие химические элементы, включая уран и торий, выделяющие теплоту за счёт реакций радиоактивного распада.

От поверхности Земли к ее центру возрастают давление, плотность и температура: давление в центре $\sim 3{,}6\cdot 10^{11}$ Н/м$^2$, плотность $\sim 12{,}5$ г/см$^3$, температура ~ 5000 °С. Поверхность Земли излучает в среднем $\sim (6{,}3 \div 7{,}5)\cdot 10^{-2}$ Вт/м$^2$, преимущественно в инфракрасном диапазоне.

По современным представлениям, теплота из недр Земли выносится и конвекцией вещества. С конвекцией связывают рождение литосферных плит, их движение и погружение в мантию [17-19].

В результате дифференциации вещества в недрах Земли и дегазации возникли ее гидросфера и атмосфера.

## 1.2 Структура и свойства атмосфер Солнца и Земли

Атмосфера планеты - это газовая оболочка, окружающая данное небесное тело. Ее характеристики зависят от размера, массы, температуры, скорости вращения и химического состава небесного тела. Они определяются историей формирования тела, начиная с момента зарождения. Для Солнца, представляющего собой вращающееся газовое тело такое определение атмосферы будет, конечно, условным.

Остановимся чуть подробнее на свойствах атмосферы Солнца и Земли, поскольку в дальнейшем мы будем обсуждать физику именно их атмосферных явлений.

**Солнечная атмосфера** состоит из нескольких различных слоев (см. Рис. 1.1, а также справочную литературу [1,8,9,16]). Самый тонкий из них - фотосфера, непосредственно наблюдаемая в видимом непрерывном спектре. Толщина фотосферы всего около 300 км. Чем глубже слои фотосферы, тем они горячее. Здесь энергия, приходящая из конвективной зоны, преобразуется в излучение. Во внешних более холодных слоях фотосферы на фоне непрерывного спектра образуются фраунгоферовы линии поглощения.

Фотосфера имеет характерную зернистую структуру - гранулы, размером около 1000 км. Гранулы окружены темными промежутками. Возникновение грануляции связано с происходящей под фотосферой конвекцией. Отдельные гранулы на несколько сотен градусов горячее окружающего их газа, и в течении нескольких минут их распределение по диску Солнца меняется.

Спектральные измерения свидетельствуют о движении газа в гранулах, похожих на конвективные движения: в гранулах газ поднимается, а между ними - опускается. Эти движения газов порождают в солнечной атмосфере акустические волны, подобные звуковым волнам в воздухе. Распространяясь в верхние слои солнечной атмосферы, волны, возникшие в конвективной зоне и в фотосфере, передают им часть механической энергии конвективных движений и производят нагревание газов последующих слоев атмосферы - хромосферы и короны. В результате верхние слои фотосферы, со средней температурой около $4500\,°\mathrm{K}$, оказываются самыми "холодными" на Солнце. Как вглубь, так и вверх от них температура газов быстро растет.

Над фотосферой расположены более разреженные слои: хромосфера и корона. Они почти прозрачны для непрерывного оптического излучения.

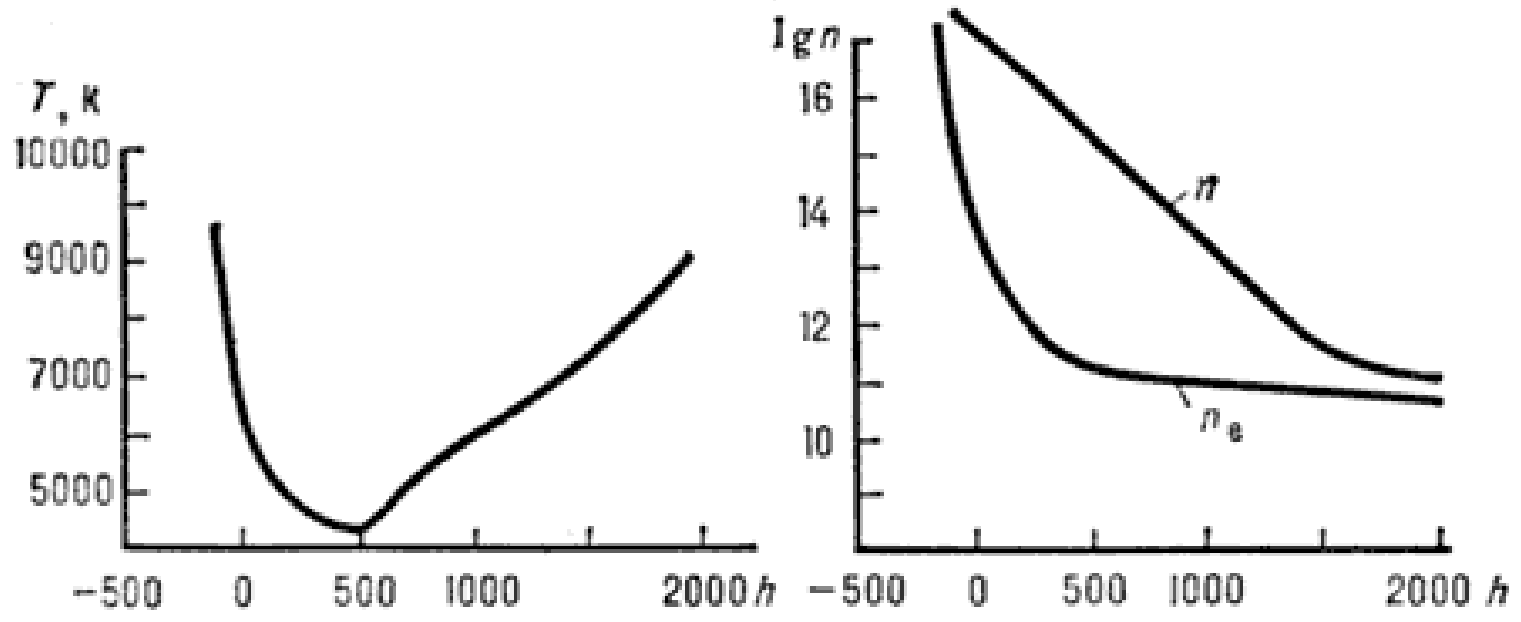

Рис. 1.2 Распределение температуры T, концентраций нейтрального водорода n и свободных электронов $n_e$ в фотосфере и нижней хромосфере (h - высота в км) [16].

В хромосфере можно наблюдать язычки пламени: вытянутые столбики из уплотненного газа - хромосферные спикулы. Спектр хромосферы состоит из ярких эмиссионных линий водорода, гелия, ионизированного кальция и других элементов, которые внезапно вспыхивают во время полной фазы затмения (спектр вспышки).

Хромосфера отличается от фотосферы значительно более неправильной и неоднородной структурой. Можно выделить два типа неоднородностей - яркие и темные. По своим размерам они превышают фотосферные гранулы. В целом распределение неоднородностей образует так называемую хромосферную сетку, особенно хорошо заметную в линии ионизированного кальция. Как и грануляция, она является следствием движения газов в подфотосферной конвективной зоне, только происходящих в более крупных масштабах.

Температура в хромосфере быстро растет, достигая в верхних ее слоях десятков тысяч градусов (Рис. 1.2).

Вдоль лимба Солнца яркость хромосферы меняется: в активных областях возрастает число спикул - отдельных газовых потоков, и усиливается излучение. В среднем излучение хромосферы в активных областях возрастает в 3-5 раз, что соответствует увеличению плотности газа примерно в 2 раза.

Хромосфера выше 1500 км представляет собой в основном набор сравнительно плотных газовых волокон, с плотно-стью частиц $n_H \approx 10^{10}$-$10^{11}$ см$^{-3}$ и температурой Т $\approx$ 6000-15000 К, и струй с гораздо более разреженным газом между ними. Выше 4-5 тыс. км остаются только спикулы.

В ячейке хромосферной сетки газ растекается от центра к периферии со скоростью 0,3-0,4 км/с. Магнитное поле на границе ячеек усилено, среднее время жизни такого образования - около суток. Горизонтальное растекание ионизованного газа от центра ячейки к периферии сгребает слабое магнитное поле (с почти вертикальными силовыми линиями). Усиление поля вызывает интенсификацию свечения хромосферы близ границ сетки, аналогично тому, как это происходит в слабых активных областях.

Интенсивность излучения хромосферы в целом невелика. Для звёзд солнечного типа установлено, что хромосферная эмиссия в линиях H, K и др. падает с уменьшением скорости вращения звёзд и их возрастом. Согласно этому критерию, Солнце - довольно старая звезда с низкой активностью.

Самая внешняя и самая разреженная часть солнечной атмосферы - корона, прослеживающаяся от солнечного лимба до расстояний в десятки солнечных радиусов и имеющая температуру около миллиона градусов. Корону можно видеть только во время полного солнечного затмения или с помощью коронографа.

Солнечная атмосфера постоянно колеблется. В ней распространяются как вертикальные, так и горизонтальные волны с длинами в несколько тысяч километров. Колебания носят резонансный характер и происходят с периодом около 5 минут.

В возникновении явлений, происходящих на Солнце, большую роль играют сложная система магнитных полей, изменяющихся как во времени, так и в пространстве. Напряженности полей могут достигать значений в тысячи Эрстед, т.е. могут в десятки тысяч раз превышать напряженности земных магнитных полей.

Вещество на Солнце всюду представляет собой намагниченную плазму, смесь электронов, ядер водорода и гелия. Иногда в отдельных областях напряженность магнитного поля быстро и сильно возрастает. Этот процесс сопровождается возникновением целого комплекса явлений солнечной активности в различных слоях солнечной атмосферы. К ним относятся факелы и пятна в фотосфере, флоккулы в хромосфере, солнечные вспышки, зарождающиеся в хромосфере, и протуберанцы (выбросы вещества) в короне.

Солнечные пятна появляются парами в тех местах, где линии искаженного магнитного поля выходят из поверхности и входят в нее. Пара пятен образует пару полюсов магнитного поля - южный и северный.

В годы повышенной солнечной активности магнитное поле искажено сильнее и пятен на Солнце больше. В годы "спокойного" Солнца пятен может и не быть. Период изменения солнечной активности приближенно принято считать равным 11,2 года. После появления пятна могут просуществовать от нескольких часов до нескольких месяцев. Форма и размеры пятен бывают различными. Их температура на 1000-1500° ниже, чем у остальной поверхности Солнца, и поэтому они кажутся темными. Холодными пятна можно считать только относительно прочих частей поверхности Солнца.

Газовая оболочка в короне и вокруг него представляет собой практически полностью ионизированную плазму, постоянно обновляющуюся. Основная часть этого вещества уходит от Солнца по всем направлениям - этот поток вещества называют солнечным ветром. Он выходит и за пределы солнечной системы в область межзвездных расстояний.

Солнечный ветер - это почти радиальный поток плазмы, движущийся от Солнца. Истечение солнечной плазмы неоднородно - выделяют, так называемые, “медленные” потоки со средней скоростью $\langle v \rangle \approx 320$ км/с и “быстрые” потоки, скорость которых может достигать 600-700 км/с и более (вплоть до $10^3$ км/с).

Плазма солнечного ветра также почти полностью ионизована. Средняя концентрация частиц в ней на орбите Земли составляет $\langle n \rangle = (5 \div 10)$ см$^{-3}$, но в солнечном ветре могут существовать неоднородности с повышенной концентрацией частиц. Средний поток частиц плазмы солнечного ветра у Земли составляет $\approx 10^8$ см$^{-2}$с$^{-1}$. Его состав в целом отражает состав солнечного вещества в короне, т.е. преобладают протоны и электроны.

Следует отметить, что ионный и изотопный состав солнечного ветра полностью не изучен, так как проведение необходимых измерений представляет собой достаточно сложную экспериментальную задачу. Механизмы формирования солнечного ветра также до конца не исследованы. Известно, что этот поток формируется в короне Солнца под действием нескольких сил различной физической природы - газового давления, постоянных и переменных электромагнитных полей, сил тяжести.

Проблема происхождения солнечного ветра тесно связана с проблемой нагрева хромосферы и короны. Их экспериментальное и теоретическое изучение в центре внимания исследований по физике Солнца.

Собран обширный наблюдательный материал, тем не менее, он пока еще не позволяет ответить на многие основные вопросы. В этом плане большое значение придается информации, получаемой на протяжении многих лет (с 1996 г.) от комплексной солнечной и гелиосферной обсерватории SOHO, расположенной вблизи первой лагранжевой точки либрации, то есть на расстоянии около 1.5 млн. км от Земли в сторону Солнца.

Сейчас активно обсуждаются различные международные космические проекты, предусматривающие измерения непосредственно в областях формирования солнечного ветра. Конкретно, речь идет о посылке космических аппаратов ближе к Солнцу вплоть до 30 или даже до 4 солнечных радиусов. Часть из этих проектов будет осуществлена в первой четверти нашего столетия. (см., например [7]).

В 2010 году предполагается запустить на вытянутую орбиту вокруг Земли Всемирную Космическую Обсерваторию - Ультрафиолет (ВКО/УФ). Участником этого проекта является Казахстан в лице Национального Космического Агентства Республики Казахстан (НКА РК), а главным научным исполнителем проектных заданий от Казахстана заявлен Астрофизический институт имени Фесенкова МОН РК. Запуск будет осуществлен с космодрома Байконур.

**Атмосфера Земли.** В настоящий исторический период атмосфера земли имеет азотно-кислородный состав: 78,1% - азота, 20,9% - кислорода. В ней также содержится от 0,3 до 3% паров воды, 0,9% аргона и 0,03% углекислого газа. Такой состав атмосфера имеет до высоты 100-120 км при общей толщине газовой оболочки 1800-2000 км.

Атмосфера имеет стратифицированное строение. До высоты 100 - 120 км вследствие активных турбулентных процессов, вызванных температурными контрастами между экватором и полюсами, неравномерным нагреванием земной поверхности солнечным теплом, происходит интенсивное перемешивание воздушных масс. Выше указанной границы происходит гравитационное разделение газов по удельному весу. От 120 до 400 км преобладают молекулярный азот и атомарный кислород. Выше (до высоты 700 км) преобладает атомарный кислород.

Внешняя часть атмосферы (до 1000 - 1500 км) имеет преимущественно гелиево-водородный состав. Легкие водород и гелий как бы всплывают над более тяжелой молекулярной оболочкой. Выделяются четыре основных слоя: тропосфера, стратосфера, мезосфера и термосфера (ионосфера).

Тропосфера. Это приземный слой атмосферы, простирается до высоты 12-18 км. В нем содержится до 80% массы всей атмосферы, водяной пар и частицы пыли антропогенного и естественного происхождения (вулканизм, пыльные бури и т.д.). На уровне моря атмосферное давление равно 760 мм ртутного столба, или $1{,}01 \cdot 10^5$ Па. С высотой давление падает и на верхней границе тропосферы не превышает 0,026 атм ($\approx 2{,}6 \cdot 10^5$ Па). Тропосфера пронизывается двумя видами солнечной энергии - световой и тепловой.

Потоки света и тепла частично рассеиваются облаками и частицами пыли и газов тропосферы, но в основном достигают земной поверхности, нагревая ее до 20 - 40 °С. Нагреваясь, Земля отдает тепло в атмосферу, но уже в длинноволновом диапазоне - инфракрасном. Это тепло поглощается парами воды и углекислого газа.

Стратосфера. От верхней границы тропосферы до высоты 50 - 55 км температура мало меняется и составляет около 220

°К. Лучистая теплопроводность стратосферы значительно выше, чем тропосферы. Этим объясняется наблюдаемая стабильность ее температуры.

Давление на верхней границе снижается до $3 \cdot 10^{-3}$ атм. Температура повышается до 270 °К (около 0 °С) за счет фотохимических реакций разложения молекулы озона, сопровождающихся выделением тепла. Озоновый слой располагается на высоте 20 - 30 км и является последним щитом на пути губительного для биосферы ультрафиолетового излучения.

В промежутке высот 50 - 85 км располагается слой низких температур атмосферы, получивший название - мезосфера. Температура здесь падает до минус 100 - 130 С. В эту область газовой оболочки уже не поступает теплое инфракрасное излучение от земной поверхности. Давление здесь падает до $7 \cdot 10^{-5}$ атм ($\approx$ 7 Па). В этой области наблюдаются так называемые серебристые облака - красивое и удивительное атмосферное явление.

В тропосфере, стратосфере и мезосфере вместе, до высоты 80 км, заключается больше, чем 99,5% всей массы атмосферы.

Верхняя часть атмосферы, над мезосферой, характеризуется очень высокими температурами и потому носит название термосферы. В ней различаются, однако, две части: ионосфера, простирающаяся от мезосферы до высот порядка тысячи километров, и лежащая над нею внешняя часть - экзосфера, переходящая в земную корону.

Воздух в ионосфере чрезвычайно разрежен, его средняя плотность $\sim 10^{-8} \div 10^{-10}$ г/м$^3$. Но и при такой малой плотности каждый кубический сантиметр воздуха на высоте 300 км еще содержит около одного миллиарда молекул или атомов, а на высоте 600 км - свыше 10 миллионов. Это на несколько порядков больше, чем содержание газов в межпланетном пространстве.

Ионосфера характеризуется очень сильной степенью ионизации воздуха - содержание ионов здесь во много раз больше, чем в нижележащих слоях, несмотря на сильную общую разреженность воздуха. Эти ионы представляют со-бой в основном заряженные атомы кислорода, заряженные молекулы окиси азота и свободные электроны.

В ионосфере выделяется несколько слоев, или областей, с максимальной ионизацией, в особенности на высотах 100-120 км (слой E) и 200-400 км (слой F). Но и в промежутках между этими слоями степень ионизации атмосферы остается очень высокой. Положение ионосферных слоев и концентрация ионов в них все время меняются. Спорадические скопления электронов с особенно большой концентрацией носят название электронных облаков.

От степени ионизации зависит электропроводность атмосферы. Поэтому в ионосфере электропроводность воздуха, в общем, в $10^{12}$ раз больше, чем у земной поверхности. Радиоволны испытывают в ионосфере поглощение, преломление и отражение. Волны длиной более 20 м вообще не могут пройти сквозь ионосферу: они отражаются уже электронными слоями небольшой концентрации в нижней части ионосферы (на высотах 70- 80 км). Средние и короткие волны отражаются вышележащими ионосферными слоями.

В ионосфере наблюдаются полярные сияния и свечение ночного неба - постоянная люминесценция атмосферного воздуха. Наблюдаются и резкие колебания магнитного поля - ионосферные магнитные бури.

Ионизация в ионосфере обязана своим существованием действию ультрафиолетовой радиации Солнца. С изменениями солнечной активности связаны изменения в потоке корпускулярной радиации, идущей от Солнца в земную атмосферу и который вызывает возмущения в ионосфере.

Температура в ионосфере растет с высотой до очень больших значений. На высотах около 800 км она достигает 1000° C. Высокие температуры ионосферы означают на самом деле то, что частицы атмосферных газов движутся там с очень большими скоростями. При этом надо иметь в виду то, что плотность воздуха в ионосфере очень мала. Поэтому тело, находящееся в ионосфере, например летящий спутник, не будет нагреваться путем теплообмена с воздухом. Температурный режим спутника будет зависеть от непосредственно-го поглощения им солнечной радиации и от отдачи его собственного излучения в окружающее пространство.

Атмосферные слои выше 800-1000 км выделяются под названием экзосферы или внешней атмосферы. Скорости движения частиц газов, особенно легких, здесь очень велики. Поскольку воздух на этих высотах имеет очень высокую степень разреженности, то частицы могут облетать Землю по эллиптическим орбитам, не сталкиваясь между собою. Отдельные частицы могут при этом иметь скорости, достаточные для того, чтобы преодолеть силу тяжести.

Для незаряженных частиц критической скоростью будет 11,2 км/сек. Частицы с такой или большей скоростью могут, двигаясь по гиперболическим траекториям, вылетать из атмосферы в космическое пространство.

Из наблюдений с помощью ракет и спутников было выяснено, что водород, ускользающий из экзосферы, образует вокруг Земли так называемую земную корону (протоносферу), простирающуюся более чем до 20 000 км. Плотность газа в земной короне ничтожно мала $\sim 10^3$ частиц на см$^3$.

Но в межпланетном пространстве концентрация частиц (преимущественно протонов и электронов) еще на порядок или два меньше.

С помощью спутников и геофизических ракет установлено существование в верхней части атмосферы и в околоземном космическом пространстве радиационных поясов Земли, начинающихся на высоте нескольких сотен километров и простирающихся на десятки тысяч километров от земной поверхности. Они состоят из электрически заряженных частиц, захваченных магнитным полем Земли. Кинетическая энергия этих частиц - порядка сотен кэВ. Радиационный пояс постоянно теряет частицы в земной атмосфере и пополняется потоками солнечной корпускулярной радиации.

**Солнечно-земные связи.** Обратимся к вопросам солнечно-земных связей. Воздействие солнечного и космического излучения на Землю и ее атмосферу является огромным и определяющим. Обратное влияние Земли на Солнце ничтожно мало.

Для биосферы главным фактором будет состояние тонкого слоя земной поверхности и нижних слоев атмосферы, поскольку именно в этих слоях сохраняются приемлемые для жизни условия и сосредоточена основная масса воздуха и воды.

Состояние верхних слоев атмосферы, расположенных на высотах от 60 до 300 и даже 1000 км от поверхности Земли, также изменяется. Здесь развиваются сильные ветры, штормы и проявляются такие удивительные электрические явления, как полярные сияния. Большинство из этих феноменов связано с потоками солнечной радиации и космического излучения, а также действием магнитного поля Земли.

Высокие слои атмосферы - это также и химическая лаборатория, поскольку там, в условиях, близких к вакууму, некоторые атмосферные газы под влиянием мощного потока солнечной энергии вступают в химические реакции.

Сейчас установлено, что потоки энергии из глубоких слоев Солнца проникают в космическое пространство далеко за орбиту Земли, и даже за пределы Солнечной системы. Этот солнечный ветер обтекает магнитное поле Земли, формируя удлиненную «полость», внутри которой и сосредоточена земная атмосфера (Рис. 1.3).

Магнитное поле Земли сужено с обращенной к Солнцу дневной стороны и образует длинный язык, вероятно выходящий за пределы орбиты Луны, - с противоположной, ночной стороны. Граница магнитного поля Земли называется магнитопаузой. С дневной стороны эта граница проходит на расстоянии около семи земных радиусов от поверхности, но в периоды повышенной солнечной активности оказывается еще ближе к поверхности Земли. Магнитопауза является одновременно границей земной атмосферы, внешняя оболочка которой называется также магнитосферой, так как в ней сосредоточены заряженные частицы (ионы), движение которых обусловлено магнитным полем Земли.

Установлена связь между активностью Солнца и геомагнитной активностью. Отмечена циклическая зависимость климатических изменений и многих биологических процессов - динамика популяций, эпидемий и т.п. От колебаний геомагнитного поля зависит состояние человека и животных. Низкочастотные колебания электромагнитных полей вызывают ответную реакцию живых организмов. Они воздействуют на нервную, эндокринную и кроветворную системы, влияют на психическое состояние.

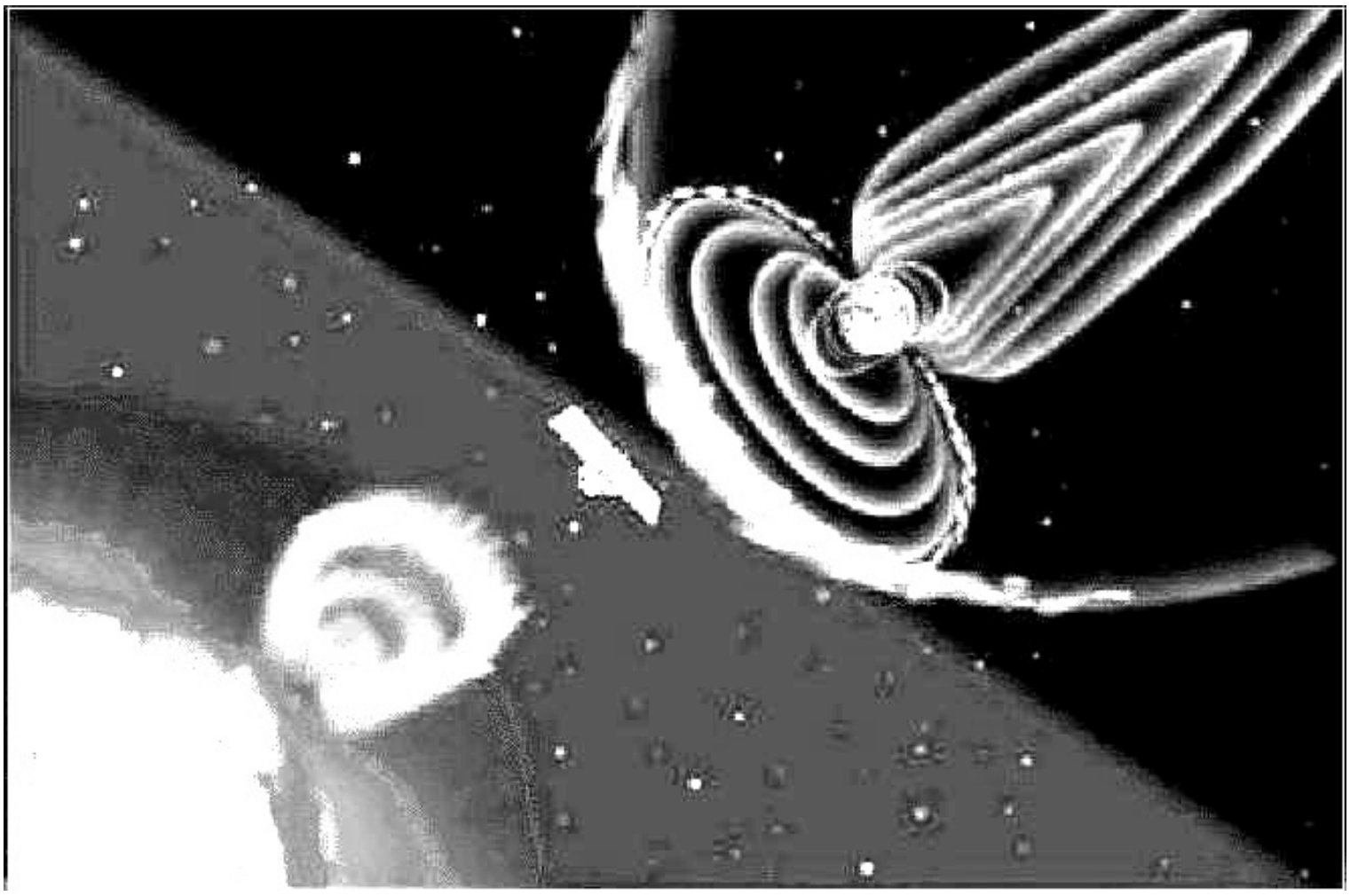

Рис.1.3 Схематичное изображение выброса коронального вещества и воздействия ударной волны на магнитосферу Земли.

Электрическое состояние атмосферы также сильно меняется во времени и в пространстве, что обязано воздействию Солнца и космических лучей.

Ясно, что Землю от межпланетного пространства отделяет мощный защитный слой. Космическое пространство пронизано мощным ультрафиолетовым и рентгеновским излучением Солнца и еще более жестким космическим излучением от далеких звезд и галактик. Эти виды радиации губительны для человека и всего живого.

Однако, значительная часть солнечного и космического излучения задерживается атмосферой Земли. Поглощением этого излучения объясняются многие свойства высоких слоев атмосферы.

Самый нижний, приземный слой атмосферы особенно важен для человека. Человек обитает в месте контакта твердой, жидкой и газообразной оболочек Земли.

Верхняя оболочка «твердой» Земли называется литосферой. Около 72% поверхности Земли покрыто водами океанов, составляющими большую часть гидросферы. Атмосфера граничит как с литосферой, так и с гидросферой. Человек живет на дне воздушного океана и вблизи или выше уровня океана водного. Взаимодействие этих океанов является одним из важных факторов, определяющих состояние атмосферы.

**Сделаем некоторые выводы.** Атмосферы Солнца и Земли являются сложными структурными газово - плазменными образованиями. Их характеризует наличие различных по физическим характеристикам атмосферных слоев, перемещение больших потоков масс между слоями и внутри них, температурные перепады и неоднородности. Важным является регулирующее действие магнитных полей.

Основные отличия физических условий атмосфер Солнца и Земли можно сформулировать в следующем:

- масса Солнца в треть миллиона раз больше массы Земли,
- центробежные силы на поверхности Солнца, локальные извержения вещества и энергии значительно больше, чем на Земле,
- температуры в атмосфере Солнца более, чем на 5000 °С выше атмосферных температур Земли,
- внешние воздействия на солнечную атмосферу идут, в основном, из глубин Солнца,
- внешние воздействия на земную атмосферу идут из космоса и, в первую очередь, от Солнца, и т.д. и т.п.

Некоторое сходство можно найти в следующих явлениях:

1. Атмосфера Солнца - это, в основном, разреженная плазма. Атмосфера Земли - это, в средних и нижних слоях, тоже плазма, но весьма разреженная и холодная (криогенная) плазма, состоящая, в основном, из отрицательных и положи-тельных ионов, в высоких слоях, состоящая из положительных ионов и электронов, а в самом верхнем слое - из протонов и электронов.

2. На Солнце встречные газо-плазменные потоки с разной плотностью и температурой, создают ионизацию трением и огромные по масштабам объемные электрические заряды и поля.

В земной атмосфере встречные (или пронизывающие друг друга) потоки воздуха также создают ионизацию трением и огромные по земным масштабам объемные электрические заряды и поля (например, многочисленные грозовые облака).

3. На Солнце происходят электроразрядные явления (солнечные молнии) и наблюдаются выбросы огромных масс вещества и излучения в верхние слои атмосферы (корональные выбросы, спикулы).

В атмосфере Земли происходят грозы, сопровождающиеся молниями (электрические разряды), выбросы масс вещества и излучения также в верхние слои атмосферы (спрайты, джеты и т.д.).

4. Молниевые процессы в атмосфере Земли и Солнца провоцируются или инициируются, как правило, внешними воздействиями:

- в случае Солнца перестройкой магнитных полей или инжекцией частиц с поверхности Солнца в его атмосферу,
- в случае Земли космическими лучами, например, ШАЛ - широкими атмосферными ливнями.

# 2. ФОРМИРОВАНИЕ ОБЪЕМНЫХ ЗАРЯДОВ В АТМОСФЕРЕ СОЛНЦА

Перейдем к более детальному анализу атмосферных процессов. Рассмотрим сначала солнечную атмосферу.

Конечно, процессы, происходящие в атмосфере Солнца, многообразны и зависят от многих важных факторов. Например, от магнитных полей, их внезапных изменений, от выбросов вещества из глубин Солнца, распространения ударных волн и плазменных струй и т.п. Учесть все эти факторы является грандиозной и пока неподъемной задачей. Но многие важные закономерности уже известны, а некоторые достаточно простые явления установлены и могут быть легко объяснены.

Остановимся на изучении периодически повторяющиеся и более или менее устойчивых процессах, которые можно считать равновесными или квазиравновесными в определенных рамках.

Рассмотрим модель формирования заряженных объемов газа в атмосфере Солнца, которые назовем солнечными «грозовыми облаками». Модель будет базироваться на теории ионизационного равновесия и уравнениях Саха. Они описывают равновесные и локальные квазиравновесные процессы ионизации газов [21, 22, 23].

Рассмотрим, например, что происходит со встречными потоками газа на границе спикул в области хромосферы. Хромосферные спикулы - отдельные потоки газа, которые поднимаются или опускаются со скоростью ~ 20 км./с, отличаются еще и тем, что газовые потоки движутся горизонтально со скоростью более 0,2 км./с от центра спикул к их периферии. Таким образом, на границе спикул происходит практически постоянное столкновение газовых потоков. Более того, сталкивающиеся потоки имеют разные температуры - это следует из данных анализа интенсивности и спектра излучения хромосферной сетки.

Во встречных потоках газо - плазменного вещества на границе хромосферных ячеек, в пограничной зоне между потоками происходит столкновение атомов и молекул этих встречных потоков, что приводит к повышенной диссоциации молекул и ионизация атомов и молекул.

В этой области - области «трения» газовых потоков, образуются положительные и отрицательные ионы, в основном это водородные радикалы $H_2^+, H_2^-$, $H^-$, и некоторая часть свободных электронов.

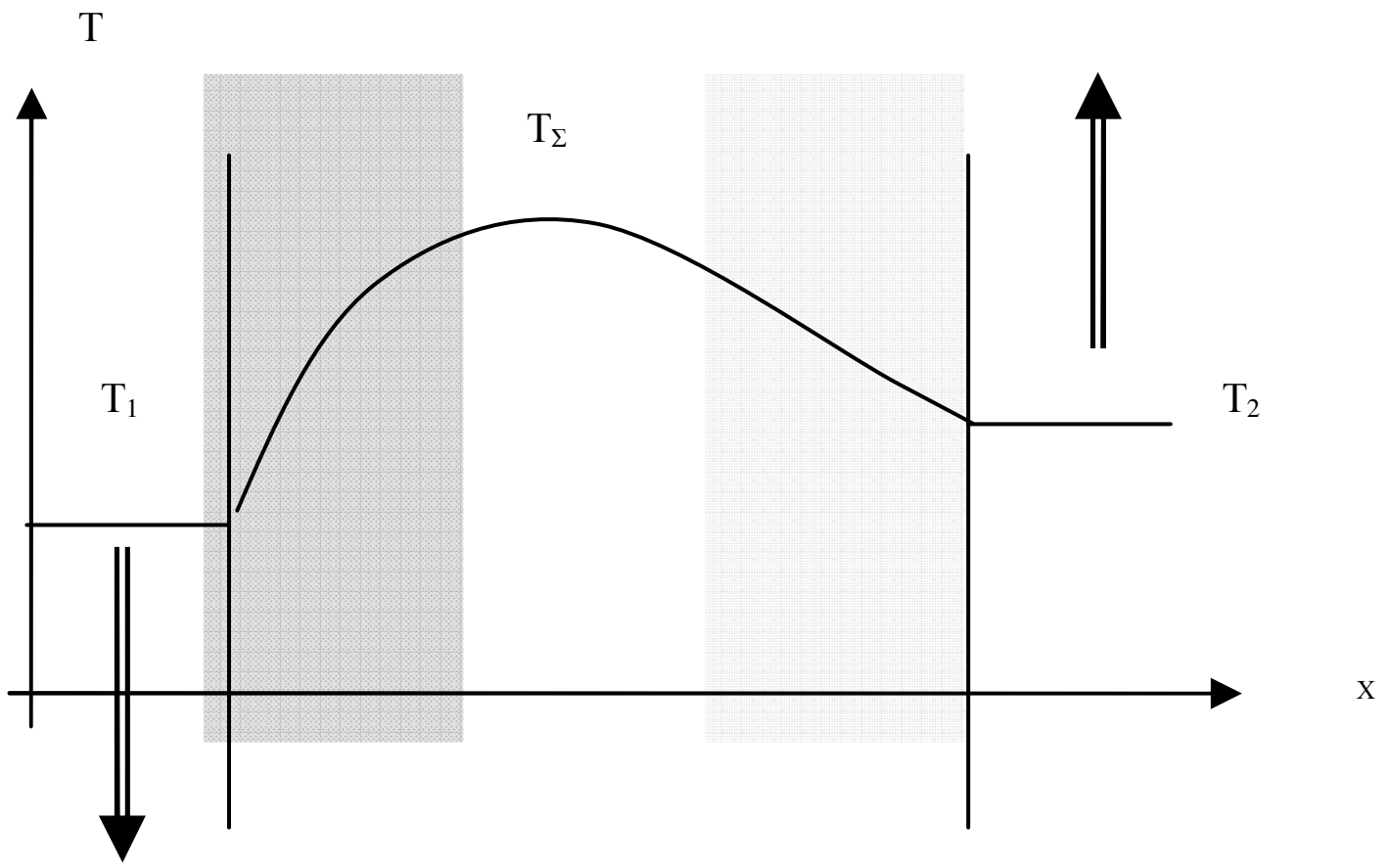

Рис. 2.1 Температурные характеристики области столкновения двух встречных потоков с разными температурами $T_1$ и $T_2$.

Где $T_\Sigma =< E_\Sigma >$ , $E_\Sigma = E_1 + E_2 + 2E_1^{1/2}E_2^{1/2}$,

$E_1$ и $E_2$ - кинетические энергии частиц потоков.

Ось x перпендикулярна движению потоков. Двойные стрелки указывают на противодвижение потоков. В зоне «трения» более темным квадратом обозначена холодная область, и более светлым - горячая область.

Если встречные потоки имеют разную температуру, то зона трения приобретает температурный градиент от центра зоны, где температура максимальна к его границам (Рис. 2.1). Это типичная картина, связанная с движением и взаимодействием двух потоков холодного и теплого слоев атмосферы. Неоднородности и турбулентные искажения в зоне «трения» для простоты рассмотрения опускаются.

В силу разных температурных градиентов от центра области «трения» по направлению к области 1 и по направлению к области 2 концентрации ионов разного сорта будут разными. Действительно, концентрация ионов каждого сорта имеет свою, вполне определенную зависимость от температуры. Это является следствием баланса прямых и обратных реакций, интенсивность которых зависит от температуры. Такой баланс описывается формулами Саха, которые часто называют теорией ионизационного равновесия [23].

Отметим, что, из данных по энергиям ионизации атомов и молекул водорода и формул Саха, следует, что в более холодных потоках будут концентрироваться преимущественно положительные ионные молекулы $H_2^+$, а в более теплых массах - отрицательно заряженные ионы $H^-$.

Что касается свободных протонов, то они образуются в основном в короне, где температуры на два порядка выше. Свободных протонов в хромосфере еще ничтожно мало.

## 2.1 Метод ионизационного равновесия

Рассмотрим общее термодинамическое описание реакций, происходящих в газовых средах. Общие условия химического равновесия реакций следуют из закона действующих масс

$$\sum \nu_i \mu_i = 0 \qquad (2.1)$$

где $\mu_i$ – химические потенциалы реагирующих веществ, а $\nu_i$ – балансовые коэффициенты. Например, для реакции образования молекул воды

$$2H_2 + O_2 - 2H_2O = 0$$

балансовые коэффициенты равны $\nu(H_2) = 2$, $\nu(O_2) = 1$, $\nu(H_2O) = -2$. Эта реакция имеет в обычных условиях взрывной характер - взрыв водородно-кислородной газовой смеси.

В общем случае реагирующих компонент может быть несколько. Химический потенциал каждого из газов смеси равен [21]

$$\mu_i = T \ln P_i + \chi_i(T), \qquad (2.2)$$

где $P_i$ - парциальное давление газа номера i, $P_i = C_i \cdot P$, $P$ - общее давление газовой смеси, $C_i = N_i / N$ - его концентрация, связанная с отношением числа частиц сорта $N_i$ к полному числу частиц газовой смеси: $N = \sum N_i$ .

Константа химического равновесия определяется выражением

$$K_p(T) = \exp(-1/T \sum \nu_i \mu_i). \qquad (2.3)$$

Здесь и далее, где это не оговорено особо, выбрана система единиц: $c = 1, \hbar = 1$ и постоянная Больцмана $k = 1$. Это выражение прямо связано с известной формулой закона действующих масс

$$\prod C_i^{\nu_i} = P^{-\sum \nu_i} K_P(T) \equiv K_C(P,T). \qquad (2.4)$$

Отметим, закон действующих масс выполняется и для ре-акций между растворенными веществами, если их концентрации в растворителе малы.

Закон справедлив и для реакций ионизации и рекомбинации атомов. Именно такая ситуация и имеет место для процессов в слабоионизованных газах.

**Уравнения Саха.** Задача «ионизационного равновесия» одноатомной плазмы сводится к следующему. Будем считать, что при низких температурах газ состоит только из нейтральных атомов. Затем с повышением температуры происходит ионизация атомов за счет столкновений и рекомбинационных излучений: однократная - $A_0 \to A_1 + e^-$ двукратная - $A_1 \to A_2 + e^-$, трехкратная - $A_2 \to A_3 + e^-$, и т.д. Здесь, как обычно, введены обозначения: $A_0$ для нейтрального атома, $A_1$ - однократно ионизованного атома, $A_2$ - двукратно ионизованного, и т.д.

Фактически, ионизационное равновесие есть частный случай химического равновесия. В применении к этим реакциям закон действующих масс приводит к системе уравнений

$$\frac{C_{n-1}}{C_n C} = PK_p^{(n)}(T) \quad , \qquad n = 1, 2, \ \ldots \tag{2.5}$$

где $C_0$ - концентрация нейтральных атомов, $C_1$, $C_2$,... - концентрация ионов различной кратности, а $C$ - концентрация свободных электронов.

Равенство $C = C_1 + 2 \cdot C_2 + 3 \cdot C_3 + \cdots$ отражает электрическую нейтральность ионизованного газа в целом. Константы равновесия для одноатомных газов легко определяются. Теплоемкости частиц одинаковы и равны $c_P = 5/2$, и тогда

$$K_P^n = \frac{g_{n-1}}{2g_n} \frac{1}{T^{5/2}} \left( \frac{2\pi}{m_e} \right)^{3/2} \exp(I_n / T) \ , \tag{2.6}$$

где $g = (2L+1)(2S+1)$ - статистический вес атомов или ионов

(L,S- их орбитальный момент и спин), $m_e$- масса электрона, а $I_n = \varepsilon_{0,n} - \varepsilon_{0,n-1}$ -энергия n-ой ионизации.

Приведенная система уравнений (2.5) определяет концентрации различных ионов и дает их функциональную зависимость с изменением температуры (формулы Саха).

Важно, что константы равновесия не зависят от деталей реакций, а определяются лишь температурой среды, а также начальными и конечными энергиями и другими характеристиками квантовых состояний частиц, участвующих в реакции.

С повышением температуры растет число ионизированных атомов, а при падении температуры ионы и электроны рекомбинируют и образуют нейтральные атомы. Концентрации обычно определяются как отношение числа частиц данного сорта к полному числу частиц среды.

Важно, что газ будет ионизован даже при температурах, малых по сравнению с энергией ионизации. Хотя число возбужденных атомов в газе будет еще крайне мало. Когда же температура среды сравнивается с энергией ионизации, то газ будет уже практически полностью ионизован. При температурах близкой к энергии отрыва последнего электрона атомарный газ можно считать состоящим из одних только электронов и голых ядер.

Изложенные здесь кратко положения теории ионизационного равновесия могут быть полностью перенесены на процессы диссоциации и рекомбинации молекул, а также на процессы, происходящие с ионизацией молекул и молекулярных комплексов [24].

**Процессы ионизации слабо ионизованной водородной плазмы.** Рассмотрим процессы ионизации молекул газа в атмосфере Солнца при различных температурах.

При столкновениях молекул водорода имеют место, как процессы диссоциации, так и ионизации. Запишем, например, реакции, требующие значительной, по величине, энергии

$$H + Q \rightarrow H^+ + e \qquad Q \geq 13{,}6\ eV\ ;$$

$$H_2 + Q \to H_2^+ + e \qquad Q \ge 15{,}4\,eV;$$
$$H_2 + Q \to H + H \qquad Q \ge 4{,}5\,eV; \qquad (2.7)$$
$$H_2 + H_2 + Q \to H_2^+ + H_2^- \qquad Q \ge 3.922\,eV;$$

где $Q$ - энергия, получаемая молекулой при столкновении с частицами среды или γ-квантами.

Запишем также реакции с участием молекулы водорода, осуществление которых требует меньших энергий

$$H_2^+ + Q \to H + H^+ \qquad Q \ge 2.652\,eV;$$
$$H_2 + H + Q \to H_2^+ + H^- \qquad Q \ge 1.062\,eV; \qquad (2.8)$$

и реакцию с подхватом второго электрона

$$H + e + Q \to H^- \qquad Q \le -0.754\,eV. \qquad (2.9)$$

Последняя из реакций является эндотермической. В процессах ионизации, рассматриваемых как химические равновесные реакции, температурные зависимости концентраций ионов определяются похожими константами равновесия - $K^{(n)}(T)$. Следует учесть, конечно, дополнительные внутренние степени свободы у двухатомной молекулы водорода и у молекулярных ионов.

Для концентраций $N_{H^-}(T)$ отрицательного иона атома водорода $H^-$ следует

$$\frac{N_{H^-}}{N_H \cdot N_e} = \frac{g_{H^-}}{g_H \cdot g_e}\left(\frac{2\pi}{m_e T}\right)^{3/2} \exp\left(\frac{\varepsilon(H^-)}{T}\right), \qquad (2.10)$$

где $\varepsilon(H^-)$ - энергия связи электрона в состоянии $(1s^2\,1S_0)$ иона $H^-$, $g_{H^-} = 1$ - статистический вес этого состояния, $g_H = 2$ - статистический вес атома водорода в состоянии $(1s)$, $g_e = 2$ - статистический вес электрона.

Для концентраций $N_{H_2^+}(T)$ положительного молекулярного иона $H_2^+$ формула типа Саха будет иметь вид

$$\frac{N_{H_2^+}}{N_H \cdot N_{H^+}} = \frac{Z(H_2^+)}{g_H \cdot g_{H^+}} \left( \frac{2\pi}{m_e T} \right)^{3/2} \exp\left( \frac{D}{T} \right), \tag{2.11}$$

где $D = 2{,}65$ эВ - энергия диссоциации $H_2^+ \to H + H^+$, $g_{H^+} = 1$, а $Z(H_2^+)$ - есть внутренняя статистическая сумма для молекулярного иона $H_2^+$, учитывающая сумму всех вращательных и колебательных уровней $H_2^+(X\ {}^2\Sigma_g^+)$ [25-27].

Здесь выписаны только основные реакции, приводящие к ионизации молекул и атомов водорода. Полное описание требует, конечно, учета всех ветвей и цепочек реакций, включая реакции с возбуждением электронных состояний молекул и атомов, переходами в спектрах колебательных и вращательных уровней молекул, а также учета реакций с атомами гелия и других примесных компонент.

Наша задача, однако, сводится к тому, чтобы указать физический механизм - каким образом может происходить электризация встречных потоков газа, имеющих разную среднюю температуру.

Главным условием остается возможность установления в системе термодинамического равновесия, пусть даже локального, т.е. квазиравновесия. Это условие мы считаем выполненным, в частности, в пределах как самих движущихся потоков, так и в тонких слоях пограничной области трения, сталкивающихся потоков газов (см. Рис. 2.1).

Слои на границе самих потоков и области «трения» являются областями частичного взаимного проникновения заряженных частиц, т.е. ионов, ионных молекул и электронов.

## 2.2 Модель формирования «грозовых облаков» в атмосфере Солнца

Обратимся к потокам в хромосфере Солнца. Анализ наблюдений и расчеты показали, что даже при температуре 6000 ºK отношение концентраций отрицательного иона водорода и нейтрального атома водорода равно $N_{H^-} / N_H \approx 10^{-8}$ [25], т.е. газ слабо ионизован.

Интересным является также то, что при температурах T ~ 5000 - 15000 ºK полная концентрация молекулярных ионов водорода $H_2^+$ оказывается того же порядка, что и полная концентрация ионов $H^-$. Однако, с уменьшением температуры, в частности, в области T < 3000-3500 ºK, концентрация $N_{H_2^+}$ становится много большим, чем $N_{H^-}$ (Рис.2.2). Следует отметить, что даже при достаточно высоких температурах концентрация нейтральных молекул водорода $H_2$ в основном электронном состоянии на несколько порядков превышает концентрацию ионов $H_2^+$. Например, в фотосфере

$$N_{H_2} / N_{H_2^+} \approx 10^4 - 10^5 \ .$$

Данные по зависимостям от температуры относительных концентраций различных ионов водорода приведены в работе Собельмана с соавторами [25].

Важным для нас является то, что даже при слабой ионизации температурные различия в концентрациях ионов разных зарядов, в частности, ионов $H_2^+$ и $H^-$, ведут к обогащению ионами $H_2^+$ слоев в области трения, имеющих более низкую температуру, а ионами $H^-$ слоев этой области, имеющих более высокую температуру.

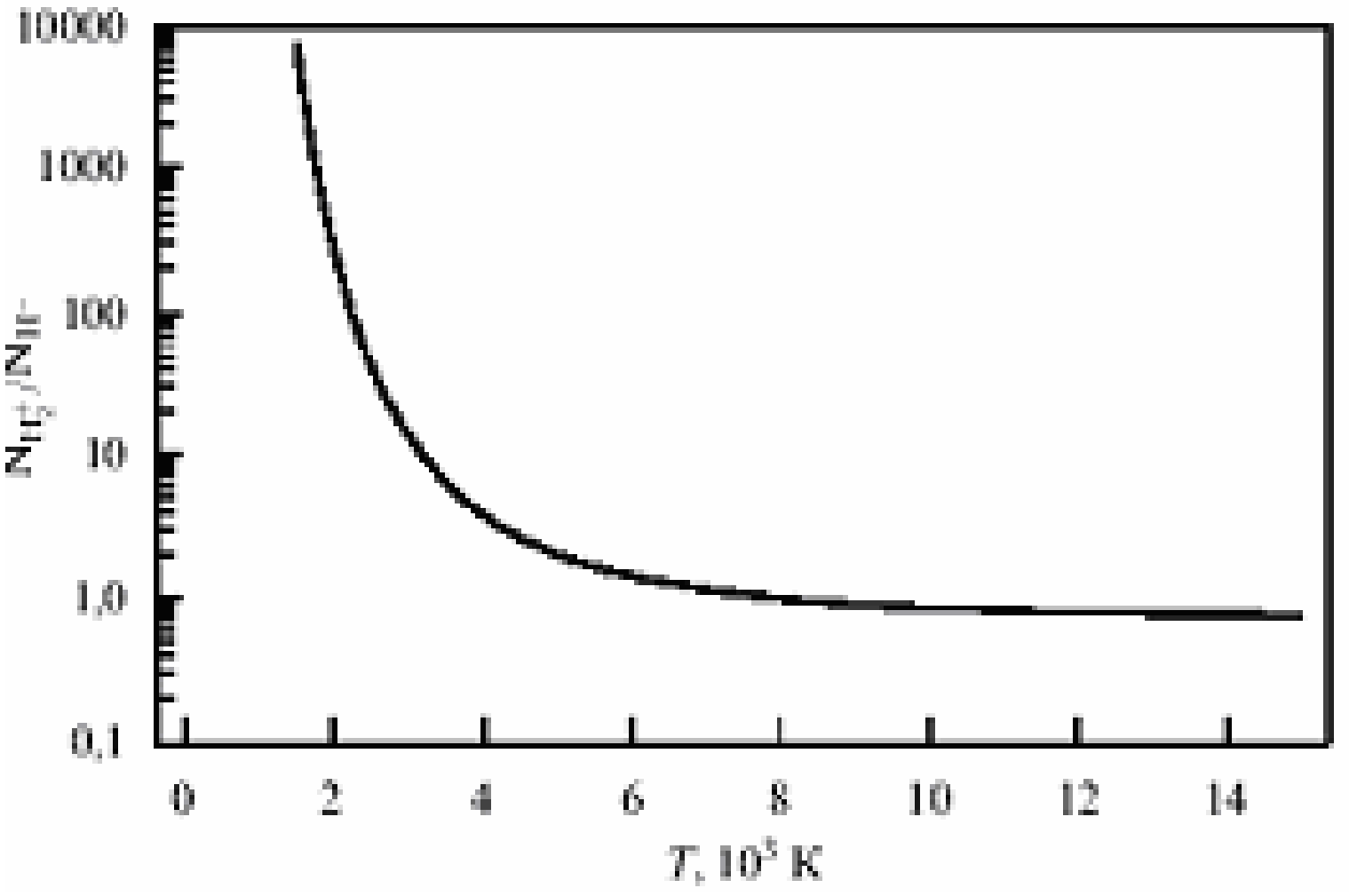

Рис. 2.2 Отношение концентраций $N_{H_2^+} / N_{H^-}$ [25].

На Рис. 2.3 изображены потоки и локальные области, имеющие разные температуры. Поток, идущий слева направо, имеет, например, температуру Т1 много меньшую, чем температура Т2 потока, идущего справа налево.

В зоне ионизации и трения могут возникать турбулентные движения и плазменные колебания свободных электронов относительно положительных ионов. Часть этих электронов будет захватываться ионами или атомами в той части зоны, где условия для их захвата будут преимущественными.

Так, в схеме рисунка 2.3, в нижней части зоны трения и ионизации температура выше и более эффективно идет образование Н- , которые будут проникать и скапливаться в потоке с температурой Т2 .

В верхней части зоны трения температура ниже и более эффективно образуются молекулярные водородные ионы $H_2^+$, которые будут проникать и скапливаться в потоке с более низкой температурой Т1 .

Обсудим физическую причину электризации солнечных «грозовых облаков». В центральной области трения темпера-тура будет самой высокой и степень ионизации также самой высокой. Здесь газ будет состоять из молекул и атомов водорода и ионов: $H^-$, $H_2^+$, $H_2^-$, а также и свободных электронов, являясь в целом электрически нейтральным. Свободных электронов будет еще очень мало, и они будут быстро захватываться атомами и положительными ионами.

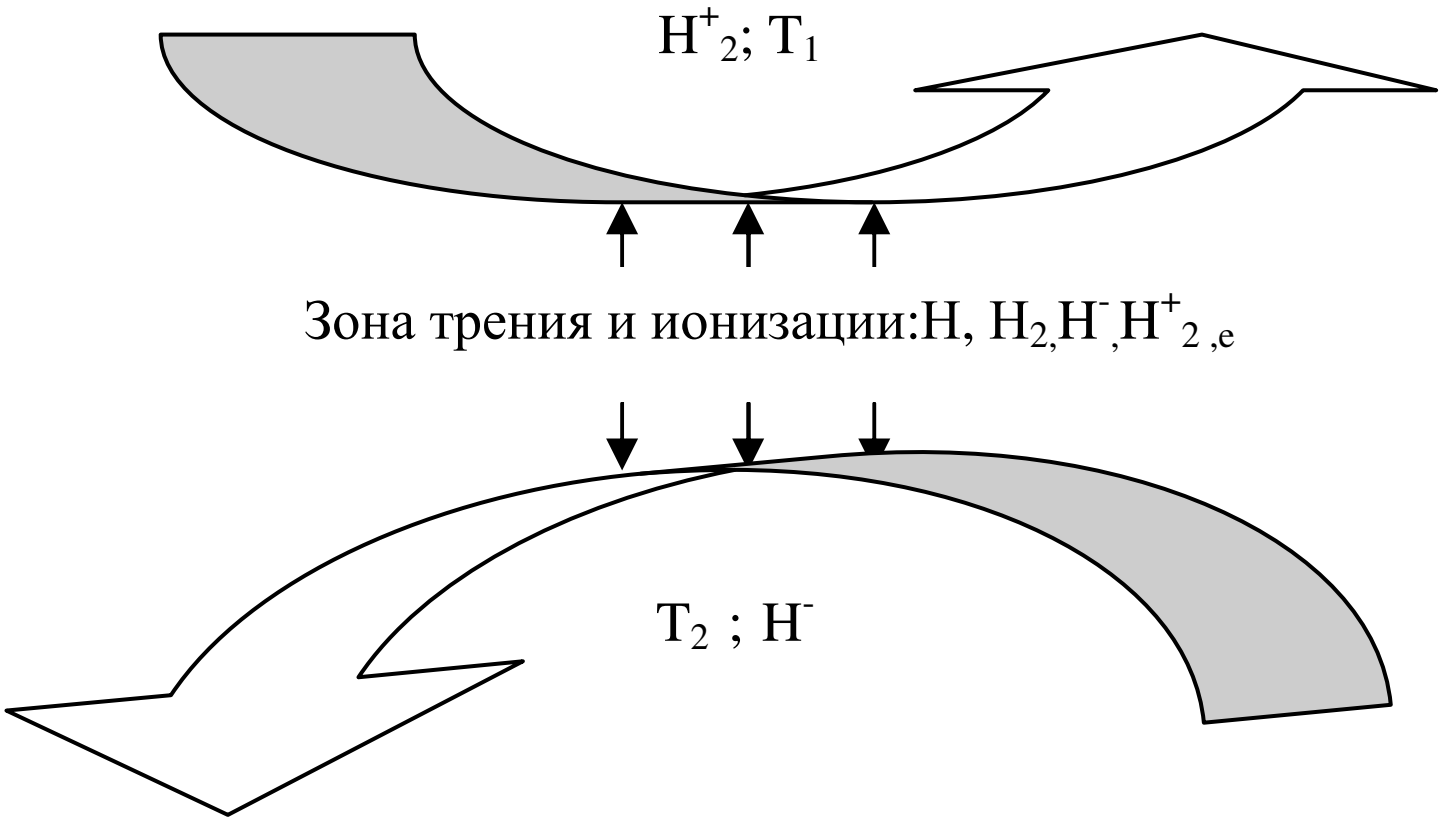

Рис. 2.3 Схема движения встречных потоков с разной температурой. Показано направление повышения концентраций ионов, имеющих разные темепратурные зависимости (обогащение слоев разноимен-но заряженными ионами в области «трения»).

В слоях ближе к области 2 с температурой $T_2 > T_1$ атомы и ионы водорода $H^-$ будут увлекаться потоком немного больше, чем фракция из тяжелых молекул.

В слоях ближе к области 1 будет больше фракции из тяжелых частиц - молекул водорода и молекулярных ионов $H_2^+$ и $H_2^-$. Отметим, в этой связи, что согласно энергетическому балансу реакций в (2.8) и (2.9) преимущество в концентрации ионной молекулы $H_2^+$ будет явным перед ионной молекулой $H_2^-$.

Таким образом, противопоточная масс - диффузия и термодиффузный и термодинамические эффекты приведут к постепенному разделению заряженных фракций, и создавая своеобразное газоплазменное «электрическое динамо».

В процессе движения встречные потоки будут обогащаться ионами противоположных зарядов и, удаляясь друг от друга, будут создавать большие по величине объемные некомпенсированные заряды.

Сформировавшиеся «заряженные облака» образуют в атмосфере Солнца сложную систему электрических и магнитных полей, дополняющих электрические и магнитные поля от других структур Солнца и его атмосферы.

Пользуясь данными по температурным зависимостям ионных концентраций можно оценить суммарный некомпенсированный объемный заряд солнечных «грозовых облаков». Они формируются в сталкивающихся потоках газов на границах спикул в хромосфере Солнца. Считаем, что эти некомпенсированные заряды в расходящихся потоках равны между собой, но противоположны по знаку.

Объемный некомпенсированный заряд в облаке равен произведению единичного заряда на полное число ионов (двух и многократную ионизацию будет считать малым по величине)

$$Q = e \cdot N_{H^-}, \tag{2.12}$$

где e - единичный заряд, $N_{H^-}$ полная сумма ионов в газовом потоке. Эту сумму можно определить, зная размеры и температуру потока, и концентрацию ионов данного сорта.

Из данных наблюдений и спектрального анализа следует, что

$$b_H = \frac{N_{H^-}}{N_H} = 10^{-8}$$

при температуре $T = 6000K$. Полное число нейтральных атомов в потоке $N_H = n \cdot V_O$, где n - число атомов в единице объема (концентрация атомов), $V_O$ - объем облака (потока) водородного газа. На высотах хромосферы примем, для примера, что концентрация частиц газа равна $n \approx 2{,}7 \cdot 10^{17}$ м$^{-3}$ , а размер газового облака равен

$$V_O = 10^4 \cdot 500 \cdot 500\ km^3 = 2{,}5 \cdot 10^{18}\ м^3 ,$$

т.е. это небольшое облако по солнечным масштабам.

При разности температур потоков

$$\Delta T = T_2 - T_1 \approx 3000\ ^\circ K ,$$

оценки дают громадное число отрицательных ионов в «грозовом облаке» (Рис. 2.3)

$$N_{H^-} \approx 10^{-8} \cdot N_H \approx 6{,}7 \cdot 10^{35} \qquad (2.13)$$

и огромные по величине некомпенсированные объемные заряды, уносимые каждым из потоков

$$Q = e \cdot N_{H^-} \approx 10^{17}\ Кл . \qquad (2.14)$$

Эти «грозовые облака» будут образовывать мощную сеть электрических полей в солнечной атмосфере, а с учетом движения потоков - формировать сильные магнитные поля. Первопричиной энергии этих полей является, конечно, гигантская энергии вращения Солнца и тепловая энергия, исходящая из его глубин.

Для проведения точных расчетов, конечно, следует принять во внимание несколько важных факторов. Это турбулентность потоков, влияние магнитных полей, воздействие колебаний плотности, связанных с ударными и периодическими волнами в атмосфере Солнца, действие потока излучения и т.п.

Неравномерности извержения вещества и излучения из

Солнца в его атмосферу, сложная структура меняющихся магнитных полей и конвективных потоков создают условия для постоянного формирования разноименно заряженных газовых потоков. Их разряды - «солнечные молнии», представляет собой перманентно проходящие процессы, которые служат фоном хорошо известным вспышкам и активным областям в солнечной атмосфере.

# 3. ГЕНЕРАЦИЯ НЕЙТРОНОВ В АТМОСФЕРЕ СОЛНЦА

Касаясь солнечных атмосферных процессов, отметим их скоротечность и энергичность, например, изменение активных областей, направлений и форм движений газовых потоков и плазмы, магнитных полей и т.п. Известно, что солнечные вспышки захватывает области не только хромосферы, но и фотосферы и короны. При этом выделяется огромная по величине энергия в виде электромагнитного излучения, спектр которого простирается от радиоволн до высокоэнергичных γ-квантов, выбросов огромных масс вещества и солнечных космических лучей (СКЛ).

Рассмотрим солнечные процессы формирования «грозовых облаков» и развития «молний» с целью поиска новых ветвей и типов ядерных реакций, характерных именно для атмосферных сред. Действительно, в отличие от плотной плазмы в глубине Солнца, в ее относительно разреженной и прозрачной атмосфере влияние на слабо заряженные газоплазменные потоки со стороны внешних электромагнитных полей и жесткого γ излучения может быть определяющим для атмосферных ядерных реакций.

Речь идет, прежде всего, о фотоядерных реакциях и коллективных эффектах ускорения частиц, например, электронов.

Напомним в этом плане некоторые особенности солнечной атмосферы. Солнечная атмосфера содержит практически весь состав известных химических элементов, но подавляющим по числу атомов является водород, почти на порядок меньшим гелий и еще почти на порядок - остальные элементы. Изучение солнечного спектра позволяет определить не только средние концентрации, но и временные их изменения [25, 26].

**Солнечные атмосферные ядерные реакции.** Результаты определения химического состава атмосферы Солнца показывают, что отношение концентраций бериллия и лития меняется со временем, причем это, наверное, связано с циклами солнечной активности. При высокой активности и, соответственно, при большой мощности вспышек, это отношение больше единицы. При уменьшении солнечной активности это отношение значительно уменьшается.

По существующим представлениям об обилии (распространенности) химических элементов следует, что бериллия в атмосфере Солнца должно быть существенно меньше чем лития.

В работах Кужевского Б.М. (см. [3, 28] и приведенные там ссылки) было указано, что такое возможно, если этими элементами являются радиоактивный изотоп ${}^{7}_{4}\mathrm{Be}$ и стабильный элемент ${}^{7}_{3}\mathrm{Li}$.

Тогда, если наиболее распространенные после водорода изотопы гелия ${}^{4}_{2}\mathrm{He}$ и более редкий ${}^{3}_{2}\mathrm{He}$ вступают в реакцию синтеза, то образуется ${}^{7}_{4}\mathrm{Be}$

$${}^{4}_{2}\mathrm{He}+{}^{3}_{2}\mathrm{He}\to{}^{7}_{4}\mathrm{Be}+\gamma. \qquad (3.1)$$

Более тяжелые изотопы бериллия, в частности, ${}^{9}_{4}\mathrm{Be}$ в таких реакциях появиться не могут. В результате электронного захвата изотоп ${}^{7}_{4}\mathrm{Be}$ превращается в элемент ${}^{7}_{3}\mathrm{Li}$. Время жизни такого процесса $\approx 53$ дней.

Т.е. ядра бериллия интенсивно возникают во время активных периодов и распадаются за время сравнимое с циклом активности.

Проблема здесь остается в достаточной по величине концентрации атмосферного гелия - ${}^{3}_{2}\mathrm{He}$. Если он образуется в результате реакций, происходящих в атмосфере

$$p + d \rightarrow {}^3_2He + \gamma \ , \qquad (3.2)$$

тогда, соответственно, возникает вопрос о достаточной концентрации атмосферного дейтерия.

Для наработки дейтерия необходимо иметь достаточное количество атмосферных нейтронов

$$p + n \rightarrow d + \gamma \, . \qquad (3.3)$$

Вопрос об атмосферных нейтронах остается открытым.

Если же ${}^3_2He$ инжектируются с поверхности Солнца, то возникает другой вопрос - как они там образовались?

Существует интересная проблема, связанная с образованием этого изотопа внутри Солнца и его влияния на характер солнечных реакций синтеза (см., например, [29-31]).

Что касается атмосферных реакций синтеза элементов, то отмечается, что они происходят почти непрерывно, усиливаясь в областях и периоды активности [3]. В этих реакциях могут вырабатываться нейтроны, которые, приходя в тепловое равновесие со средой, эффективно захватываются протонами. В результате этого радиационного захвата - (3.3), образуется дейтерий и высокоэнергичный γ-квант с энергией 2,223 МэВ. Сделанные оценки показали, что в одной солнечной вспышке синтезируется в среднем около тонны дейтерия [3].

## 3.1 Водород и его ионы

Известно, что основной вклад в солнечные спектры поглощения и излучения света в инфракрасной, видимой и УФ областях дают атомы водорода, его ионы и молекулы. Выше уже говорилось, что в квазиравновесной плазме, характерной для фотосферы Солнца и при эффективной температуре ≈ 6000 K, отношение концентрации отрицательного иона водорода $H^-$ к концентрации нейтрального атома водорода H оказывается еще очень малым $\sim 10^{-8}$.

Концентрация молекулярного иона $H_2^+$ в области температур $T \sim 5000-15000K$ будет такого же порядка. В то же время, при более низких температурах $T < 3000-3500$ K отношение концентраций этих ионов меняется

$$N(H_2^+)/N(H^-) >> 1.$$

Причем даже в области высоких температур нейтральные молекулы водорода остаются в подавляющем большинстве, например, в фотосфере

$$(T \sim 6000 \text{ K}): \ N(H_2)/N(H_2^+) \sim 10^4 - 10^5 \ ,$$

т.е. газовая плазма является еще слабо ионизированной.

Еще одна, достаточно очевидная, особенность атмосферы Солнца связана с движением огромных газоплазменных потоков, имеющих отличающиеся внутренние средние температуры (см. предыдущий раздел).

Столкновение таких потоков вызывает не только повышение степени ионизации газовых масс, но и электризацию потоков. Расходящиеся потоки осуществляют пространственное разделение разноименных зарядов. Образование пространственно разделенных объемных зарядов разного знака ведет к накоплению сильных электрических напряженностей, и, затем, к развитию «солнечных молний» - мощных электрических разрядов.

Примером таких процессов являются грозовые облака, молнии, спрайты и иные электрические явления, происходящие в атмосфере Земли, но во много меньших масштабах.

Здесь нужно отметить интереснейшее физическое явление, предсказанное Гуревичем и сопутствующее земным атмосферным молниям, - это генерация энергичных потоков гамма излучения от «убегающих электронов» [32].

Унесенные высотными электрическими полями электроны, лавина которых образовалась в молниевых разрядах, ускоряются полем и образуют поток релятивистских электронов (названные Вильсоном, в свое время, «убегающими» электронами). Эти электроны дают мощное рентгеновское и гамма излучение.

Другое интересное явление в земной атмосфере связано со случаями регистрации нейтронов во время грозовых молний. Это вызвало повышенное внимание физиков - ядерщиков.

Так, при анализе вторичных явлений, вызванных сильными грозовыми процессами, были обнаружены не только сильные радиопомехи, ударные волны и скачки электрических напряженностей, но и излучение $\gamma$-квантов достаточно высоких энергий, и другие аномалии. В нескольких случаях был зафиксирован поток нейтронов, сопутствующих земным молниевым разрядам [33].

Были высказаны гипотезы по возможности образования таких нейтронов. Был предложен механизм стимулированного ядерного синтеза с участием атмосферного дейтерия [34]. В других работах обосновывается механизм генерации нейтронов в фотоядерных реакциях, связанных с потоком $\gamma$-квантов, порожденных гигантскими высотными атмосферными разрядами от убегающих релятивистских электронов [35, 36]. Оценки потока таких нейтронов, сделанные Бабичем, дали значения $N_n \sim 10^{15}$ на одно такое грозовое событие [35].

Указывается на случаи регистрации нейтронов в лабораторных экспериментах, имитирующих грозовые условия [35, 36].

Ясно, что процессы, аналогичные грозовым процессам в земной атмосфере, могут происходить и в активных областях атмосферы Солнца. По интенсивности они будут много мощнее и, соответственно, на много продуктивнее, чем на Земле. Имеющиеся снимки таких областей в УФ показывают их интересную динамику, исследовать которую представляется весьма важным.

В солнечных «грозовых» процессах основными носителями зарядов будут, конечно, электроны и ионы водорода, в то время как в земных грозовых облаках носителями зарядов являются ионы, образующиеся при диссоциации молекул воды.

**Молекулярный ион** $H_2^+$**.** Здесь мы коснемся наиболее интересного объекта в солнечной атмосфере - ионной молекулы $H_2^+$. Мощные электрические напряженности в активных областях атмосферы Солнца приведут к молниевым разрядам и потоку «убегающих» электронов. Тогда на треке прошедшей молнии должна остаться слабо заряженная плазма, которая будет состоять в основном из ионных молекул $H_2^+$.

Речь дальше пойдет о механизме стимулированного захвата электрона одним из протонов, входящих в состав трехчастичной системы $H_2^+$.

Во-первых, здесь важным является то, что внешнее, по отношению к плазменному следу молнии, слабое электромагнитное поле будет индуцировать дипольный момент у ионных молекул. Т.е. слабое внешнее поле будет деформировать электронную орбиту, например, у ионной молекулы $H_2^+$.

Внешнее поле может быть от оставшегося некомпенсированного заряда в газоплазменном облаке, либо от имеющейся системы внешних электромагнитных полей. Оно может возникнуть и в самом плазменном шнуре, так как в треке прошедшей молнии остается избыток положительного заряда, а собственные плазменные колебания также могут наводить переменный дипольный момент. Обозначим такую индуцированную ионную молекулу символом: $H_2^+(d)$.

Из данных экспериментов следует, что индуцированный (наведенный) дипольный момент $\vec{d}$ уже при незначительных по силе внешних электрических полях становится равным $\vec{d} \approx e\vec{b}/2$, где $\vec{b}$ есть межъядерной расстояние. И он почти не меняется с ростом напряженности поля. Причем в переменных полях дипольный момент меняет направление синхрон-но полю, что означает, что электрон периодически смещается то к одному, то к другому из протонов. Более того, сама ионная молекула постепенно ориентируется по направлению внешнего поля [37].

Во-вторых, близость смещенного электрона к одному из протонов означает, что интеграл перекрытия электрона с этим протоном становится существенно большим в отличие от случая, когда поле отсутствует или когда оно очень мало. Но это еще не означает, что может произойти электронный захват, поскольку главным запретом остается дефицит энергии - реакция эндоэнергетическая.

Ситуация кардинально меняется при облучении такой системы, т.е. $H_2^+(d)$, потоком жестких $\gamma$-квантов достаточных энергий: $E_\gamma \geq 0.8$ МэВ. Тогда процесс захвата электрона становится возможным, и произойдет излучение нейтрино с превращением одного из протонов в нейтрон

$$\gamma + H_2^+(d) \rightarrow n + p + \nu . \qquad (3.4)$$

Реакция (3.4) соответствует реакции фотопоглощения и вынужденного электронного захвата в молекулярном ионе $H_2^+(d)$, имеющем наведенный дипольный момент.

Влияние сильных электромагнитных полей на скорость $\beta$ - процессов изучается уже много лет [38,39,40]. Уже имеются экспериментальные результаты, и примеры стимулирования таких процессов воздействием сильных полей [41].

Свободный нейтрон, образовавшийся в реакции (3.4), сталкиваясь затем с ядрами или протонами среды, будет ими захвачен. Связываясь между собой, протон и нейтрон будут образовывать дейтрон и излучать $\gamma$-квант с энергией 2.23 МэВ.

Если число нейтронов, порождаемых таким $\gamma$ - стимулированием, будет значительным, то это станет дополнительным источником выделяемой ядерной энергии.

## 3.2 Основное электронное состояние $H_2^+$ - системы. Внешнее поле и наведенный дипольный момент

Нужно отметить, что $H_2^+$ - система является уникальным по свойствам объектом. $H_2^+$ представляет собой систему трех частиц - один легкий электрон и два тяжелых протона. Но $H_2^+$ - система является простейшей трех частичной системой, в состав которой входит только один электрон.

Общее корректное решение трех тельной задачи можно определить из уравнений Фаддеева [42]. Решения находятся, как правило, численным путем. Это создает определенные трудности анализа и интерпретации решений.

Существуют, однако, модельные задачи, в которых решения могут быть найдены в аналитической форме [43, 44]. Это создает базу для поиска приближенных решений, описывающих в главном свойства многих реальных физических систем.

В приближении Борна-Оппенгеймера система уравнений Фаддеева может быть так же, как и уравнение Шредингера разделена на электронную и ядерную части. Такое разделение автоматически возникает в классе точно решаемых модельных задач квантовой механики трех тел [43]. К этому классу относится задача рассеяния легкой частицы на двух очень тяжелых частицах, в случае, если парные t-матрицы имеют сепарабельную форму. Решение такой задачи проводится до конца и выражается в аналитической форме [44].

В приближении бесконечно тяжелых масс протонов по сравнению с массой электрона решение задачи может быть существенно упрощено. Напомним, что отношение масс электрона и протона $\frac{m_e}{M_p} \approx \frac{1}{1836}$, т.е. является очень малой величиной. Таким образом, решение задачи взаимодействия электрона с двумя бесконечно тяжелыми заряженными центрами является достаточно хорошим приближением.

Еще одна важная особенность $H_2^+$-системы заключается в том, что для одного единственного электрона области, близкие к любому из протонов, другими электронами не экранируются - таких электронов попросту нет.

В более сложных ионных молекулах наличие внутренних электронов на замкнутых атомных орбитах вокруг ядер, препятствует молекулярному электрону, приблизится к этим ядрам. Т.е. внутренние электроны на атомных орбитах могут блокировать приближение валентных (молекулярных) электронов к любому из атомных ядер.

Очевидно, что расчет молекулярной орбиты электрона будет в этом случае более простым, а анализ более ясным, чем в случае многоэлектронных молекул.

Обратимся к модельной задаче, позволяющей провести решения до конца и получить их в аналитической форме. Отсылая за подробностями к следующему разделу и работам [43,44,45], приведем здесь исходные положения и основные результаты.

Исходным считается, что парные амплитуды нам известны. Т.е. амплитуды рассеяния электрона на протоне нам должны быть известны и представлены в аналитической форме.

Далее, считается, что парные амплитуды могут быть представлены в «расщепленном» виде, что иначе называют сепарабельной формой. Такое представление может быть и приближенным - требуется, однако, чтобы сепарабельное приближение было главным, а поправки к нему малыми.

И, наконец, отношение масс легкой и тяжелых частиц должно быть тоже очень малой величиной. Как уже отмечалось выше, масса электрона почти в две тысячи раз легче массы протона, т.е. это положение выполняется.

Парные амплитуды рассеяния электрона на каждом из протонов в области низких энергий известны и могут быть записаны в аналитическом виде [46-51].

Отличительной чертой таких парных состояний является наличие спектра, т.е. связанных состояний электрона в поле ядра (протона). Для парных амплитуд это означает, что они имеют полюса при энергиях соответствующих этим связанным состояниям [46-48].

Однако, полюсные члены парных амплитуд уже имеют сепарабельную, т.е. «расщепленную», форму. Более того, в области низких энергий они вносят основной вклад в решение трех тельной задачи. Таким образом, и это исходное положение выполняется. Остается оценить величину поправок.

Для оценки поправок оставшиеся (неполюсные) члены парных амплитуд можно аппроксимировать средним эффективным полем, и взять его так же в сепарабельной форме. После этой процедуры поправка оказывается действительно очень малой - ее величина зависит от ранга сепарабельного приближения.

Проще говоря, если после введения среднего поля в форме однократного сепарабельного члена, остаток дает вклад, который желательно уменьшить, т.е. увеличить точность оценки, то этот остаток можно также аппроксимировать другим сепарабельным слагаемым и снова определить решение. Это будет аппроксимация среднего поля уже двухранговым сепарабельным членом, точность которого будет выше.

При желании процедуру можно повторить еще несколько раз и добиться очень высокой точности.

Главное, что сепарабельное представление позволяет получить решение трех частичной задачи в аналитической форме, т.е. точно.

Запишем решение задачи трех частиц в символьной форме. В рамках отмеченных допущений, амплитуда рассеяния электрона на двух тяжелых заряженных центрах принимает вид

$$\mathrm{M}(\vec{\mathrm{b}}) = \frac{1}{1-\mathrm{B}(\vec{\mathrm{b}})}\left[\mathrm{J}(\vec{\mathrm{b}}) + \mathrm{J}(\vec{\mathrm{b}})\eta\mathrm{J}(-\vec{\mathrm{b}})\right] , \qquad (3.5)$$

где

$$\mathrm{B}(\mathrm{E},\vec{\mathrm{b}}) = \mathrm{J}(\vec{\mathrm{b}})\eta\mathrm{J}(-\vec{\mathrm{b}})\eta .$$

Матрица $J(\vec{b})$ соответствует переходу электрона от одного протона к другому и содержит форм - факторы полюсных членов парных амплитуд

$$J(\vec{b}) = \int d\vec{k}\, \Lambda(E,\vec{k}) \exp(i\vec{k}\vec{b}) \ , \qquad (3.6)$$

где

$$\Lambda(E,\vec{k}) \equiv \Lambda_{ij} = \nu_i(\vec{k}) \frac{1}{E - k^2/2m_e + i\cdot 0} \nu_j(\vec{k}) \ , \qquad (3.7)$$

причем индексы $i \neq j$ отвечают номеру соответствующего тяжелого центра. $\nu(\vec{k})$ есть форм-фактор полюсного члена парной амплитуды, структура которого для уровня номера $n$ имеет вид: $\nu(\vec{k}) \cdot \eta(E) \cdot \nu(\vec{k})$ , где $\eta = 1/(E - E_n)$, а $E_n < 0$.

Используемый подход позволяет выразить энергии уровней электрона в молекуле в виде аналитических функций от атомных характеристик и параметра $\vec{b}$ - радиус-вектора, соответствующего расстоянию между тяжелыми центрами. Энергиям уровней электрона в молекуле $H_2^+$ отвечают полюса трехчастичной амплитуды рассеяния в (3.5).

Амплитуда $M(E,\vec{b})$ из (3.5) имеет полюса в точках $E = E(b) < 0$. В этих точках детерминант матрицы $D(E,b) = 1 - B(E,b)$ обращается в нуль. Нулям детерминанта в области отрицательных энергий соответствуют связанные состояния системы. Эти состояния называют связывающими молекулярными орбитами, а их энергии - энергиями связи молекулярных орбит. В нашем случае это электронные молекулярные орбиты в $H_2^+$-системе.

Если параметр $b = |\vec{b}|$ меняется, то меняется и значение $E(b)$, при котором детерминант $\| D(E,b) \| = 0$. Зависимость энергии связывающего состояния электрона от величины расстояния между двумя тяжелыми центрами представляет собой явное отражение трех частичной динамики.

Отметим, что в области положительных энергий $E > 0$, полюса амплитуды $M(E, \vec{b})$ располагаются в комплексной плоскости энергий. Соответствующие состояния системы представляют собой квазистационарные состояния. Такие состояния обычно называют антисвязывающими состояниями, поскольку их время жизни конечно.

На рис.3.1 приведена кривая для энергии основного связывающего состояния молекулы $H_2^+$ за минусом энергии связи электрона в атоме водорода $H$.

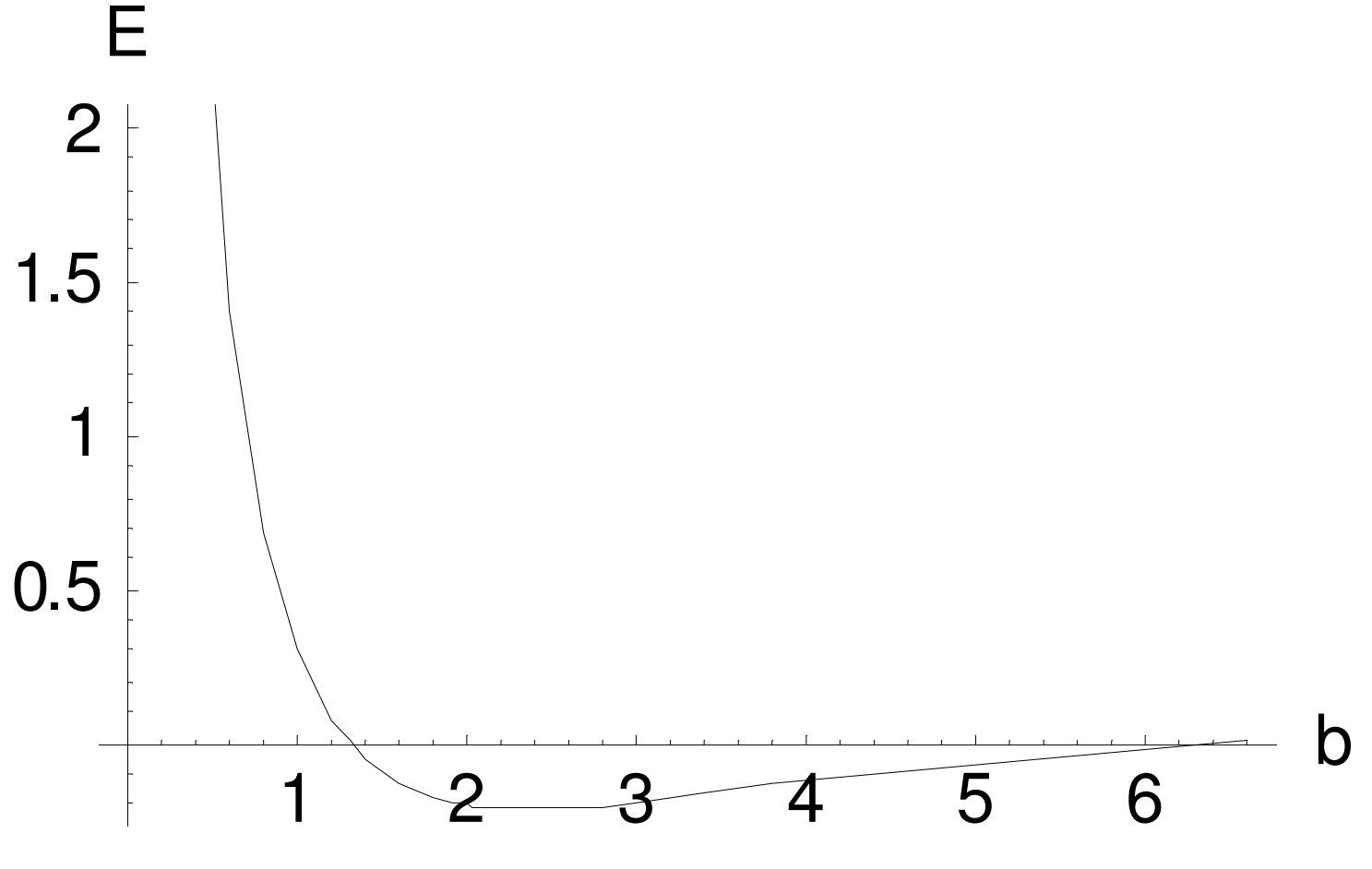

Рис. 3.1 Энергии связывающего состояния ионной молекулы $H_2^+$ как функция $b$ - расстояния между протонами.

В этом случае асимптотике при $b \to \infty$ будет отвечать развал молекулы на протон и нейтральный атом водорода. Минимум кривой на рисунке 3.1 соответствует равновесному значению межъядерного расстояния в молекуле $H_2^+$: $E_{min} = -2{,}92\ eV$ при $b = 1{,}32$ A. Экспериментальные значе-ния: $E_{exp} = -2{,}79\,eV$ и $b_{exp} = 1{,}058\,A$ [37, 46].

На рисунке 3.1 расстояние берется в единицах Боровского радиуса электрона, энергия в кулоновских единицах. В силу симметрии системы $H_2^+$ относительно перестановки протонов, волновая функция основного связывающего состояния электрона будет также обладать такой симметрией. Наибольшая плотность вероятности местоположения электрона будет точка центра масс системы, т.е. точка $\vec{r} = 0$, когда положение протонов отвечает значениям

$$\vec{r}_1 = \vec{b}/2\,, \quad \vec{r}_2 = -\vec{b}/2\,,$$

так что $\vec{r}_1 - \vec{r}_2 = \vec{b}$ есть межъядерное расстояние.

Определим теперь действие слабого внешнего поля на $H_2^+$ систему, которое должно привести к смещению электрона из центра молекулы, и создать, тем самым, наведенный дипольный момент.

Введем внешнее поле непосредственно в выражения для парных амплитуд, аппроксимируя его сепарабельными членами вида: $\pm\xi(\vec{k})\cdot x\cdot\xi(\vec{k})$, где разные знаки отличают вклады в разные по номеру парные амплитуды, форм-факторы $\xi(\vec{k})$ и параметр $x$ - подбираются таким образом, чтобы удовлетворительно описать слабо меняющееся внешнее поле $\vec{E}_{ext}$. Решения задачи можно опять получить в аналитическом виде.

Параметр $x = \vec{d}\cdot\vec{E}_{ext}/E_B$ приведен к безразмерной относительной величине, где $\vec{d}$ - индуцированный дипольный момент.

Поскольку у внешнего поля имеется направление действия, то симметрия относительно замены центров местами положений исчезает. Такой эффект всегда возникает при действии направленных внешних полей. Кривая сдвига энергии связывающего состояния как функция параметра $x$ приведе-на на рис.3.2.

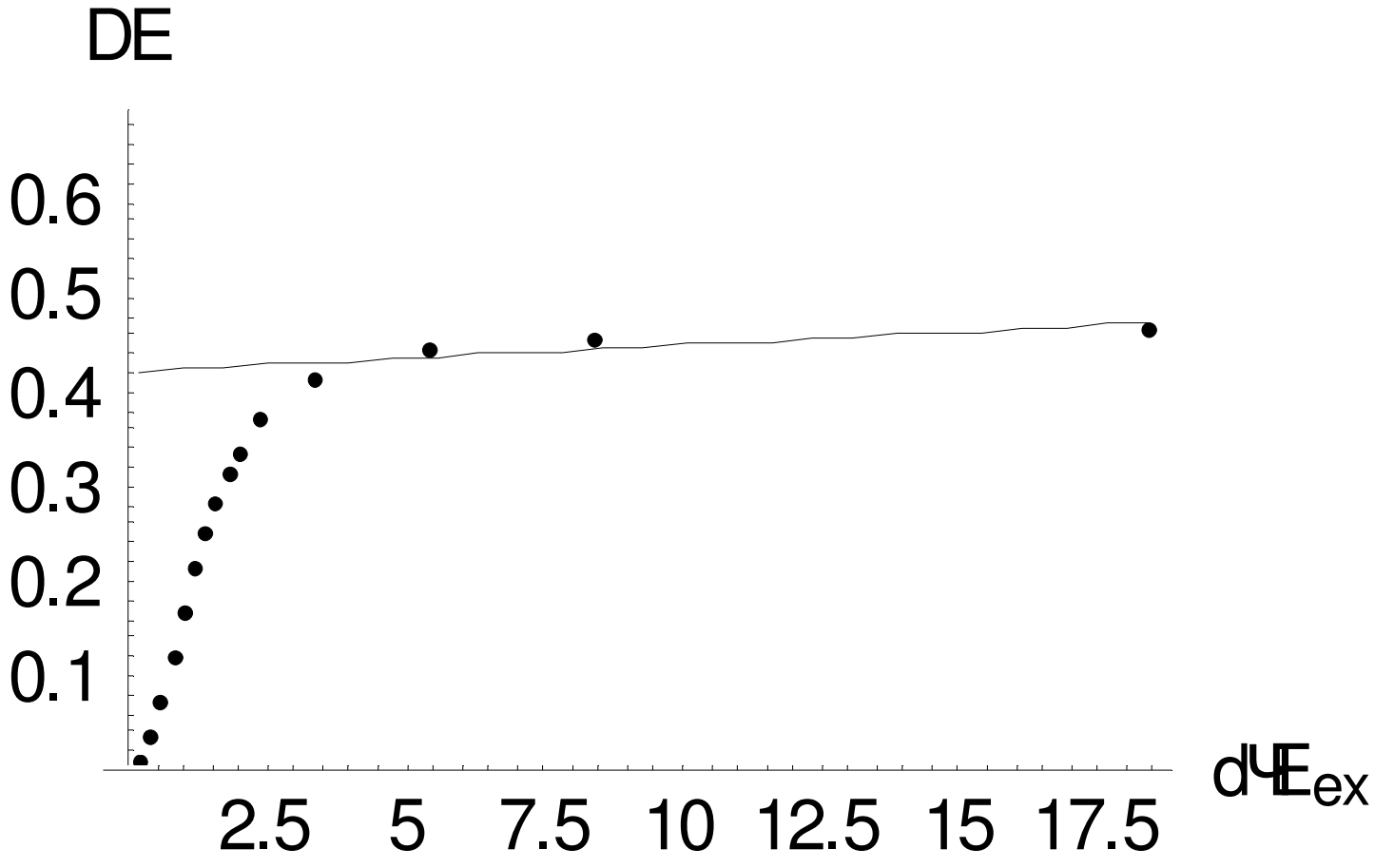

Рис. 3.2 Сдвиг энергии связывающего состояния ионной молекулы $H_2^+(d)$ как функция параметра $x = \vec{d} \cdot \vec{E}_{ext}$.

Отметим, что хотя внешнее поле нарушает симметрию относительно перестановки протонов в системе $H_2^+(d)$, кривая сдвига электронной энергии оказывается все равно симметричной относительно обращения направления поля.

Существенно, что кривая имеет излом, отвечающий определенному значению индуцированного дипольного момента. Это находится в согласии с опытными и расчетными данными [37].

Причина излома довольно проста. Легкий электрон быстрее реагирует на внешнее поле, при этом межъядерное расстояние практически не меняется. При очень малых значениях внешнего поля сдвиг энергии электрона будет связан в основном с его отклонением от равновесной точки, т.е. со смещением электрона от центра масс системы. Поскольку смещение электрона почти пропорционально величине поля, то при малых $x$ наклон кривой будет большим.

Говоря о «положении» электрона в системе $H_2^+(d)$, нуж-но, конечно, понимать, что это есть лишь мера вероятности найти электрон в соответствующей точке пространства внутри этой системы.

Далее, с ростом величины внешнего поля, наступает момент, когда смещение становится невозможным - система должна либо развалиться на несвязные подсистемы, либо, если энергии поля не достаточно для такого развала, то она должна медленно расширяться за счет роста межъядерного расстояния. Именно такое медленное расширение, и отвечает слабому наклону кривой при $x > 2{,}5$.

Кривая связывающего состояния (Рис. 3.1) будет сдвигаться вверх внешним полем и, как только произойдет ее пересечение с кривой антисвязывающего состояния, система распадется.

В случае слабо меняющихся переменных внешних полей, например, в поле лазерного облучения, описанная ситуация повторяется, но с периодическим смещением электрона то к одному, то к другому протону. Известно, что в этом случае происходит медленная «раскачка» молекулярной системы с постепенным увеличением межъядерного расстояния [37].

Интеграл перекрытия электрона и протона в $H_2^+(d)$ системе вблизи излома будет отличен от нуля и его величина может составлять десятые доли процента и более. Учитывая громадные по интенсивности потоки жесткого γ-излучения в атмосфере Солнца, можно говорить о достаточной величине выхода нейтронов, рожденных в «грозовых» солнечных облаках и молниях.

## 3.3 Решение электронного уравнения для ионной системы $H_2^+$

В этом параграфе мы рассмотрим более подробно решение задаче трех тел, о которой говорилось в предыдущем разделе.

В работах [43-45] был предложен способ получения точ-ных решений в задачах квантовой механики трех тел, в которых рассматривается рассеяние легкой частицы на паре двух очень тяжелых частиц, а парные взаимодействия имеют сепарабельную форму.

Получение точных решений означает, что решения проводятся до конца и записываются в аналитическом виде. Это оказывается возможным вследствие того, что сепарабельная (или «расщепленная») форма парных сил позволяет записать парные решения точно, а предел $\zeta = m / M \to 0$, где $m, M$ - масса легкой и тяжелой частиц, ведет к развязке кинематических переменных для разных пар частиц. В результате решения трехчастичной задачи тоже записываются точно.

В силу того, что сепарабельные потенциалы представляют собой простой класс среди многообразия различных типов взаимодействий, то в практических целях важно расширить диапазон приложений и обобщить метод получения точных решений для различных задач квантовой механики трех тел.

Ключом к реализации таких обобщений может служить «пола логия» - подход, в котором в качестве главного приближения берется вклад ближних и наиболее сильных особенностей парных амплитуд рассеяния, т.е. вклад от наиболее близких к физической области полюсов этих амплитуд [47]. Дополнительным инструментом может служить метод сепарабилизации парных $t$-матриц или парных потенциалов [48,49].

Задача $H_2^+$ имеет большую историю и исключительное значение, особенно, в квантовой теории двухатомных молекул и квантовой химии [46, 50].

В задачах по определению спектра, т.е. в области отрицательных полных энергий, для такой трехчастичной системы, состоящей из одного электрона и двух малоподвижных протонов, проведено огромное число расчетов разнообразными способами и методами. Анализ касался, прежде всего, классификации уровней молекулярного спектра и решений ядерного уравнений Шредингера, когда электронное уравнение считается уже определенным [46].

Такое разделение автоматически возникает в пределе

$$\zeta = m/M \to 0,$$

где $m$ -масса электрона, а $M$ - масса протона. Это предел называют приближением Борна-Оппенгеймера. Отметим, что полное решение задачи является весьма сложным, так как задача рассеяния трех заряженных частиц имеет свои специфические трудности, которые связаны с характером кулоновских сил, недостаточно быстро спадающих на больших расстояниях [48,51].

В общем случае, решение электронного уравнения требует преодоления этих трудностей, причем даже для определения спектра, т.е. дискретных уровней электрона в системе двух протонов, решения находятся лишь численно, например, с использованием метода линейной комбинации независимых атомных орбиталей, взятых в качестве исходного базиса [46,50].

Мы здесь рассмотрим решение электронного уравнения, записанного в интегральной форме, которое следует из уравнений Фаддеева и адекватно электронному уравнению Шредингера, имеющему дифференциальную форму.

Введем необходимые упрощения, чтобы обозначить рамки, рассматриваемой здесь модели и указать на ее определенные ограничения.

Будем считать, что в задаче на определение спектра:

1) разбиение на парные взаимодействующие подсистемы является оправданным [51];

2) главным приближением для парной t-матрицы взаимодействия электрона с протоном будут полюсные члены, отвечающие вкладу от дискретного спектра пары;

3) неполюсной «остаток» парной t-матрицы может быть аппроксимирован сепарабельным слагаемым (или суммой таких слагаемых), характеризующей эффективное «остаточное» потенциальное поле;

4) поправки к решениям в низшем приближении могут

быть определены по обычной теории возмущений.

Выше уже отмечалось, что полюсные члены парных t-матриц соответствуют вкладу от дискретного спектра, а их форма имеет сепарабельный, т.е. «расщепленный» вид. Существенно, что вычеты в этих полюсных членах выражаются через парные волновые функции этого спектра.

Будем считать, что искажения от экранировки для этих функций несущественны, по крайней мере, для волновых функций низших уровней этого дискретного спектра.

Полюсные члены парных кулоновских t-матриц возникают, если в парной системе действуют кулоновские силы притяжения, т.е. в парах, состоящих из электрона и любого из двух протонов.

Мы сохраним такие полюсные члены в неизменной форме. Остальные, т.е. неполюсные (и локальные) члены парных t-матриц заменим их сепарабельной аппроксимацией. Такая аппроксимация будет оправдана в задачах определения спектра.

В альтернативном случае, например, в задачах на рассеяние, нужно корректно учесть кулоновские асимптотики и показать малость вклада поправочных членов.

Объектом решения задачи трех тел будет амплитуда рассеяния электрона на двух протонах. В задаче на определение спектра достаточно найти полюса этой величины. Полюса амплитуды будут отвечать связанным, виртуальным или квазистационарным состояниям электрона в системе двух протонов.

Мы определим положение этих состояний, в частности, энергию уровней дискретного спектра и, затем, эффективный потенциал взаимодействия двух тяжелых центров. Этот потенциал будет входить в ядерное уравнение, определяющее вращательные и колебательные уровни самой ионной молекулы [46,50].

**Аналитическое решение модельной задачи.** Будем использовать следующие общепринятые обозначения. Запишем парную t-матрицу в виде (i - номер пары, $V_i$ - ее кулоновский потенциал и $G_i(Z)$ - ее полная функция Грина)

$$t_i = V_i + V_i G_i (Z) V_i = t_i^{sep} + t_i' \ , \qquad (3.8)$$

где сепарабельные члены t - матрицы есть сумма по всему спектру

$$t_i^{sep} = \sum_n V_i \mid \Psi_n^i > \frac{1}{Z_i + \mid E_n \mid} < \Psi_n^i \mid V_i \ , \qquad (3.9)$$

и именно они будут приниматься во внимание для определения решений задачи трех тел в низшем приближении, т.е. на первом этапе. Остальные слагаемые парных t-матриц, включая интеграл по континууму,

$$t_i' = V_i + \int d\vec{p}_S \ V_i \mid \Psi_S^i > \frac{1}{Z_i - E_S + i0} < \Psi_S^i \mid V_i \ , \qquad (3.10)$$

будем учитывать в данной схеме как поправочные, например, определять по теории возмущений. На первом этапе положим $t_i' = 0$.

В нашей задаче $\zeta << 1$, а t-матрицы в низшем приближении берутся в виде суммы сепарабельных членов

$$t_i^{sep} = \sum_n \mid \nu_n^i > \eta_n^i < \nu_n^i \mid ,$$

$$(3.11)$$

$$< \nu_n^i \mid = < \Psi_n^i \mid V_i \ , \quad \eta_n^i = \frac{1}{Z_i + \mid E_n \mid}$$

Дадим необходимые определения. Электрон обозначим номером 1, а протоны - номерами 2 и 3.

Выражение для трехчастичной T-матрицы, где

$$T = \sum T_{ij} \ , \quad i, j = 2,3 \ ,$$

будет иметь вид

$$T_{ij} = |\nu_i > (\eta_i \delta_{ij} + \eta_i M_{i,j} \eta_j) < \nu_j| \quad , \qquad (3.12)$$

где величина $M_{ij}$ есть решение уравнения

$$M_{ij} = \Lambda_{ij} + \sum_{l=2,3} \Lambda_{il} \eta_l M_{lj} \ . \qquad (3.13)$$

Потенциал $\Lambda_{ij}$ при $\zeta \to 0$ сводится к следующей форме

$$\Lambda_{ij} = 2m \frac{\nu_i(\vec{p}) \cdot \nu_j(\vec{p})}{(p_0^2 - p^2 + i\gamma)} \ , \qquad i \neq j, \quad \vec{p} = \vec{p}_1 \ , \qquad (3.14)$$

причем $\vec{p}_1 + \vec{p}_2 + \vec{p}_3 = 0$. Связь с внешними импульсами $\vec{p}_2$ и $\vec{p}_3$ можно определить в более удобной записи [44]

$$\Lambda_{ij} = \int d\vec{r} \exp(i\vec{r}\vec{p}_2) J_{ij}(\vec{r}; p_0) \exp(i\vec{r}\vec{p}_3) \ , \qquad (3.15)$$

где $J_{ij}(\vec{r}; p_0)$ есть Фурье-образ потенциала в (3.15)

$$J_{ij}(\vec{r}) = J_{ij}(\vec{r}; p_0) = \int d\vec{p} \exp(i\vec{r}\vec{p}) \Lambda_{ij}(\vec{p}; p_0) \quad . \qquad (3.16)$$

Фурье преобразование уравнения (3.13) и амплитуды

$$M_{ij} = M_{ij}(\vec{p}_2, \vec{p}_3)$$

с использованием представления

$$M_{ij}(\vec{p}_2, \vec{p}_3) = \int d\vec{r} \int d\vec{r}' \exp(i\vec{r}\vec{p}_2) M_{ij}(\vec{r}, \vec{r}') \exp(i\vec{r}'\vec{p}_3) \ ,$$

приведет тогда к уравнению для амплитуды $M_{ij}(\vec{r},\vec{r}')$ в конфигурационном пространстве

$$M_{ij}(\vec{r},\vec{r}') = J_{ij}(\vec{r})\delta(\vec{r}+\vec{r}') + J_{il}(\vec{r})\eta_l(p_0) * \\ * M_{lj}(-\vec{r},\vec{r}') \qquad (3.17)$$

Введем выражение

$$M_{ij}(\vec{r},\vec{r}') = M^{+}{}_{ij}(\vec{r})\delta(\vec{r}+\vec{r}') + M^{-}{}_{ij}(\vec{r})\delta(-\vec{r}+\vec{r}')$$

и обозначим

$$B_{ij}(\vec{r}) = \left[J(\vec{r})\eta J(-\vec{r})\eta\right]_{ij}.$$

Отметим, что по номерам пар матрица

$$B_{ij}(\vec{r}) \equiv B_{ii}(\vec{r})$$

диагональна, т.е. $j = i$. В то время, как матрица $J_{ij}$ всегда недиагональна, где $j \neq i$ (см. (3.16) и (3.14)).

Запишем точные решения задачи в виде

$$M^{+}{}_{ij}(\vec{r}) = \left[\frac{1}{1-B(\vec{r})}\right]_{ii} J_{ij}(\vec{r}), \qquad (3.18)$$

$$M^{-}{}_{ii}(\vec{r}) = \left[\frac{1}{1-B(\vec{r})}\cdot B(\vec{r})\right]_{ii} \eta_i^{-1}. \qquad (3.19)$$

Интегрирования по координате промежуточного состояния в (3.17) уже нет - значение $\vec{r}$ одно и тоже, как в начальном состоянии, так и в промежуточном и конечном со-стояниях. Отсюда следует, что $\vec{r}$ от номера канала уже не зависит, а суммирование в (3.18) и (3.19) может быть лишь по дискретным состояниям внутри пар.

Выделим далее парциальные компоненты, записывая величину $J_{ij}(\vec{r})$ в виде

$$J_{ij}(\vec{r}\,) = \sum_{LM} Y_L^{*M}(\hat{r}) J^L{}_{ij}(r\,) C^{l_j m_j}_{LM;l_i m_i} \,, \tag{3.20}$$

где $Y_L^M(\hat{r})$ - сферические функции углового момента $L$ и его проекции $M$ на выделенную ось квантования. Получим выражения для этих компонент

$$J^L{}_{ij}(r\,) = (-1)^{L/2} \frac{\Pi(Ll_i)}{\sqrt{4\pi}\Pi(l_j)} C^{l_j 0}_{L0;l_i 0} \frac{m}{\pi^2} *$$

$$* \int_0^\infty p^2 dp \frac{\nu_{l_i}(p)\nu_{l_j}(p)}{p_0^2 - p^2 + i\cdot\gamma} j_L(pr) \tag{3.21}$$

$\Pi(L\,l) = \sqrt{(2L+1)(2l+1)}$, $j_L(x)$ - функции Бесселя, а $C^{l_j m_j}_{LM;l_i m_i}$ - коэффициенты Клебша - Гордана. Далее, для краткости записи будем объединять номера каналов с номерами их парциальных компонент (в частности, $l_j \equiv j$ и $l_i \equiv i$), что не должно вызвать особой путаницы.

Аналогично (3.20) представим величины $M^{\pm}{}_{ij}(\vec{r})$ в форме

$$M^{+}{}_{ij}(\vec{r}\,) = \sum_{LM} Y_L^{*M}(\hat{r}) M^L{}_{ij}(r\,) C^{jm_j}_{LM;\,im_i} \,, \tag{3.22}$$

и получим, следуя (3.13) уравнения для компонент [45]

$$M^{L}{}_{ij}(r) = J^{L}{}_{ij}(r) + B^{LK}_{ii}(r) M^{K}{}_{ij}(r) \ . \qquad (3.23)$$

Матричная величина $B^{LK}_{ii}$ будет равна

$$B^{LK}{}_{ii}(r) = \frac{(-1)^{L-j}}{4\pi} \sum (-1)^{L_2+\tilde{L}-k} \Pi(L_1 L_2 K \tilde{L} ik) C^{\tilde{L}0}_{L_1 L_2} C^{L0}_{\tilde{L}K} *$$

$$* \begin{Bmatrix} L_1 L_2 \tilde{L} \\ iik \end{Bmatrix} \begin{Bmatrix} \tilde{L} KL \\ jii \end{Bmatrix} J^{L_1}_{ik}(r) \eta_k J^{L_{21}}_{ki}(r) \eta_i \qquad (3.24)$$

где $\eta_k \equiv \eta_k(p_0)$ и приняты стандартные обозначения для $6j$-символов в фигурных скобках.

При ограничении S-волновыми компонентами решение $M_{ij}(r)$ принимает очень простую форму

$$M(r) = \frac{J(r)}{D(r, p_0)} \ , \quad D = 1 - [J(r)\eta(p_0)]^2 \ . \qquad (3.25)$$

При учете высших парциальных волн возникает система уравнений (3.23), но если число волн ограниченно, то решения могут быть проведены до конца, и анализ особенностей амплитуд не представляет особых трудностей.

**Решения задачи рассеяния на двух фиксированных центрах.** Рассмотрим волновую функцию системы двух тяжелых частиц, фиксированных в конфигурационном пространстве. Будем считать, что частица 2 фиксирована в точке $\vec{R}_2$, а частица 3 в точке $\vec{R}_3 = \vec{R}_2 + \vec{b}$. Частицы 2 и 3 будем считать одинаковыми (протоны).

Исходя из нормировки $< \Psi_2 | \Psi_2 > = 1$ для частицы лока-лизованной в определенной ограниченной области, введем волновую функцию $\Psi_2$ в виде

$$\Psi_2(\vec{r}) = C \cdot \exp[-\frac{(\vec{r} - \vec{R}_2)^2}{2\Delta^2}] , \tag{3.26}$$

где $C^2 = \Delta^{-3}\pi^{-3/2}$. Аналогичное выражение примем и для волновой функции $\Psi_3$. Здесь предполагается, что в конечных выражениях будет взят предел: $\Delta \to 0$, с тем, чтобы «сжать» эту область и фиксировать тяжелые центры в точках $\vec{R}_2$ и $\vec{R}_3$.

Взаимодействие между тяжелыми центрами можно на этом этапе считать выключенным благодаря множителю $\zeta \to 0$. Напомним, что мы решаем здесь только электронное уравнение.

Чтобы определить физическую амплитуду рассеяния, нужно взять выражение для Т-матрицы из (3.12) в обкладках волновых функций начального $< \Psi_{in} |$ и конечного $| \Psi_f >$ состояний системы. Структура этих волновых функций очевидна, например, для $< \Psi_{in} | = < \chi_1 \Psi_2 \Psi_3 |$, где $\chi_1$ - свободная волновая функция легкой частицы (электрона).

Отделяя движение центра масс трехчастичной системы и выделяя формфакторы начального - $\nu_i(\vec{p})$ и конечного - $\nu_j(\vec{p}')$ взаимодействий легкой частицы, получим выражение для трехчастичной части амплитуды рассеяния

$$< M_{ij} > \equiv < \Psi_{\vec{R}_i}(\vec{r}) | M_{ij}(\vec{r}, \vec{r}') | \Psi_{\vec{R}'_j}(\vec{r}') > .$$

Затем нужно перейти к пределу $\Delta \to 0$, и $\delta$-функциональная зависимость величины $M_{ij}(\vec{r}, \vec{r}')$ позволит получить окончательное решение задачи. При $\Delta \to 0$ радиус вектор $\vec{r} \to -\vec{b}/2$, если $i = j$, и $\vec{r} \to \vec{b}/2$, когда $i \neq j$. Учи-тывая это тривиальное условие, можно в выражениях (3.18) - (3.25) с самого начала положить $|\vec{r}|=|\vec{b}/2|$ и исследовать свойства амплитуд $M^{+}{}_{ij}(-\vec{b}/2)$, где $i \neq j$, и $M^{-}{}_{ii}(\vec{b}/2)$, когда $i = j$.

Важно помнить, что в полученных решениях координаты и импульсы частиц задаются в системе центра инерции для трех частичной системы в целом. Поэтому радиус-вектор $\vec{r}$ будет отвечать положению той из тяжелых частиц, с которой легкая частица взаимодействует первой при входе в систему. Соответственно, радиус-вектор $\vec{r}'$ будет отвечать координате тяжелой частицы, которая взаимодействует последней с легкой частицей на ее выходе из системы. Очевидно, тогда, что

$\vec{R}_2 = -\vec{b}/2$, а $\vec{R}_3 = \vec{b}/2$.

Формальное решение в данной модельной задачи дает для полной амплитуды рассеяния выражение

$$f(\vec{p},\vec{p}') = \sum_{i=2,3} \nu_i(\vec{p})\eta_i(p_0)\nu_i(\vec{p}') + \\ + \sum_{i,j=2,3} \nu_i(\vec{p})\cdot\eta_i(p_0) < M_{i,j} > \eta_j(p_0)\cdot\nu_j(\vec{p}')| \quad . \qquad (3.27)$$

Здесь первая сумма в правой части соответствует сумме независимых парных амплитуд рассеяния легкой частицы на каждом из двух рассеивающих тяжелых центров, вторая - амплитуде рассеяния легкой частицы на двух центровой системе. Можно сказать, что во втором случае происходит многократное перерассеяние легкой частицы на двух фиксированных центрах.

В первой сумме справа в (3.27) зависимости от параметра $\vec{b}$ нет и, конечно, быть не может. Во второй сумме зависимость от параметра $\vec{b}$ есть и она является важной характери-стикой системы.

**Решение электронного уравнения в низшем приближении.** Будем пользоваться кулоновскими переменными с единицей массы равной массе электрона и энергиями уровней дискретного спектра, записанными в форме $E_n = -E_B(2n^2)^{-1}$, где $E_B = e^2 / r_B$, $r_B$ - Боровский радиус электрона.

Помня о том, что в конечных выражениях переменная $\vec{r}$ должна быть приравнена фиксированной величине $\pm\vec{b}/2$, сохраним для удобства восприятия в ниже следующих выкладках эту переменную в привычном обозначении, т.е. как $\vec{r}$.

Для простоты ограничимся $n = 1$. Форм-факторы $< \nu^i_{n=1} | = \nu_i(\vec{p})$ в (3.11) будут равны

$$\nu_i(\vec{p}) =< \Psi^i_{n=1} | V_i = \int d\vec{r} R_{n=1}(r) Y_{L=0}(\hat{r}) \frac{e^2}{r} =$$

$$= E_B r_B^{3/2} \frac{8\pi}{1 + p^2 r_B^2} Y_{L=0}(\hat{p}) \tag{3.28}$$

где $R_n(r)$ - радиальная часть кулоновской волновой функции n-го уровня дискретного спектра. Вводя безразмерные переменные соотношениями $r \equiv r / r_B$, $p \equiv p r_B$ и $p_0 \equiv p_0 r_B$, получим, следуя (3.14) - (3.16), выражение

$$J(r\ ; p_0) = 4 E_B \left[ \frac{2}{r} \cdot \frac{e^{-r} - e^{i p_0 r}}{(1 + p_0^2)^2} + \frac{e^{-r}}{1 + p_0^2} \right] . \tag{3.29}$$

При этом из (3.11) следует

$$\eta_i(p_0) = \eta_n(p_0) = 2 \cdot E_B^{-1} / (p_0^2 + n^{-2}) ,$$

что дает для $n = 1$

$$\eta(p_0) = 2 \cdot E_B^{-1} / (1 + p_0^2) .$$

Отметим, что произведение $J(r; p_0) \cdot \eta(p_0)$ будет безразмерным.

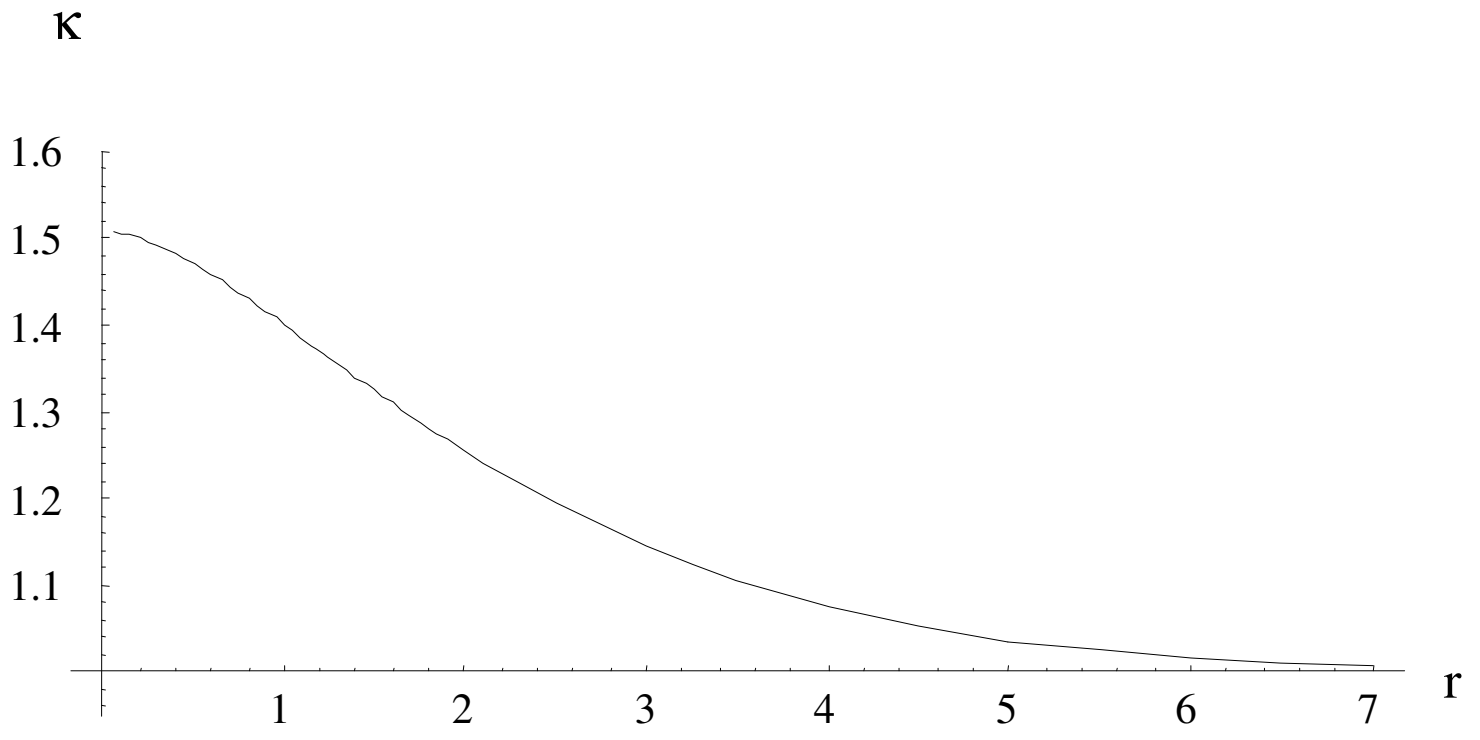

Рис. 3.3 Зависимость $\kappa = \kappa(r)$, где $r$ есть расстояние между протонами. Расстояние берется в единицах Боровского радиуса. Волновое число нормировано на волновое число основного состояния электрона в атоме водорода.

Определим нули функции $D(r, p_0)$ из (3.18)-(3.19) и (3.25). На мнимой положительной полуоси комплексной плоскости $p_0$, т.е. при $p_0 = i\kappa, \kappa \geq 0$, нули функции $D(r, i\kappa)$ будут отвечать связанным состояниям электрона в системе двух протонов. На рисунке 3.3 приведена кривая $\kappa = \kappa(r)$, отвечающая траектории движения особенности $D(r, i\kappa) = 0$. Она соответствует основному состоянию - «связывающей» орбитали электрона. Кривая демонстрирует, что энергия связи электрона в системе двух протонов больше ее энергии связи в атоме водорода, для которого $\kappa_H = 1$.

Особенности $D(r, i\kappa) = 0$ имеются и при меньшей связи:

$0 \le \kappa < 1$. В рамках данного приближения эти слабосвязанные состояния возникают лишь в области $r > r_{crit} = 3.49$ и соответствуют «антисвязывающим» орбиталям или возбужденным состояниям электрона в системе.

Нули функции $D(r, p_0)$ возникают и при комплексных значениях: $p_0 = \pm p_R + i \cdot p_I$, причем, если $p_R = \mathrm{Re}\, p_0 \ne 0$, то $p_I = \mathrm{Im}\, p_0 < 0$. Такие нули отвечают квазистационарным (или резонансным) состояниям электрона в системе двух протонов.

Нули функции $D(r, p_0)$ образуют в плоскости $p_R$ и $r$ ряд отрезков (траекторий), ограничивающих допустимые значения $p_R$ и $r$. Для резонансных состояний условие $\mathrm{Re}\, D = 0$ ограничена областью значений $0 < p_R < 0.87$, и есть несколько точек, где $\mathrm{Im}\, D$ становится равным нулю.

Можно сказать, что система $H_2^+$ имеет значительно более богатый спектр состояний по сравнению со спектром атома водорода.

Рассмотрим теперь, что дает основное «связывающее» решение, для которого функция $D(r, i\kappa) = 0$. Если считать, что положение протонов фиксировано и расстояние между ними равно $b$, то, следуя (3.26), мы должны положить $r = b/2$ (см., например, [43-45]) и перейти к решению ядерного уравнения.

Ядерное уравнение будет уже определять состояния двух протонов в эффективном поле быстродвижущегося связующего электрона [46,50]. Энергия электрона $E_e = -E_B \cdot \kappa^2 / 2$ в сумме с потенциалом кулоновского отталкивания между протонами создает эффективную потенциальную функцию для этого ядерного уравнения.

Минимум этой функции по величине $b$ отвечает равновесной конфигурации молекулы и определяет равновесное расстояние между протонами в молекуле $H_2^+$ и ее энергию диссоциации на атом водорода и свободный протон.

В рассмотренном здесь приближении равновесное рас-стояние оказывается равным $b_{eq} = 1.59$ A°, а энергия диссоциации - $E_{dis} = 1.21$ eV, что по порядку величин согласуется с данными эксперимента: $b_{exp} = 1.058$ A° и $\Delta E_{exp} = 2.79$ eV. Но результаты хуже, чем в стандартных расчетах, например, в методе линейной комбинации атомных орбиталей [46].

Включение эффективного поля и определение равновесного состояния молекулярного иона $H_2^+$. Результаты низшего приближения можно существенно улучшить, если учесть в парных $t$-матрицах вклады неполюсных слагаемых из (3.10) и вклады высших уровней дискретного спектра с $n \ge 2$.

Сначала оценим соответствующие вклады по теории возмущений, следуя соотношению Гельмана - Фейнмана [52,53] (см. также [54,55,47])

$$\Delta E = \frac{<\Psi_0 \mid \Delta V \mid \Psi_0>}{<\Psi_0 \mid \Psi_0>}\,, \qquad (3.30)$$

где $\Psi_0$ - волновая функция связанного состояния электрона, а $\Delta V$ - соответствующий поправочный член. Здесь неизвестной величиной является $\Psi_0$, которую мы определим из уравнения

$$\Psi_0 = G_0(E_0) V \Psi_0\,, \quad \text{при} \quad E_0 = -E_B \frac{\kappa_0^2}{2}\,, \ \text{а}$$

$$V\Psi_0 = \text{Const} \cdot M(p, i\kappa_0)\,. \qquad (3.31)$$

Хотя точный вид волновой функции $\Psi_0$ - электрона в системе двух протонов, нам неизвестен, но известна амплитуда $M(r, i\kappa_0)$ как решение (3.25). Это дает возможность записать в импульсном представлении выражение

$$\Psi_0(p) = \mathrm{Const} \cdot G_0(E_0) \int d\vec{r}\, \frac{\sin(pr)}{pr} M(r, i\kappa_0) \;, \qquad (3.32)$$

и из условия $< \Psi_0 \mid \Psi_0 >= 1$ найти нормировочную константу $\mathrm{Const}$.

Действительно, амплитуда вблизи полюсной точки представима в виде

$$M(r; p_0 \to i\kappa_0) \approx 4 \cdot \frac{J(r; i\kappa_0) \cdot J(r; i\kappa_0)}{p_0^2 + \kappa_0^2} \;. \qquad (3.33)$$

С другой стороны, можно записать общую форму для полюсного члена, отвечающего вкладу в t-матрицу от основного связанного уровня

$$t_{pol} = V \mid \Psi_0 > \frac{1}{E - E_0} < \Psi_0 \mid V \;. \qquad (3.34)$$

Сравнивая вычеты полюсов в (3.32) и (3.33) при $E = E_0$, получим, что $J(r; i\kappa_0) = \sqrt{2} \cdot V \mid \Psi_0 >$. Это дает нам дополнительный критерий для проверки сделанных оценок.

Вклад уровня с $n = 2$ дает поправку к энергии связи низшего состояния $H_2^+$ менее 3%, в то время, как неполюсная часть из (3.10), т.е. когда берется $\Delta V = t'$, вносит более ощутимый вклад ~ 10%. Поэтому представляется важным учесть вклад неполюсных слагаемых (3.10) с самого начала, т.е. уже в исправленном низшем приближении с тем, чтобы получить новые более точные решения задачи.

Для этого представим $\Delta V = t'$ в сепарабельном виде. Используем хорошо разработанную процедуру сепарабилизации потенциалов произвольной формы (см., например, [48,49])

$$\Delta V = \sum \Delta V \mid \chi_m > D_{mk} < \chi_k \mid \Delta V \;,$$

$$D^{-1}_{mk} = <\chi_m | \Delta V | \chi_k> , \qquad (3.35)$$

где набор функций и точки $m, k$ подбираются так, чтобы обеспечить выполнение определенных условий, например, из условия минимума вариаций энергии основного уровня относительно смещения положения протонов [45,46]. Из условия минимума равновесной энергии, следуя соотношению Гельмана - Фейнмана, определим параметры, введенные в процедуре сепарабилизации.

Функцию $\chi(r)$ возьмем в простой форме

$$\chi(r) = A \cdot \exp(-\tau * r),$$

и произведем замену

$$t_i' \Rightarrow \tilde{t}_i^{sep} = \frac{V_i | \chi >< \chi | V_i}{<\chi | V_i | \chi>} . \qquad (3.36)$$

Здесь принято во внимание, что доминирующим в (3.10) будет первое слагаемое справа, т.е. кулоновский потенциал, поэтому в правой части (3.36) берется $t_i' \approx V_i$.

Теперь парная амплитуда будет суммой двух сепарабельных членов $t_i^{sep} + \tilde{t}_i^{sep}$, полюсного слагаемого $t_i^{sep}$, отвечающего основному уровню электрона в атоме водорода (3.11), и эффективного поля «расщепленного» вида - $\tilde{t}_i^{sep}$, описываемого правой частью (3.36). Соответственно, в выражениях для матриц $J_{ij}, B_{ii}, M_{ij}$ нужно учесть, что в каждом парном канале число функций $\eta_{i,k}$ и форм-факторов $\nu_{i,k}(p)$ будет равно 2 ($k = 1,2$)

$$\nu_{i,1} = <\Psi_{n=1} | V_i \ , \quad \eta_{i,1} = 2 \cdot E_B / (1 + p_0^2) \ ,$$

$$\nu_{i,2} = <\chi | V_i \ , \quad \eta^{-1}_{i,2} = <\chi | V_i | \chi> , \qquad (3.37)$$

и ранг матриц $J, \Lambda, B$ и $M$ должен быть удвоен.

Отметим, что у определяемой величины в (3.36) зависимости от нормировочной константы $A$ не будет, и подгоночным параметром будет только величина $\tau$. Значение $\tau = 2.2$ хорошо аппроксимирует опытные данные. Расчеты дают оценки для равновесного расстояния между протонами значение - $b_{eq} = 1.248\ A°$, и для энергии диссоциации - $E_{dis} = 2.92\ eV$, которые уже удовлетворительно согласуются с экспериментальными данными.

Выше, на рисунке 6 уже был приведен результат расчетов - график энергии основного «связывающего» состояния для $H_2^+$ как функции от b, где b - расстояние между протонами (см. предыдущий раздел).

Отметим, что одна из важнейших задач квантовой химии и молекулярной физики - система $H_2^+$, давно ставшая эталонной для многих расчетных методов, может быть решена в классе точно решаемых моделей [44,45]. Это дает новые возможности анализа энергетических уровней системы и оценки влияния на нее внешних сил, например, воздействия сильных переменных и постоянных электромагнитных полей.

**Интеграл перекрытия.** Обсудим важный вопрос об интеграле перекрытия волновых функций электрона с ближайшим протоном в $H_2^+$ и $H_2^+(d)$ системах.

Следуя работам [38,39], отметим, что область формирования $\beta$-процесса должна быть порядка длины волны $E_\gamma$. Поглощение энергии электроном от поля облучения будет своеобразным фотоэффектом в этом процессе. И т.к. $E_\gamma \geq 0.8$ МэВ, то эти два условия и будут определять то, что мы называем интегралом перекрытия в данном случае.

Действительно, вероятность процесса в области невысоких энергий, т.е. < 10 МэВ, может быть представлена в форме [38,39,56]

$$\lambda = \frac{2\pi}{\hbar} | M_{fi} |^2 \rho_f \rho_i \quad , \qquad (3.38)$$

где $\rho_{f,i}$ - плотности конечного и начального состояний, а $M_{fi}$ - матричный элемент, который в первом приближении может быть записан в виде

$$M_{fi} = G_F \int d\Omega\, \Psi_f^* \Psi_i \ , \tag{3.39}$$

$G_F$ - константа Ферми слабого взаимодействия.

Интеграл перекрытия в (3.39) быстро убывает с ростом относительного расстояния между электроном и ядром [56]. Поэтому если для атома ${}^7_4\mathrm{Be}$ вероятность электронного захвата еще достаточно велика, то для самого атома водорода такая вероятность уже очень мала, так что даже в потоке мощного γ-облучения электронный захват в такой системе остается маловероятным (экспоненциально подавленным).

Аналогичная ситуация имеет место и для $H_2^+$ - системы, в котором относительное расстояние между электроном и любым из протонов по ядерным масштабам велико $\sim b/2$. При этом интеграл перекрытия конечного - $\overline{\Psi}(p,n,\nu)$ и начального $\Psi(H_2^+,\gamma)$ состояний будет пропорционален величине I

$$I = \int d\vec{r}_e\ \overline{\Phi}_{n,\nu;p}(\vec{R}_e) \cdot \Psi_e(\vec{r}_e) \ , \tag{3.40}$$

где приняты обозначения: $\overline{\Phi}(\vec{R}_e) = < \overline{\Psi}_{n,\nu} \mid \Psi_p >$ и $\vec{R}_e = \vec{b}/2 - \vec{r}_e$. Величина I и, соответственно, интеграл перекрытия оказываются экспоненциально малыми.

В случае же $H_2^+(d)$ -системы из-за смещения электрона к одному из протонов величина $I_d$, подобная (3.40), будет уже отлична от нуля, т.к.

$$I_d = \int d\vec{r}_e\ \overline{\Phi}_{n,\nu;p}(\vec{R}_{d,e}) \cdot \Psi_e(\vec{r}_e) =$$

$$= \int d\vec{r}_e \; \overline{\Phi}_{n,\nu;p}(-\vec{r}_e) \cdot \Psi_e(\vec{r}_e) \;, \qquad (3.41)$$

где $\vec{R}_{d,e} = \vec{R}_e - \vec{b}/2 \approx -\vec{r}_e$.

Нужно сказать, что величина интеграла перекрытия зависит от конкретного вида волновой функции связывающего электрона $\Psi_e(\vec{r}_e)$ . Эту волновую функцию удается определить лишь приближенно, например, используя метод линейной комбинации атомных орбиталей (ЛКАО).

Такая неоднозначность связана с тем, что точные решения (3.22) и (3.23) получаются лишь для амплитуд рассеяния, взятых на массовой (энергетической) поверхности. Как известно, аналитическое продолжение, т.е. сход с массовой поверхности, является процедурой не однозначной. Поэтому точное построение внемассовых амплитуд и, соответственно, точное определение молекулярных волновых функций нуждается в дополнительном анализе или использовании физически обоснованных приближений и методов. Таким, очевидно, является широко используемое в квантовой химии приближение или метод ЛКАО [46].

В нашем методе, использующем сепарабельные приближения для парных амплитуд, точные решения для амплитуд рассеяния трех частичной системы находятся точно (см. формулы в (3,18), (3,19) и (3,25)). Но, сделанные приближения задают, в свою очередь, форму решения и для волновой функции электрона, захваченного системой двух тяжелых центров - протонов. Эту волновую функцию можно символически записать в следующем виде (см., например, (3.31))

$$\Psi_e(\vec{r}_e) \equiv |\,\Psi> = G_0 V_{eff} \,|\,\Psi> \quad . \qquad (3.41)$$

Эффективный потенциал $V_{eff}$ нам неизвестен, но мы можем определить его, следуя формальным соотношениям (3.31) - (3.34), возникающего для амплитуды трех частиц при энергии $E \to E_0$, т.е. вблизи полюса, где доминирует лишь полюсной член амплитуды. Тогда $\Psi_e(\vec{r}_e)$ можно записать в

матричной форме

$$|\Psi(\vec{r}_e)>=|\varphi>\{\eta J\}. \qquad (3.42)$$

Здесь $|\varphi>$ - есть столбец, состоящий из элементов $\varphi_j$, $j=0,1,2,...,n$. Эти элементы соответствуют волновым функциям атомных электронных орбиталей - $\varphi_n$, когда $j\geq 1$, и волновой функции эффективного среднего поля $\varphi_0 \equiv \chi$ (см. (3.36) и (3.37)), когда $j=0$.

В свою очередь, матрица $\eta \cdot J$ определяется вычетом амплитуды в (3.33), элементы которой можно найти из матричных выражений (3.18) и (3.19). Норма функции (3.42) близка к единице.

Найденная таким образом волновая функция (3.42) дает значение для интегралов перекрытия: $I \sim 10^{-10}$, но $I_d \sim 10^{-2}$. Это означает, что процессы электронного захвата возможны в $H_2^+(d)$ - системе, но они не возможны в обычной $H_2^+$ -системе.

# 4. ГРОЗЫ И МОЛНИИ В ЗЕМНОЙ АТМОСФЕРЕ

Атмосферные процессы на Земле исследованы и известны нам, конечно, намного лучше, чем на Солнце. Ясно при этом, что сами физические явления и процессы, протекающие в атмосферах Солнца и Земли, остаются во многом схожими. Отличие же будет связано с величиной их физических характеристик в целом, в первую очередь, масс этих объектов и температур, как на их поверхности, так и в атмосфере. Отличается структура атмосфер и состав вещества, интенсивность протекающих процессов.

Важным для атмосферы Земли является состав воздуха (~ 80% азота и 20% кислорода) и наличие в нем паров воды. Именно молекулы воды являются главными участниками процессов ионизации и формирования грозовых облаков в нижних слоях атмосферы.

Облачные образования являются одним из наиболее интересных и многогранных явлений, характерных для атмосферы Земли. Их научным исследованиям в последние годы придают все большее значение и важность. Они представляют собой, с одной стороны, природный индикатор состояния атмосферы и изменений климата: как в локальном, так и в глобальном масштабе - т.е. как для географических регионов, так и для всей планеты в целом. С другой стороны, они являют собой инструмент воздействия природы на биосферу - т.е. воздействия Солнца и космоса, изменений в глубинах Земли влекут за собой масштабные сдвиги в характере атмосферных процессов.

Физика атмосферных явлений находится сейчас в стадии быстрого развития. Изучаются периодичность глобальных климатических изменений, локальные возмущения нижней и верхней атмосферы, эволюция во времени и связи с солнечной и космической активностью. За последние годы получено очень много опытных данных [10-13, 57-59].

Однако, на многие вопросы, касающихся облаков, их образования и динамики, до сих пор исчерпывающих ответов нет.

В данной работе мы рассмотрим ряд частных вопросов в проблеме динамики облачных образований в земной атмосфере. Это вопросы формирования объемных электрических зарядов, которые происходят при образовании и росте грозовых облаков и вопросы формирования и эволюции серебристых облаков в нижних слоях мезосферы.

В основу исследований будут положены методы термодинамического описания равновесных и квазиравновесных процессов, развивающихся в этих атмосферных явлениях.

Следует отметить, что сами термодинамические методы описания атмосферных газовых сред базируются на квантовой теории взаимодействия частиц - элементарных составляющих этих газовых сред. Т.е. в основе описания лежат квантовые свойства атомов, ионов и молекул, входящих в состав облачных образований, и характер их взаимодействий.

Таким образом, как, возможно, и следовало ожидать, именно квантовая механика способна объяснить многие важные явления и процессы, происходящие в таких макроскопических объектах, каким является газ и слабо ионизированная газовая плазма. Иными словами, квантовая механика опять выходит на первый план с тем, чтобы расставить все точки над «i».

Конечно, для полного и исчерпывающего описания атмосферных процессов, включая грозовые облака и молнии, потребуется исследовать и сами квантовомеханические системы, особенно, такие сложные, как молекулы, ионы и ионные комплексы. И, кроме того, дополнить рассмотрение теоретическими методами газодинамики, теплофизики и физики плазмы. Учесть осо-бенности кинетики и динамики турбулентных движений и ударных волн, и т.п.

Мы ограничимся низшим уровнем описания - описанием при учете характеристик квантовых состояний ионов, атомов и молекул. При небольших плотностях газов и скоростях потоков воздуха такое описание можно считать состоятельным.

Отличительной особенностью земной атмосферы является насыщенность ее нижних слоев парами воды. При этом, отличия в локальных термодинамических и кинетических характеристиках воздушных объемов приводят к существенным отличиям в скорости гидратации (захвате молекул воды) и кластеризации зарядов разного знака (т.е. образовании заряженных капелек воды с ионом или молекулярным радикалом в центре).

Разные частицы - элементарные составляющие воздуха, имеют отличающиеся энергетические уровни возбуждения, т.е. энергетическую емкость, и разную скорость реакций ионизации, диссоциации, перезарядки и рекомбинации.

Более того, термодинамические характеристики, например, удельные концентрации, различаются для разных типов частиц и по зависимости от изменений температуры среды. Таким образом, микроскопические характеристики создают условия для макроскопической дифференциации (т.е. пространственного разделения) частиц в некоторых средах.

Например, в более холодных воздушных массах образуются преимущественно отрицательно заряженные кластеры - т.е. заряженные отрицательно капельки воды, а в более теплых массах - положительно заряженные кластеры [60].

Хотя в каждом частном случае ситуация с образованием заряженных кластеров и капелек воды зависит от конкретных свойств ионов - центров гидратации и кластеризации, а также от макро характеристик воздушных потоков: температуры, скорости их движения, влажности, состава примесей и т.д.

## 4.1 Особенности грозовых облаков

Грозовыми облаками называют облака, которые сопровождаются молниями. Наличие молний говорит о процессах электрического разряда, происходящих внутри облака, или между облаками, или между облаком и землей.

Каким образом в облаке накапливаются мощные электрические заряды, как эти заряды разделяются в пространстве, и как накапливается такая гигантская энергия - эти вопросы сейчас интенсивно исследуются [6,10-13,59].

Грозовое облако обычно представляет собой множество одновременно "работающих" гроз (до полутора тысяч), распределенных в нижней части атмосферы - тропосфере.

Грозовое облако живет от часа до нескольких часов. На смену одним грозам приходят другие, формирующиеся в тропосфере по соседству. Сейчас проводятся спутниковые измерения и имеются наземные системы регистрации, на основе которых созданы базы данных по грозовым молниям [10-13].

Частота вспышек над поверхностью океана в среднем на порядок ниже, чем над континентами в тропиках. Одна из причин это интенсивная конвекция в континентальных областях, где суша эффективно прогревается солнечным излучением. Быстрый подъем прогретого насыщенного влагой воздуха способствует образованию мощных конвективных облаков вертикального развития, в верхней части которых температура ниже -40°С [10].

Над океанами высота облаков в среднем ниже, чем над континентами, и процессы электризации менее эффективны. В последнее время обсуждается и другой фактор - различие в концентрациях аэрозолей над океаном и континентами. Так как аэрозоли служат ядрами конденсации, необходимыми для образования частиц в переохлажденном воздухе, их обилие над сушей повышает вероятность сильной электризации облака. Количественный анализ этого фактора требует детальных экспериментов, которые только начинаются [6,10].

Земная атмосфера представляет собой исключительно хороший диэлектрик, расположенный между двумя провод-никами: поверхностью земли снизу и ионосферы, сверху. Эти слои являются пассивными компонентами земной глобальной электрической цепи. Между отрицательно заряженной поверхностью земли и положительно заряженной верхней атмосферой поддерживается постоянная разность потенциалов величиной ~ 300 тысяч вольт. Считается, что этот «ионосферный потенциал» является результатом заряда, получаемого от гроз, которые создают глобальную электрическую «батарею».

Обычно, нижняя часть облака, обращённая к земле, заряжена отрицательно, а верхняя часть - положительно. Космические лучи сталкиваются с молекулами воздуха и ионизируют их. Положительные заряды двигаются вниз к отрицательно заряженной земле и скапливаются под облаком, а отрицательные заряды - притягиваются к верхней части облака, заряжая его отрицательно.

При накоплении достаточного заряда происходит электрический пробой атмосферы - молния. Разряд молнии характеризуется чрезвычайно быстрым нарастанием тока до пикового значения, как правило, достигаемого за время от 1 до 80 мкс (миллионных долей секунды), и последующим падением тока обычно за 3-200 мкс после пикового значения [10,12].

Такова одна из моделей молний, которая активно исследуется в последнее время. Сложность проблемы объясняется тем, что в формировании молнии и грозы задействованы сразу несколько явлений, характерные длины которых изменяются на 15 порядков величины. Это и разделение зарядов на молекулярном уровне, и вспышки молний несколько километров длиной, и конвекционные потоки воздуха, которые могут охватывать континенты. Все эти факторы нужно рассматривать совместно, чтобы понять, как работает глобальная электрическая цепь нашей планеты.

Отмечается, что в грозовой ячейке молниевые разряды происходят в среднем каждые 15-20 секунд, т.е. действующий в облаке механизм зарядки очень эффективен, хотя средняя плотность электрического заряда редко превышает несколько нКл/м$^3$ [6,10].

Измерения электрического поля на поверхности земли и внутри облаков показали, что в типичном грозовом облаке "основной" отрицательный заряд - в среднем несколько десятков кулон - занимает интервал высот, соответствующий температурам от - 10 до -25°С. "Основной" положительный заряд составляет также несколько десятков кулон, но располагается выше основного отрицательного.

Большая часть молниевых разрядов облако-земля отдает земле отрицательный заряд. Но в нижней части облака часто обнаруживается меньший по величине (~10 Кл) положительный заряд [6, 10-13].

Для объяснения такой трипольной структуры поля и заряда в грозовом облаке рассматривается множество механизмов разделения зарядов. Они зависят от таких факторов, как температура, фазовый состав среды, спектр размеров облачных частиц.

Очень важна зависимость величины передаваемого за одно соударение заряда $\Delta q$ от электрического поля. По этому параметру принято подразделять все механизмы на индукционные и безындукционные. Для первого класса механизмов заряд $\Delta q$ зависит от величины и направления внешнего электрического поля и связан с поляризацией взаимодействующих частиц. Безындукционный обмен зарядами между сталкивающимися частицами в явном виде от напряженности поля не зависит [6, 10,13].

Несмотря на обилие различных микрофизических механизмов электризации, сейчас многие авторы считают главным безындукционный обмен зарядами при столкновениях мелких (с размерами от единиц до десятков микрометров) кристаллов льда и частиц снежной крупы (с размерами порядка нескольких миллиметров).

В лабораторных экспериментах было установлено наличие характерного значения температуры, при которой меняется знак заряда - точки реверса, лежащей обычно между -15 и -20°С. Именно эта особенность сделала данный механизм столь популярным, так как с учетом типичного профиля температуры в облаке она объясняет трипольную структуру распределения плотности заряда.

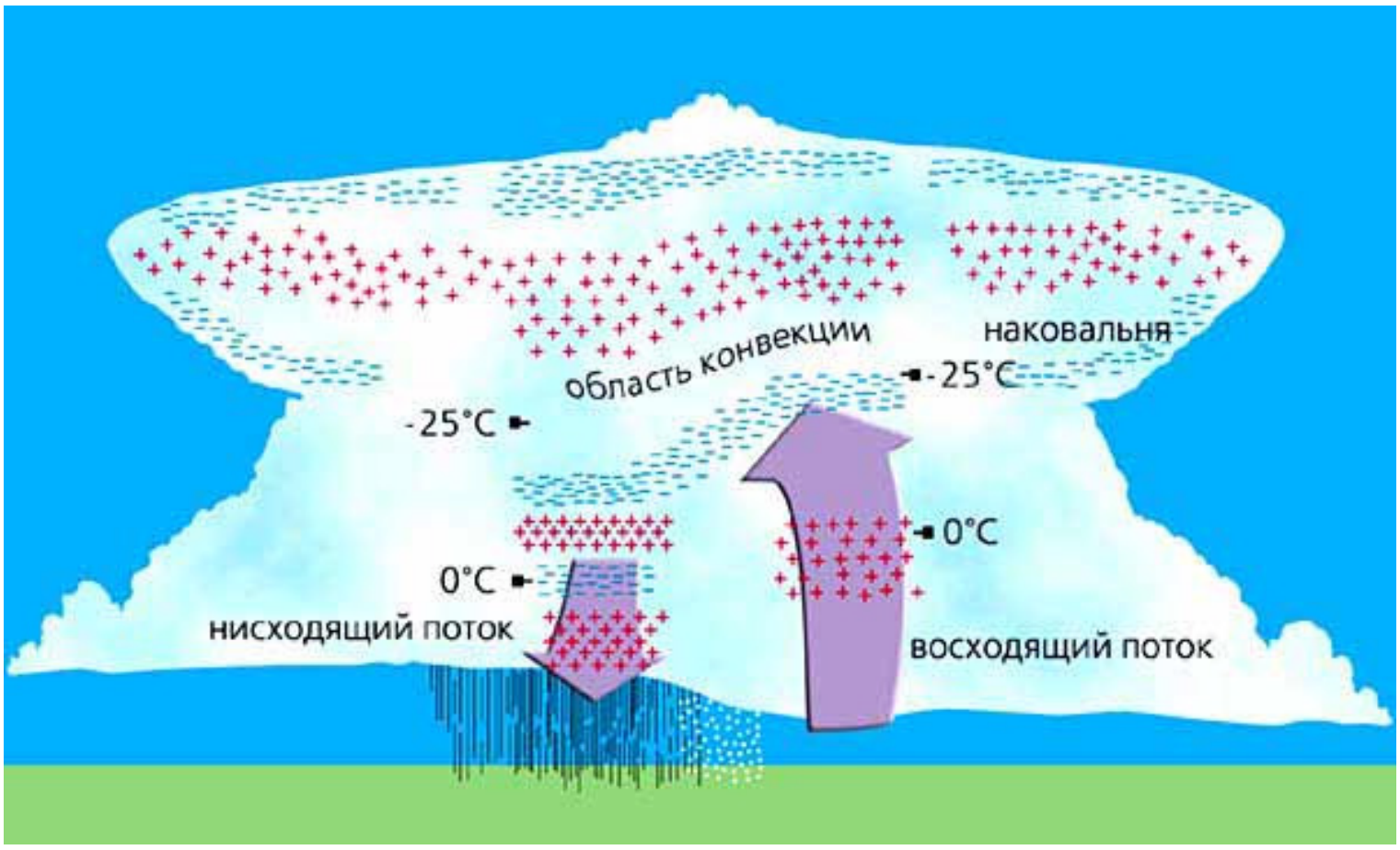

Рис. 4.1. Схематическое изображение типичного конвективного грозового облака (По: Stolzenburg M. et al. // J. Geophys. Res. 1998. V.103. P.14097-14108).

Недавние эксперименты показали [10-13,61], однако, что многие грозовые облака обладают еще более сложной структурой пространственного заряда (до шести слоев, рис.4.1). Особенно интересны мезомасштабные (с горизонтальными масштабами от десятков до сотен километров) конвективные системы, служащие важным источником грозовой активности. Их характерная черта - наличие единой электрической структуры, включающей область интенсивной конвекции и протяженную (до нескольких сотен километров) стратифицированную область.

В области стратификации восходящие потоки достаточно слабые, но электрическое поле имеет устойчивую многослойную структуру. Вблизи нулевой изотермы здесь формируются достаточно узкие (толщиной в несколько сотен метров) и стабильные слои пространственного заряда, во многом ответственные за высокую молниевую активность мезомасштабных конвективных систем. Вопрос о механизме и закономерностях образования слоя положительного заряда в окрестности нулевой изотермы остается дискуссионным. Мо-дель, основанная на механизме разделения зарядов при таянии ледяных частиц, удовлетворительно согласуется с экспериментальными данными [10-13].

Интересный механизм, где ключевую роль играют ионизационные излучения, был предложен в работе [6, 62]. К этим излучениям относятся радиоактивные газы, выносимые в атмосферу из недр земли, и поступающие из космоса галактические лучи, порождающие широкие атмосферные ливни (ШАЛ), а также солнечный ветер и рентгеновское и ультрафиолетовое излучение от Солнца.

Это излучение является основным поставщиком разноименно заряженных частиц грозового облака на стадии его зарождения. Масштабное разделение объемных зарядов происходит в облаке из-за того, что процессы конденсации водяного пара начинаются на отрицательно заряженных частицах при меньших температурах, чем они начинаются на положительных частицах. Молнии в грозовых облаках инициируются ШАЛ [6, 62].

Подчеркнем некоторые общие черты грозовых процессов, которые подтверждены метеорологическими наблюдениями и их прямыми исследованиями, лабораторными и модельными экспериментами, и присутствуют во многих теоретических моделях грозового облака или предлагаемых механизмах:

- грозовым облакам сопутствует движение больших масс воздуха, насыщенного водяным паром;
- потоки воздушных масс имеют разные средние температуры - теплые потоки пронизывают или сталкиваются с холодными потоками, конвекция потоков имеет место и внутри грозового облака;
- грозы почти не случаются на высоких широтах или в зимнее время;
- молнии инициируются внешними воздействиями (космическими лучами - ШАЛ);
- сами молнии являются мощным ионизатором воздуха;
- грозовые облака могут иметь сложную структуру, как по температуре, так и по конфигурации электрических полей;

- грозовые облака являются важной частью глобальной электрической сети планеты Земля и т.д.

## 4.2 Молекула воды. Энергии диссоциации, ионизации и гидратации

Как уже было отмечено выше, в процессах формирования облаков исключительная роль принадлежит водной компоненте воздуха. Вода может присутствовать в потоках воздуха и в облачных образованиях в форме водяного пара, водных капелек или кристалликов льда - снежинок или градинок.

Рассмотрим физические свойства молекулы воды. Молекула воды $H_2O$ имеет вид треугольника. Угол между двумя связками: кислорода с каждым из двух атомов водорода, примерно равен 104 градусов. Атомы водорода в молекуле расположены по одну сторону от кислорода, т.е. электрические заряды в молекуле рассредоточены и молекула имеет дипольный момент.

Молекула воды полярная, что является причиной особого взаимодействия между разными молекулами. Атомы водорода в молекуле $H_2O$ имеют наведенный положительный заряд, из-за смещения их электронных орбит в сторону ядра атома кислорода, и взаимодействуют с электронами атомов кислорода соседних молекул. Такая химическая связь называется водородной. Она объединяет молекулы $H_2O$ в своеобразные полимеры пространственного строения.

Плоскость, в которой расположены водородные связи, перпендикулярна плоскости атомов той же молекулы $H_2O$. Взаимодействием между молекулами воды и объясняются в первую очередь незакономерно высокие температуры её плавления и кипения. Нужно подвести дополнительную энергию, чтобы расшатать, а затем разрушить водородные связи. И энергия эта очень значительна. Это объясняет, почему так велика теплоёмкость воды.

В молекуле воды имеются две полярные ковалентные связи между атомами водорода и кислорода. Они образованы за счёт перекрывания двух одноэлектронных р - облаков атома кислорода и одноэлектронных S - облаков двух атомов водорода. Обычно, эти связи обозначают графически в виде

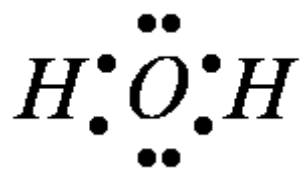

В молекуле воды атом кислорода имеет четыре электронных пары. Две из них участвуют в образовании ковалентных связей, т.е. являются связывающими. Две другие электронные пары являются не связывающими.

В молекуле имеются четыре полюса зарядов: два - положительные и два - отрицательные. Положительные заряды сосредоточены у атомов водорода. Два отрицательных полюса приходятся на две не связывающие электронные пары кислорода.

Подобное представление о строении молекулы позволяет объяснить многие свойства воды, в частности структуру льда. В кристаллической решётке льда каждая из молекул окружена четырьмя другими.

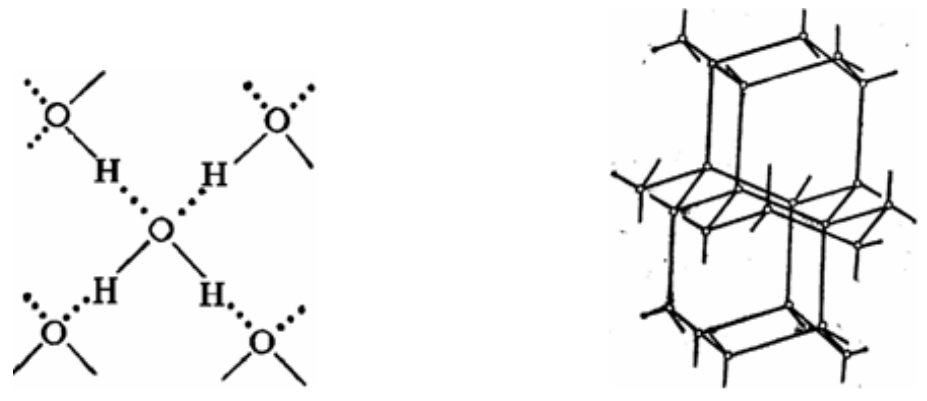

а б

Рис. 4.2 Плоскостное изображение связей молекул воды - а), и трехмерное - б).

Отметим, что связь между молекулами осуществляет атом водорода. Положительно заряженный атом водорода одной молекулы воды притягивается к отрицательно заряженному атому кислорода другой молекулы воды. Такая связь называют водородной и графически ее обозначают точ-ками. По прочности водородная связь примерно в 15 - 20 раз слабее, чем ковалентная связь. И водородная связь легко разрывается, что наблюдается, например, при испарении воды. Вследствие теплового движения молекул воды одни водородные связи разрываются, другие быстро образуются.

## 4.3 Термодинамическое описание образования водных капель в облаках

Важным фактом наблюдений и экспериментов по формированию водных капель в газовых средах является температурная зависимость этих процессов, а именно то, что отрицательно заряженные ионы образуют водные капли при более низких температурах среды [6,62].

Рассмотрим в этой связи термодинамику этих процессов гидратации. Будем считать, что в газовой среде образуется в результате столкновений достаточное количество ионов, которые могут либо рекомбинировать, либо стать центрами образования водных капель и просуществовать в таком состоянии значительно время.

Ионизационное равновесие в такой среде будем описывать методом Саха (см., например [21-23]). В классическом варианте этот метод определяет степень ионизации слаборазреженной плазмы в зависимости от температуры. Будем определять аналогичным образом степень комплексности (массивности) кластера, к примеру, кластера типа: $O^- \cdot (H_2O)_n$, и рассматривать процессы последовательного отделения (присоединения) молекул воды от ионов и ионных кластеров [60].

При термодинамически равновесных процессах в системе константы равновесия - $K^{(n)}(T)$ не зависят от деталей химических реакций, и определяются лишь энергиями соответствующих квантовых состояний и температурой. В этом случае, можно записать

$$\frac{C_{n+1}}{C_n \cdot C} = PK^{(n)}(T), \qquad (4.1)$$

где $C_n$ есть концентрация кластеров с определенным значением n, а $C$ - концентрация молекул воды в целом - т.е. как свободных, так и захваченных ионными комплексами. Величина $C$ может быть нормирована условием

$$C = \sum_{n=1}^{N} n \cdot C_n \qquad (4.2)$$

где $N$ - максимально возможное значение n.

Масса кластера, т.е. величина N, ограничена условием баланса между энергией связи и поверхностным натяжением молекул воды в кластере. Величина С - концентрация молекул паров воды, которая определяется начальными условиями задачи и считается независящей от последующих локальных изменений температуры.

Константы равновесия $K^{(n)}(T)$ могут быть определены из выражений [21,60]

$$K^{(n)}(T) = \frac{g_{n+1}}{g_n \cdot g}\left(\frac{2\pi \cdot m_{n+1}}{m_n \cdot m}\right)^{3/2} T^{-5/2} \exp\left(\frac{I_n}{T}\right), \qquad (4.3)$$

где $g_n, m_n$ есть статистический вес и масса соответствующего кластера, g, m - статистический вес и масса молекулы $H_2O$, соответственно. Величина $I_n = \varepsilon_{0,n+1} - \varepsilon_{0,n}$ есть энергия n-ого присоединения (прилипания) к кластеру молекулы $H_2O$.

Мы будем пользоваться системой единиц, в которой для простоты записи положено: $c = 1, \hbar = 1$, и постоянная Больцмана также положена равной единице: $k = 1$. Это отражено в записи формулы (4.3).

Полагая n >> 1, будем считать, что в довольно широких пределах значений n энергия прилипания молекулы к кластеру не зависит от этой величины. Действительно, в процессах роста многих кластеров такая закономерность имеет место, включая процессы гидратации [63-65]. Таким образом, можно принять, что $I_n = \Delta\varepsilon_n \approx \varepsilon_0$.

Полагая, что масса кластера растет монотонно с ростом числа n, и статистические веса кластеров при этом почти не меняются, получим, что правые части (4.3) при больших n от самой величины n перестают зависеть.

Тогда выражение (4.3) можно упростить [24,60]. Введем для этого обозначение

$$\alpha(T) = CPK^{(n)}(T) \approx CPK(T) \approx$$

$$\approx \frac{C}{g}\left(\frac{2\pi}{mT}\right)^{3/2} \exp\left(\frac{\varepsilon_0}{T}\right). \qquad (4.4)$$

Уравнение (4.1) можно тогда привести к виду

$$\frac{C_{n+1}}{C_n} = \frac{C_n + \Delta C_n}{C_n} = CPK^{(n)}(T) \approx CPK(T) = \alpha(T)\,, \qquad (4.5)$$

и, определяя функцию $C(n) \equiv C_n$, записать (4.5) в дифференциальной форме:

$$\frac{d}{dn}\ln\{C(n)\} = \alpha(T) - 1 \qquad (4.6)$$

Его решение имеет простой вид:

$$C(n) = B \cdot \exp\{n \cdot (\alpha(T) - 1)\}\,, \qquad (4.7)$$

где $B = B(T)$ можно определить из нормировочного условия (4.2), записанного в интегральной форме

$$C = \int_1^N dx\; x \cdot C(x)\,. \qquad (4.8)$$

Если параметр $\alpha = \alpha(T) < 1$, то с точностью до членов $\sim N^{-1/3}$ получим

$$B(T) \approx C \cdot \frac{(1-\alpha)^2}{(2-\alpha)} \exp(1-\alpha)\,, \qquad (4.9)$$

и, соответственно,

$$C(n) = C \frac{(1-\alpha)^2}{2+\alpha} \exp\{-(n-1)\cdot(1-\alpha)\}\ . \qquad (4.10)$$

Отсюда следует, что концентрации $C(n)$ при $n >> 1$ будут экспоненциально подавлены. Основную роль в этом случае будут играть только малые кластеры.

В ситуации, когда $\alpha > 1$, решения для $C(n)$ принимают вид

$$C(n) = \frac{C\cdot(\alpha-1)}{N} \exp\{-(N-n)\cdot(\alpha-1)\}\ . \qquad (4.11)$$

Из (4.11) видно, что будут подавлены почти все кластеры, за исключением самых массивных, т.е. тех, у которых $n \approx N$. Величина $\alpha(T)$ может быть найдена из выражения (4.4) и записана в приближенной форме:

$$\alpha(T) \approx \exp\{\varepsilon_0 / T - \beta(T)\}\,. \qquad (4.12)$$

Критическим является значение $\alpha_c = \alpha(T_c) = 1$. Когда $\alpha(T) > \alpha_c$ идет быстрое образование массивных кластеров, а когда $\alpha(T) < \alpha_c$ - их распад.

Если температура $T$ велика, то величина $\varepsilon_0 / T$ будет мала так, что $\alpha(T) < 1$. При низких температурах величина $\varepsilon_0 / T$ может стать, наоборот, очень большой и $\alpha(T) > 1$. Такие температуры и приводят к образованию массивных кластеров, т.е. крупных капелек воды, а из них - кристалликов льда.

Оценим значения параметров $\alpha$ и $\beta$ в области формирования грозовых облаков. Грубые оценки дают:

$$\beta(T) \approx 9 + 1{,}5 \cdot \ln(\frac{T}{300}) - \ln \tilde{C} \quad , \qquad (4.13)$$

где $\tilde{C}$ - концентрация паров воды, взятая в процентах. Сама эта концентрация, конечно, зависит от температуры воздуха.

В выражении (4.13) за точку отсчета взята температура 300 K.

Полагая, например, энергию прилипания равной $\varepsilon_0 \approx 0.09$ eV, получим, что для температур $T_1 \approx 283\,^\circ K$ следует $\varepsilon_0 / T \approx 3.69$. В области $T_2 \approx 303\,^\circ K$ должно быть $\varepsilon_0 / T \approx 3.45$. Тогда $\alpha$ не сможет превысить свою критическую величину, даже если концентрации паров воды будет высокой. В этом случае крупные водные капли вообще не будут образовываться.

Для ионов $H^+$, которые имеют наибольшую энергию гидратации среди легких ионов (1076 кДж/моль или 11,15 эВ), получим $\varepsilon_0 / T \approx 411 >> 1$ даже при нулевых температурах по Цельсию. При этом, величина $\alpha$ будет всегда превышать свою критическую величину и вызывать быстрое образование массивных ионных кластеров. Таким образом, свободные протоны могут практически всегда образовывать крупные водные капли.

Однако, температуры, требуемые для ионизации атомов водорода и образования свободных протонов в самой нижней атмосфере не достижимы, свободные протоны могут появиться только из-за внешних воздействий, например, от галактических космических лучей. Концентрация таких частиц в нижней атмосфере очень мала.

Аналогичные свойства, как активных центров гидратации, имеют практически все ионы средних и тяжелых элементов (см., например, таблицу 4.1).

Энергия гидратации ионного комплекса $OH^-$ равна - 339 кДж/моль (или 3.51 эВ). Это дает в области $T_1$ значение $\varepsilon_0 / T \approx 130$, что указывает на большую способность комплекса образовать крупный ионный кластер, т.е. такие ионные комплексы тоже являются центрами образования крупных водных капель.

Большое значение для процессов формирования водных капель имеет начальная концентрация молекул воды в потоке воздуха.

Конечно, сила прилипания молекулы воды к иону, имеющему большую энергию гидратации, экранируется первыми нижними слоями молекул воды, но не компенсируются полностью. Это является следствием высокого дипольного момента у молекулы воды и ее сильными водородными связями. Энергия гидратации остается на верхнем уровне энергии диссоциации кластерных ионов даже при больших значениях $n$, т.е. для комплексов $A^{\pm}(H_2O)_n$, где A обозначает ион или радикал [63,65].

Оценки показывают, что в рассматриваемых диапазонах температур и концентраций паров воды в атмосфере, процессы гидратации могут активно реализоваться, что имеет место в области трения, т.е. в полосе встречи горячих и холодных потоков воздуха.

Если исходить из модели молекулярной ионизации паров воды, т.е. образования молекулярных ионов $H_2^+$ и $O^-$, $OH^-$ и даже $H^+$ в области трения двух потоков воздуха, то полная концентрация ионов будет пропорциональна полной исходной концентрации молекул воды в этих потоках. Присутствие примесей может только увеличить эту концентрацию.

Отметим, что ионизация (или ионизационная диссоциация) молекул кислорода $O_2$ и, тем более, азота $N_2$ является практически маловероятной при атмосферных температурах, которая для этих процессов еще не достаточно высока. Так, известно, что азот часто используется как инертный наполнитель для предотвращения электрического пробоя во многих физических приборах и установках.

Относительно других примесей, например, указанных в Таблице 4.1, можно сказать, что их концентрации в атмосфере ничтожны по сравнению с парами воды и, поэтому, их вкладом можно пренебречь. Однако, при высоких концентрациях эти примеси могут значительно ускорить процессы гидратации и вызвать быстрое выпадение осадков. Это часто используют для принудительного выпадения осадков, т.е. устранения облаков и создания «хорошей погоды».

Таблица 4.1 Значение энергий гидратации в кДж/моль.

| | | | |
|---|---|---|---|
| $Fe^{3+}$ | 4707 | $Cu^{2+}$ | 2063 |
| $Al^{3+}$ | 4548 | $Mg^{2+}$ | 1887 |
| $Cr^{3+}$ | 4142 | $Fe^{2+}$ | 1874 |
| $La^{3+}$ | 3339 | $Cr^{2+}$ | 1883 |
| $Zn^{2+}$ | 2130 | $Ca^{2+}$ | 1569 |

## 4.4 Проблемы образования грозовых облаков

Необходимым условием для возникновения грозового облака является конвекция воздушных потоков, обогащенных парами воды. Рассмотрим столкновение двух встречных воздушных потоков, имеющих разные температуры.

Обратимся к схеме, изображенной на рисунке 2.1. Схема изображает области разных температур по линии перпендикулярной движению двух газовых потоков в атмосфере Солнца. Эта схема может быть использована и для описания встречных воздушных потоков в атмосфере Земли. Хотя в условиях земной атмосферы градиенты температур и сами их значения существенно меньше солнечных, но масштабность явлений остается.

Итак, область «трения» имеет самое высокое значение температуры. В сторону к области 2 падение температуры меньше, чем в сторону потока 1. Отличие в температурном градиенте будет существенным для разделения сортов заряженных частиц (кластеров) в пространстве.

Следует отметить, что в земной атмосфере среда, т.е. облачные образования, и проходящие в ней процессы являются более сложными, чем в атмосфере Солнца, из-за разнообразия молекулярных соединений, кластерных образований, сосуществования воды в разных агрегатных состояниях, и т.п.

В тоже время, в солнечной атмосфере сама среда имеет простую структуру из-за высоких температур, но более сложными являются внешние воздействия, например, магнитные поля, меняющиеся во времени и пространстве.

Однако, общей характеристикой атмосфер будет их разреженность и наличие сильных конвективных явлений, разнообразных по характеру и силе.

Конвекция, приводящая к развитию гроз, возникает в следующих случаях:

- при неравномерном нагревании приземного слоя воздуха над различной подстилающей поверхностью. Например, над водной поверхностью и сушей из-за различий в температуре воды и почвы. Над крупными городами и индустриальными центрами интенсивность конвекции значительно выше, чем в их окрестностях.
- при подъеме или вытеснении теплого воздуха холодным на атмосферных фронтах. Конвекция на атмосферных фронтах значительно интенсивнее и чаще, чем при внутримассовой конвекции. Часто фронтальная конвекция развивается одновременно со слоисто-дождевыми облаками и обложными осадками, что маскирует образующиеся кучево-дождевые облака.
- при подъеме воздуха вблизи горных массивов. Даже небольшие возвышенности на местности приводят к усилению образования облаков - рельеф местности создает вынужденную конвекцию.

Высокие горы являются причиной конвекции и усложняют их развитие. В таких районах имеет место регулярность облачных образований и их большая интенсивность.

Восходящие и нисходящие потоки в изолированных грозах обычно имеют диаметр от 0.5 до 2.5 км и высоту от 3 до 8 км. Иногда диаметр восходящего потока может достигать 4 км. Вблизи поверхности земли потоки обычно увеличиваются в диаметре, а скорость в них падает по сравнению с выше расположенными потоками. Характерная скорость восходящего потока лежит в диапазоне от 5 до 10 м/с, и доходит до 20 м/с в верхней части крупных гроз.

Исследовательские самолеты, пролетающие сквозь грозовое облако на высоте около 10 км, регистрируют скорость восходящих потоков свыше 30 м/с. Наиболее сильные восходящие потоки наблюдаются в организованных грозах [10-13].

Движение грозового облака относительно земли определяется, прежде всего, взаимодействием восходящего и нисходящего потоков облака с несущими воздушными потоками в средних слоях атмосферы в которых развивается гроза. Скорость перемещения изолированной грозы обычно порядка 20 км/час, но некоторые грозы двигаются гораздо быстрее. В экстремальных ситуациях супер ячейковое облако может двигаться со скоростями 65 - 80 км/час.

В большинстве гроз по мере рассеивания старых грозовых ячеек последовательно возникают новые грозовые ячейки. При слабом ветре отдельная ячейка за время своей жизни может пройти совсем небольшой путь, меньше чем пара километров; однако в более крупных грозах новые ячейки запускаются нисходящим потоком, вытекающим из зрелой ячейки, что дает впечатление быстрого движения не всегда совпадающего с направлением ветра.

В больших многоячейковых грозах существует закономерность, когда новая ячейка формируется справа по направлению несущего воздушного потока в северном полушарии и слева от направления несущего потока в южном полушарии [12,13].

Распределение и движение электрических зарядов внутри и вокруг грозового облака является сложным непрерывно меняющимся процессом. Можно, однако, представить обобщенную картину распределения электрических зарядов на стадии зрелости облака. Доминирует положительная дипольная структура, где положительный заряд находится в верхней части облака, а отрицательный заряд в нижней.

Физические параметры насыщения воздуха парами воды известны. Для оценок мы будет пользоваться усредненными характеристиками:

- абсолютная влажность воздуха над поверхностью воды при температуре ~ 20÷25° C дается величиной $a_В = 17 \div 22$ г$\cdot$м$^{-3}$ (для оценок мы возьмем 20 г$\cdot$м$^{-3}$ ). Это дает значение для концентрации молекул воды в воздухе: $C \approx 2{,}5 \cdot 10^{-2}$ ;

- область трения потоков холодного и теплого воздуха, конечно, зависит от размеров этих потоков, морских и воздушных течений, географических особенностей и т.п. (для оценок мы выберем модель области трения, имеющей протяженность 5 км, ширину 1 км и высоту 20 м);

- будем считать, что в самих воздушных потоках быстро устанавливается локальное термодинамическое равновесие, молекулярная ионизация происходит преимущественно в области трения потоков, а результатом локального равновесия является образование ионных кластеров - заряженных капелек воды.

Температурные различия во встречных потоках воздуха могут быть в пределах $20 \div 50\ ^{\circ}C$. Большой температурный градиент возможен в предгорных и экваториальных регионах, где восходящие теплые потоки могут сталкиваться с потоками нисходящего холодного воздуха. Воздух над твердой поверхностью прогревается быстрее, а вращение Земли и высотные ураганные потоки создают в тропосфере очень сложную картину завихрения и передвижения воздушных масс.

Обратимся теперь к рисунку 2.3. Воспроизведем его снова и укажем типы образующихся ионов в области трения (Рис.4.3). Отметим, что в области трения образуются, в результате столкновения молекул встречных потоков, различные молекулярные ионы. Ионы, имеющие более сильную способность к гидратации, становятся центрами образования крупных водных капель. Это относится, в первую очередь, к отрицательно заряженным кластерам.

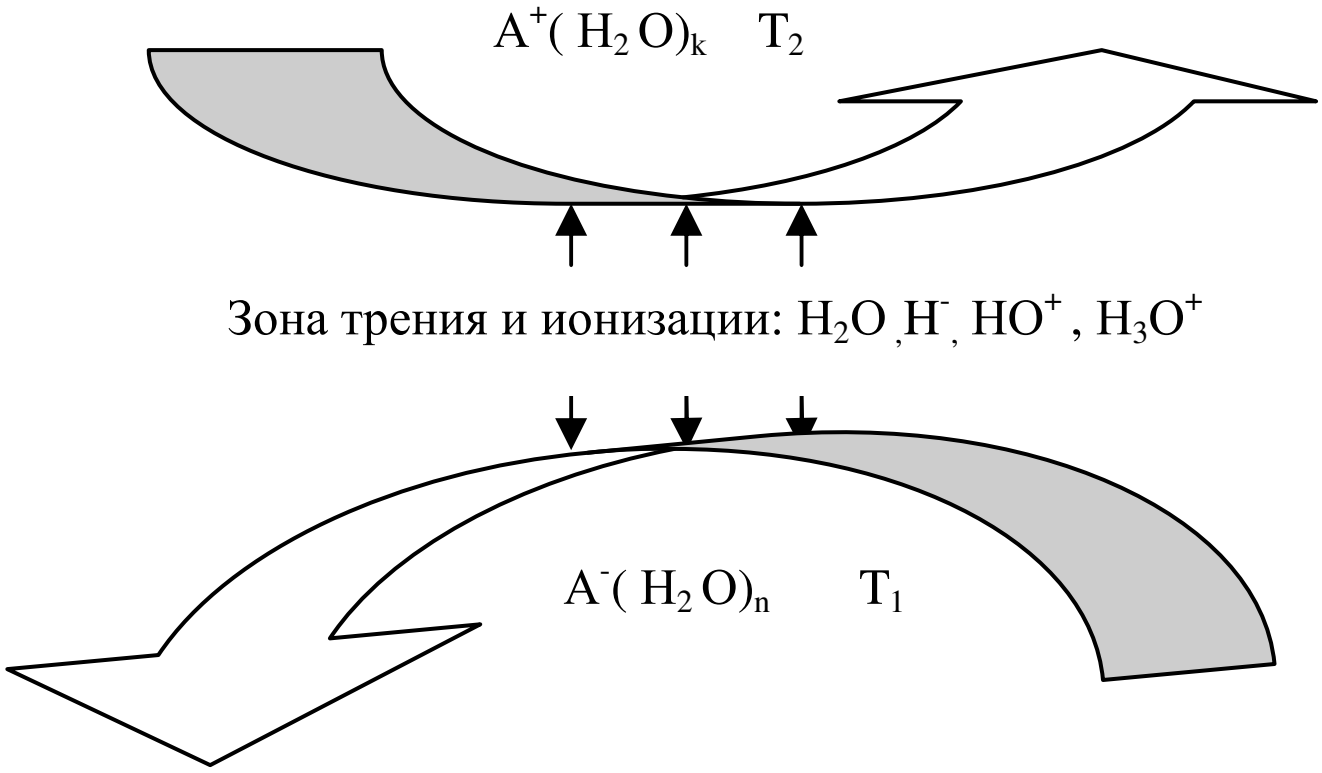

Рис. 4.3 Схема движения встречных потоков воздуха, имеющих разные температуры. Показано расслоение концентраций водных ионных кластеров с n>>k, имеющих противоположные заряды, $A^{\pm}$ обозначают ионы - центры гидратации.

Итак, преимущество в формировании крупных водных капель (или заряженных кластеров) остается за отрицательно заряженными ионами молекул воды. И эти же образования имеют преференции дальнейшего роста в области более низких температур.

Более мелкие капли оказываются, в основном, заряженными положительно и будут оставаться в области более высоких температур.

Эти два фактора - масса кластера и температурная преференция по росту и концентрации, создают основу для пространственного расслоения кластеров с разными массами и зарядами. Это расслоение дополняется общей кинетикой противодвижения потоков, что приводит к увлечению воздушным потоком заряженных капель определенного сорта, и разделению в пространстве заряженных грозовых ячеек.

Действительно, поток воздуха, имеющий более высокую температуру и кинетическую энергию частиц, будет увлекать более легкие кластеры, имеющие в основном положительные заряды.

В тоже время, поток воздуха с более низкой температурой будет обогащаться более массивными кластерами, имеющими отрицательные заряды.

Подчеркнем, что в результате диффузии (противопоточная масс - диффузия) и термодиффузии ионные кластеры разных типов пространственно разделяются внутри области трения. Более крупные кластеры, имеющие отрицательный заряд, скапливаются ближе к потоку Т1, а более легкие кластеры, имеющие положительный заряд, скапливаются у потока Т2.

Таким образом, холодные потоки будут обогащаться отрицательным объемным зарядом и более крупными водными каплями, а теплые потоки - положительным зарядом и более мелкими каплями.

Следует отметить, что потоки воздуха у земной твердой поверхности будут накапливать в основном отрицательный заряд, т.е. земля будет положительной по заряду, что связано со свойствами твердых тел, являющихся хорошими проводниками тепла и электричества.

Оценим грубо величину заряда в небольшом облаке. Считая, что ионизируется одна молекула воды на миллион (это очень низкое значение), получим суммарное число заряженных частиц, образующихся в области трения: $N_{ион} = C \times V_T \times N_Л / 10^6 \approx 7 \cdot 10^{25}$ . Здесь $V_T$ - объем области трения, а $N_Л$ - число Лошмидта. Некомпенсированный заряд будет равен: $Q \approx 5.4 \cdot 10^6$ Кл. Таким образом, в нашем небольшом модельном облачке уже будет запасен огромный потенциал энергии.

Источником энергии, как уже отмечалось, является в этом случае громадная по величине кинетическая энергия вращения Земли и энергия конвективных потоков, которые преобразуются в потенциальную энергию электрических полей грозовых облаков.

Известно, что накоплению заряда будет противодействовать пробойный разряд, который во влажном воздухе составляет ~ 5÷10 кВ/см . Размеры облака будут во многом определяться динамикой атмосферных смен различных потоков, восходящих и нисходящих, географическими особенностями и т.д.[10,12]

На самом деле, картина, конечно, значительно более сложная и более богатая, чем здесь была рассмотрена. Нашей задачей было максимально упростить модель грозового облака, для того, чтобы показать возможности микроскопического рассмотрения физических процессов.

Рассмотрение атмосферных явлений с привлечением термодинамических методов, теорий реакций, столкновений и перерассеяния, использование данных по исследованию плазмы и физики космических лучей представляется важным и полезным. Такое рассмотрение способно раскрыть физические детали многих интересных явлений.

## 4.5 Модель формирования мезосферных серебристых облаков.

В атмосфере Земли на больших высотах происходят процессы, которые пока не имеют однозначного объяснения. Это относится к спрайтам, джетам и так называемым мезосферным серебристым облакам (МСО).

Известно, что в промежуточном слое между стратосферой и термосферой - мезосфере - на высоте около 85 км образуются серебристые облака. Характер рассеяния солнечного света серебристыми облаками позволил установить, что они представляют собой скопления частиц размером 0,1-0,7 мкм. О природе этих частиц высказывались самые разные гипотезы - это ледяные кристаллы, мелкие частицы вулканической пыли, кристаллы в ледяной "шубе", космическая пыль, частицы метеорного или кометного происхождения.

Одна из последних гипотез связывает серебристые облака с возникновением озоновых дыр в стратосфере (см. литературу [66-70]).

Процесс формирования этих облаков изучается все активнее в связи с запусками ракет. Дело в том, что запуски космических аппаратов с водородо - кислородными двигате-лями служат причиной заноса водяного пара в мезосферу и, следовательно, стимулируют формирование облаков. Появление в этой области облаков, в свою очередь, создает проблемы при возвращении космических аппаратов на Землю. Поэтому необходимо создание надежной теории серебристых облаков, дающей возможность прогнозировать и даже управлять этим явлением природы [69].

**Физические особенности мезосферы в области образования МСО.** Как известно, в ионосфере непрерывно протекают процессы ионизации и рекомбинации. Наблюдаемые концентрации электронов и ионов есть результат баланса между скоростью их образования и нейтрализации (в частности, рекомбинации, захвата и т.д.).

Причины и процессы ионизации и нейтрализации разные в различных областях. В верхней части области D на высотах 85-100 км ионизацию вызывает в основном солнечное рентгеновское излучение с $\lambda < 85$ Å, а ниже 60-70 км днем и ниже 80-90 км ночью ионизация осуществляется космическими лучами галактического происхождения.

Существенный вклад в ионизацию вносят корпускулярные потоки, например электроны с энергией < 40 кэВ, а также солнечное излучение первой линии серии Лаймана с $\lambda =$ 1215,7 Å [16].

Скорость исчезновения ионов характеризуется эффективным коэффициентом рекомбинации $\alpha$, который определяет величину концентрации электронов $n_e$ и ее изменение во времени. На малых высотах значение $\alpha$ на несколько порядков выше, чем ее значение на больших высотах. Поэтому область D оказывается в целом слабо ионизированной, причем существенным становится образование комплексных ионов-гидратов типа $(H_2O)_n\ H^+$ , а также отрицательных ионов типа $O_2^-$ , $NO_3^-$ и других. Важно отметить, что отрицательные ионы наблюдаются лишь в D области.

Концентрация электронов в области D при переходе от дня к ночи еще больше уменьшается, поскольку ионизирующее излучение от Солнца закрывается Землей. Наконец, в области D и в области E ионосферы иногда наблюдаются кратковременные, очень узкие слои повышенной ионизации, состоящие из многозарядных ионов Mg, Fe, Ca и др. Они представляют собой следы метеоритов и других тел, проходящих через толщу атмосферы.

Приведенные здесь особенности протекания процессов в мезосфере позволяют описывать их как процессы в слабо ионизированной ионной плазме без участия свободных электронов. При этом, среда будет представлять собой почти идеальный разреженный газ, постоянный по составу: 80% $N_2$ и 20% $O_2$ .

Указанная область имеет интересные физические свойства, отличающиеся от других слоев атмосферы. Ионизационные и температурные характеристики здесь таковы, что этот слой естественным образом детектирует треки инородных тел. Можно сказать, что эта область мезосферы представляет собой природную камеру Вильсона, функционирующую, конечно, в иных временных, пространственных и температурных режимах, в отличие от известных лабораторных устройств.

**Химическое равновесие в криогенной плазмы.** Рассмотрим состояние области D нижней ионосферы, например, в дневной период времени. Поток солнечной радиации, очевидно, будет ионизировать атомы и молекулы среды, но ионизация здесь уже будет значительно меньшей, чем в более высоких слоях ионосферы. При этом плотность воздуха в этой области будет существенно более высокой, чем на больших высотах, но еще малой и сильно разреженной по сравнению с нижними слоями атмосферы.

Отличия будут касаться интенсивности и спектра радиационного излучения, а также температуры среды. Среда в области D будет слабо ионизирована, причем электронная компонента будет быстро поглощаться нейтральными молекулами, которые начнут пополнять число отрицательно заряженных ионов - важную компоненту такой квази - криогенной плазмы. В ночное время температура резко падает, и начинаются процессы, устанавливающие в среде тепловое и ионизационное равновесие. Это ведет к изменению качественного и количественного состава квази - криогенной плазмы и, соответственно, в самой среде.

В лабораторных условиях рекомбинирующая криогенная плазма помещается во внешнее электрическое поле для того, чтобы поддержать температуру электронной компоненты $T_e$ на заданном уровне [63-65]. В нашем случае, поток радиации от Солнца в дневное время естественным образом будет поддерживать существование квази - криогенной плазмы в верхних слоях мезосферы, и этот поток играет роль внешнего поля. Солнечное излучение и солнечные космические лучи будут при этом постоянно поставлять или порождать достаточное количество заряженных частиц, в частности, свободных электронов и положительных ионов.

Важными процессами, происходящими в рассматриваемой области мезосферы, являются процессы коагуляции капелек воды. Причиной коагуляции могут явиться как акустические волны, так и плазменные колебания. В последнем случае, особый интерес представляют процессы роста комплексных или кластерных ионов [64].

Кластерные ионы выступают как центры конденсации паров воды и других веществ тогда, когда давление паров превышает давление насыщенного пара при данной температуре. (Заметим, что именно такой механизм является основой работы камеры Вильсона.) При этом энергия диссоциации кластерных ионов ($\varepsilon_c = 0{,}09 \div 1{,}7\,\mathrm{eV}$) в среднем меньше, чем энергия химической связи молекул ($\varepsilon_m = 0{,}75 \div 11{,}1\,\mathrm{eV}$), но значительно выше энергии диссоциации ван-дер-ваальсовых молекул ($\varepsilon_V = 0{,}9 \cdot 10^{-3} \div 0{,}105\,\mathrm{eV}$).

Большие кластеры, как макроскопические системы, могут находиться в твердом и жидком агрегатном состояниях. Описывая кластер в рамках капельной модели можно считать, что его плотность равна плотности макроскопической жидкости. Тогда радиус кластера [63]

$$r(n) = r_W \cdot n^{1/3}, \tag{4.14}$$

где n - число атомов в кластере, т.е. элементарных составляющих - молекул или собственно атомов, $r_W$ - радиус Виг-

нера - Зейтца

$$r_W = \left(\frac{3m}{4\pi\rho}\right)^{1/3}, \qquad (4.15)$$

где m - масса элементарной составляющей (далее атома), ρ - плотность материала кластера.

В отличие от твердых параметры жидких кластеров монотонно зависят от их размера. Так, полная энергия связи атома кластера определяется формулой

$$E(n) = \varepsilon_0 n - A \cdot n^{2/3}, \qquad (4.16)$$

где $\varepsilon_0$ - удельная энергия сублимации макроскопической системы, приходящаяся на один атом, а второе слагаемое в правой части соответствует поверхностной энергии, выраженной через поверхностное натяжение. Здесь предполагается, что энергия связи $\varepsilon_0$ много больше тепловой энергии элемента кластера $T_m$ и что энергия связи $\varepsilon_0$ при температуре плавления и нулевой температуре одинаковы.

Пренебрегая температурной зависимостью параметров $\varepsilon_0$ и A, получим энергию связи атома в кластере

$$\varepsilon_n = dE(n)/dn = \varepsilon_0 - \Delta\varepsilon / n^{1/3}, \quad \Delta\varepsilon = (2/3)A. \qquad (4.17)$$

Модель жидкой капли позволяет анализировать кинетические параметры кластера, находящегося в газе или плазме. Например, сечение прилипания атома к поверхности большого кластера равно

$$\sigma(n) = \pi r^2, \qquad (4.18)$$

где $r = r(n)$ - его радиус. Для скорости прилипания следует

$$\nu(n) = N v \sigma(n) = N k_0 n^{1/3}, \qquad (4.19)$$

где N - плотность атомов, а приведенная константа прилипания будет равна

$$k_0 = \sqrt{\frac{8T}{\pi m}} \pi r_W^2, \qquad (4.20)$$

где T - температура газа.

Переходя к равновесию отдельного кластера в атомном паре и используя экспоненциальную зависимость для равновесной плотности атомов, получаем для частоты испарения атомов $\nu_{ev}(n)$ с поверхности кластера, содержащего n атомов

$$\nu_{ev}(n) = \nu(n) \frac{p_0}{TN} \exp(-\varepsilon_n / T). \qquad (4.21)$$

Таким образом, частота испарения кластера выражается через $p_0$ - давление насыщенного пара, энергию связи атомов кластера и частоту прилипания атомов к кластеру. Соответственно, уравнение баланса для описания эволюции размера кластера имеет вид [63-65]

$$\frac{dn}{dt} = k_0 n^{2/3} N - \nu_{ev}(n). \qquad (4.22)$$

Итак, модель жидкой капли позволяет описать свойства больших жидких капель и их поведение в собственном атомном паре.

**Плазменная модель образования мезосферных серебристых облаков.** Рассмотрим модель образования мезосферных серебристых облаков, основываясь на качественных особенностях верхних слоев мезосферы, смене ее температурных и иных физических режимов, а также на процессах образования больших молекулярных кластеров и особенностях

квази - криогенной плазмы [71].

1-й этап, дневной период времени. Солнечное излучение и солнечные космические лучи создают в области D нижней ионосферы и верхних слоях мезосферы относительно небольшую степень ионизации, т.е. слабо разреженную плазму. В этой части атмосферы присутствует небольшое, но достаточное для дальнейших процессов количество паров воды. Поэтому наличие молекулярных ионов различного типа, особенно отрицательных, создает условия для образования комплексных или кластерных ионов типа $O^{-}\cdot(H_2O)_n$ при n >> 1. Среднее число и характеристики кластерных ионов могут быть определены в соответствии с приведенными выше формулами и общими оценками.

2 этап, ночной период времени. Поток частиц и излучения от Солнца закрывается телом Земли, генерация свободных электронов и ионизация в разреженной среде прекращаются, температура среды резко падает. Квази - криогенная плазма переходит в стадию рекомбинации с образованием некоторого числа кластерных ионов, рассеянных хаотично по всей рассматриваемой области.

3 этап, пробуждение - переход к дневному времени. Появившийся поток солнечных лучей и частиц приводит к распаду кластеров, появлению новых ионизированных частиц и слаборазреженной плазмы, т.е. наступает 1-й этап, и далее весь процесс повторяется.

Рассмотрим теперь ситуацию, когда происходит неординарное внешнее воздействие - например, мезосферу пронизывает метеорных дождь или происходит выброс большого количества вещества, происходящего при запуске ракетно-космических комплексов. В этом случае в мезосфере возникают большие по протяженности и плотности ионизационные треки, аналогично трекам в камерах Вильсона, ионизационных счетчиках и в других аналогичных приборах. Главным в нашем случае будет большой масштаб ионизированных участков и их локализация в определенной области мезосферы. В этой области и областях, прилежащих к ним (назовем всю эту область - метеорной областью М), будут воз-никать специфические явления.

1-й этап, дневной период времени. В области М присутствует значительная по степени ионизация. Распадающееся квази - криогенная плазма будет испытывать переход в квазиравновесное состояние. Солнечная ионизация может в этом случае рассматриваться как фоновая и незначительная по величине.

Как и ранее, здесь возникают условия образования комплексных и кластерных ионов типа $O^{-} \cdot (H_2O)_n$, но также и новых, например, $Fe^{3+} \cdot (H_2O)_n$, $Al^{3+} \cdot (H_2O)_n$ и других. Скоротечность процессов будет обусловлена быстрым падением температуры при адиабатическом расширении зоны повышенной ионизации. При этом степень роста ионных кластеров будет существенно большей, чем в ранее рассмотренном случае, т.е. средняя величина n будет значительно большей.

Кроме того, в этой области степень ионизации плазмы будет также значительно выше, чем в других обычных по режиму областях. При этом свободные электроны будут быстро захватываться молекулами среды, и плазма приобретет свойства чисто ионной, т.е. без электронной компоненты. В такой плазме имеют место так называемые плазменные колебания, участниками которых становятся уже тяжелые заряженные частицы (отрицательные и положительные ионы) и даже массивные ионные кластеры.

В силу огромной по величине массы ионных кластеров в плазме возникнут длинноволновые плазменные колебания. Сами эти колебания будут создавать преимущественные условия для обычного роста кластеров

$$M_n + m_A \to M_{n+1} \ , \qquad (4.23)$$

с одной стороны, и развития процессов коагуляции

$$M_n + M_q \to M_{n+q} \ , \qquad (4.24)$$

с другой стороны.

Нужно отметить, что, в принципе, типы, массы и заряды отрицательных и положительных кластеров могут различаться между собой. Поэтому приведенные формулы имеют оценочный характер.

2 этап, ночной период времени. Квази – криогенная плазма переходит в стадию адиабатического остывания, рекомбинации и коагуляции кластерных ионов. В отличие от дневного режима, в области M будут формироваться более крупные по размеру кластеры (жидкие капли), с достаточно упорядоченной пространственной структурой вокруг или вблизи первоначальной области макро - ионизационных треков. Дальнейшее падение температуры приведет к образованию в области M длинных облаков, состоящих из ледяных кристалликов и растянутых по трекам прошедших инородных тел.

3 этап, пробуждение - переход к дневному времени. Солнечные лучи высвечивают облака ледяных кристалликов, называемых мезосферными серебристыми облаками. Усиливающийся поток солнечных лучей и частиц приводит к их постепенному испарению, затем к появлению новых ионизированных частиц и слаборазреженной плазмы, т.е. наступает 1-й этап, и т.д.

Приведенная здесь модель образования МСО дает однозначную их связь с неординарными внешними воздействиями по отношению к мезосфере и системе Солнце - Земля - Атмосфера. Однако воздействие МСО на атмосферные и погодные явления остаются пока не исследованными, и находятся вне проведенного здесь анализа.

**Термодинамическое описание образования МСО.** Рассмотрим низкотемпературную плазму по составу и характеристикам, близким к условиям мезосферы, где образуются серебристые облака. Ионизационное равновесие в такой плазме будем снова описывать методом Саха. Как и выше (смотри предыдущий раздел), будем определять степень комплексности (массивности) кластера и рассматривать процессы последовательного отделения (или присоединения) молекул воды от ионных кластеров [24,60].

Приведенные в разделе 4.4 формулы и выкладки (см.

(4.1)-(4.13)) будут в еще большей степени правомерны для рассмотрения процессов в МСО. Обращаясь к этим формулам, выразим величину $\beta(T)$ с привязкой к условиям специфичным для высот мезосферы. Выберем область температур вблизи температурной точки $T(^\circ K) = 300\ ^\circ K$, а значения концентраций вблизи концентраций $C = 10^{-6}$, характерных для областей МСО.

Отметим, что $\beta(T)$ является функцией, слабо зависящей от температуры

$$\beta(T) \approx 18.19 + \frac{3}{2}\ln\left\{\frac{T(^\circ K)}{300}\right\} - \ln\left\{C \cdot 10^{6}\right\}. \qquad (4.25)$$

Как и ранее, критическим является значение $\alpha_c = \alpha(T_c) = 1$.

Когда $\alpha(T) > 1$ идет быстрое образование массивных кластеров, а когда $\alpha(T) < 1$ - их распад.

Если температура $T$ велика, то величина $\varepsilon_0 / T$ будет мала так, что $\alpha(T) < 1$. При низких температурах величина $\varepsilon_0 / T$ может стать, наоборот, очень большой и $\alpha(T) > 1$. Такие температуры и приводят к образованию массивных кластеров, а из них - кристалликов льда. Их скопление и будет представлять собой серебристые облака.

Оценим критическое значение параметра $\alpha_c$. Тепловая энергия атомов при комнатной температуре равна $\approx 0.025$ eV. В D области мезосферы температуры меняются в интервале ~ 170÷230 $^0$K, т.е. соответствуют энергиям $\sim 0.0147 \div 0.0198$ eV.

Полагая, например, энергию диссоциации равной $\varepsilon_0 \approx 0.09$ eV, получим, что для ночных температур в D области должно быть $\varepsilon_0 / T \approx 6.14$. В дневное же время должно быть $\varepsilon_0 / T \approx 4.5$. В случае $\varepsilon_0 \approx 0.423$ eV, получим $\varepsilon_0 / T \approx 29$ для ночного времени и $\varepsilon_0 / T \approx 18$ для дневного времени. Если же, $\varepsilon_0 \approx 1.7$ eV, то для ночных температур будет $\varepsilon_0 / T \approx 116$, а для дневных - $\varepsilon_0 / T \approx 86$.

Очевидно, что комплексы на основе ионов, приведенных в таблице 1, особенно ионы $Fe^{3+}$, $Al^{3+}$, $Cr^{3+}$, будут являться более мощными центрами конденсации, чем многие другие.

Оценки, данные выше, показывают, что уже в рассматриваемых диапазонах температур $T = 230 \div 170\ ^{\circ}K$ и концентраций $C \approx 10^{-6} \div 10^{-7}$, процессы гидратации могут реализоваться, если ионные комплексы, будут иметь $\varepsilon_0 > 0.26$ eV.

Из (4.25) ясно видно, что величина C является очень важной. Имеющиеся опытные данные по концентрациям паров воды в верхних слоях атмосферы имеют неопределенности и достаточно большой разброс. Но даже в рамках полученных здесь грубых оценок уже можно говорить об удовлетворительном согласии предлагаемой плазменной модели образования МСО и опытных данных.

# ЗАКЛЮЧЕНИЕ

Нами были рассмотрены некоторые из многообразных и интересных явлений, происходящих в атмосферах Солнца и Земли. Мы коснулись лишь тех их них, которые были связаны с формированием и развитием объемных атмосферных электрических зарядов и их особенностями.

На Солнце мы назвали их «грозовыми облаками» по аналогии с земными грозовыми облаками. Они являются носителями больших по величине электрических зарядов, в которых происходят молнии - электрические разряды.

Был проведен анализ и сравнение грозовых процессов в атмосфере Земли и Солнца. Молнии на Земле обычно сопровождаются акустическими эффектами - громами. Иногда молниевые процессы инициируют (или им сопутствуют) выбросы в атмосферу больших масс вещества (плазмы), достигающих больших высот (спрайты) или даже ионосферы (джеты). В атмосфере Солнца также происходят молнии, т.е. электрические разряды, выбросы плазмы (спикулы, корональные выбросы и вспышки и т.п.). Показано, что многие атмосферные процессы на Солнце и Земле во многом имеют сходные физические особенности.

Наш интерес касался вопроса - как образуются «грозовые облака»? Откуда, где и каким образом облака набирают громадные по величине электрические заряды (причем, объемные и пространственно разделенные некомпенсированные заряды)?

Были сделаны оценки величины электрического заряда в громадных потоках газовых масс в атмосферах Солнца и Земли, имеющих разные температуры и, часто, противоположные по направлению движения. Было ясно, что гигантская энергия движения этих масс пре-образуется при столкновениях в огромный потенциал электрического поля, т.е. некомпенсированные электрические заряды, собранные в объемные структуры - газоплазменные облака на Солнце и грозовые облака на Земле.

Был проведен анализ процессов гидратации и ионизационной диссоциации в методе термодинамического описания слаборазреженной плазмы.

Важным здесь является то, что концентрации различных ионов имеют разную температурную зависимость. Причиной этого является их электронная структура, т.е. особенности внутреннего квантовомеханического строения энергетических уровней этих ионов и ионных молекул. Но важным для развития макроскопических атмосферных процессов является не только микроскопические характеристики ионов (таких, как энергии ионизации или диссоциации, их массы и молекулярная структура, моменты, колебательные и вращательные спектры, и т.п.), но и макроскопические характеристики газовых потоков, внешних полей и излучений.

Поскольку с изменением температуры концентрации одних ионов могут значительно превосходить концентрации других ионов, даже если эти концентрации при определенной температуре были равны между собой, то это дает ключ к пониманию того, почему объемные заряды могут разделиться в пространстве.

Здесь становятся важными особенности движения потоков относительно друг друга, которые имеют разные температуры. Во-первых, если потоки не имеют относительного движения, то разность температур в них не может долго сохраняться. Во-вторых, если относительное движение мало, то большие объемные заряды разных знаков не могут быть удержаны сколько-нибудь продолжительное время вдали друг от друга. И только, если относительное движение потоков достаточно по величине, то оно способно разнести объемные заряды далеко друг от друга и сохранить собственную температуру в каждом из потоков почти неизменной.

Локально изолированный объемный и электрически заряженный объект - «грозовое облако» может разрядиться, встретившись с другим облаком, имеющим противоположный заряд и возникшим аналогичным образом в другом месте. Примером этому, на наш взгляд, могут служить земные грозовые облака и поведение газовых потоков в спикулах в хромосфере Солнца.

Расчеты показывают, что даже в небольших по размеру облаках могут содержаться гигантские по величине некомпенсированные заряды. Становится ясным, каким мощным является «электрическое динамо» Земли. И поражает сила атмосферных явлений на Солнце - на сколько она является гигантской, по сравнению с земными.

Отметим особую роль внешних воздействий на описываемые атмосферные явления. Это касается солнечно-земных связей и, в частности, космических лучей, идущих от Солнца и из глубин галактик.

В механизме «грозовых облаков» внешние воздействия играют исключительную роль. В случае земной атмосферы космические лучи и порождаемые ими широкие атмосферные ливни являются «детонаторами» молниевых разрядов.

В случае солнечной атмосферы определяющая роль принадлежит плазменным выбросам из солнечных глубин и магнитным полям.

К задачам атмосферной динамики Солнца и звезд относится и предлагаемый механизм генерации атмосферных нейтронов. Процессы, определяющие этот механизм, восходят к явлениям земной атмосферы - потоку γ-излучения от «убегающих электронов», задачам квантовой механики трех частиц - ионной молекуле $H_2^+$, и физике частиц - $\beta$-процессам, стимулированным сильными внешними полями.

**Проведенный анализ позволяет констатировать следующее:**

-многое в физике атмосферных явлений Земли еще не исследовано, но все они являются интересными и важными для науки будущего - науки «погода Земли»;

-используя методы квантовой теории и квантовой статистики, можно достичь определенного понимания в механизмах формирования грозовых облаков, и понять некоторые из очень интересных физических явлений, происходящих в атмосферах Солнца и Земли;

-атмосферу Земли можно рассматривать, как своеобразную природную лабораторию - «детектор» по изучению электрических атмосферных процессов. Такое изучение может стать ключом к пониманию многих солнечных и звездных «грозовых молний» и атмосферных явлений.

Важно при этом понимать, что существует огромное отличие между земными, солнечными и звездными процессами: по мощности, температурам, массе, величине магнитных полей, размерам, моменту вращения и даже химическому составу атмосфер.

# ЛИТЕРАТУРА

**Часть 3**

# АСТРОФИЗИЧЕСКИЕ S - ФАКТОРЫ ЛЕГЧАЙШИХ ЯДЕР

*Дубовиченко С.Б.*

***Автор посвящает этот раздел 80-летию Владимира Германовича Неудачина***

В этой части рассмотрены ядерные фотопроцессы, включающие фоторазвал ядра в двухчастичный канал гамма квантом и радиационный захват двух кластеров с образованием ядра в основном состоянии с испусканием такого кванта. Основной характеристикой таких процессов являются астрофизические S - факторы, которые определяют поведение сечения ядерных реакций при энергии, стремящейся к нулю.

Астрофизические S - факторы выражаются через полные сечения процесса радиационного захвата и энергию взаимодействующих частиц

$$S = \sigma E_{cm} \exp\left( \frac{31.335\ Z_1 Z_2 \sqrt{\mu}}{\sqrt{E_{cm}}} \right)$$

где $\sigma$ - полное сечение процесса радиационного захвата в барн, $E_{cm}$ - энергия частиц в кэВ для системы центра масс, $\mu$ - приведенная масса в а.е.м. и Z - заряды частиц.

Для выполнения подобных расчетов в ядерной астрофизике [1], для легких атомных ядер и низких энергий, требуется умение решать уравнение Шредингера или связанную систему этих уравнений. Результатом решения является волновая функция, которая описывает квантовое состояние некоторой системы ядерных частиц и, в принципе, содержит всю информацию об этом состоянии. Такая волновая функция позволяет выполнять и любые расчеты характеристик фотоядерных процессов при низких энергиях.

Существует довольно много различных математических методов решения дифференциальных уравнений или их систем второго порядка, которым является уравнение Шредингера [2]. Некоторую проблему обычно составляет выбор наиболее оптимального математического численного метода, применимого для рассмотрения определенного круга задач ядерной физики, основанных на решениях уравнения Шредингера или системы таких уравнений.

Решению некоторых из этих проблем посвящена первая глава этой части данной книги, которая описывает несколько математических и численных методов, непосредственно применимых для нахождения волновых функций, т.е. решений уравнения Шредингера или системы таких уравнений в задачах ядерной физики низких энергий и ядерной астрофизики.

Во второй главе приводятся расчеты полных сечений фотоядерных реакций в двухкластерной модели легких ядер. В качестве межкластерных потенциалов взаимодействия выбран особый класс сравнительно новых потенциалов с запрещенными состояниями, параметры которых предварительно согласованы с фазами их упругого рассеяния.

Присутствие таких состояний позволяет эффективно учитывать принцип Паули без выполнения полной и явной антисимметризации волновых функций ядерной системы, что существенно упрощает всю вычислительную процедуру, не приводя, по-видимому, к заметному ухудшению результатов по сравнению с точными методами.

# 1. УРАВНЕНИЯ ШРЕДИНГЕРА С ТЕНЗОРНОЙ КОМПОНЕНТОЙ

В данной главе приведены численные математические методы для решения системы уравнений Шредингера при наличии в действительном потенциале взаимодействия тензорной компоненты. Описаны общие и вычислительные методы решения системы уравнений Шредингера для задачи рассеяния с тензорными силами, когда начальные и асимптотические условия записываются в наиболее общем виде. В матричной форме дан весь основной математический аппарат для решения поставленной задачи.

Далее рассматриваются математические численные методы решения связанной системы уравнений Шредингера с тензорными силами для дискретного спектра собственных значений. При решении такой задачи предложена комбинация численных и вариационных методов нахождения отрицательных собственных значений, т.е. энергии связи, которая позволяет определять их с большой точностью, контролируемой на основе невязок.

## 1.1 Общие методы решение системы уравнений Шредингера в задачах рассеяния

Использование ядерных потенциалов с тензорной компонентой приводит нас к системе связанных уравнений Шредингера.

Будем исходить в дальнейшем из обычных уравнений [3], которые учитывают действительные центральную и тензорную часть ядерных потенциалов

$$u''(r) + [\, k^2 - V_c(r) - V_{cul}(r)]u(r) = \sqrt{8}\, V_t(r)w(r)\,, \tag{1.1}$$

$$w''(r) + [\, k^2 - V_c(r) - 6/r^2 - V_{cul}(r) + 2V_t(r)\,]w(r) = \sqrt{8}\, V_t(r)u(r)\,,$$

где u(r) и w(r) - волновые функции (ВФ), r - скалярное расстояние между частицами, $V_{cul}(r)= 2\mu/\hbar^2\, Z_1Z_2/r$ - кулоновский потенциал, $Z_1$, $Z_2$ - заряды частиц, $\mu$ - приведенная масса частиц.

Константа $\hbar^2/M_N$ принималась равной 41.4686 (или 41.47 в нуклон - нуклонной или NN задаче) МэВ Фм$^2$, $M_N$ - масса нуклона, $k^2 = 2\mu E/\hbar^2$ - волновое число относительного движения частиц, E - энергия относительного движения частиц, $V_c = 2\mu/\hbar^2 V_{cn}(r)$ - центральная часть потенциала, $V_t = 2\mu/\hbar^2 V_{tn}(r)$ - тензорная часть потенциала взаимодействия, $V_{cn}(r)$ и $V_{tn}(r)$ - радиальные части центрального и тензорного потенциалов, которые обычно берутся в виде гауссойды или экспоненты вида $V_{c(t)n}(r) = V_{c(t)0}\exp(-\alpha r)$, здесь $V_{c0}$ - глубина потенциала, $\alpha$ – его ширина.

Решением системы (1.1) являются четыре волновые функции, получающиеся с двумя типами начальных условий вида [3]

$$1.\; u_1(0)=0\,,\quad u'_1(0)=1\,,\quad w_1(0)=0\,,\quad w'_1(0)=0\,, \qquad (1.2)$$

$$2.\; u_2(0)=0\,,\quad u'_2(0)=0\,,\quad w_2(0)=0\,,\quad w'_2(0)=1\,,$$

которые для состояний рассеяния ($k^2 > 0$) образуют линейно независимые комбинации, представляемые в форме [3]

$$u_\alpha = C_{1\alpha} u_1 + C_{2\alpha} u_2 \longrightarrow \mathrm{Cos}(\varepsilon)\,[F_0\, \mathrm{Cos}(\delta_\alpha) + G_0\, \mathrm{Sin}(\delta_\alpha)]\,,$$
$$w_\alpha = C_{1\alpha} w_1 + C_{2\alpha} w_2 \longrightarrow \mathrm{Sin}(\varepsilon)\,[F_2\, \mathrm{Cos}(\delta_\alpha) + G_2\, \mathrm{Sin}(\delta_\alpha)]\,,$$

$$u_\beta = C_{1\beta} u_1 + C_{2\beta} u_2 \longrightarrow -\mathrm{Sin}(\varepsilon)\,[F_0\, \mathrm{Cos}(\delta_\beta) + G_0\, \mathrm{Sin}(\delta_\beta)]\,,$$
$$w_\beta = C_{1\beta} w_1 + C_{2\beta} w_2 \longrightarrow \mathrm{Cos}(\varepsilon)\,[F_2\, \mathrm{Cos}(\delta_\beta) + G_2\, \mathrm{Sin}(\delta_\beta)]\,, \qquad (1.3)$$

где $F_L$ и $G_L$ - кулоновские функции рассеяния [3,4], $\delta$ - фазы рассеяния, $\varepsilon$ - параметр смешивания состояний с разными орбитальными моментами.

Пары функций $u_\alpha$ и $w_\alpha$ , и $u_\beta$ и $w_\beta$ являются наиболее общими решениями уравнений (1.1) и при больших расстояниях порядка R = 15-20 Фм. стремятся к своим асимптотическим значениям, определяемым правой частью выражений (1.3).

Сами начальные условия (1.2) строятся исходя из того, что волновые функции при r = 0 всегда равны нулю, а их производные u' и w' должны принимать не нулевые значения, равные некоторой заданной константе, причем не обязательно 1. Величина этой константы определяет амплитуду ВФ, не влияя на ее форму, и в реальных численных расчетах принимается равной 0.01. Амплитуда ВФ внутри обрасти решения уравнения (1.1), т.е. при $r < R$ определяется из сшивки ее численных значений со своей аналитической асимптотикой (1.3) при $r = R$.

При отсутствии тензорной части потенциала, параметр смешивания состояний с различным орбитальным моментом $\varepsilon$ становится равен нулю, уравнений (1.1) превращается в два не связанных уравнения и функции $u_\alpha$ и $w_\beta$ переходят в решения $u_0$ и $w_2$, которые определяют волновые функции рассеяния для частиц с относительным орбитальным моментом L, равным 0 и 2.

В случае нейтрон - протонной (np) задачи рассеяния, когда заряд одной из частиц равен нулю, кулоновские функции $F_L$ и $G_L$ превращаются в обычные сферические функции Бесселя [5].

Здесь мы рассмотрим более общий случай заряженных частиц и вынесем Cos(δ) в правой части выражений (1.3), тогда они преобразуются к виду

$$u_{1\alpha} = C'_{1\alpha} u_1 + C'_{2\alpha} u_2 \longrightarrow \mathrm{Cos}(\varepsilon)\, [F_0 + G_0\, \mathrm{tg}(\delta_\alpha)] ,$$
$$w_{1\alpha} = C'_{1\alpha} w_1 + C'_{2\alpha} w_2 \longrightarrow \mathrm{Sin}(\varepsilon)\, [F_2 + G_2\, \mathrm{tg}(\delta_\alpha)] ,$$

$$u_{2\beta} = C'_{1\beta} u_1 + C'_{2\beta} u_2 \longrightarrow -\mathrm{Sin}(\varepsilon)\, [F_0 + G_0\, \mathrm{tg}(\delta_\beta)] ,$$
$$w_{2\beta} = C'_{1\beta} w_1 + C'_{2\beta} w_2 \longrightarrow \mathrm{Cos}(\varepsilon)\, [F_2 + G_2 \mathrm{tg}(\delta_\beta)] ,$$

где $C' = C/\mathrm{Cos}(\delta)$ и $u_{i\alpha} = u_\alpha/\mathrm{Cos}(\delta)$.

Более компактно можно записать эти выражения для $u_{1\alpha}$ и $w_{1\alpha}$, и $u_{2\beta}$ и $w_{2\beta}$ в матричном виде [6]

$$V = XC' \longrightarrow FU + GU\sigma ,$$

где

$$V = \begin{pmatrix} u_{1\alpha} & u_{2\beta} \\ w_{1\alpha} & w_{2\beta} \end{pmatrix}, \qquad X = \begin{pmatrix} u_1 & u_2 \\ w_1 & w_2 \end{pmatrix}, \qquad G = \begin{pmatrix} G_0 & 0 \\ 0 & G_2 \end{pmatrix},$$

$$C' = \begin{pmatrix} C'_{1\alpha} & C'_{1\beta} \\ C'_{2\alpha} & C'_{2\beta} \end{pmatrix}, \qquad F = \begin{pmatrix} F_0 & 0 \\ 0 & F_2 \end{pmatrix},$$

$$U = \begin{pmatrix} \mathrm{Cos}\varepsilon & -\mathrm{Sin}\varepsilon \\ \mathrm{Sin}\varepsilon & \mathrm{Cos}\varepsilon \end{pmatrix}, \quad \sigma = \begin{pmatrix} \mathrm{tg}\delta_\alpha & 0 \\ 0 & \mathrm{tg}\delta_\beta \end{pmatrix} . \qquad (1.4)$$

Аналогичное уравнение можно написать и для производных волновых функций

$$V' = X'C' \longrightarrow F'U + G'U\sigma .$$

Исключая из этих уравнений C', после несложных преобразований, для K матрицы рассеяния, определяемой в виде

$$K = U\sigma U^{-1},$$

окончательно будем иметь

$$K = - [ X(X')^{-1}G' - G]^{-1} [X(X')^{-1}F' - F] .$$

Тем самым, K матрица рассеяния оказывается выраженной через кулоновские функции, численные решения исходных уравнений и их производные при некотором $r = R$.

Как известно, K матрица рассеяния в параметризации

Блатта - Биденхарна выражается через фазы рассеяния и параметр смешивания следующим образом [3]

$$K = \begin{pmatrix} \mathrm{Cos}^2\varepsilon \mathrm{tg}\delta_\alpha + \mathrm{Sin}^2\varepsilon \mathrm{tg}\delta_\beta & \mathrm{Cos}\varepsilon \mathrm{Sin}\varepsilon(\mathrm{tg}\delta_\alpha - \mathrm{tg}\delta_\beta) \\ \mathrm{Cos}\varepsilon \mathrm{Sin}\varepsilon(\mathrm{tg}\delta_\alpha - \mathrm{tg}\delta_\beta) & \mathrm{Sin}^2\varepsilon \mathrm{tg}\delta_\alpha + \mathrm{Cos}^2\varepsilon \mathrm{tg}\delta_\beta \end{pmatrix} . \quad (1.5)$$

Тогда, приравнивая соответствующие элементы, получим для матричных элементов K матрицы следующие выражения

$$K_{12} = K_{21} = 1/2\ (\mathrm{tg}\delta_\alpha - \mathrm{tg}\delta_\beta)\ \mathrm{Sin}(2\varepsilon)\ ,$$
$$K_{11} + K_{22} = \mathrm{tg}\delta_\alpha + \mathrm{tg}\delta_\beta\ ., \quad (1.6)$$
$$K_{11} - K_{22} = (\mathrm{tg}\delta_\alpha - \mathrm{tg}\delta_\beta)\ \mathrm{Cos}(2\varepsilon)\ .$$

Откуда имеем

$$\mathrm{tg}(2\varepsilon) = 2K_{12}/(K_{11}-K_{22})\ ,$$
$$\mathrm{tg}\delta_\alpha = (A+B)/2\ ,$$
$$\mathrm{tg}\delta_\beta = (A - B)/2\ ,$$

$$A = K_{11} + K_{22}\ ,$$
$$B = (K_{11} - K_{22})/\mathrm{Cos}(2\varepsilon)\ . \quad (1.7)$$

Здесь

$$a = f\ (u_1w'_2 - u_2w'_1)\ , \quad b = f\ (u'_1u_2 - u_1u'_2)\ , \quad (1.8)$$
$$c=f\ (w_1w'_2-w'_1w_2)\ , \quad d=f\ (u'_1w_2-u'_2w_1)\ , \quad f=(u'_1w'_2-u'_2w'_1)^{-1}\ ,$$

$$A = aG'_0 - G_0\ , \quad B = bG'_2\ , \quad E = cG'_0\ ,$$
$$D = dG'_2 - G_2\ , \quad F = PD\ , \quad G = - PB\ ,$$
$$N = - PE\ , \quad M = PA\ , \quad P = - (AD - BE)^{-1}\ ,$$
$$R = aF'_0 - F_0\ , \quad S = bF'_2\ , \quad T = cF'_0\ ,$$
$$Z = dF'_2 - F_2\ , \quad K_{11} = FR + GT\ , \quad K_{12} = FS + GZ\ ,$$
$$K_{21} = NS + MZ\ , \quad K_{22} = NR + MT\ .$$

Таким образом, получаются сравнительно простые вы-ражения для определения фаз рассеяния $\delta_\alpha$ и $\delta_\beta$, и параметров смешивания ε, которые требуется определить для процессов рассеяния квантовых частиц, через значения численных волновых функций на асимптотике, т.е. при R = r и известные кулоновские функции. При отсутствии тензорной компоненты параметр смешивания оказывается равен 0 и фазы $\delta_\alpha$ и $\delta_\beta$ переходят в $\delta_0$ и $\delta_2$, т.е. фазы с орбитальным моментом 0 и 2.

Для численных решений, производные и функции в одной точке R можно заменить только на волновые функции в двух точках $R_1$ и $R_2$ - при этом вид полученных выражений не меняется. При этом можно считать, что величины, например, без штриха находятся в первой точке, а со штрихом во второй. Расстояние между этими двумя точками обычно выбирается равным 5-10 шагов численной схемы [4].

По определенным фазам рассеяния легко можно найти в матричном виде и коэффициенты C'

$$C' = X^{-1}(FU + GU\sigma) ,$$

где $X^{-1}$ - обратная к X матрица.

Расписывая это матричное выражение, имеем

$$C'_{1\alpha} = a(A + F) + b(E + H) ,$$
$$C'_{2\alpha} = c(A + F) + d(E + H) ,$$

$$C'_{1\beta} = a(B + G) + b(D + K) ,$$
$$C'_{2\beta} = c(B + G) + d(D + K) , \qquad (1.9)$$

где

$$a = fw_2 , \quad b = - fu_2 , \quad c = - fw_1 , \quad d = fu_1 ,$$
$$f = (u_1w_2 - u_2w_1)^{-1},$$

$$A = F_0Cos(\varepsilon) , \quad B = - F_0Sin(\varepsilon) , \quad E = F_2Sin(\varepsilon) ,$$
$$D = F_2Cos(\varepsilon) , \quad F = G_0Cos(\varepsilon)tg(\delta_\alpha) , \quad G = - G_0Sin(\varepsilon)tg(\delta_\beta) ,$$
$$H = G_2Sin(\varepsilon)tg(\delta_\alpha) , \quad K = G_2Cos(\varepsilon)tg(\delta_\beta) . \qquad (1.10)$$

В результате, можно получить полный вид ВФ во всей области при r < R, а радиус сшивки R обычно принимается равным 15 - 20 Фм. Для численного решения исходного уравнения можно использовать метод Рунге - Кутта четвертого порядка [4,7] с автоматическим выбором шага при заданной точности результатов по фазам рассеяния и параметру смешивания.

Фазы рассеяния для NN задачи обычно принято выражать в параметризации Сака, а не в использованном выше представлении Блатта - Биденхарна. Между этими представлениями фаз существует простая связь [8]

$$\theta_J^{J-1} + \theta_J^{J+1} = \delta_\alpha + \delta_\beta \ ,$$

$$\mathrm{tg}(\theta_J^{J-1} - \theta_J^{J+1}) = \cos(2\varepsilon)\,\mathrm{tg}(\delta_\alpha - \delta_\beta) \ , \qquad (1.11)$$

$$\sin(2E) = \sin(2\varepsilon)\sin(\delta_\alpha - \delta_\beta) \ ,$$

где $\theta_J^{J\pm1}$, E - фазы и параметр смешивания в параметризации Сака.

Отсюда при J = 1 находим

$$\sin(2E)=\sin(2\varepsilon)\sin(\delta_\alpha-\delta_\beta) , \qquad \mathrm{tg}(2E)=\sin(2E)/(1-\sin(2E)^2)^{1/2} ,$$

$$E=1/2\ \mathrm{atn}(\mathrm{tg}(2E)) , \qquad \theta^0=1/2\ (\delta_\alpha+\delta_\beta+\mathrm{atn}(\cos(2\varepsilon)\mathrm{tg}(\delta_\alpha-\delta_\beta))) ,$$

$$\theta^2=\delta_\alpha+\delta_\beta-\theta^0 . \qquad (1.12)$$

Это и есть нужные нам выражения для фаз упругого нуклон-нуклонного рассеяния в параметризации Сака, которые обычно определяются при фазовом анализе экспериментальных данных.

## 1.2 Численные методы решения системы уравнений Шредингера в задачах рассеяния

Решения системы (1.1) связанных уравнений Шредингера с начальными условиями (1.2) образуют линейно независимые комбинации, представляемые в форме [8]

$u_\alpha = C_{1\alpha} u_1 + C_{2\alpha} u_2$ ,
$w_\alpha = C_{1\alpha} w_1 + C_{2\alpha} w_2$ ,

$u_\beta = C_{1\beta} u_1 + C_{2\beta} u_2$ ,
$w_\beta = C_{1\beta} w_1 + C_{2\beta} w_2$ .

Система уравнений (1.1) может решаться методом Рунге - Кутта [9] с автоматическим выбором шага по заданной точности вычисления фаз и параметра смешивания. Для нахождения решения системы двух уравнений второго порядка, перепишем уравнение (1.1) в следующем виде

$u''(x) + A(x)u(x) = B(x)w(x)$ ,

$w''(x) + C(x)w(x) = B(x)u(x)$ , (1.13)

или

$u''(x) = B(x)w(x) - A(x)u(x) = F(x,u,w)$ ,

$w''(x) = B(x)u(x) - C(x)w = G(x,u,w)$ . (1.14)

Введем две новые переменные

$y = u'$ ,

$z = w'$ .

Тогда система (1.13) или (1.14) перепишется в виде четырех связанных уравнений первого порядка

$$u' = y \ ,$$
$$y' = F(x,u,w) \ ,$$
$$w' = z \ ,$$
$$z' = G(x,u,w) \ , \qquad (1.15)$$

с двумя начальными условиями

1. $u_1(0) = 0$ , $y_1(0) = const$ , $w_1(0) = 0$ , $z_1(0) = 0$ , (1.16)

2. $u_2(0) = 0$ , $y_2(0) = 0$ , $w_2(0) = 0$ , $z_2(0) = const$ .

Решение системы (1.15), записанной в общем виде

$$u' = f(x,y,z,u,w) \ ,$$
$$y' = g(x,y,z,u,w) \ ,$$
$$w' = d(x,y,z,u,w) \ ,$$
$$z' = s(x,y,z,u,w) \ , \qquad (1.17)$$

можно представить в форме [7]

$$y_{n+1} = y_n + \Delta y_n \ ,$$
$$z_{n+1} = z_n + \Delta z_n \ ,$$

$$u_{n+1} = u_n + \Delta u_n \ ,$$
$$w_{n+1} = w_n + \Delta w_n \ , \qquad (1.18)$$

где

$$\Delta y_n = 1/6(k_1 + 2k_2 + 2k_3 + k_4) \ ,$$
$$\Delta z_n = 1/6(m_1 + 2m_2 + 2m_3 + m_4) \ ,$$

$$\Delta u_n = 1/6(v_1 + 2v_2 + 2v_3 + v_4) \ ,$$
$$\Delta w_n = 1/6(b_1 + 2b_2 + 2b_3 + b_4) \ , \qquad (1.19)$$

и

$$k_1 = hg(x_n,y_n,z_n,u_n,w_n) \ ,$$

$$m_1 = hs(x_n,y_n,z_n,u_n,w_n) \ ,$$
$$v_1 = hf(x_n,y_n,z_n,u_n,w_n) \ ,$$
$$b_1 = hd(x_n,y_n,z_n,u_n,w_n) \ ,$$

$$k_2 = hg(x_n+h/2,\ y_n+k_1/2,\ z_n+m_1/2,\ u_n+v_1/2,\ w_n+b_1/2) \ ,$$
$$m_2 = hs(x_n+h/2,\ y_n+k_1/2,\ z_n+m_1/2,\ u_n+v_1/2,\ w_n+b_1/2) \ ,$$
$$v_2 = hf(x_n+h/2,\ y_n+k_1/2,\ z_n+m_1/2,\ u_n+v_1/2,\ w_n+b_1/2) \ ,$$
$$b_2 = hd(x_n+h/2,\ y_n+k_1/2,\ z_n+m_1/2,\ u_n+v_1/2,\ w_n+b_1/2) \ ,$$

$$k_3 = hg(x_n+h/2,\ y_n+k_2/2,\ z_n+m_2/2,\ u_n+v_2/2,\ w_n+b_2/2) \ , \qquad (1.20)$$
$$m_3 = hs(x_n+h/2,\ y_n+k_2/2,\ z_n+m_2/2,\ u_n+v_2/2,\ w_n+b_2/2) \ ,$$
$$v_3 = hf(x_n+h/2,\ y_n+k_2/2,\ z_n+m_2/2,\ u_n+v_2/2,\ w_n+b_2/2) \ ,$$
$$b_3 = hd(x_n+h/2,\ y_n+k_2/2,\ z_n+m_2/2,\ u_n+v_2/2,\ w_n+b_2/2) \ ,$$

$$k_4 = hg(x_n+h,\ y_n+k_3,\ z_n+m_3,\ u_n+v_3,\ w_n+b_3) \ ,$$
$$m_4 = hs(x_n+h,\ y_n+k_3,\ z_n+m_3,\ u_n+v_3,\ w_n+b_3) \ ,$$
$$v_4 = hf(x_n+h,\ y_n+k_3,\ z_n+m_3,\ u_n+v_3,\ w_n+b_3) \ ,$$
$$b_4 = hd(x_n+h,\ y_n+k_3,\ z_n+m_3,\ u_n+v_3,\ w_n+b_3) \ .$$

Поскольку

$$f(x,y,z,u,w) = y \ ,$$
$$g(x,y,z,u,w) = F(x,u,w) \ ,$$
$$d(x,y,z,u,w) = z \ ,$$
$$s(x,y,z,u,w) = G(x,u,w) \ ,$$

то формулы (1.20) преобразуются к виду

$$k_1 = hF(x_n,u_n,w_n) \ ,$$
$$m_1 = hG(x_n,u_n,w_n) \ ,$$
$$v_1 = hy_n \ ,$$
$$b_1 = hz_n \ , \qquad (1.21)$$

$$k_2 = hF(x_n+h/2,\ u_n+v_1/2,\ w_n+b_1/2) \ ,$$
$$m_2 = hG(x_n+h/2,\ u_n+v_1/2,\ w_n+b_1/2) \ ,$$
$$v_2 = h(y_n+k_1/2) \ ,$$
$$b_2 = h(z_n+m_1/2) \ ,$$

$k_3 = hF(x_n+h/2,\ u_n+v_2/2,\ w_n+b_2/2)$ ,
$m_3 = hG(x_n+h/2,\ u_n+v_2/2,\ w_n+b_2/2)$ ,
$v_3 = h(y_n+k_2/2)$ ,
$b_3 = h(z_n+m_2/2)$ ,

$k_4 = hF(x_n+h,\ u_n+v_3,\ w_n+b_3)$ ,
$m_4 = hG(x_n+h,\ u_n+v_3,\ w_n+b_3)$ ,
$v_4 = h(y_n+k_3)$ ,
$b_4 = h(z_n+m_3)$ .

Тогда выражения (1.19) для определения функций заметно упрощаются

$$\Delta u_n = 1/6h(6y_n + k_1 + k_2 + k_3)\ ,$$
$$\Delta w_n = 1/6h(6z_n + m_1 + m_2 + m_3) \qquad (1.22)$$

и в формулах (1.21) нужно вычислять только две величины k и m.

## 1.3 Компьютерная программа решения уравнения Шредингера для потенциалов с тензорной компонентой в задачах рассеяния

Компьютерная программа для поиска ядерных фаз рассеяния в системах с тензорными силами написана на алгоритмическом языке “Basic” и использовалась для расчетов в среде компилятора “Turbo Basic-1.0” фирмы “Borland International Inc.” [6,10].

Ниже приведены результаты контрольного счета по этой программе для случая классического NN потенциала Рейда с мягким кором [11]. Вычислительная точность в компьютерной программе задавалась на уровне 0.5%, начальное число шагов 1000, а величина начального шага принималась равной 0.02. Для определения фаз NN рассеяния сшивка численной волновой функции с ее асимптотикой выполнялась на рас-стояниях 20 Фм. Результаты наших расчетов фаз рассеяния и их сравнение с результатами Рейда [11] приведены в табл.1.1.

Табл. 1.1 - Результаты расчета ядерных фаз рассеяния. Здесь: E - энергия частиц, $\delta_{\alpha,\beta}$ - фазы рассеяния, ε - параметр смешивания.

| E, MeV | $\delta_\alpha$, rad Рез-ты [11] | $\delta_\alpha$, rad (Наш расчет) | $\delta_\beta$, rad Рез-ты [11] | $\delta_\beta$, rad (Наш расчет) | Sin(2ε), Рез-ты [11] | Sin(2ε) (Наш расчет) |
|---|---|---|---|---|---|---|
| 24 | 1.426 | 1.426 | -0.050 | -0.050 | 0.064 | 0.064 |
| 48 | 1.105 | 1.105 | -0.115 | -0.116 | 0.081 | 0.081 |
| 96 | 0.749 | 0.748 | -0.215 | -0.216 | 0.114 | 0.114 |
| 144 | 0.521 | 0.520 | -0.281 | -0.282 | 0.152 | 0.152 |
| 208 | 0.300 | 0.299 | -0.340 | -0.341 | 0.203 | 0.203 |
| 304 | 0.057 | 0.056 | -0.403 | -0.404 | 0.269 | 0.269 |

Как видно из этой таблицы отличие наших расчетов и результатов, приведенных в работе Рейда, составляет величину порядка 0.001 радиана и находится на уровне ошибок округления, что демонстрирует полную работоспособность, как описанных выше математических методов, правильность выбора численных способов решения уравнения Шредингера, так и составленной компьютерной программы.

В дальнейшем эти численные методы и компьютерная программа использовались для вычисления ядерных фаз $^2H^4He$ и NN упругого рассеяния для потенциалов с тензорной компонентой и запрещенным только в S - волне состоянием [12,13].

В этих работах впервые была показана возможность построения потенциалов с тензорной компонентой для этих ядерных систем, которые полностью удовлетворяли общетеоретическим выводам о структуре их запрещенных и разрешенных состояний, т.е. не содержат запрещенного связанного уровня в D - волне.

Предложенные потенциалы хорошо согласованы с фазами упругого рассеяния при низких и средних энергиях и ха-рактеристиками низкоэнергетического NN рассеяния. Они позволяют правильно описать практически все рассмотренные характеристики связанных состояний ядер $^2$H и $^6$Li [14], включая квадрупольный момент $^6$Li, который не описывался ранее в рамках других моделей или каких-либо иных подходов.

Использование концепции запрещенных состояний позволило внести определенную структуру в те области ядра, где ранее предполагалось наличие только феноменологического отталкивающего кора [14], что позволило сократить число подгоночных параметров, например, для NN потенциала с нескольких десятков до, всего лишь, трех. Они определяют глубину и ширину потенциальной ямы, и фактор обрезания потенциала однопионного обмена.

Причем, первые два, однозначно фиксируются по низкоэнергетическим характеристикам рассеяния и, только третий, варьируется для достижения наилучшего описания S и D фаз NN рассеяния и параметра смешивания в области до 500 МэВ.

Кроме того, было показано, что наилучшие результаты по описанию фаз NN рассеяния достигаются при значении константы πNN связи равной 0.074. Это полностью подтверждает полученные ранее результаты Нимегенской группы [15] для некоторых типов NN взаимодействий с отталкивающим кором и лучше всего согласуются с экспериментальными данными, выполненными группой политехнического института Виржинии [16], которая получила значение 0.076.

## 1.4 Постановка задачи для решения системы уравнений Шредингера на связанные состояния

Для расчетов энергии и волновых функций связанных состояний ядерной системы с тензорными потенциалами исходим из обычных уравнений Шредингера (1.1).

Решением этой системы являются четыре волновые функции, получающиеся с начальными условиями типа (1.2), которые образуют линейно независимые комбинации, представляемые в виде

$$\chi_0 = C_1 u_1 + C_2 u_2 = \exp(-kr) \ , \qquad (1.23)$$

$$\chi_2 = C_1 w_1 + C_2 w_2 = [1 + 3/kr + 3/(kr)^2]\exp(-kr) \ ,$$

или с учетом кулоновских сил

$$\chi_0 = C_1 u_1 + C_2 u_2 = W_{\eta,0}(2kr) \ , \qquad (1.24)$$

$$\chi_2 = C_1 w_1 + C_2 w_2 = W_{\eta,2}(2kr) \ ,$$

где $W_{\eta,L}(2kr) = W_{\eta L}(Z)$ - функция Уиттекера [4] для связанных состояний, которая является решением исходных уравнений (1.1) при $k^2<0$ без ядерных потенциалов, которая определяет асимптотику ВФ при $r \geq R$ = 15-20 Фм; $Z = 2kR$ и

$$\eta = \frac{\mu Z_1 Z_2}{k\hbar^2}$$

кулоновский параметр.

Волновые функции связанных состояний нормированы на единицу следующим образом

$$\int_0^\infty \left(\chi_0^2 + \chi_2^2\right) dr = 1$$

а интеграл от квадрата волновой функции D состояния определяет ее вес.

Орбитальные состояния системы при наличии тензорных потенциалов смешиваются, и сохраняется только полный момент, который определяется векторной суммой орбиталь-

ных и спиновых моментов

**J** = **S** + **L** .

Откуда для орбитального момента можно получить выражение

/J - S/ ≤ L ≤ /J + S/ .

В частности, для дейтрона или $^6$Li в двухкластерной $^2$H$^4$He модели, полный момент равен единице, спин также единица, и орбитальный момент может принимать два значения 0 и 2.

## 1.5 Численные методы решения системы уравнений Шредингера в задачах на связанные состояния

Для нахождения энергий и ВФ связанных состояний системы (1.1) с тензорным потенциалом использовалась комбинация численных и вариационных методов, которая заключается в последовательном выполнении следующих шагов:

1. При некоторой изначально заданной энергии связанного состояния (которая не является собственным значением данной задачи), численным методом находилась ВФ системы (1.1). Для этого использовался обычный метод Рунге - Кутта [9].
2. Затем система уравнений (1.1) представлялась в конечно - разностном виде с центральными разностями [9]

$$u_{i+1} - 2u_i + u_{i-1} + h^2[\, k^2 - V_c - V_{cul}]u_i = h^2\sqrt{8}\, V_t w_i \ , \qquad (1.25)$$

$$w_{i+1} - 2w_i + w_{i-1} + h^2[\, k^2 - V_c - 6/r^2 - V_{cul} + 2\, V_t\, ]w_i = h^2\sqrt{8}\, V_t u_i \ ,$$

или

$$u_{i+1} + h^2[-2/h^2 + k^2 - V_c - V_{cul}]u_i + u_{i-1} - h^2\sqrt{8}\,V_t w_i = 0\ ,$$

$$w_{i+1} + h^2[-2/h^2 + k^2 - V_c - 6/r^2 - V_{cul} + 2V_t]w_i + w_{i-1} - h^2\sqrt{8}\,V_t u_i = 0$$

и полученная численная ВФ подставлялась в эту систему уравнений.

3. Левая часть этих уравнений будет равна нулю только в случае, когда энергия и ВФ являются собственными решениями такой задачи. При произвольной энергии и найденной по ней ВФ левая часть будет отлична от нуля, и можно говорить о методе невязок [17], который позволяет оценить степень точности нахождения собственных функций и собственных значений.

4. Из уравнений

$$N_{si} = u_{i+1} + h^2[-2/h^2 + k^2 - V_c - V_{cul}]u_i + u_{i-1} - h^2\sqrt{8}\,V_t w_i\ , \quad (1.26)$$

$$N_{ti} = w_{i+1} + h^2[-2/h^2 + k^2 - V_c - 6/r^2 - V_{cul} + 2V_t]w_i + w_{i-1} - h^2\sqrt{8}\,V_t u_i$$

вычислялась сумма невязок в каждой точке численной схемы

$$N_s = \sum_i N_{si}\ ,$$

$$N_t = \sum_i N_{ti}\ .$$

5. Варьируя энергию связи (или $k^2<0$), проводилась минимизация значений всех невязок

$$\delta\left[\left|N_s(k^2)\right| + \left|N_t(k^2)\right|\right] = 0\ . \quad (1.27)$$

6. Энергия ($k^2$), дающая минимум невязок, считалась собственной энергией $k_0^2$, а функции $\chi_0$ и $\chi_2$, приводящие к этому минимуму - собственными функциями задачи, т.е. ВФ связанного состояния ядерной системы [4].

Рассмотренный вариационный метод сходится достаточно быстро, позволяет получать практически любую реальную точность, при использовании в программе двойной точности, и может применяться при решении любых задач на собственные значения для системы двух дифференциальных уравнений, типа уравнения Шредингера.

Этот метод прекрасно показал свою работоспособность, как для контрольных задач, в качестве которых выбиралась нуклон - нуклонная система с классическим потенциалом Рейда [11], так и реальных расчетов физических характеристик связанных состояний кластеров в легких атомных ядрах [4].

## 1.6 Программа решения уравнения Шредингера для потенциалов с тензорной компонентой в задачах на связанные состояния

На основе приведенных выражений (1.25)-(1.27), на алгоритмическом языке “Basic” в среде компилятора “Turbo Basic-1.0” фирмы “Borland International Inc.”, была написана компьютерная программа [18], которая использовалась для вычисления ядерных характеристик дейтрона и связанных состояний $^4He^2H$ кластерной системе ядра $^6Li$.

Программа тестировалась на нуклон - нуклонном потенциале Рейда [11] и сравнение результатов, полученных в работе [11] и по разработанной здесь программе приведены в табл.1.2.

Здесь приняты следующие обозначения: $E_d$ - энергия связи дейтрона; $R_d$ - среднеквадратичный радиус дейтрона; $Q_d$ - квадрупольный момент дейтрона; $P_d$ - вероятность D - со-стояния в дейтроне; $A_s$ - асимптотическая константа S - волны; η - отношение асимптотических констант D и S волн; $a_t$ - триплетная длина нуклон - нуклонного рассеяния; $a_s$ - синглетная длина нуклон - нуклонного рассеяния; $r_t$ - триплетный эффективный радиус нуклон - нуклонного рассеяния; $r_s$ - синглетный эффективный радиус нуклон - нуклонного рассеяния.

Табл.1.2. Характеристики дейтрона и np рассеяния.

| Характеристики дейтрона | Расчет Рейда [11] | Наш Расчет [14] |
|---|---|---|
| $E_d$, МэВ | 2.22464 | 2.22458 |
| $Q_d$, $Фм^2$ | 0.2762 | 0.2757 |
| $P_d$ , % | 6.217 | 6.217 |
| $A_S$ | 0.87758 | 0.875(2) |
| $\eta=A_D/A_S$ | 0.02596 | 0.0260(2) |
| $a_t$, Фм | 5.390 | 5.390 |
| $r_t$, Фм | 1.720 | 1.723 |
| $a_s$, Фм | -17.1 | -17.12 |
| $r_s$, Фм | 2.80 | 2.810 |
| $R_d$, Фм | 1.956 | 1.951 |

Из этих результатов видно, что совпадение наших расчетов и расчетов Рейда [11] по энергии связанного состояния дейтрона имеет величину порядка нескольких тысячных процента или 60 эВ. Ошибки в асимптотических константах получены усреднением их значений в области 10-15 Фм и в пределах этих ошибок согласуются с результатами работы [11]. Низкоэнергетические np характеристики, по сути, совпадают между собой с точностью до ошибок округления. Величина квадрупольного момента и среднеквадратичного радиуса несколько меньше значений, полученных в работе [11]. Это обусловлено тем, что в наших расчетах не учитывался очень длинный хвост ВФ, и интегрирование проводилось только до 20 Фм. Но это отличие для $Q_d$ составляет величину меньше 0.2%, а для среднеквадратичного радиуса меньше

0.3%.

Далее этот метод использовался для рассмотрения характеристик связанных состояний кластеров в легких атомных ядрах, в частности, связанного состояния $^2H^4He$ кластеров с тензорными силами в атомном ядре $^6Li$ [10], и позволил получить хорошие результаты по описанию квадрупольного момента этого ядра.

Оказалось, что на основе простых гауссовых потенциалов в качестве центральной и тензорной частей, на базе единых параметров можно правильно описать не только фазы упругого рассеяния, но и среднеквадратичный радиус ядра, квадрупольный момент и асимптотические константы связанного состояния в этом канале.

Правильно получается не только отрицательный знак, но и величина $\eta_D$, определяющая отношение асимптотических констант в D и S волнах [14]. Причем только при ее отрицательных значениях можно получить правильный по величине и знаку, отрицательный квадрупольный момент $^6Li$, равный -0.064 Фм$^2$, который хорошо согласуется с его экспериментальным значением -0.0644(7) Фм$^2$ [19]. Для величины $\eta_D$ было получено -0.0120(10) при известной экспериментальной величине -0.0125(25) [20].

Во всех этих расчетах задавались целые значения масс частиц [21], а константа $\hbar^2/m_0$ принималась равной 41.47 МэВ Фм$^2$.

Кулоновский параметр

$$\eta = \mu Z_1 Z_2 e^2/(k\hbar^2),$$

представлялся в виде

$$\eta = 3.44476\ 10^{-2} Z_1 Z_2 \mu/k,$$

где k - волновое число в Фм$^{-1}$, $\mu$ - приведенная масса в а.е.м.

Кулоновский потенциал с $R_{cul} = 0$ записывался

$V_{cul}$ (МэВ) = 1.439975 $Z_1Z_2/r$,

где r - расстояние в Фм.

# Заключение

Таким образом, предложенная комбинация численных и вариационных методов математической модели, используемой для решения системы уравнений Шредингера в задаче на связанные состояния с тензорными силами, позволяет получить полностью устойчивые решения, контролируемые с помощью невязок, при нахождении собственных значений рассмотренной системы квантовых частиц для дискретного спектра.

Вариационный процесс сходится сравнительно быстро и позволяет получить практически любую точность при вычислении энергии связи. Найденные, в результате таких вычислений, ВФ имеют правильную асимптотику на расстояниях порядка 10-15 Фм, что подтверждается устойчивостью асимптотических констант в этой области расстояний.

Использование многопараметрического вариационного метода для математической модели обратной задачи квантовой теории рассеяния, рассматриваемой на основе системы уравнений Шредингера с тензорными силами, который применяется для построения ядерных потенциалов, позволяет вполне однозначно определять значения параметров таких потенциалов по фазам упругого рассеяния.

Этому способствует и классификация связанных состояний по орбитальным схемам Юнга, которая позволяет четко определить статус уровня - запрещенный или разрешенный. От этого зависит форма потенциала, а значит и количество узлов в каждой парциальной волне ВФ рассматриваемой ядерной системы.

Такая методика построения ядерных потенциалов избавляет нас от дискретной и непрерывной неоднозначности получения параметров взаимодействий, присутствующих в стандартной оптической модели и делает процедуру нахождения параметров потенциалов вполне однозначной [22].

# 2. АСТРОФИЗИЧЕСКИЕ S - ФАКТОРЫ

Переходя к рассмотрению астрофизических S - факторов заметим, что для проведения всех расчетов использована микроскопическая двухкластерная модель легких ядер с классификацией кластерных состояний по орбитальным симметриям, которая выполнена с помощью схем Юнга и потенциалы, параметры которых заранее фиксированы по фазам упругого рассеяния соответствующих кластеров.

Полученные результаты согласуются с имеющимися экспериментальными данными при энергиях 0.5 - 20 МэВ [23,24]. Поэтому вполне естественно проводить вычисления и для более низких, астрофизических энергий порядка 1 кэВ - 0.5 МэВ.

Такой подход дает возможность избавиться от традиционной экстраполяции экспериментальных данных в область астрофизических энергий, который приводит к большей неоднозначности получаемых результатов [14].

Радиационный $p^2H$ захват при сверхнизких энергиях входит в протон - протонный термоядерный цикл и дает существенный вклад в энергетический выход ядерных реакций [25], которые обуславливают горение солнца и звезд нашей Вселенной. Поскольку взаимодействующие ядерные частицы водородного цикла имеют минимальную величину потенциального барьера, он является первой цепочкой ядерных реакций, которые могут происходить при самых низких энергиях и звездных температурах.

В этой цепочке процесс радиационного $p^2H$ захвата является основным для перехода от первичного

$$p+p \rightarrow {}^2H+e^-+\nu_e$$

слияния протонов, идущего почти со 100% вероятностью, до финального, в р-р - цепочке, процесса вероятность которого около 85% [26]. Поэтому детальное изучение реакции $p^2H$ захвата с теоретической и экспериментальной точки зрения представляет существенный интерес не только для ядерной астрофизики, но и вообще для всей ядерной физики сверхнизких энергий и легчайших атомных ядер.

$^3He+^3He \rightarrow ^4He+2p$,

Процесс радиационного $p^3H$ захвата при звездных энергиях, мог по-видимому, происходит на дозвездной стадии развития Вселенной [1] и приводит к образованию стабильного ядра $^4He$.

В рассматриваемой модели находятся как полные сечения фотоядерных реакций, входящих в водородный термоядерный цикл, так и астрофизические S - факторы в области сверхнизких энергий. Полученные результаты разумно согласуются и имеющимися экспериментальными данными для S - факторов при нулевой энергии.

## 2.1 Радиационный $p^2H$ захват

Рассмотрим возможность описания астрофизических S-факторов на основе потенциальной кластерной модели, в которой учитывается супермультиплетная симметрия волновой функции с разделением орбитальных состояний по схемам Юнга, позволяющая анализировать структуру межкластерных взаимодействий, определять наличие и положение разрешенных и запрещенных состояний в межкластерных потенциалах, как это было сделано в работах [23,27].

В рамках этой концепции были выполнены расчеты дифференциальных сечений фотопроцессов в $N^2H$, $N^3H$ и многих других кластерных системах для потенциалов с запрещенными состояниями и разделением состояний по орби-тальным симметриям [27]. Такой подход позволяет хорошо описать имеющиеся экспериментальные данные, и дает возможность рассматривать структуру межкластерных взаимодействий на тех расстояниях, где обычно предполагается присутствие отталкивающего кора.

Полные сечения фотопроцессов для легчайших ядер в потенциальной кластерной модели с запрещенными состояниями и разделением орбитальных состояний по схемам Юнга рассматривались в нашей работе [23]. В этих расчетах процессов фоторазвала ядер $^3$He и $^3$H в p$^2$H и n$^2$H каналы учитывались E1 переходы, обусловленные орбитальной частью электрического оператора $Q_{Jm}(L)$ [14].

Сечения E2 процессов и сечения, зависящие от спиновой части электрического оператора, оказались на несколько порядков меньше. Далее предполагалось, что электрические E1 переходы в N$^2$H системе возможны между основным чистым $^2$S состоянием ядер $^3$H и $^3$He и дублетным $^2$P состоянием рассеяния [14].

Для выполнения расчетов фотоядерных процессов в рассматриваемых системах ядерная часть межкластерного потенциала p$^2$H, p$^3$H и p$^3$He взаимодействий представляется в виде

$$V(r)=V_0\exp(-\alpha r^2)+V_1\exp(-\beta r) \qquad (2.1)$$

с обычным кулоновским потенциалом при нулевом радиусе, гауссовой притягивающей с $V_0$ и экспоненциальной отталкивающей с $V_1$ частью. Потенциал каждой парциальной волны строился так, чтобы правильно описывать соответствующую парциальную фазу упругого рассеяния [28]. Используя эти данные, были получены потенциалы p$^2$H взаимодействия для процессов рассеяния, параметры которых приведены в табл.2.1 [29].

Они позволяют хорошо описать экспериментальные данные по фазам рассеяния в обоих спиновых каналах, но в дублетном состоянии приводят к неправильной величине энергии связи ядер $^3$He и $^3$H, т.к. эти состояния оказываются смешанными по схемам Юнга [14,27].

Табл. 2.1. Дублетные потенциалы взаимодействия $p^2H$ [23] системы, смешанные по схемам Юнга и синглетные потенциалы $p^3He$ [24] системы чистые по орбитальным симметриям и изоспину с T = 1.
Здесь: $E_{cc}$ - энергии связанных состояний. В скобках приведены значения энергии для $n^2H$ и $n^3H$ систем.

| Система | L | $V_0$, (МэВ) | $\alpha$, (Фм$^{-2}$) | $V_1$, (МэВ) | $\beta$, (Фм$^{-1}$) | $E_{cc}$, (МэВ) |
|---|---|---|---|---|---|---|
| $p^2H$ | Чет. | -35.0 | 0.1 | -- | -- | -9.3 (-10.1) |
| | Нечет. | -10.0 | 0.16 | +0.6 (+0.1) | 0.1 | --- |
| $p^3He$ | Чет. | -110 | 0.37 | +45 | 0.67 | -9.0 (-11.4) |
| | Нечет. | -15 | 0.1 | --- | --- | --- |

Экспериментальные дублетные фазы в $N^2H$ системе, смешанные по схемам Юнга {3} и {21}, могут быть представлены в виде полусуммы чистых, с одной определенной схемой Юнга, фаз рассеяния [14,27]

$$\delta_L^{\{f_1\}+\{f_2\}} = \frac{1}{2}\delta_L^{\{f_1\}} + \frac{1}{2}\delta_L^{\{f_2\}}.$$

В данном случае имеем $\{f_1\}=\{3\}$ и $\{f_2\}=\{21\}$, и дублетные фазы оказываются смешанными по схемам Юнга с {3}+{21}.

Если допустить, что в качестве дублетной фаз с {21} могут быть использованы квартетные фазы той же симметрии {21}, то легко определить чистые дублетные фазы с {3} - это и было сделано на основе экспериментальных данных работы [28] по фазовому анализу упругого $p^2H$ рассеяния.

Таким образом, в дублетном канале были выделены чистые фазы и на их основе построены чистые по схемам Юнга потенциалы межкластерного взаимодействия, параметры которых приведены в табл.2.2 [14,23].

Табл. 2.2. Чистые по схемам Юнга потенциалы $p^2H$ [23] и $p^3H$ [24] взаимодействий в дублетном и синглетном каналах. Здесь: $E_{cc}$ - расчетная энергия связанных состояний, $E_{эксп.}$ - ее экспериментальное значение [30]. В скобках указаны значения энергии связи и глубины потенциала для $n^3He$ системы.

| Система | L | $V_0$, (МэВ) | $\alpha$, ($Фм^{-2}$) | $E_{cc}$, (МэВ) | $E_{эксп.}$ (МэВ) |
|---|---|---|---|---|---|
| $p^2H$ | Чет. | -34.76305 | 0.15 | -5.49400 | -5.494 |
| | Нечет. | +2.4 | 0.01 | --- | --- |
| $p^3H$ | Чет. | -62.90714 (-62.4069) | 0.17 | -19.81400 (-20.87063) (-20.57800) | -19.814 (-20.578) |
| | Нечет. | +8 | 0.03 | --- | --- |

С этими потенциалами были выполнены расчеты полных сечений радиационного $p^2H$ захвата и астрофизических S - факторов при энергиях до 10 кэВ [23], хотя на тот момент нам были известны только экспериментальные данные по S - фактору в области выше 150-200 кэВ [31].

Сравнительно недавно появились новые экспериментальные данные при энергиях до 2.5 кэВ [32,33,34]. Поэтому представляется интересным выяснить - способна ли потенциальная кластерная модель описать новые данные на основе полученных ранее потенциалов.

Наши предварительные результаты показали, что для расчетов S - фактора при энергиях порядка нескольких кэВ требуется существенно повысить точность нахождения энергии связи $p^2H$ системы в ядре $^3He$, которая находилась на уровне 1-2 кэВ, и более строго контролировать поведение «хвоста» волновой функции (ВФ) связанного состояния (СС) на больших расстояниях.

Кроме того, требовалось повысить точность нахождения кулоновских волновых функций [4], определяющих поведение асимптотики волновой функции рассеяния в дублетной $^2P$ волне. Для этой цели была несколько модифицирована программа расчета полных сечений E1 захвата в $p^2H$ канале [4]. Теперь относительная точность вычисления кулоновских функций, контролируемая по величине Вронскиана, и точность поиска корня детерминанта [4], определяющая точность определения энергии связи ядра, находятся на уровне $10^{-15}$.

Для контроля поведения ВФ СС на больших расстояниях вычислялись асимптотические константы $C_0$ и $C_W$ - для первой из них асимптотика представлялась в виде экспоненты, для второй в виде точной функции Уиттекера [4]. Величина первой константы в интервале 8-16 Фм оказалась равна 2.0(1), а второй - 2.4(1). Известные нам экспериментальные данные по этим константам приводят к значениям 1.8-3.0 [35].

Кроме того, для более правильного описания экспериментального значения энергии связи ядра $^3$He в p$^2$H канале, были уточнены параметры чистого дублетного потенциала взаимодействия. Полученное значение параметров потенциала и расчетная энергия связи, вместе с экспериментом, приведены в табл.2.2. Такой потенциал стал глубже на 0.01305 МэВ [23] и приводит к полному совпадению расчетной - 5.49400 МэВ и экспериментальной энергий связи -5.494 МэВ [30].

Все расчеты выполнялись конечно-разностным методом (КРМ) [4], который учитывает кулоновское взаимодействие. Точность вычисления КРМ энергии, и ее сводимость в зависимости от числа шагов конечно-разностной сетки N и расстояния R, которое определяет область поиска энергия связи, [4] приведены в табл.2.3.

Табл.2.3. Сходимость энергии связи в p$^2$H канале ядра $^3$He. В скобках показана величина шага H конечно-разностной сетки в Фм.

| N | $E_{cc}$, МэВ | | | |
|---|---|---|---|---|
| | R = 10 Фм | R = 13 Фм | R = 16 Фм | R = 19 Фм |
| 1000 | -5.49404 (0.001) | -5.49407 (0.0013) | -5.49411 (0.0016) | -5.49416 (0.0019) |
| 2000 | -5.49401 (0.005) | -5.49401 (0.0065) | -5.49402 (0.008) | -5.49404 (0.0095) |
| 3000 | -5.49400 (0.00333) | -5.49400 (0.00433) | -5.49401 (0.00533) | -5.49401 (0.00633) |
| 4000 | -5.49400 (0.0025) | -5.49400 (0.00325) | -5.49400 (0.004) | -5.49400 (0,00475) |

Видно, что для такого потенциала насыщение по энергии связи -5.49400 МэВ достигается при числе шагов 4000 и уже не зависит от R, что говорит о насыщении вычислительного процесса при шаге меньше 0.005 Фм.

Для дополнительного контроля правильности вычисления энергии связи использовался вариационный метод [4], который уже на сетке с размерностью 10 и независимом варьировании параметров позволил получить энергию -5.49399 МэВ.

Асимптотическая константа $C_w$ вариационной волновой функции, параметры которой приведены в табл.2.4, на расстояниях 6-15 Фм сохраняется на уровне $C_W$ = 2.3(1), а величина невязок не превышает 5.8 $10^{-15}$ [4].

Табл. 2.4. Вариационные параметры $\alpha_i$ и коэффициенты разложения $C_i$ ВФ связанного состояния $p^2H$ системы.

| I | $\alpha_i$ | $C_i$ |
|---|---|---|
| 1 | 1.32946E-01 | -1.41373E-01 |
| 2 | 2.42134E-02 | -2.16041E-02 |
| 3 | 1.03213E-02 | -1.29669E-03 |
| 4 | 1.00370E-01 | -8.14873E-02 |
| 5 | 5.33156E-02 | -8.47630E-02 |
| 6 | 8.16886E+00 | +2.48810E-03 |
| 7 | 2.69096E-01 | -5.05526E-02 |
| 8 | 2.27912E-01 | -1.17652E-01 |
| 9 | 1.05537E+00 | -6.79733E-02 |
| 10 | 1.09123E+00 | +7.04227E-02 |

Поскольку вариационная энергия при увеличении раз-мерности базиса уменьшается и дает верхний предел истинной энергии связи, а конечно-разностная энергия при уменьшении величины шага увеличивается, то для реальной энергии связи в таком потенциале можно принять величину -5.493995(5) МэВ.

Во всех этих расчетах задавались точные значения масс частиц [21,36], а константа $\hbar^2/m$ принималась равной 41.4686 МэВ Фм$^2$. Кулоновский параметр и кулоновский потенциал определены в конце первой главы.

Далее, для расчетов S - фактора радиационного $p^2H$ захвата использовалось известное выражение [37], определяющее его через энергию частиц и полные сечения фотопроцессов

$$S = \sigma E_{cm} \exp\left( \frac{31.335\ Z_1 Z_2 \sqrt{\mu}}{\sqrt{E_{cm}}} \right)$$

где $\sigma$ - полное сечение процесса радиационного захвата в барн, $E_{cm}$ - энергия частиц в кэВ для системы центра масс, $\mu$ - приведенная масса в а.е.м. и Z - заряды частиц. Численный коэффициент 31.335 получен на основе современных значений фундаментальных констант [36].

Ранее проведенные расчеты E1 перехода показали [23], что лучшие результаты по описанию полных сечений радиационного захвата удается получить, если использовать потенциал $^2P$ волны $p^2H$ рассеяния с периферическим отталкиванием (табл.2.1 - глубина отталкивающей части потенциала +0.6 МэВ). Эти результаты вполне описывают новые данные по S - фактору до 20 кэВ. Однако, величина S - фактора при 1 кэВ находится на уровне 1.1(1) $10^{-4}$ кэВ бн., т.е. несколько ниже новых данных, если рассматривать полный S - фактор, без разделение его на $S_s$ и $S_p$ части, обусловленные M1 и E1 переходами, как это было сделано в работе [38], где получено $S_s(0)$ = 1.09(10) $10^{-4}$ кэВ бн, и $S_p(0)$ = 0.73(7) $10^{-4}$ кэВ бн, что дает для полного S - фактора значение 1.82(17) $10^{-4}$ кэВ бн.

Поскольку величина $^2P$ фазы $p^2H$ рассеяния параметризована с некоторой неоднозначностью [14], мы всегда можем варьировать глубину отталкивающей части потенциала, которая определяет поведение сечений и фаз рассеяния при малых энергиях. Наилучшие результаты по описанию полного S - фактора только на основе E1 перехода получаются с высотой отталкивающей части +0.1 МэВ (табл.2.1).

Результаты расчета S - фактора $p^2H$ захвата с модифицированным потенциалом $^2P$ волны при энергиях до 1 кэВ приведены на рис.2.1 и при энергиях выше 30-50 кэВ практически не отличаются от наших прежних результатов [23]. Теперь вычисленный S - фактор хорошо воспроизводит экспериментальные данные при энергиях до 10 кэВ, а при более низких энергиях расчетная кривая идет практически в полосе экспериментальных ошибок работы [34].

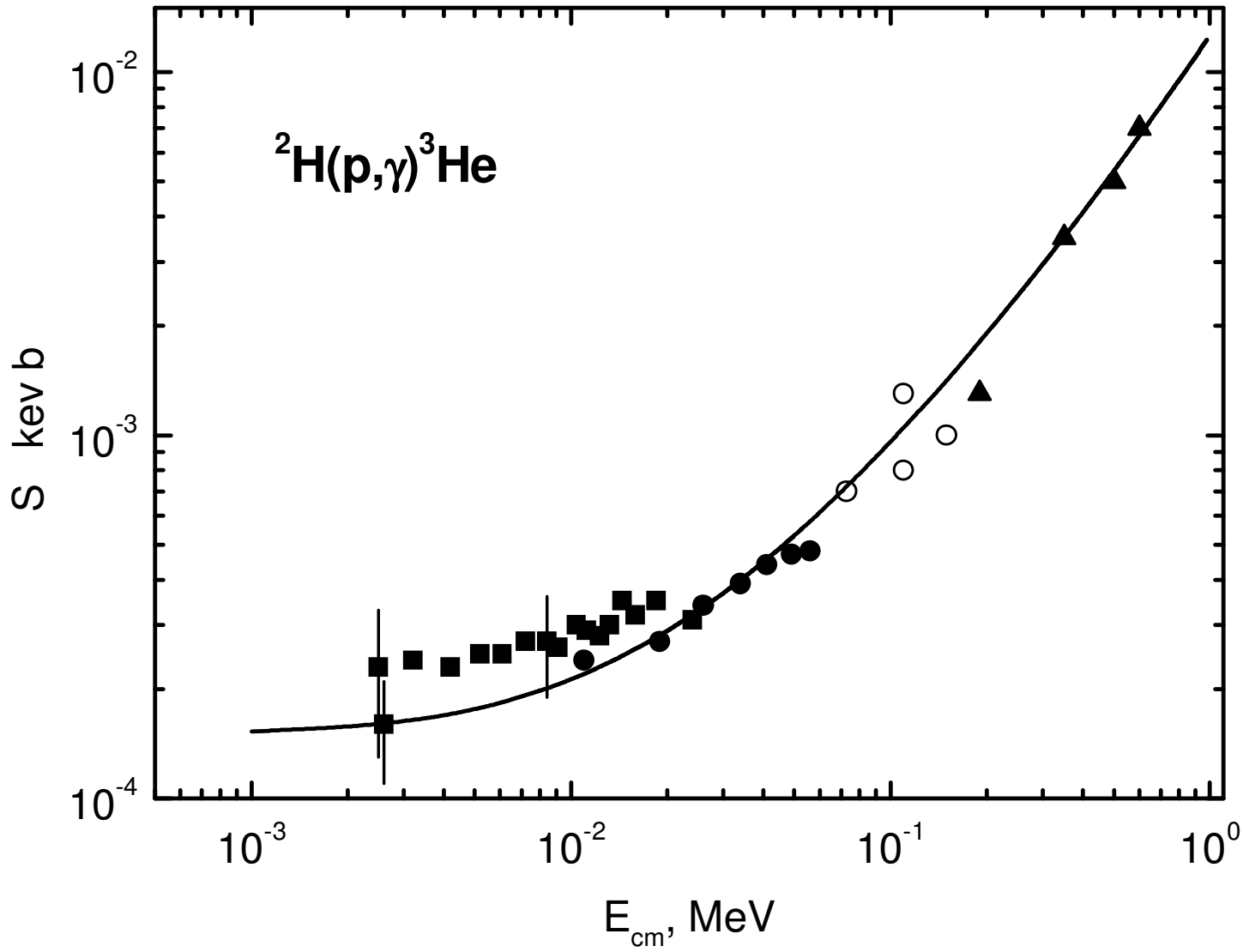

Рис.2.1. Астрофизический S-фактор радиационного $p^2H$ захвата. Треугольник - эксперимент из работы [31], кружки - [32], точки - [33], квадраты - [34].

Из рисунка видно, что ниже 3-4 кэВ S - фактор практиче-ски выходит на плато, которое при 1 кэВ приводит к значению 1.55(10) $10^{-4}$ кэВ бн. Здесь дана возможная суммарная ошибка всех вычислительных процедур, связанная с использованием в расчетах определенных численных методов, а также неопределенностей при нахождении фазы рассеяния [39].

Одно из последних экспериментальных измерений S - фактора при нулевой энергии дает 1.66(14) $10^{-4}$ кэВ бн [40], а предыдущие измерения этих же авторов 1.21(12) $10^{-4}$ кэВ бн [41]. В работе [34] для S(0) получено 2.16(10) $10^{-4}$, в [42] приводится величина 1.85(5) $10^{-4}$, а в [38] найдено 1.82(17) $10^{-4}$ кэВ бн. Среднее между этими экспериментальными измерениями приводит к значению 1.7(6) $10^{-4}$ кэВ бн, которое вполне согласуется с полученной здесь величиной 1.55(10) $10^{-4}$ кэВ бн.

В тоже время, если принять величину отталкивающей части $^2$P потенциала p$^2$H рассеяния +1.1 МэВ, для $S_p$ при 1 кэВ можно получить значение 0.8(1) $10^{-4}$ кэВ бн, согласующееся с данными [38]. Все эти изменения отталкивающей части P взаимодействия, по сути, находятся в пределах неоднозначностей различных результатов по фазовому анализу p$^2$H рассеяния [28].

Однако, потенциал с отталкиванием +0.1 МэВ более правильно описывает $^2$P фазу при наиболее низких энергиях, а в работе [23], мы строили потенциал так, чтобы он в целом описывал фазы при энергиях 5-15 МэВ, где наблюдается пик в полных сечениях p$^2$H захвата.

Отметим, что все эти изменение отталкивающей части потенциала слабо сказывается на полных сечениях p$^2$H захвата при энергиях 5-15 МэВ, полученных нами в работе [23, 43].

## 2.2 Радиационный p$^3$H захват

Рассмотрим теперь возможность описания S - фактора p$^3$H захвата в потенциальной кластерной модели с запрещен-ными состояниями. В рамках такого подхода выполнялись расчеты дифференциальных сечений $p^3H$ развала ядра $^4He$ [27,44]. Полные сечения этого процесса на основе потенциальной кластерной модели рассчитывались нами в работе [24], где считалось, что основной вклад в сечения E1 фоторазвала ядра $^4He$ в $p^3H$ канал или радиационного $p^3H$ захват дают процессы с изменением изоспина $\Delta T = 1$ [45]. Поэтому можно использовать $^1P_1$ потенциал рассеяния чистого по изоспину с $T = 1$ синглетного состояния $p^3He$ системы, параметры которого приведены в табл.2.1 и потенциал основного чистого состояния ядра $^4He$ с $T = 0$ для системы $p^3H$ (табл.2.2) [24].

Параметры потенциалов (табл.2.1) строятся так, чтобы правильно описать экспериментальные фазы упругого рассеяния в $p^3He$ системе с $T = 1$. Поскольку имеется несколько различных вариантов фазовых анализов [46] для синглетной $^1P_1$ и триплетной $^3P_1$ волн, параметры потенциала, приведенные в табл.2.1, подбирались так, чтобы получить определенный компромисс между разными фазовыми анализами [14].

Фазы рассеяния $p^3H$ системы находятся вполне однозначно и приводят к определенным параметрам смешанных по изоспину и схемам Юнга взаимодействий [14]. На основе известных фаз в $p^3He$ и $p^3H$ системах строятся чистые фазы $p^3H$ рассеяния и получены чистые потенциалы $p^3H$ взаимодействия, параметры которых приведены в табл.2.2. Такие взаимодействия правильно описывают канальную энергию связи $p^3H$ системы и среднеквадратичный радиус ядра $^4He$ [24].

Результаты расчета астрофизического S - фактора при энергиях до 10 кэВ, выполненные нами в работе [24], приведены на рис.2.2. Экспериментальные данные взяты из работ [47] и на момент этих расчетов были известны нам только для энергий выше 700 кэВ. Впоследствии, в работах [48,49,50], появились новые экспериментальные данные по S - фактору при энергиях до 12 кэВ, которые также показаны на рис.2.2. Из рисунка видно, что расчеты сделанные нами около 15 лет назад хорошо воспроизводят новые данные по S - фактору при энергиях до 50 кэВ [24].

Здесь мы продолжили эти расчеты, и вычислили S - фактор при энергиях до 1 кэВ, вид которого также показан на рис.2.2 [51,52]. При энергии 1 кэВ его величина оказалась равна 1.1(1) $10^{-3}$ кэВ бн, а результаты его расчета при энергии меньше 50 кэВ лежат несколько ниже новых данных работы [50], где для S(0) получено 2.0(2) $10^{-3}$ кэВ бн.

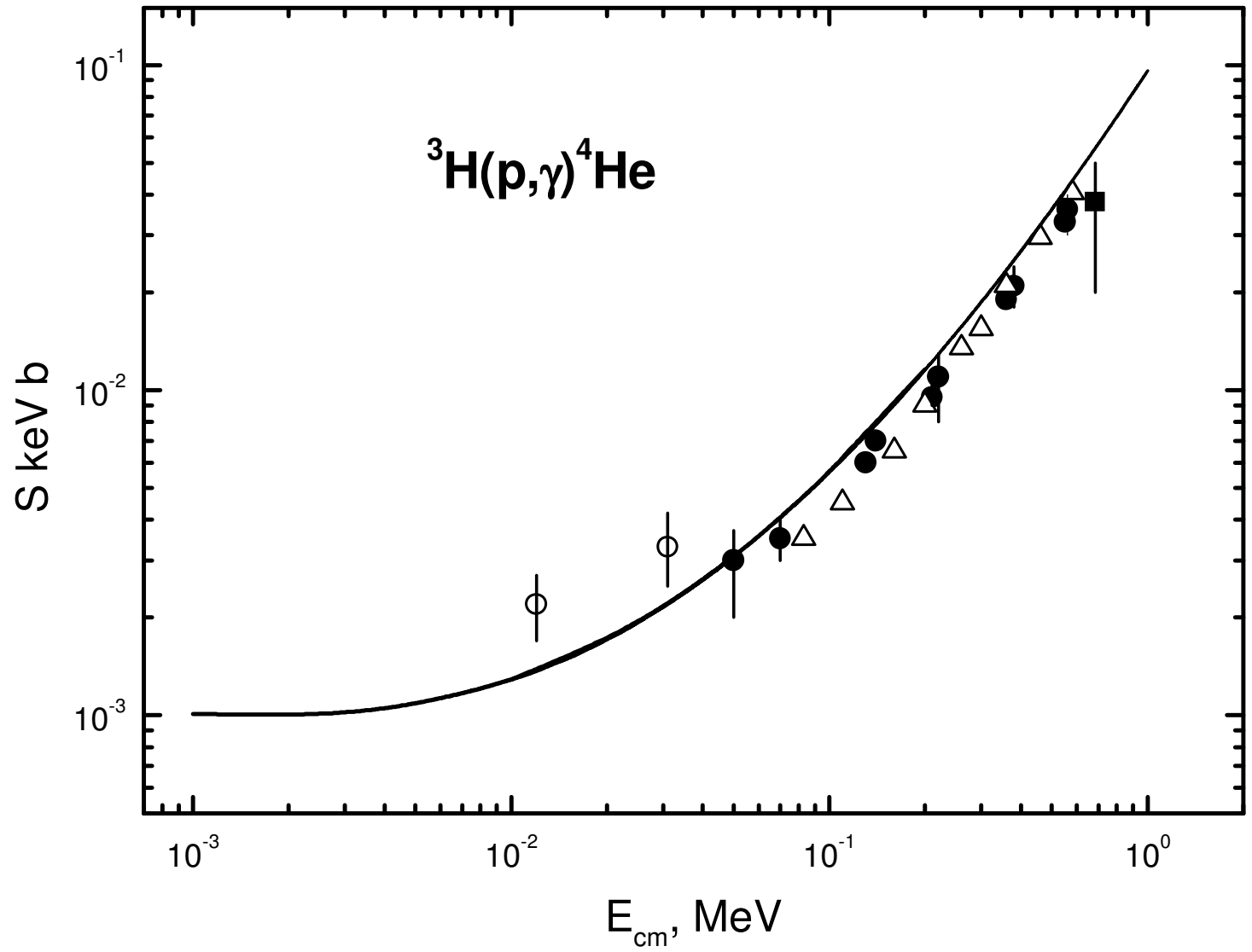

Рис.2.2. Астрофизический S-фактор радиационного $p^3H$ захвата. Квадрат - экспериментальные данные из работ [47], точки - [48], треугольники [49], кружки - [50].

Заметим, что простая экстраполяция имеющихся экспериментальных данных к 1 кэВ по трем последним точкам работ [48,49] приводит к его значению 0.6(2) $10^{-3}$ кэВ бн, т.е. в три раза меньше результатов [50].

Для проведения этих расчетов были уточнены параметры потенциала основного состояния $p^3H$ системы в ядре $^4He$, которые отличаются от приведенных в работе [24] на 0.19289 МэВ (табл.2.2). В основном, это отличие связано с использованием здесь точных значений масс частиц p и $^3H$ [36].

Сходимость энергии к экспериментальной величине -

19.814 МэВ [53] для конечно-разностного метода, как и в $p^2H$ системе, достигается при шаге меньше 0.005 Фм. Точность вычисления «хвоста» ВФ СС $p^3H$ системы проверялась по асимптотической константе, которая для двух видов асимптотик на интервале 8-16 Фм оказалась равна $C_0$= 4.1(1) и $C_w$=4.6(1), а известные нам экспериментальные данные дают значения от 4.2 до 5.1 [35].

Для контроля энергии связи $p^3H$ системы также использован вариационный метод, который позволил получить величину -19.81398 МэВ с асимптотическими константами $C_0$ = 4.1(1) и $C_w$ = 4.5(1) в области 5-12 Фм. Тем самым, истинной энергией для такого потенциала можно считать -19.81399(1) МэВ.

# Заключение

Расчеты S - фактора $p^2H$ радиационного захвата при энергии до 10 кэВ, выполненные нами около 15 лет назад [23], когда для S - факторов были известны только экспериментальные данные выше 150-200 кэВ, хорошо согласуются с новыми данными из работ [32,33] в области 20 - 150 кэВ.

Тем самым, использованная нами потенциальная кластерная модель с запрещенными состояниями и классификацией их по схемам Юнга оказалась способна правильно ***предсказать*** поведение S - фактора $p^2H$ захвата при низких энергиях, вплоть до 20 кэВ [23,54].

Выполненные здесь вычисления S - фактора при более низкой энергии, хотя и содержат большую ошибку, все же демонстрируют определенную тенденцию к его постоянству в области энергий 1-4 кэВ. Эти новые результаты вполне укладываются в полосу ошибок работы [34], где S - фактор определялся при энергиях до 2.5 кэВ.

Удалось ***предсказать*** поведение S - фактора $p^3H$ захвата при энергии до 50 кэВ, т.к. наши расчеты до 10

кэВ также были выполнены около 15 лет назад [24], когда нам были известны только экспериментальные данные выше 700 кэВ [47]. Хотя результаты расчета при более низких энергиях и лежат несколько ниже новых данных работы [50], эти данные содержат сравнительно большую ошибку и, по-видимому, подлежат уточнению.

Возможность предсказания поведения S - фактора при энергиях до 20-50 кэВ в $p^2H$ и $p^3H$ системах вполне можно рассматривать, как очередное свидетельство в пользу потенциального подхода в кластерной модели, когда межкластерные взаимодействия строятся на основе фаз упругого рассеяния кластеров.

Каждая парциальная волна описывается своим потенциалом, например, гауссового вида, а в некоторых случаях к нему добавляется периферическое отталкивание, что приводит к общей форме парциального взаимодействия, представленного выражением (2.1) [55].

Такое разделение общего взаимодействия по парциальным волнам позволяет детализировать его структуру, а классификация орбитальных состояний по схемам Юнга, которая дает возможность определить наличие и число запрещенных состояний, приводит к вполне определенной глубине взаимодействия, что позволяет избавиться от дискретной неоднозначности глубины потенциала, присущей оптической модели.

Форма каждой парциальной фазы рассеяния может быть правильно описана только при определенной ширине этого потенциала, что избавляет нас от непрерывной неоднозначности, которая также имеет место в классической оптической модели.

В результате, все параметры такого потенциала фиксируются вполне однозначно, а чистая по схемам Юнга компонента взаимодействия позволяет правильно описать многие характеристики связанного состояния легчайших кластеров, которое с большой вероятностью реализуется в легких атомных ядрах [27,56].

Конечно, все сказанное верно только в случае точ-ного определения фаз рассеяния из экспериментальных данных по упругому рассеянию. Однако, до настоящего времени, для большинства легчайших ядерных систем, фазы рассеяния найдены с довольно большими ошибками, доходящими, например для $^2H^4He$ системы, до 20-30%.

Все это затрудняет построение потенциалов межкластерного взаимодействия, и приводят, в итоге, к большим неоднозначностям в конечных результатах, получаемых в потенциальной кластерной модели легких атомных ядер.

Тем не менее, во многих случаях и для различных кластерных систем, удается по фазам упругого рассеяния построить потенциалы их взаимодействия, которые позволяют в целом правильно воспроизвести различные характеристики связанных состояний кластерных ядер и некоторые фотоядерные процессы на таких ядрах [14].

# ЗАКЛЮЧЕНИЕ

Таким образом, зная методы расчета волновых функций ядра в непрерывном и дискретном спектрах можно рассматривать любые модельные задачи для выполнения различных вычислений и решения любых проблем ядерной физики низких энергий и ядерной астрофизики.

Конечно, при этом нужно знать потенциалы взаимодействия между легкими ядерными частицами. Такие потенциалы были получены в работах [14,57] для систем ядерных частиц np, $p^2H$, $p^3H$, $p^3He$, $p^4He$, $p^6Li$, $^2H^3He$, $^2H^3H$, $^3H^3He$, $^3H^3H$, $^2H^4He$, $^2H^6Li$, $^4He^4He$ и т.д.

Для построения этих потенциалов использовалась концепция запрещенных и разрешенных состояний в относительном движении указанных выше кластеров, что позволило избавиться от наличия отталкивающего кора в таких взаимодействиях [57].

В частности, полученные взаимодействия оказываются способными правильно воспроизвести астрофизические S - факторы в фотоядерных реакциях для систем частиц $p^2H$, $p^3H$ и т.д. при низких, астрофизических энергиях.

И в заключение всего материала, изложенного в данной части, подчеркнем, что использованный здесь метод расчетов ядерных процессов при низких и астрофизических энергиях можно назвать микроскопическим подходом в ядерной астрофизике, поскольку он использует как имеющиеся экспериментальные данные, так и микроскопически обоснованные теоретические модели атомного ядра, а именно, потенциальную кластерную модель легких атомных ядер.

Автор выражает огромную признательность Неудачину В.Г. (НИИЯФ МГУ, Москва), Узикову Ю.Н. (ОИЯИ

Дубна), Бурковой Н.А. и Неронову В.С. (Каз.НУ им. аль-Фараби, Алматы), Буртебаеву Н.Т., Дуйсебаеву А.Д. и Ибраевой Е.Т. (ИЯФ НЯЦ РК, Алматы).

Выражаю также благодарность Казахской Академии Труда и Социальных Отношений (Алматы), где автор работал в течение ряда лет.

# ЛИТЕРАТУРА

# ЗАКЛЮЧЕНИЕ

***К книге***

В книге рассмотрен ряд актуальных вопросов космологии ранней Вселенной, физики атмосфер Солнца и планет, а также ядерной астрофизики. Все они имеют преимущественно теоретический характер. Поэтому по их содержанию можно судить об уровне казахстанской теоретической физики.

Резюмируем в главном наши новые результаты:

-впервые дано обоснование принципа давление - доминантности в космологии ранней Вселенной; показано, что формирование первичных возмущений барионного вещества может происходить и вследствие антигравитационной неустойчивости субстрата; предложен новый механизм выпрямления космических струн;

-исследованы квантово-статистические процессы формирования «грозовых облаков» в атмосфере Солнца и грозовых и мезосферных серебристых облаков в атмосфере Земли; предложен новый механизм генерации нейтронов в атмосфере Солнца;

-развиты альтернативные численные методы решения системы уравнений Шредингера в задачах на связанные состояния; рассчитан ряд астрофизических S-факторов звездных ядерных реакций.

Все эти результаты являются важным вкладом в познание физических процессов, происходящих в дальнем и ближнем космосе.

*В заключение авторы выражают искреннюю благодарность рецензентам книги:*

*академику НАН РК Омарову Т.Б. и академику НАН РК Боосу Э.Г.*

*за внимательное ознакомление с ее содержанием и ряд критических замечаний, которые были учтены в процессе окончательного редактирования монографии.*

**Дубовиченко С.Б., Такибаев Н.Ж., Чечин Л.М.**

# ФИЗИЧЕСКИЕ ПРОЦЕССЫ В ДАЛЬНЕМ И БЛИЖНЕМ КОСМОСЕ

В книге рассматриваются актуальные вопросы современной космологии, физики атмосферных явлений, физики квазичастиц и ядер. Она рассчитана на специалистов в области астрофизики и космологии, атмосферной физики, ядерной астрофизики и физики частиц, а также на преподавателей высших учебных заведений, аспирантов и студентов, интересующихся достижениями отечественной теоретической физики и физики космоса.

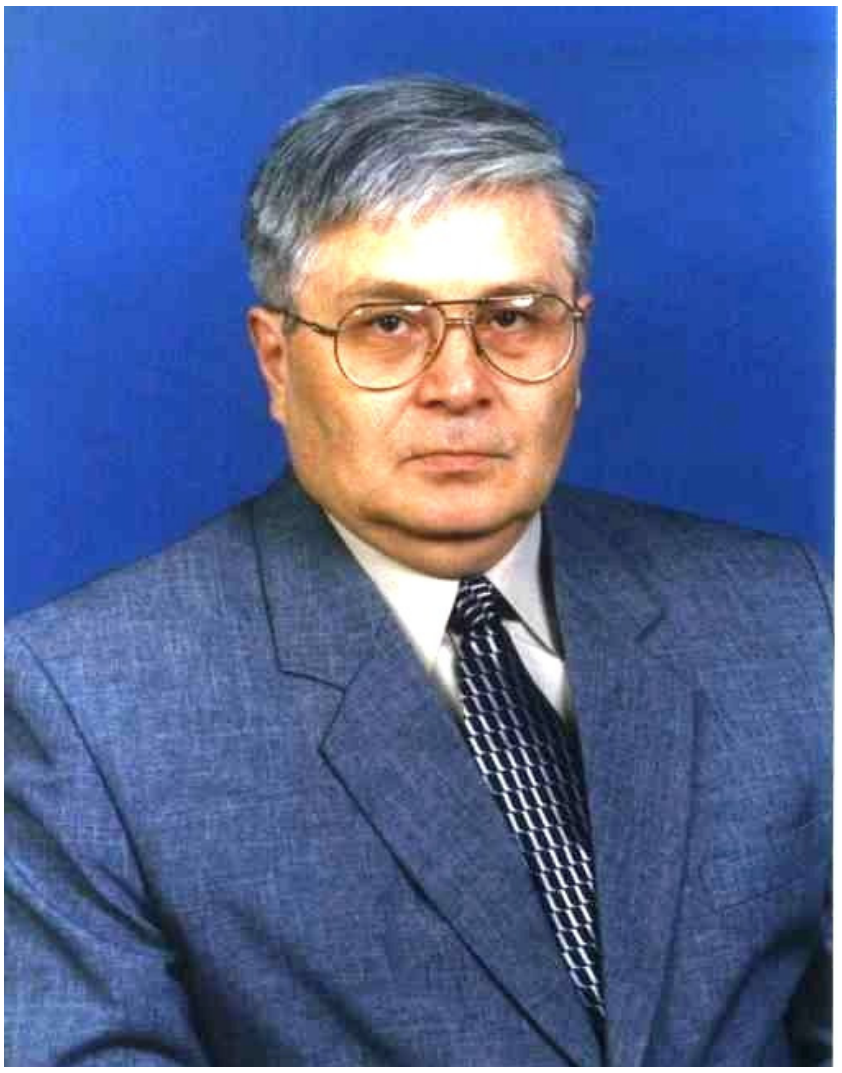

# Такибаев Нургали Жабагаевич

Академик
Национальной Академии
Наук
Республики Казахстан

***Заведующий лабораторией ядерной астрофизики Каз.НПУ им. Абая***

***Доктор физико - математических наук***

***Профессор Кафедры теоретической физики и численного моделирования Каз.НПУ им. Абая***

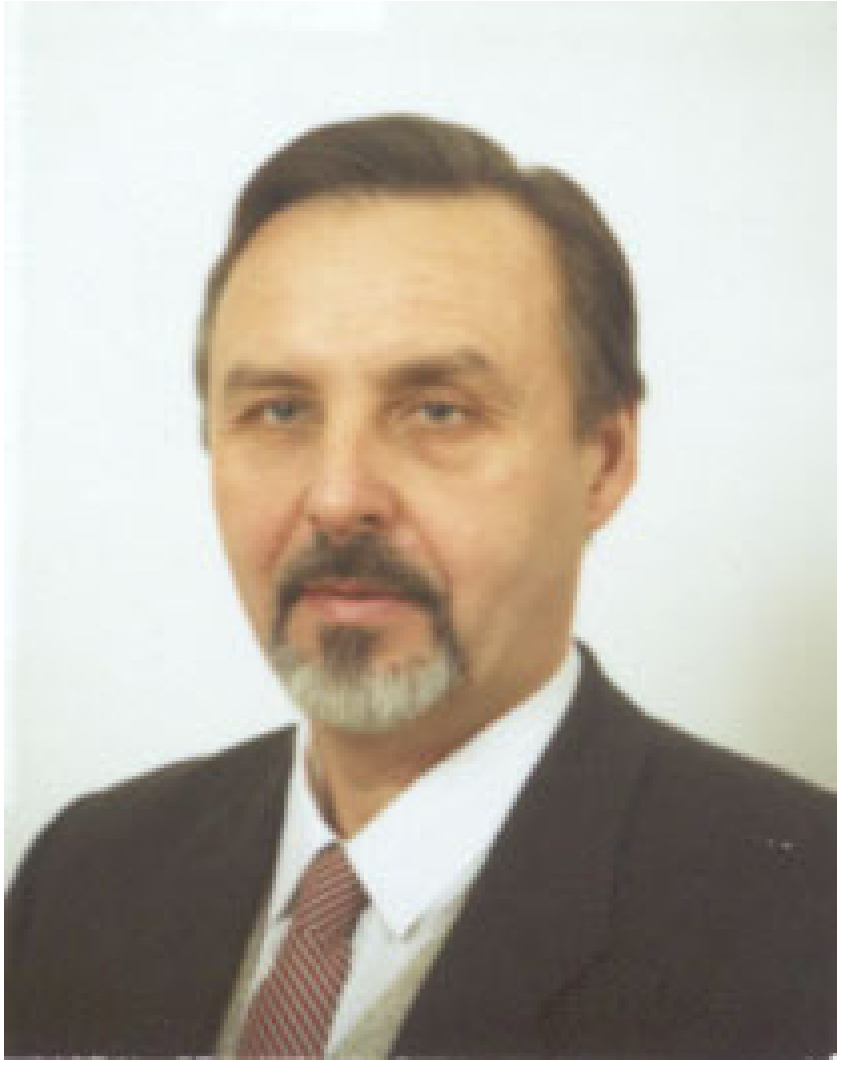

# Чечин Леонид Михайлович

*Академик Международной Академии Информатизации Республики Казахстан*

***Директор Астрофизического института им. В.Г. Фесенкова***

***Доктор физико - математических наук***

***Профессор Кафедры теоретической физики и численного моделирования Каз.НПУ им. Абая***

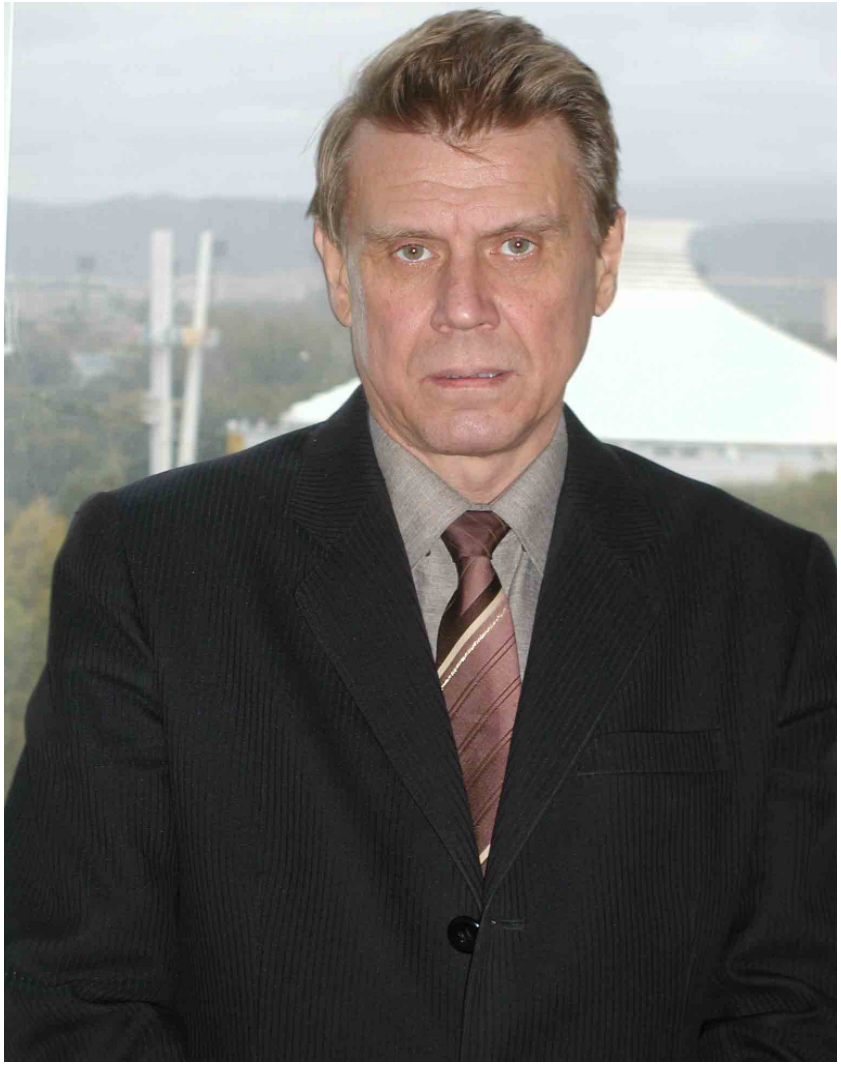

# Дубовиченко Сергей Борисович

*Член - корреспондент Международной Академии Информатизации Республики Казахстан*

***Главный научный сотрудник Астрофизического института им. В.Г. Фесенкова***

***Доктор физико - математических наук***

***член Европейского Физического Общества***

***член Нью - Йоркской Академии Наук***

***Лауреат премии ЛКСМ Казахстана***

***Лауреат гранты международного фонда Сороса***

***Профессор Каз.АТиСО***